\documentclass[twoside,fleqn]{ActaStyle}
\usepackage{amssymb,amsthm,amsmath}

\usepackage[dvips]{graphicx}
\usepackage{times,cite}
\usepackage{multirow}
	
\def\authorheading{J. Stre\v{c}ka and M. Ja\v{s}\v{c}ur}
\def\shorttitle{A brief account of the Ising and Ising-like models}
	
\begin{document}

\pagerange{1}{133}

\title{A BRIEF ACCOUNT OF THE ISING AND ISING-LIKE MODELS: \\ MEAN-FIELD, EFFECTIVE-FIELD AND EXACT RESULTS}

\author{Jozef Stre\v{c}ka\email{jozef.strecka@upjs.sk}, Michal Ja\v{s}\v{c}ur \email{michal.jascur@upjs.sk}}
{Department of Theoretical Physics and Astrophysics, Faculty of Science, \\ P. J. \v{S}af\'{a}rik University, Park Angelinum 9, 040 01 Ko\v{s}ice, Slovakia}

\abstract{The present article provides a tutorial review on how to treat the Ising and Ising-like models within the mean-field, effective-field and exact methods. 
The mean-field approach is illustrated on four particular examples of the lattice-statistical models: the spin-1/2 Ising model in a longitudinal field, 
the spin-1 Blume-Capel model in a longitudinal field, the mixed-spin Ising model in a longitudinal field and the spin-$S$ Ising model in a transverse field. 
The mean-field solutions of the spin-1 Blume-Capel model and the mixed-spin Ising model demonstrate a change of continuous phase transitions to discontinuous 
ones at a tricritical point. A continuous quantum phase transition of the spin-$S$ Ising model driven by a transverse magnetic field is also explored within the mean-field method. 
The effective-field theory is elaborated within a single- and two-spin cluster approach in order to demonstrate an efficiency of this approximate method, which affords superior
approximate results with respect to the mean-field results. The long-standing problem of this method concerned with a self-consistent determination of the free energy 
is also addressed in detail. More specifically, the effective-field theory is adapted for the spin-1/2 Ising model in a longitudinal field, the spin-$S$ Blume-Capel model in a longitudinal field and the spin-1/2 Ising model in a transverse field. The particular attention is paid to a comprehensive analysis of tricritical point, continuous and discontinuous phase transitions of the spin-$S$ Blume-Capel model. Exact results for the spin-1/2 Ising chain, spin-1 Blume-Capel chain and mixed-spin Ising chain in a longitudinal field are obtained using the transfer-matrix method, the crucial steps of which are also reviewed when deriving the exact solution of the spin-1/2 Ising model on a square lattice. The critical points of the spin-1/2 Ising model on several planar (square, honeycomb, triangular, kagom\'e, decorated honeycomb, etc.) lattices are rigorously obtained with the help of dual, star-triangle and decoration-iteration transformations. The mapping transformation technique {\phantom{is subsequently}}}

\vspace*{-0.6cm}
{\hfill{\itshape\small [Continued on next page]\qquad\quad}}

\vspace{0.3cm}
\pacs{05.50.+q; 05.70.Jk; 75.10.-b; 75.10.Hk; 75.10.Jm; 75.30.Kz; 75.40.Cx}

\vspace{-0.1cm}
\begin{minipage}{2.5cm}
\quad{\small {\sf KEYWORDS:}}
\end{minipage}
\begin{minipage}{10.0cm}
Ising model, Blume-Capel model, Mean-field theory, Effective-field theory, Exact results, Mapping transformations, Critical behavior, Mixed spins, Transverse field
\end{minipage}

\abstract{is subsequently adapted to obtain exact results for the critical temperature and spontaneous magnetization of the mixed-spin Ising model on decorated planar lattices and three-coordi\-nated archimedean (honeycomb, bathroom-tile and square-hexagon-dodecagon) lattices. It is shown that an increase in the coordination number of the mixed-spin Ising model on decorated planar lattices gives rise to reentrant phase transitions, while the critical temperature of the mixed-spin Ising model on a regular honeycomb lattice is always greater than the critical temperature of the same model on two semi-regular archimedean lattices with the same coordination number. The effect of selective site dilution of the mixed-spin Ising model on a honeycomb lattice upon phase diagrams is also examined in detail. Last but not least, the review affords a brief account of the Ising-like models previously solved within the mean-field, effective-field and exact methods along with a few comments on their future applicability.}

{
\tableofcontents
}

\setcounter{equation}{0} \setcounter{figure}{0} \setcounter{table}{0}\newpage
\section{Introduction}
\label{sec:intr}

The Ising model represents one of the simplest but at the same time also one of the most frequently studied models of cooperative phenomena in statistical 
mechanics. An early history of the Ising model dates back to pioneering works by Lenz \cite{lenz20} and Ising \cite{isin25}, which have introduced and exactly solved one-dimensional (1D) version of this lattice-statistical model originally aimed at a theoretical description of a magnetic phase transition of ferromagnetic materials. However, the exact solution has refuted an existence of finite-temperature phase transition in the spin-1/2 Ising linear chain, whereas the same conclusion was erroneously drawn also for its higher-dimensional versions \cite{isin25}. This faulty conjecture has significantly suppressed an investigation of the Ising model over subsequent two decades during which it became of marginal research interest. The situation has dramatically changed after Onsager found the exact solution for the spin-1/2 Ising model on a rectangular lattice \cite{onsa44}, which has rigorously proved an intriguing phase transition at finite temperature notwithstanding of an extremely short-ranged character of the interaction potential. The two-dimensional (2D) Ising model has accordingly gained a prominent place in statistical mechanics as paradigmatic example of exactly solved lattice-statistical model with a remarkable phase transition. It is beyond the scope of the present work to review here all crucial aspects of historical development of the Ising model and its various extensions, which have been exhaustively described in several previous review papers \cite{brus67,niss05} to which the interested readers are referred to for further details. By contrast, the primary goal of the present work is to provide a comprehensive description of mean-field, effective-field and exact methods when applying them to different variants of the Ising model. Before doing this, it might be quite useful to recall some elementary facts about possible applications of the Ising model. 

The Ising model and its various extensions have an extraordinary capability of elucidating cooperative phenomena of seemingly diverse physical origin in spite of the fact that the model was originally proposed for a theoretical description of insulating magnetic materials. In this regard, let us at first briefly comment on traditional applications of the Ising model in magnetism. It is quite curious that it took almost a half century from the early development of the Ising model until the first experimental realizations of it were found in magnetism. For many years, the Ising model was therefore regarded more as a mathematical curiosity without any detailed correspondence to a specific magnetic material, because this model totally disregards quantum effects closely related to a non-commutability of spin operators. There are three principal requirements for a magnetic substance in order to be identified as the Ising-like material:
\begin{enumerate}
\item The dominant interaction between magnetic ions, which governs the overall magnetic behavior, must be extremely short-ranged.
\item The ground state of magnetic ion must be a doublet, which is well separated in energy from all other excited states
(typically magnetic ions with Kramer's doublet ground state). 
\item There should be no quantum-mechanical coupling between the states of mutually interacting magnetic ions. 
\end{enumerate}
At present, two families of insulating magnetic materials are known, which conform with greater or lesser success all three aforementioned requirements. The first class of the Ising-like magnetic materials are \textit{rare-earth compounds}, which contain as the only carriers of magnetic moment rare-earth ions (such as Dy$^{3+}$, Ho$^{3+}$, Tb$^{3+}$, etc.) interacting among themselves almost entirely through dipolar forces. Although the magnetic dipole-dipole interaction is in its essence long-ranged, the dipolar interactions between more distant rare-earth ions can occasionally cancel out. Under this circumstance, one may merely consider the dipole-dipole interaction between nearest-neighbor rare-earth ions and to neglect all other dipolar interactions between more distant rare-earth ions. In spite of a certain oversimplification, the agreement between theoretical predictions of the Ising model and relevant experimental data is found to be very satisfactory \cite{jong74,stry77,wolf00}. 

The second class of the Ising-like magnetic materials represent some particular examples of \textit{coordination compounds}. The most important interaction between magnetic centers (usually transition-metal ions) within this class of magnetic compounds is a superexchange coupling, which is mediated via intervening non-magnetic ligand and is therefore extremely short-ranged. However, the highly anisotropic Ising interaction between the magnetic centers can be obtained just for the magnetic ions with a strong enough spin-orbit interaction and/or crystal-field anisotropy such as for instance Co$^{2+}$ or low-spin Fe$^{3+}$ ions. This specific condition represents the most important limitation for transition-metal coordination compounds to be identified as experimental realizations of the Ising-like materials \cite{jong74,stry77,wolf00}. For illustration, a few typical experimental realizations of the Ising-like spin systems from both aforementioned families of magnetic materials are listed in Tab. \ref{tabex}.      
\begin{table}[!b]
\caption{The list of Ising-like magnetic materials compiled from Refs. \cite{jong74,stry77,wolf00}.} 
\bigskip
\centering 
\begin{tabular}{llll} 
\hline\hline 
Chem. formula &  Type of ordering & $T_{\rm c} (K)$\\ 
\hline 
Dy(C$_2$H$_5$SO$_4$)$_3$.9H$_2$0     & dipolar ferromagnet & 10\\
LiTbF$_4$, LiHoF$_4$  & dipolar ferromagnet    &  29\\
Dy$_2$Ti$_2$O$_7$  &  spin ice  & $<$ 0.05\\
Dy$_3$Al$_5$O$_{12}$   &  antiferromagnet  &  25\\
DyPO$_{4}$    &  antiferromagnet  &  34\\
Rb$_2$CoF$_4$, K$_2$CoF$_4$& 2D antiferromagnet   &  101\\
Rb$_3$CoCl$_5$, Cs$_3$CoCl$_5$& 2D antiferromagnet   &  1.1, 0.5\\
\hline \hline
\end{tabular}
\label{tabex}
\end{table}

Apart from the traditional applications in magnetism, the Ising-like models have also found manifold applications closely related to cooperative phenomena of interacting many-particle systems from seemingly diverse research areas. As a matter of fact, there are several physical motivations for exploring the Ising model and its different variants, which have proved their usefulness in the realm of statistical physics for modeling of order-disorder phase transitions in metal alloys \cite{brag34,ylee52,leey52}, vapour-liquid coexistence curves at liquid-gas transition \cite{ylee52,leey52,whee77} or phase separation of liquid mixtures \cite{whee77,whee68,blum71}. The Ising-like models also reveal a profound essence of various collective phenomena in biophysics such as for instance the saturation curve of hemoglobine \cite{mono65,thom68,thom79}, the initial reaction rate of allosteric enzymes \cite{thom79,thom68}, the melting curve of helix-coil transition in DNA \cite{mont66,goel68,wart72,wart85} or the critical behavior of lipid membranes \cite{fraz07,hone08,mach11,mach12,meer15}. More recently, the Ising-like models have opened up a brand new research field focused on several socio-economical aspects of society. Among these unconventional applications of the Ising model one could mention an investigation of the role of socio-economic interactions in relation to business confidence indicators \cite{hohn05,stau06,stau08}, the analysis of financial time series in business markets \cite{zhao13,zhan15}, urban segregation \cite{sche71,stau07,mull08} or linguistic change of language \cite{stau08,itoh04,schu08}. Another remarkable applications of the Ising model have been recently found in the medicine when affording promising modeling of a tumor growth \cite{torq11} or electromechanical alternans of cardiac cells \cite{rest09,alva15}. The latter undesirable effect is caused by an instability of calcium cycling closely connected with the order-disorder phase transition from the Ising universality class  \cite{rest09,alva15}. Nowadays, the Ising model is also being considered as a useful framework for the chronification of pain due to unknown causes, which represents challenging up-to-date topic of current medicine aimed at a more thorough understanding and treating of long-standing pain \cite{gran15}. Of course, the aforementioned list of applications is far from being completed, because the main purpose of this list is to demonstrate versatility of the Ising-like models rather than to provide comprehensive description of all possible applications reported hitherto.

It has been already mentioned that the primary goal of this article is to provide a tutorial review on mean-field, effective-field and exact theories when applied to various versions of the Ising-like models. It is our hope that the present review will afford a fairly good background even for graduate students or non-experts how to treat the Ising-like models within all three aforementioned approaches widely used for diverse lattice-statistical spin models. Because of manifold applications reaching far beyond typical problems of physics (e.g. economics or medicine) we have decided to present a theoretical treatment of the Ising-like models in its full details preferably on a few selected textbook examples for which a lot of (even exact) results can be derived after relatively straightforward and modest calculations. The extension to other more complex Ising-like models is of course possible and is briefly commented on at the end of each section. Last but not least, a comparison between exact and approximate theories should contribute to a capability of distinguishing reliability of the obtained approximate results, because a danger of over-interpretation is inherent in any approximation. 
 
The organization of this short review is as follows. The section \ref{mft} is devoted to the mean-field theory, which is still frequently used due to its conceptual simplicity as a useful benchmark for more precise approximate and exact theories. Among other matters, we will address in this section a possible change of continuous phase transitions to discontinuous ones at a so-called tricritical point, as well as, a quantum phase transition driven by a transverse magnetic field. The basic ideas of the effective-field theory, which is superior with respect to the mean-field theory, are reviewed in the section \ref{eft}. This section provides a more fundamental understanding of how magnetic response functions are influenced when short-range correlations are accounted for. In addition, the section \ref{eft} also brings insight into a long-standing problem of the effective-field theory concerned with a self-consistent determination of the Gibbs free energy. The section \ref{exact} provides a lot of exact results for 1D and 2D Ising models. The presented exact results shed light on a strong universality of the critical behavior of the simple spin-1/2 and mixed-spin Ising model on several planar lattices, whereas our particular attention is paid to the dependence of critical temperature on the lattice geometry. Finally, the review ends up with a brief summary of our findings and conclusions mentioned in the section \ref{conclusion}.

\setcounter{equation}{0} \newpage
\section{Mean-field theory}
\label{mft}

In this section we will provide a comprehensive description of the calculation procedure based on the molecular-field (mean-field) approximation \cite{wei90,wei70}, which is often used due to its simplicity as the first primary tool for rough estimate of a magnetic behavior of more complex lattice-statistical spin systems. All crucial steps of this approximate method will be illustrated on four essential lattice-statistical spin models: the spin-1/2 Ising model in a longitudinal magnetic field, the spin-1 Blume-Capel model in a longitudinal magnetic field, the mixed-spin Ising model in a longitudinal magnetic field and the spin-$S$ Ising model in a transverse magnetic field.

\subsection{Spin-1/2 Ising model in a longitudinal field}
\label{mfa12l}

Let us begin with by considering the simplest example of the spin-1/2 Ising model in a longitudinal magnetic field, which can be defined through the Hamiltonian:
\begin{equation}
{\cal H} = - J \sum_{\langle i j \rangle}^{Nq/2} \sigma_i \sigma_j - h \sum_{k=1}^{N} \sigma_k.
\label{ham12}
\end{equation}
In above, $\sigma_i = \pm 1/2$ is the Ising spin placed at the $i$th lattice point, the first summation accounts for the Ising coupling $J$ between the nearest-neighbor spin pairs, the second summation accounts for the Zeeman's energy of individual magnetic moments, $N$ denotes the total number of the lattice sites (spins) and $q$ is their coordination number (i.e. the number of nearest neighbors). It worth mentioning that the Ising spin variable $\sigma_i = \pm 1/2$ can be alternatively viewed as the $z$-component of the standard spin-1/2 operator (Pauli matrix) with the eigenvalues $\sigma_i^z = \pm 1/2$, whereas its canonical ensemble average directly equals to the magnetization as the Land\'e $g$-factor and Bohr magneton $\mu_{\rm B}$ were adsorbed into the definition of the magnetic field ($h = g \mu_{\rm B} H$) within this convention. The Ising model given by the Hamiltonian (\ref{ham12}) is semi-classical, because it correctly takes into account a quantization of the magnetic moment (spin), but it disregards quantum fluctuations connected to a non-commutability of other spatial components of the same spin operator completely left out from the Hamiltonian (\ref{ham12}).  

An exact treatment of the Ising model given by the Hamiltonian (\ref{ham12}) demands (notwithstanding of its conceptual simplicity) an utilization of very sophisticated mathematical methods (see Section \ref{exact}), since degrees of freedom of individual spins cannot be accounted for independently of each other due to mutual nearest-neighbor spin-spin interactions. To avoid this complexity of the cooperative nature one performs within the mean-field approximation a summation over spin degrees of freedom of just one individual (central) spin and replaces an expectation value for all other spins by the mean value. Hence, it follows that the Hamiltonian of a single spin is at the mean-field level given by: 
\begin{equation}
{\cal H}_i = - J \sigma_i \sum_{j= 1}^{q} \langle \sigma_j \rangle - h \sigma_i.
\label{hami12}
\end{equation}
The keystone of the mean-field approximation is to neglect all local fluctuations in spin-spin correlations, which substantially simplifies further treatment as the considered model essentially becomes a set of non-interacting spins in a presence of the effective field $h_{\rm m.f.} = h + J q \langle \sigma_j \rangle$. In the following, we will elaborate the mean-field theory starting from the Gibbs-Bogoliubov inequality \cite{bog47,fey55,fal70}, because this approach allows to obtain 'self-consistent' expressions for all basic thermodynamic quantities unlike a few other alternative formulations. 

The Gibbs-Bogoliubov variational principle \cite{bog47,fey55,fal70} is based on a validity of the inequality:
\begin{equation}
G \leq \Phi \equiv G_{0} + \left \langle {\cal H} - {\cal H}_0 \right \rangle_{0},
\label{gbi}
\end{equation}
which is proved in full details in Appendix. Above, $G$ and ${\cal H}$ represent the true Gibbs free energy and the full Hamiltonian of a considered lattice-statistical model, whereas $G_0$ and ${\cal H}_0$ represent the trial Gibbs free energy and the trial Hamiltonian of a simplified lattice-statistical model for which the relevant calculations can be performed exactly (the symbol $\langle \cdots \rangle_{0}$ denotes canonical ensemble average within the simplified model defined by the Hamiltonian ${\cal H}_0$). It can be understood from the inequality (\ref{gbi}) that the expression $\Phi$ represents the variational Gibbs free energy, which provides an upper bound for the true Gibbs free energy $G$. In this respect, the variational Gibbs free energy  $\Phi$ is subject to minimalization achieved by convenient choice of variational parameters involved in the trial Hamiltonian ${\cal H}_0$. An accuracy of the developed variational approach will of course depend on how  concisely the simplified lattice-statistical model describes the original lattice-statistical model.

It has been argued previously that the mean-field theory presumes a virtual independence of the Ising spins, which is achieved by the renormalization of the actual magnetic field to the effective field taking into account the contribution from the averaged nearest-neighbor interactions. According to this, the mean-field approximation of the spin-1/2 Ising model in a longitudinal magnetic field can be accomplished by the following choice of the trial Hamiltonian: 
\begin{equation}
{\cal H}_0  = -\gamma \sum_{k=1}^{N} \sigma_k,
\label{th12}
\end{equation}  
which depends on a single variational parameter $\gamma$ with the physical meaning of the effective field. The partition function $Z_0$, the Gibbs free energy $G_0$ and the magnetization $m_0 = \langle \sigma_i \rangle_{0} $ of the spin-1/2 Ising model defined by the trial Hamiltonian (\ref{th12}) are given by: 
\begin{eqnarray}
Z_0  \!\!\!&=&\!\!\! \sum_{\{ \sigma_k \}} \exp(- \beta {\cal H}_0) = \prod_{k = 1}^N \sum_{\sigma_k = \pm \frac{1}{2}} \exp(\beta \gamma \sigma_k) 
                   = \left[2 \cosh \left(\frac{\beta \gamma}{2} \right) \right]^N, \nonumber \\
G_0 \!\!\!&=&\!\!\! - k_{\rm B} T \ln Z_0  = - N k_{\rm B} T \ln 2 - N k_{\rm B} T \ln \cosh \left(\frac{\beta \gamma}{2} \right), \nonumber \\
m_0 \!\!\!&=&\!\!\! - \frac{1}{N} \frac{\partial G_0}{\partial \gamma} = \frac{1}{2} \tanh \left( \frac{\beta \gamma}{2} \right),
\label{tpf12}
\end{eqnarray}
where $\beta = 1/(k_{\rm B} T)$, $k_{\rm B}$ is Boltzmann's constant, $T$ is the absolute temperature and the symbol $\sum_{\{ \sigma_k \}}$ labels summation over all possible states of a set of non-interacting Ising spins. It can be easily proved with the help of Eq. (\ref{tpf12}) that the expectation value of the spin-spin correlation can be decoupled within the mean-field theory to 
a second power of the magnetization: 
\begin{equation}
\langle \sigma_i \sigma_j \rangle_0 = \langle \sigma_i \rangle_0 \langle \sigma_j \rangle_0 = m_0^2,
\label{cor12}
\end{equation}
and subsequently, one in turn obtains the mean value for a difference between the true and trial Hamiltonians entering the right-hand-side of the Gibbs-Bogoliubov inequality (\ref{gbi}):   
\begin{equation}
\left \langle {\cal H} - {\cal H}_0 \right \rangle_{0} = - \frac{Nq}{2} J m_0^2 - N (h - \gamma) m_0.
\label{mv12}
\end{equation} 
The final explicit form of the variational Gibbs free energy $\Phi$, which determines an upper bound for the true Gibbs free energy, is then given by: 
\begin{equation}
\Phi= - N k_{\rm B} T \ln 2 - N k_{\rm B} T \ln \cosh \left(\frac{\beta \gamma}{2} \right) - \frac{Nq}{2} J m_0^2 - N (h - \gamma) m_0.
\label{vp12}
\end{equation}
At this stage, one may minimize the variational Gibbs free energy (\ref{vp12}) in order to get the mean-field estimate for the true Gibbs free energy from the extremum condition:
\begin{equation}
\frac{\partial \Phi}{\partial \gamma} = 0 \quad \Longleftrightarrow \quad \left(q J m_0 + h - \gamma \right) \frac{\partial m_0}{\partial \gamma} = 0,
\label{ec12}
\end{equation} 
which implies the optimal value of the variational parameter $\gamma =  q J m_0 + h$ in agreement with the previous phenomenological expectation (\ref{hami12}). The mean-field estimate for the true Gibbs free energy of the spin-1/2 Ising model in a longitudinal field can be now obtained by substituting the gained optimal value of the variational parameter $\gamma =  q J m_0 + h$ into Eq. (\ref{vp12}):   
\begin{equation}
G = - N k_{\rm B} T \ln 2 - N k_{\rm B} T \ln \cosh \left[\frac{\beta}{2} \left(q J m + h \right) \right] + \frac{Nq}{2} J m^2,
\label{g12}
\end{equation}
whereas the magnetization $m = m_0$ should obey according to Eqs. (\ref{tpf12}) and (\ref{g12}) the following 'self-consistent' equation:
\begin{equation}
m = \frac{1}{2} \tanh \left[\frac{\beta}{2} \left(q J m + h \right) \right].
\label{m12}
\end{equation}
In a zero magnetic field $h=0$, the 'self-consistent' formula for the magnetization (\ref{m12}) reduces to a simpler form:
\begin{equation}
m = \frac{1}{2} \tanh \left(\frac{\beta q J m}{2} \right),
\label{m12zf}
\end{equation}
which indicates the non-zero spontaneous magnetization related to the ferromagnetic long-range order at low enough temperatures. 

It is well known that the spin-1/2 Ising model may exhibit in an absence of the magnetic field ($h=0$) a continuous (second-order) phase transition from a spontaneously ordered ferromagnetic state to a disordered paramagnetic state upon rising temperature. The critical temperature of the order-disorder phase transition can be accordingly acquired by expanding the right-hand-side of Eq. (\ref{m12zf}) and keeping just the first linear term from the relevant expansion ($\tanh_{x \to 0} x \approx x$) as soon as the spontaneous magnetization continuously vanishes at a critical point. This straightforward procedure gives the following critical condition: 
\begin{equation}
m = \frac{1}{4} \beta_{\rm c} q J m \quad \Longleftrightarrow \quad q \beta_{\rm c} J = 4 \quad \Longleftrightarrow \quad \frac{k_{\rm B} T_{\rm c}}{q J} = \frac{1}{4},
\label{cc12}
\end{equation}
where $T_{\rm c}$ and $\beta_{\rm c}  = 1/(k_{\rm B} T_{\rm c})$ is the critical temperature and inverse critical temperature, respectively. The precise nature of how the spontaneous magnetization disappears in a vicinity of the critical point can be obtained from the expansion of the right-hand-side of Eq. (\ref{m12zf}) when keeping two first non-zero terms ($\tanh_{x \to 0} x \approx x - x^3/3$), which serve in evidence of the power-law behavior of the spontaneous magnetization with the critical exponent $\beta_m = 1/2$ close to but slightly below the critical temperature: 
\begin{equation}
m (T \leq T_c, h = 0) \approx \left(1 - \frac{T}{T_{\rm c}} \right)^{\frac{1}{2}}. 
\label{ce12}
\end{equation}
It should be stressed that the critical point of a continuous phase transition can be alternatively attained according to the Landau theory of phase transitions from the expansion of the zero-field Gibbs free energy with respect to the relevant order parameter. The formula (\ref{g12}) for the Gibbs free energy reduces in a zero magnetic field to the following expression:
\begin{equation}
G (h = 0) = - N k_{\rm B} T \ln 2 - N k_{\rm B} T \ln \cosh \left(\frac{\beta q J m}{2} \right) + \frac{Nq}{2} J m^2,
\label{g12zf}
\end{equation}
whereas its expansion with respect to the magnetization in a vicinity of the critical point reads:
\begin{eqnarray}
G (h \!\!\!&=&\!\!\! 0, m \to 0) = N \left( g_0 + g_2 m^2 + g_4 m^4 + \ldots \right) \nonumber \\
g_0 \!\!\!&=&\!\!\! - k_{\rm B} T \ln 2, \nonumber \\
g_2 \!\!\!&=&\!\!\! \frac{qJ}{2} \left(1  - \frac{\beta q J}{4} \right), \nonumber \\
g_4 \!\!\!&=&\!\!\! \frac{qJ}{24} \left(\frac{\beta q J}{2} \right)^3. 
\label{g12ex}
\end{eqnarray}
The continuous phase transition is given by the constraint $g_2 = 0, g_4 > 0$ within the Landau theory of phase transitions, which is indeed in accordance with the critical condition (\ref{cc12}) reported previously from the 'self-consistent' relation for the spontaneous magnetization. The Landau theory thus unambiguously verifies an existence of the continuous phase transition at a critical point of the spin-1/2 Ising model independently of the lattice coordination number and the lattice dimensionality. It should be also pointed out that a rather naive expression for the critical temperature (\ref{cc12}) obtained within the mean-field theory depends exclusively upon the coordination number of the underlying lattice, while it does not depend neither on the lattice dimensionality nor the lattice topology. From this point of view, the non-zero critical temperature of the spin-1/2 Ising linear chain ($k_{\rm B} T_{\rm c}/J = 1/2$ for $q = 2$) represents the most principal artifact of the mean-field approximation, whereas the mean-field estimate of the critical temperature (\ref{cc12}) generally becomes more reliable for the spin-1/2 Ising model on 2D and 3D lattices with sufficiently high coordination numbers. As a matter of fact, the mean-field theory becomes exact in the limit $q \to \infty$ due to diminishing importance of local fluctuations.

\subsection{Spin-1 Blume-Capel model in a longitudinal field}
\label{mfa1l}

The foremost advantage of the mean-field theory lies in that it can be readily generalized to more complex lattice-statistical models as for instance the spin-1 Blume-Capel model \cite{blu66,cap66,cap67}, which may even display a greater versatility of the critical behavior including a change of continuous phase transitions to discontinuous ones at a tricritical point. The Hamiltonian of the spin-1 Blume-Capel model in a longitudinal magnetic field reads:
\begin{equation}
{\cal H} = - J \sum_{\langle ij \rangle}^{Nq/2} S_i S_j - D \sum_{k=1}^{N} S_k^2 - h \sum_{k=1}^{N} S_k,
\label{ham1}
\end{equation}
where $S_i = \pm 1, 0$ is the spin-1 particle placed at the $i$th lattice point, the first summation takes into account the bilinear nearest-neighbor Ising interaction, the second summation   
the uniaxial single-ion anisotropy and the third one the Zeeman's energy in a longitudinal magnetic field. 

The trial Hamiltonian for the spin-1 Blume-Capel model in a magnetic field should include within the mean-field approximation two different variational parameters $\gamma$ and $\eta$:
\begin{equation}
{\cal H}_0  = -\gamma \sum_{k=1}^{N} S_k -\eta \sum_{k=1}^{N} S_k^2,
\label{th1}
\end{equation} 
which have physical meaning of the effective field and the effective single-ion anisotropy acting on a spin and quadrupolar moment, respectively. With regard to this, the Hamiltonian (\ref{ham1}) of interacting many-particle system is approximated by the Hamiltonian  (\ref{th1}) of the non-interacting system of the spin-1 particles, for which the partition function $Z_0$, the Gibbs free energy $G_0$, the magnetization $m_0 = \langle S_i \rangle_0$ and the quadrupolar moment $q_0 = \langle S_i^2 \rangle_0$ can easily be calculated: 
\begin{eqnarray}
Z_0  \!\!\!&=&\!\!\! \sum_{\{ S_k \}} \exp(- \beta {\cal H}_0) = \prod_{k = 1}^N \sum_{S_k = \pm 1,0} \!\!\! \!\!\! \exp(\beta \gamma S_k + \beta \eta S_k^2) 
                   = \left[1 + 2 \exp(\beta \eta) \cosh \left(\beta \gamma \right) \right]^N\!\!\!\!\!, \nonumber \\
G_0 \!\!\!&=&\!\!\! - k_{\rm B} T \ln Z_0  = - N k_{\rm B} T \ln \left[1 + 2 \exp(\beta \eta) \cosh \left(\beta \gamma \right) \right], \nonumber \\
m_0 \!\!\!&=&\!\!\! - \frac{1}{N} \frac{\partial G_0}{\partial \gamma} = \frac{2 \exp(\beta \eta) \sinh \left(\beta \gamma \right)}{1 + 2 \exp(\beta \eta) \cosh \left(\beta \gamma \right)}, \nonumber \\
q_0 \!\!\!&=&\!\!\! - \frac{1}{N} \frac{\partial G_0}{\partial \eta} = \frac{2 \exp(\beta \eta) \cosh \left(\beta \gamma \right)}{1 + 2 \exp(\beta \eta) \cosh \left(\beta \gamma \right)}. 
\label{tpf1}
\end{eqnarray} 
The mean value for the difference between the true and trial Hamiltonians entering the right-hand-side of the Gibbs-Bogoliubov inequality (\ref{gbi}) is given by:   
\begin{equation}
\left \langle {\cal H} - {\cal H}_0 \right \rangle_{0} = - \frac{Nq}{2} J m_0^2 - N (D - \eta) q_0 - N (h - \gamma) m_0,
\label{mv1}
\end{equation} 
which leads in combination with Eq. (\ref{tpf1}) to the following expression for the variational Gibbs free energy $\Phi$ determining an upper bound for the true Gibbs free energy: 
\begin{equation} 
\Phi= - N k_{\rm B} T \ln \cosh \left[1 + 2 \exp(\beta \eta) \cosh \left(\beta \gamma \right) \right] - \frac{Nq}{2} J m_0^2 - N (D - \eta) q_0 - N (h - \gamma) m_0.
\label{vp1}
\end{equation}
The minimalization of the variational Gibbs free energy (\ref{vp1}) with respect to both variational parameters $\gamma$ and $\eta$ results in two equations:
\begin{eqnarray}
\frac{\partial \Phi}{\partial \gamma} \!\!\!&=&\!\!\! 0 \quad \Longleftrightarrow \quad 
                                         \left(q J m_0 + h - \gamma \right) \frac{\partial m_0}{\partial \gamma} +  (D - \eta) \frac{\partial q_0}{\partial \gamma} = 0, \nonumber \\
\frac{\partial \Phi}{\partial \eta} \!\!\!&=&\!\!\! 0 \quad \Longleftrightarrow \quad 
                                         \left(q J m_0 + h - \gamma \right) \frac{\partial m_0}{\partial \eta} +  (D - \eta) \frac{\partial q_0}{\partial \eta} = 0,                                 
\label{ec1}
\end{eqnarray} 
which are simultaneously satisfied if the variational parameters $\gamma$ and $\eta$ obey the relations:
\begin{eqnarray}
\gamma =  q J m_0 + h \quad \mbox{and} \quad \eta = D.                        
\label{vpe1}
\end{eqnarray}
The mean-field calculation thus yields, after substituting the variational parameters $\gamma$ and $\eta$ satisfying the extremum condition (\ref{vpe1}) to Eq. (\ref{vp1}), the following estimate of the Gibbs free energy:
\begin{equation}
G = - N k_{\rm B} T \ln \cosh \left \{1 + 2 \exp(\beta D) \cosh \left[\beta (q J m + h) \right] \right\} + \frac{Nq}{2} J m^2,
\label{vgfe1}
\end{equation}
whereas the magnetization ($m = m_0$) and the quadrupolar moment ($q_D = q_0$) are determined by the expressions:
\begin{eqnarray}
m \!\!\!&=&\!\!\! \frac{2 \exp(\beta D) \sinh \left[\beta (q J m + h) \right]}{1 + 2 \exp(\beta D) \cosh \left[\beta (q J m + h) \right]}, \nonumber \\
q_D \!\!\!&=&\!\!\! \frac{2 \exp(\beta D) \cosh \left[\beta (q J m + h) \right]}{1 + 2 \exp(\beta D) \cosh \left[\beta (q J m + h) \right]}.                                 
\label{mq1}
\end{eqnarray}
In a zero magnetic field $h=0$, the final formulas for the magnetization and the quadrupolar moment (\ref{mq1}) substantially simplify to the following expressions:
\begin{eqnarray}
m \!\!\!&=&\!\!\! \frac{2 \exp(\beta D) \sinh \left(\beta q J m \right)}{1 + 2 \exp(\beta D) \cosh \left(\beta q J m \right)}, \nonumber \\
q_D \!\!\!&=&\!\!\! \frac{2 \exp(\beta D) \cosh \left(\beta q J m \right)}{1 + 2 \exp(\beta D) \cosh \left(\beta q J m \right)}.                                 
\label{mq1zf}
\end{eqnarray}

On assumption that the spin-1 Blume-Capel model undergoes a continuous phase transition from the ferromagnetic state to the disordered paramagnetic state one may easily derive the critical condition:
\begin{eqnarray}
m = \frac{2 \exp(\beta_{\rm c} D) \beta_{\rm c} q J m}{1 + 2 \exp(\beta_{\rm c} D)} \quad \Longleftrightarrow \quad 
\frac{k_{\rm B} T_{\rm c}}{qJ} = \frac{2 \exp(\beta_{\rm c} D)}{1 + 2 \exp(\beta_{\rm c} D)},
\label{cc1}
\end{eqnarray}
which readily follows from the expansion of the spontaneous magnetization (\ref{mq1zf}) when retaining only the linear term. Moreover, the critical value of the quadrupolar moment can be related to the reduced critical temperature when substituting the critical condition (\ref{cc1}) to Eq. (\ref{mq1zf}):  
\begin{eqnarray}
q_{D} (T = T_{\rm c}) = \frac{k_{\rm B} T_{\rm c}}{qJ}.
\label{cq1}
\end{eqnarray}
It can be easily understood from Eq. (\ref{cc1}) that the critical temperature depends basically on a relative strength of the uniaxial single-ion anisotropy, whereas it equals to $k_{\rm B} T_{\rm c}/(qJ) = 2/3$ for the particular case without single-ion anisotropy $D=0$. The critical value of the quadrupolar moment (\ref{cq1}) for the special case with $D=0$ suggests that all three spin states $\pm 1, 0$ are equally populated at a critical temperature. It should be mentioned that the critical condition (\ref{cc1}) can be also recovered from the Landau expansion of the zero-field Gibbs free energy (\ref{vgfe1}): 
\begin{equation}
G (h = 0) = - N k_{\rm B} T \ln \cosh \left[1 + 2 \exp(\beta D) \cosh \left(\beta q J m \right) \right] + \frac{Nq}{2} J m^2,
\label{g1zf}
\end{equation}
which takes the following form in a close vicinity of the critical point:
\begin{eqnarray}
G (h \!\!\!&=&\!\!\! 0, m \to 0) = N \left( g_0 + g_2 m^2 + g_4 m^4 + \ldots \right) \nonumber \\
g_0 \!\!\!&=&\!\!\! -k_{\rm B} T \ln [1 + 2 \exp(\beta D)], \nonumber \\
g_2 \!\!\!&=&\!\!\! \frac{qJ}{2} \left(1  - \beta q J \frac{2 \exp(\beta D)}{1 + 2 \exp(\beta D)} \right), \nonumber \\
g_4 \!\!\!&=&\!\!\! \frac{qJ}{24} (\beta q J)^3 \frac{2 \exp(\beta D)}{1 + 2 \exp(\beta D)} \left[\frac{6 \exp(\beta D)}{1 + 2 \exp(\beta D)} - 1 \right]. 
\label{g1ex}
\end{eqnarray}
The necessary condition $g_2 = 0, g_4 > 0$ for a presence of the continuous phase transition within the Landau theory of phase transitions is repeatedly in concordance with the critical condition (\ref{cc1}) gained from an expansion of the spontaneous magnetization, but the latter condition $g_4 > 0$ restricts the second-order phase transitions to the single-ion anisotropies greater than the boundary value $D > D_{\rm t}$ with $D_{\rm t}/(qJ) = - \ln (4)/3$. The special constraint $g_2 = 0, g_4 = 0$ in the Landau theory of phase transitions is prerequisite for an existence of tricritical point, which occurs in the spin-1 Blume-Capel model at the coordinates:
\begin{eqnarray}
\frac{k_{\rm B} T_{\rm t}}{q J} = \frac{1}{3}, \qquad  \frac{D_{\rm t}}{q J} = - \frac{\ln 4}{3}.
\label{tcp}
\end{eqnarray}
Interestingly, the tricritical point of the spin-1 Blume-Capel model can be eventually located from the series expansion of the spontaneous magnetization (\ref{mq1zf}) in a close vicinity of the critical point:
\begin{eqnarray}
m \!\!\!&=&\!\!\! f_1 m + f_3 m^3 + \ldots  \nonumber \\
f_1 \!\!\!&=&\!\!\! \beta q J \frac{2 \exp(\beta D)}{1 + 2 \exp(\beta D)}, \nonumber \\
f_3 \!\!\!&=&\!\!\! \frac{(\beta q J)^3}{6} \frac{2 \exp(\beta D)}{1 + 2 \exp(\beta D)} \left[1 - \frac{6 \exp(\beta D)}{1 + 2 \exp(\beta D)} \right]. 
\label{m1ex}
\end{eqnarray}
According to Eq. (\ref{m1ex}), the spontaneous magnetization follows near a critical point of the continuous phase transition the power-law behavior with the mean-field critical exponent $\beta_m = 1/2$:
\begin{eqnarray}
m (T \leq T_{\rm c}, h=0) \approx \left(\frac{1 - f_1}{f_3}\right)^{\frac{1}{2}}. 
\label{m1cc}
\end{eqnarray}
It is quite evident from Eq. (\ref{m1cc}) that the critical condition for the second-order phase transition is given by the constraint $1-f_1=0$, $f_3<0$, which is actually equivalent to the critical condition (\ref{cc1}) provided that $D > D_{\rm t}$. In addition, it turns out that the constraint $1-f_1=0$, $f_3=0$ is fully consistent with the locus of the tricritical point (\ref{tcp}) afforded by the Landau expansion of the Gibbs free energy with the special values of the expansion coefficients $g_2 = 0, g_4 = 0$.

\begin{figure}[tb]
\begin{center}
\hspace{-1.0cm}
\includegraphics[clip,width=7.45cm]{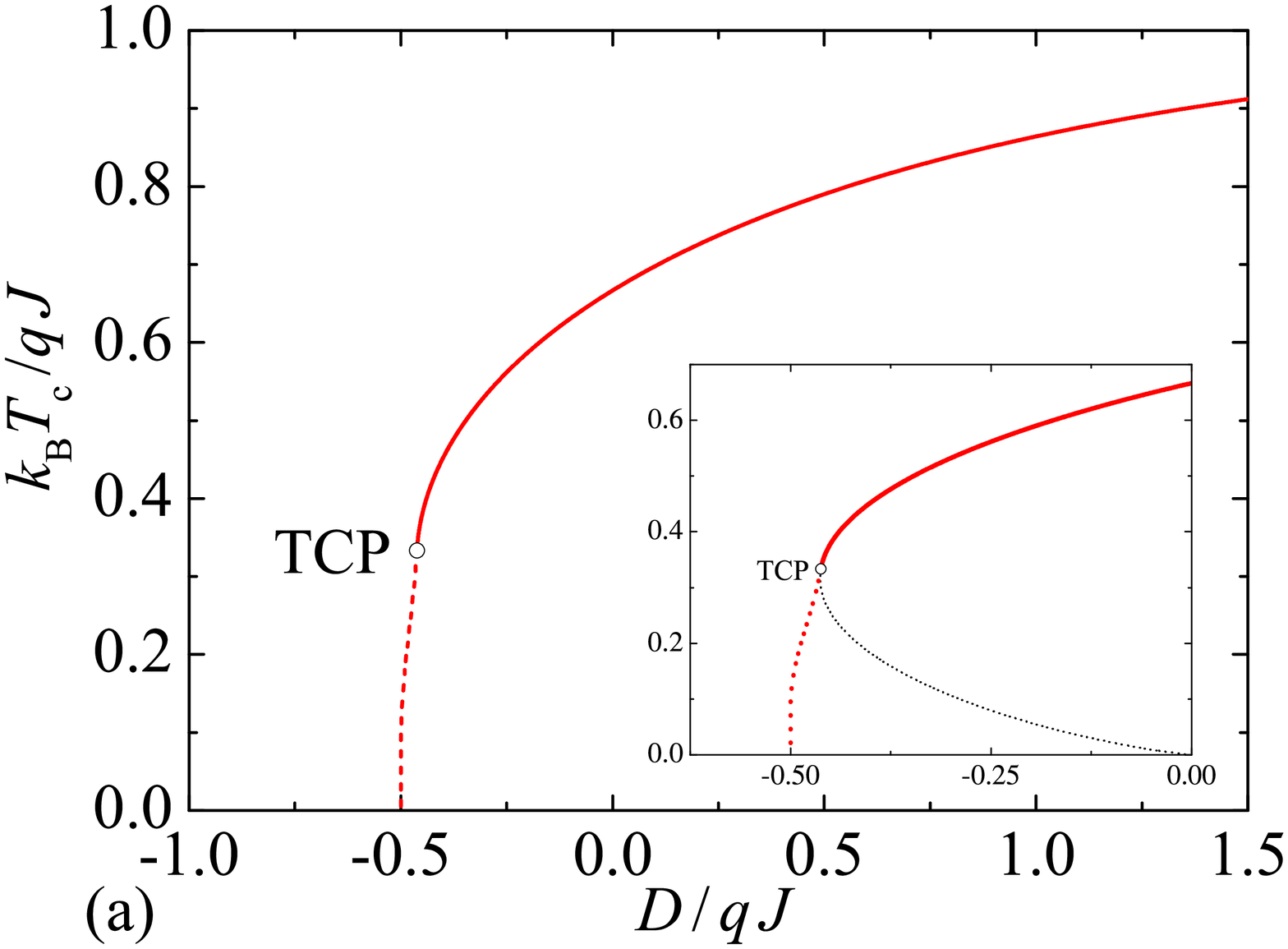}
\hspace{-1.0cm}
\includegraphics[clip,width=7.45cm]{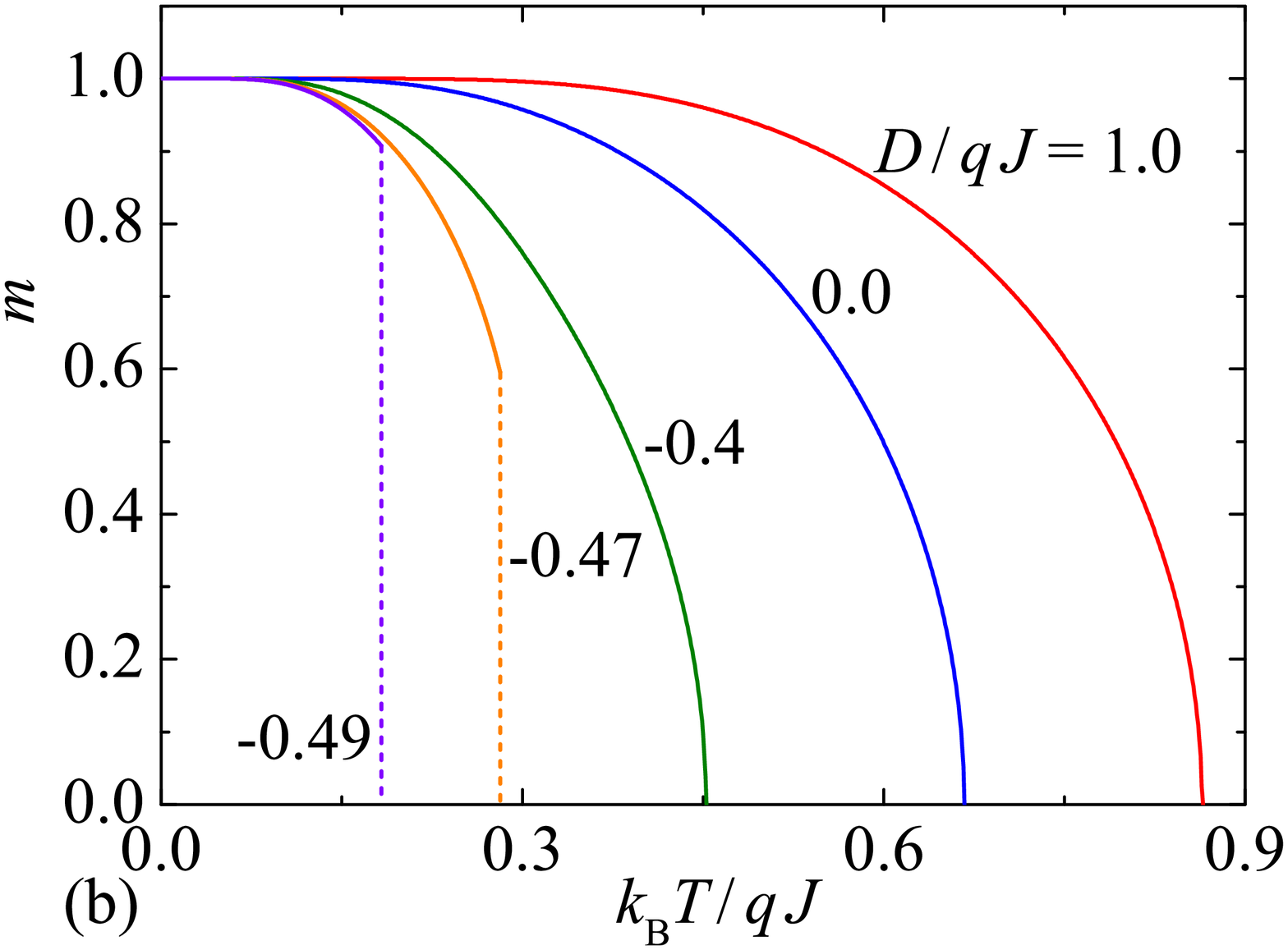}
\hspace{-1.5cm}
\vspace{-0.25cm}
\caption{(a) The critical temperature of the spin-1 Blume-Capel model as a function of the uniaxial single-ion anisotropy. The critical line of continuous (discontinuous) phase transitions is displayed by solid (broken) line, an open circle denotes a tricritical point (TCP). The inset shows detailed plot of a tricritical region including the unstable solution of the critical condition (\ref{cc1}) shown by a thin dotted line; (b) Thermal variations of the spontaneous magnetization for a few different values of the uniaxial single-ion anisotropy. Broken lines denote magnetization jumps connected with the discontinuous phase transitions.}
\label{figbc}
\end{center}
\end{figure}

It could be concluded on the basis of our mean-field calculations that the spin-1 Blume-Capel model should exhibit the second-order phase transitions with a continuous vanishing of the spontaneous magnetization at a critical temperature just for $D > D_{\rm t}$, while the first-order phase transitions with a discontinuous (abrupt) jump of the spontaneous magnetization at a critical point should emerge in the reverse case $D < D_{\rm t}$ (see Fig. \ref{figbc}(a)). The locus of discontinuous phase transitions can be found from a comparison of the Gibbs free energies (\ref{g1zf}) of the ferromagnetic state with a non-zero spontaneous magnetization $m \neq 0$ and the disordered paramagnetic state with a zero spontaneous magnetization $m=0$ both satisfying the 'self-consistent' equation (\ref{mq1zf}) for the magnetization. The lines of continuous and discontinuous phase transitions merge together at a tricritical point, which is displayed in Fig. \ref{figbc}(a) by an open circle. 
The unstable solution of the critical condition (\ref{cc1}) for the continuous phase transitions is shown in the inset of Fig. \ref{figbc}(a) by a thin dotted line. The overall correctness of the established phase diagram is demonstrated by temperature dependences of the spontaneous magnetization, which are depicted in Fig. \ref{figbc}(b) for several values of the uniaxial single-ion anisotropy. In agreement with our argumentation, the spontaneous magnetization exhibits at a critical temperature an abrupt jump to the zero magnetization for the uniaxial single-ion anisotropies $D < D_{\rm t}$ (see the curves for $D/(qJ) = -0.47$ and $-0.49$) instead of a continuous vanishing observed for $D > D_{\rm t}$. 

\subsection{Mixed-spin Ising model in a longitudinal field}
\label{mfa121l}

Another interesting example of the lattice-statistical spin system, which may display a more diverse critical behavior than the usual spin-1/2 Ising model, represents the mixed spin-1/2 and spin-1 Ising model refined by the uniaxial single-ion anisotropy and longitudinal magnetic field. The Hamiltonian of the mixed spin-1/2 and spin-1 Ising model taking into consideration the uniaxial single-ion anisotropy and longitudinal magnetic field can be defined as:
\begin{equation}
{\cal H} = - J \sum_{\langle ij \rangle}^{Nq/2} \sigma_i S_j - D \sum_{j=1}^{N} S_j^2 - h_A \sum_{i=1}^{N} \sigma_i - h_B \sum_{j=1}^{N} S_j,
\label{ham121}
\end{equation}
where $\sigma_i = \pm 1/2$ and $S_j = \pm 1, 0$ are the spin-1/2 and spin-1 Ising variables located at the $i$th and $j$th lattice point, respectively. The first summation accounts for the nearest-neighbor Ising interaction $J$, the second summation accounts for the uniaxial single-ion anisotropy $D$ acting on the spin-1 particles only, while the third and fourth summations account for the Zeeman's energy of the spin-1/2 and spin-1 particles in a longitudinal magnetic field. Note that the effective magnetic fields $h_A$ and $h_B$ acting on the spin-1/2 and spin-1 magnetic particles are chosen to be different, because different magnetic ions generally possess different Land\'e g-factors absorbed into the effective magnetic fields 
$h_A = g_A \mu_{\rm B} H$ and $h_B = g_B \mu_{\rm B} H$.

The mean-field theory for the mixed spin-1/2 and spin-1 Ising model should be developed from the trial Hamiltonian involving three different variational parameters:
\begin{equation}
{\cal H}_0  = -\gamma_A \sum_{i=1}^{N} \sigma_i -\gamma_B \sum_{j=1}^{N} S_j - \eta \sum_{j=1}^{N} S_j^2,
\label{th121}
\end{equation} 
which correspond to the effective fields $\gamma_A$ and $\gamma_B$ acting on the spin-1/2 and spin-1 particles, respectively, and the effective uniaxial single-ion anisotropy $\eta$ acting on the quadrupolar moment of the spin-1 particles. The Hamiltonian (\ref{ham121}) of the interacting mixed-1/2 and spin-1 Ising system is thus approximated within the mean-field theory by the Hamiltonian (\ref{th121}) of the non-interacting system of the spin-1/2 and spin-1 particles, for which the partition function $Z_0$, the Gibbs free energy $G_0$, both sublattice magnetizations $m_A = \langle \sigma_i \rangle_0$, $m_B = \langle S_j \rangle_0$ and the quadrupolar moment $q_B = \langle S_j^2 \rangle_0$ can readily be calculated: 
\begin{eqnarray}
Z_0  \!\!\!&=&\!\!\! \sum_{\{ \sigma_i \}} \sum_{\{ S_j \}} \exp(- \beta {\cal H}_0) 
            = \prod_{i = 1}^N \sum_{\sigma_i = \pm \frac{1}{2}} \!\!\! \exp(\beta \gamma_A \sigma_i) \prod_{j = 1}^N \sum_{S_j = \pm 1,0} \!\!\! \exp(\beta \gamma_B S_j + \beta \eta S_j^2) \nonumber \\
            \!\!\!&=&\!\!\! \left[2 \cosh \left(\beta \gamma_A \right) \right]^N \left[1 + 2 \exp(\beta \eta) \cosh \left(\beta \gamma_B \right) \right]^N\!\!\!, \nonumber \\
G_0 \!\!\!&=&\!\!\! - N k_{\rm B} T \ln 2 - N k_{\rm B} T \ln \cosh \left(\beta \gamma_A \right)
                    - N k_{\rm B} T \ln \left[1 + 2 \exp(\beta \eta) \cosh \left(\beta \gamma_B \right) \right], \nonumber \\
m_A \!\!\!&=&\!\!\! - \frac{1}{N} \frac{\partial G_0}{\partial \gamma_A} = \frac{1}{2} \tanh \left( \frac{\beta \gamma_A}{2} \right), 
\nonumber \\
m_B \!\!\!&=&\!\!\! - \frac{1}{N} \frac{\partial G_0}{\partial \gamma_B} = \frac{2 \exp(\beta \eta) \sinh \left(\beta \gamma_B \right)}{1 + 2 \exp(\beta \eta) \cosh \left(\beta \gamma_B \right)}, 
\nonumber \\
q_B \!\!\!&=&\!\!\! - \frac{1}{N} \frac{\partial G_0}{\partial \eta} = \frac{2 \exp(\beta \eta) \cosh \left(\beta \gamma_B \right)}{1 + 2 \exp(\beta \eta) \cosh \left(\beta \gamma_B \right)}. 
\label{tpf121}
\end{eqnarray} 
The right-hand-side of the Gibbs-Bogoliubov inequality (\ref{gbi}) requires calculations of the mean value for the difference between the true and trial Hamiltonians:   
\begin{equation}
\left \langle {\cal H} - {\cal H}_0 \right \rangle_{0} = - N q J m_A m_B - N (D - \eta) q_B - N (h_A - \gamma_A) m_A - N (h_B - \gamma_B) m_B,
\label{mv121}
\end{equation} 
which in combination with the trial Gibbs free energy $G_0$ given by Eq. (\ref{tpf121}) gives the following expression for the variational Gibbs free energy $\Phi$: 
\begin{eqnarray}
\Phi = \!\!\!&-&\!\!\!N k_{\rm B} T \ln 2 - N k_{\rm B} T \ln \cosh \left(\beta \gamma_A \right) - N k_{\rm B} T \ln \cosh \left[1 + 2 \exp(\beta \eta) \cosh \left(\beta \gamma_B \right) \right] \nonumber \\
\!\!\!&-&\!\!\! N q J m_A m_B - N (D - \eta) q_B - N (h_A - \gamma_A) m_A - N (h_B - \gamma_B) m_B.
\label{vp121}
\end{eqnarray} 
The minimalization of the variational Gibbs free energy (\ref{vp121}) with respect to all three variational parameters $\gamma_A$, $\gamma_B$ and $\eta$ leads to three equations:
\begin{eqnarray}
\left(q J m_B + h_A - \gamma_A \right) \frac{\partial m_A}{\partial \gamma_A} + \left(q J m_A + h_B - \gamma_B \right) \frac{\partial m_B}{\partial \gamma_A}  
+ (D - \eta) \frac{\partial q_B}{\partial \gamma_A} = 0, \nonumber \\
\left(q J m_B + h_A - \gamma_A \right) \frac{\partial m_A}{\partial \gamma_B} + \left(q J m_A + h_B - \gamma_B \right) \frac{\partial m_B}{\partial \gamma_B}  
+ (D - \eta) \frac{\partial q_B}{\partial \gamma_B} = 0, \nonumber \\
\left(q J m_B + h_A - \gamma_A \right) \frac{\partial m_A}{\partial \eta} + \left(q J m_A + h_B - \gamma_B \right) \frac{\partial m_B}{\partial \eta}  
+ (D - \eta) \frac{\partial q_B}{\partial \eta} = 0,                                
\label{ec121}
\end{eqnarray} 
which are simultaneously satisfied if the variational parameters $\gamma_A$, $\gamma_B$ and $\eta$ fulfill the relations:
\begin{eqnarray}
\gamma_A =  q J m_B + h_A, \quad \gamma_B =  q J m_A + h_B, \quad \mbox{and} \quad \eta = D.                        
\label{vpe121}
\end{eqnarray}
After substituting the variational parameters $\gamma_A$, $\gamma_B$ and $\eta$ obeying the extremum condition (\ref{vpe121}) to the variational Gibbs free energy (\ref{vp121}) one obtains the  mean-field estimate of the true Gibbs free energy:
\begin{eqnarray}
G = \!\!\!&-&\!\!\! N k_{\rm B} T \ln 2 - N k_{\rm B} T \ln \left \{\cosh \left[\frac{\beta}{2} (q J m_B + h_A) \right] \right\} \nonumber \\
  \!\!\!&-&\!\!\! N k_{\rm B} T \ln \left \{1 + 2 \exp(\beta D) \cosh \left[\beta (q J m_A + h_B) \right] \right\} + NqJ m_A m_B,
\label{vgfe121}
\end{eqnarray}
whereas the sublattice magnetizations $m_A$, $m_B$ and the quadrupolar moment $q_B$ are determined by the following expressions:
\begin{eqnarray}
m_A \!\!\!&=&\!\!\! \frac{1}{2} \tanh \left[\frac{\beta}{2} \left(q J m_B + h_A \right) \right], \nonumber \\
m_B \!\!\!&=&\!\!\! \frac{2 \exp(\beta D) \sinh \left[\beta (q J m_A + h_B) \right]}{1 + 2 \exp(\beta D) \cosh \left[\beta (q J m_A + h_B) \right]}, \nonumber \\
q_B \!\!\!&=&\!\!\! \frac{2 \exp(\beta D) \cosh \left[\beta (q J m_A + h_B) \right]}{1 + 2 \exp(\beta D) \cosh \left[\beta (q J m_A + h_B) \right]}.                                 
\label{mq121}
\end{eqnarray}
In a zero magnetic field, the derived expressions (\ref{mq121}) for both sublattice magnetizations and quadrupolar moment reduce to: 
\begin{eqnarray}
m_A \!\!\!&=&\!\!\! \frac{1}{2} \tanh \left(\frac{\beta q J m_B}{2} \right), \nonumber \\
m_B \!\!\!&=&\!\!\! \frac{2 \exp(\beta D) \sinh \left(\beta q J m_A \right)}{1 + 2 \exp(\beta D) \cosh \left(\beta q J m_A \right)}, \nonumber \\
q_B \!\!\!&=&\!\!\! \frac{2 \exp(\beta D) \cosh \left(\beta q J m_A \right)}{1 + 2 \exp(\beta D) \cosh \left(\beta q J m_A \right)}.                                 
\label{mq121zf}
\end{eqnarray}
It can be understood from Eq. (\ref{mq121zf}) that one has to solve at first a system of two coupled equations for both sublattice magnetizations $m_A$ and $m_B$ in order to get the mean-field estimate of the true Gibbs free energy according to Eq. (\ref{vgfe121}). Besides, it is quite obvious from Eq. (\ref{mq121zf}) that both spontaneous magnetizations $m_A$ and $m_B$ simultaneously vanish at a critical point of the continuous phase transition and consequently, the spontaneous magnetizations $m_A$ and $m_B$ can still be expanded up to the linear term (and subsequently eliminated) with the aim to locate the critical temperature of the second-order phase transition. Following this procedure one gains the critical condition for the continuous phase transitions: 
\begin{eqnarray}
\begin{array}{l}
m_A = \frac{\beta_{\rm c} q J}{4} m_B,  \\
m_B = \frac{2 \beta_{\rm c} q J  \exp(\beta_{\rm c} D)}{1 + 2 \exp(\beta_{\rm c} D)} m_A 
\end{array}
\Longleftrightarrow \,             
\left(\frac{\beta_{\rm c} q J}{2} \right)^2 = \frac{1 + 2 \exp(\beta_{\rm c} D)}{2 \exp(\beta_{\rm c} D)},
\label{cc121}
\end{eqnarray}
which implies a highly non-trivial dependence of the critical temperature on the uniaxial single-ion anisotropy. It is worthy to notice, moreover, that the critical condition (\ref{cc121}) of the mixed spin-1/2 and spin-1 Ising model is nothing but a geometric mean of the critical conditions (\ref{cc12}) and (\ref{cc1}) of the spin-1/2 Ising model and the spin-1 Blume-Capel model, respectively. In an absence of the uniaxial single-ion anisotropy $D = 0$, the critical condition (\ref{cc121}) allows to obtain the closed-form expression for the critical temperature of the mixed spin-1/2 and spin-1 Ising model: 
\begin{eqnarray}
\frac{k_{\rm B} T_{\rm c}}{q J} = \frac{1}{\sqrt{6}}, \qquad (D = 0).
\label{cc121d0}
\end{eqnarray}

However, one should bear in mind that the critical behavior of the mixed spin-1/2 and spin-1 Ising model may be more intricate, because an existence of the discontinuous phase transitions cannot be definitely ruled out in this more complex spin system. The tricritical point can be simply obtained from the expansion of both sublattice magnetizations when retaining the first two non-zero terms: 
\begin{eqnarray}
\label{me121}
m_A \!\!\!&=&\!\!\! \frac{\beta q J}{4} m_B - \frac{(\beta q J)^3}{48} m_B^3, \\
m_B \!\!\!&=&\!\!\! \beta q J \frac{2\exp(\beta D)}{1 + 2 \exp(\beta D)} m_A 
           + \frac{(\beta q J)^3}{6} \frac{2\exp(\beta D)}{1 + 2 \exp(\beta D)} \left[1 - \frac{6\exp(\beta D)}{1 + 2 \exp(\beta D)} \right] m_A^3. \nonumber
\end{eqnarray}
An elimination of one of the sublattice magnetizations (say $m_B$) from the system of two mutually coupled equations (\ref{me121}) yields the following expansion 
for the other sublattice magnetization ($m_A$):
\begin{eqnarray}
m_A \!\!\!&=&\!\!\! f_1 m_A + f_3 m_A^3 + \ldots \nonumber \\
f_1 \!\!\!&=&\!\!\! \frac{(\beta q J)^2}{4} \frac{2\exp(\beta D)}{1 + 2 \exp(\beta D)}   \nonumber \\
f_3 \!\!\!&=&\!\!\! \frac{(\beta q J)^4}{24} \frac{2\exp(\beta D)}{1 + 2 \exp(\beta D)} \left\{ 1 - \frac{6\exp(\beta D)}{1 + 2 \exp(\beta D)} 
           - 2 \left[ \frac{\beta q J \exp(\beta D)}{1 + 2 \exp(\beta D)} \right]^2 \right\},  
\label{mae121}
\end{eqnarray}
which gives evidence for the mean-field critical exponent $\beta_m = 1/2$ of the spontaneous magnetization close to a critical point of the second-order phase transition:
\begin{eqnarray}
m_A (T \leq T_{\rm c}, h=0) \approx \left(\frac{1 - f_1}{f_3} \right)^{\frac{1}{2}}. 
\label{mfae121}
\end{eqnarray}
The tricritical point, at which the lines of second- and first-order phase transitions of the mixed spin-1/2 and spin-1 Ising model coalesce, can be consequently found from the condition 
$1 - f_1 = 0$, $f_3 = 0$ fulfilled at: 
\begin{eqnarray}
\frac{k_{\rm B} T_{\rm t}}{q J} = \frac{1}{\sqrt{20}}, \qquad  \frac{D_{\rm t}}{q J} = - \frac{\ln 8}{\sqrt{20}}.
\label{tcpmix}
\end{eqnarray}

\begin{figure}[tb]
\begin{center}
\hspace{-1.0cm}
\includegraphics[clip,width=7.5cm]{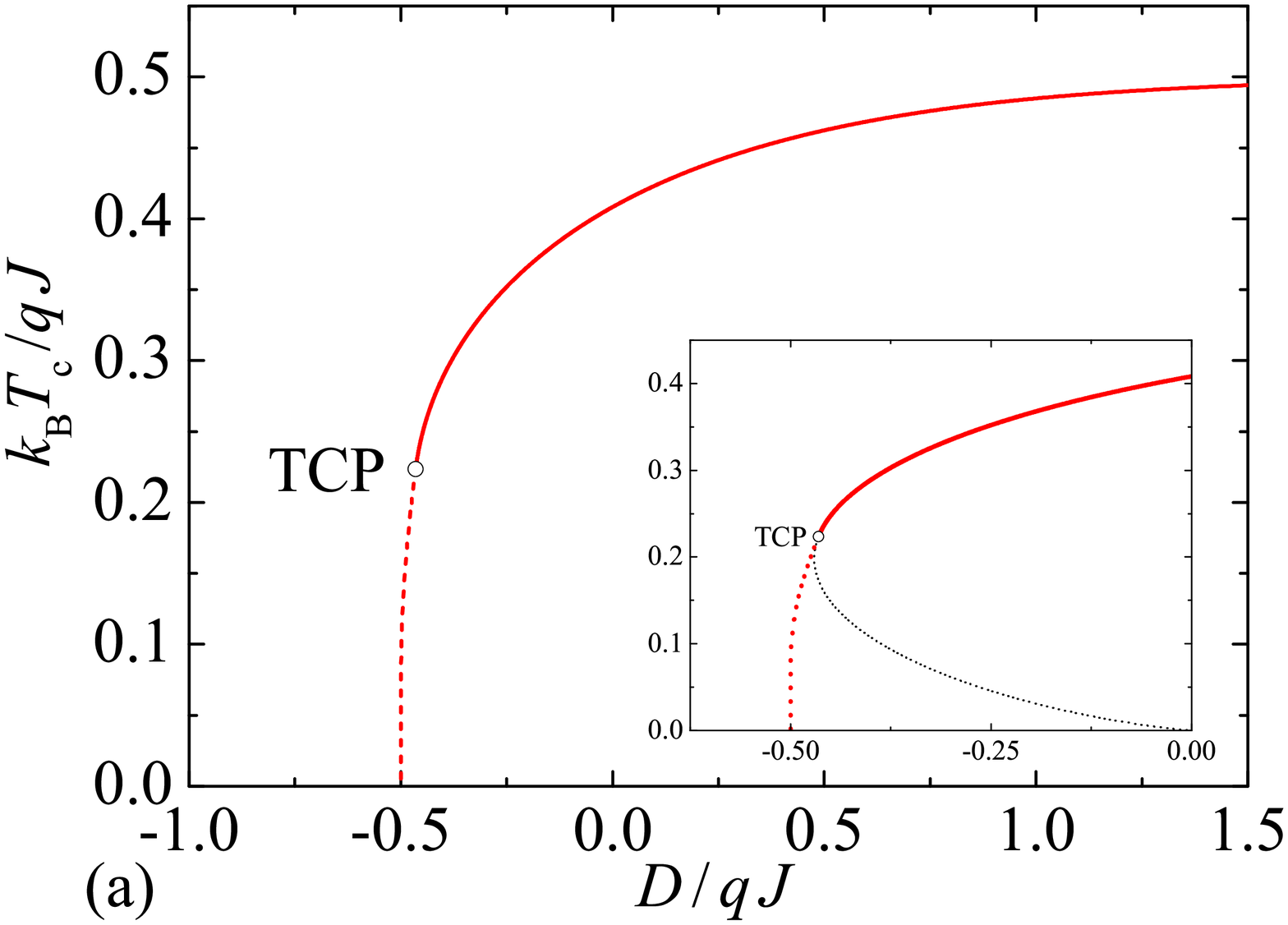}
\hspace{-1.0cm}
\includegraphics[clip,width=7.5cm]{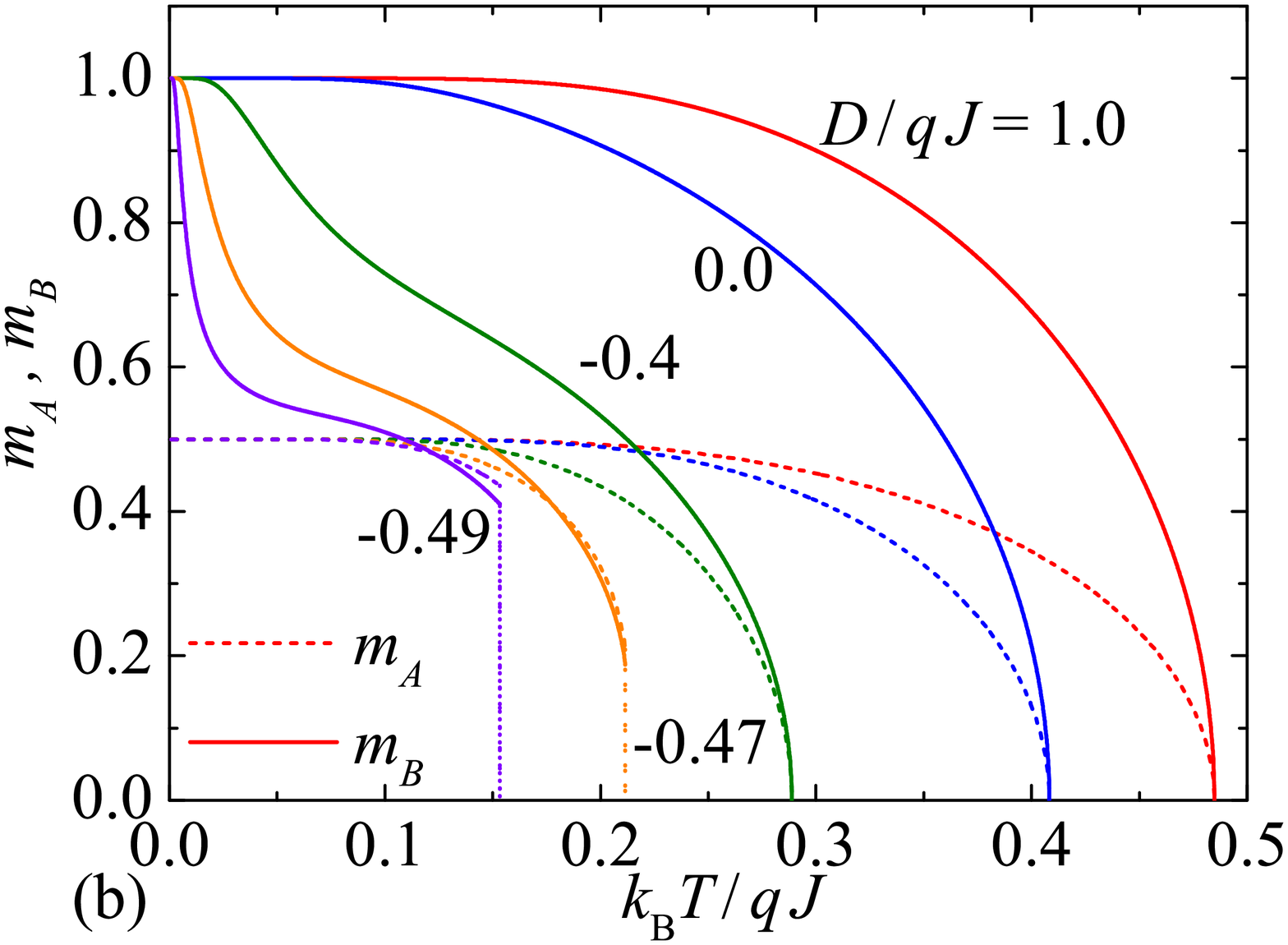}
\hspace{-1.5cm}
\vspace{-0.25cm}
\caption{(a) The critical temperature of the mixed spin-1/2 and spin-1 Ising model as a function of the uniaxial single-ion anisotropy. The critical line of continuous (discontinuous) phase transitions is displayed by solid (broken) line, an open circle denotes a tricritical point (TCP). The inset shows detailed plot of a tricritical region including the unstable solution of the critical condition (\ref{cc121}) shown by a thin dotted line;  (b) Thermal variations of the sublattice magnetizations $m_A$ (broken lines) and $m_B$ (solid lines) for a few different values of the uniaxial single-ion anisotropy. Dotted lines denote magnetization jumps connected with the discontinuous phase transitions.}
\label{figmix}
\end{center}
\end{figure}

Altogether, it can be concluded that the mean-field theory predicts for the mixed spin-1/2 and spin-1 Ising model with the uniaxial single-ion anisotropy a rather diverse phase diagram, which is quite reminiscent of that previously discussed for the spin-1 Blume-Capel model. The phase diagram of the mixed spin-1/2 and spin-1 Ising model shown in Fig. \ref{figmix}(a) indeed exhibits the second-order phase transitions for $D > D_{\rm t}$ with a continuous vanishing of the spontaneous magnetization at the critical temperature (\ref{cc121}), while the first-order phase transitions with a discontinuous (abrupt) jump of the spontaneous magnetization appear for $D < D_{\rm t}$ owing to a mutual coexistence of the ferromagnetic and paramagnetic states at a critical point. The lines of continuous and discontinuous phase transitions merge together at a tricritical point, which is unambiguously given by the coordinates (\ref{tcpmix}) and is displayed in Fig. \ref{figmix}(a) by an open circle. The inset of Fig. \ref{figmix}(a) also shows the unstable solution of the critical condition (\ref{cc121}) for the continuous phase transitions by a thin dotted line. To provide an independent check of the established phase diagram we have plotted in Fig. \ref{figmix}(b) a few typical temperature dependences of both spontaneous sublattice magnetizations. It can be seen from Fig. \ref{figmix}(b) that thermal variations of the sublattice magnetizations actually exhibit an abrupt jump at a critical temperature for $D < D_{\rm t}$ (e.g. $D/(qJ) = -0.47$ and $-0.49$) contrary to a continuous disappearance of both sublattice magnetizations at a critical point provided that $D > D_{\rm t}$. 

\subsection{Spin-$S$ Ising model in a transverse field}
\label{mfa12t}

Finally, the mean-field calculations will be exemplified on the particular example of the spin-$S$ Ising model in a transverse magnetic field, which represents a paradigmatic example of lattice-statistical spin model displaying a quantum phase transition. The spin-$S$ Ising model in a transverse magnetic field can be defined through the Hamiltonian:
\begin{equation}
\hat{\cal H} = - J \sum_{\langle ij \rangle}^{Nq/2} \hat{S}_i^z \hat{S}_j^z - \Omega \sum_{k=1}^{N} \hat{S}_k^x,
\label{hamtim}
\end{equation}
which involves two different spatial components $\hat{S}_j^x$ and $\hat{S}_j^z$ of the spin-$S$ operator at the $j$th lattice site that do not commute with each other owing to a validity of the commutation relation $[\hat{S}_j^z, \hat{S}_j^x] = {\rm i} \hat{S}_j^y$ (${\rm i}^2 = -1, \hbar = 1$). The first summation in the Hamiltonian (\ref{hamtim}) accounts for the Ising coupling $J$ between the nearest-neighbor spin pairs, while the second one accounts for the Zeeman's energy of magnetic moments in the transverse magnetic field $\Omega$. In a consequence of this fact, the magnetic behavior of the transverse Ising model given by the Hamiltonian (\ref{hamtim}) will be fundamentally affected by quantum fluctuations.

The trial Hamiltonian of the spin-$S$ Ising model in a transverse magnetic field will include within the mean-field approximation two different variational parameters $\gamma_z$ and $\gamma_x$:
\begin{equation}
\hat{\cal H}_0  = -\gamma_z \sum_{i=1}^{N} \hat{S}_i^z -\gamma_x \sum_{i=1}^{N} \hat{S}_i^x, 
\label{thtim}
\end{equation} 
which correspond to effective fields acting on each individual spin along the longitudinal and transverse direction, respectively. For further convenience, it is possible to rewrite the trial Hamiltonian (\ref{thtim}) as a sum of single-spin Hamiltonians $\hat{\cal H}_i$: 
\begin{equation}
\hat{\cal H}_0  =  \sum_{i = 1}^N \hat{\cal H}_i, \qquad \hat{\cal H}_i = -\gamma_z \hat{S}_i^z -\gamma_x \hat{S}_i^x, 
\label{thtimi}
\end{equation}
since the individual spins are treated independently of each other within the framework of the mean-field theory. The single-spin Hamiltonian $\hat{\cal H}_i$ represents in general $(2S+1)\times(2S+1)$ matrix, which must be further diagonalized in order to calculate the trial partition function:
\begin{equation}
Z_0  =  \mbox{Tr} \exp(-\beta \hat{\cal H}_0) = \prod_{i = 1}^N \mbox{Tr}_i \exp( -\beta \hat{\cal H}_i).  
\label{tpftim}
\end{equation}
Here, the symbols $\mbox{Tr}$ and $\mbox{Tr}_i$ stand for a trace over degrees of freedom of the full set of the Ising spins and the $i$th Ising spin, respectively. The most straightforward diagonalization of the single-spin Hamiltonian $\hat{\cal H}_i$ can be accomplished by the spin rotation transformation:
\begin{eqnarray}
\hat{S}_i^x \!\!\!&=&\!\!\!  \hat{S}_i^{x}{'} \cos \phi + \hat{S}_i^{z}{'} \sin \phi, \nonumber \\
\hat{S}_i^z \!\!\!&=&\!\!\! -\hat{S}_i^{x}{'} \sin \phi + \hat{S}_i^{z}{'} \cos \phi,
\label{srt}
\end{eqnarray}
which brings the single-spin Hamiltonian $\hat{\cal H}_i$ to the following form:
\begin{equation}
\hat{\cal H}_i = -\hat{S}_i^{z}{'} (\gamma_z \cos \phi + \gamma_x \sin \phi) - \hat{S}_i^{x}{'} (-\gamma_z \sin \phi + \gamma_x \cos \phi). 
\label{thitim}
\end{equation}
A suitable choice of the rotation angle $\tan \phi = \gamma_x/\gamma_z$ consequently ensures a diagonal form of the single-spin Hamiltonian $\hat{\cal H}_i$: 
\begin{equation}
\hat{\cal H}_i = -\hat{S}_i^{z}{'} \sqrt{\gamma_z^2 + \gamma_x^2}, 
\label{thitimd}
\end{equation}
which allows a straightforward calculation of the trial partition function $Z_0$, the Gibbs free energy $G_0$, the longitudinal and transverse magnetizations $m_z = \langle \hat{S}_i^z \rangle_0$ and $m_x = \langle \hat{S}_i^x \rangle_0$:
\begin{eqnarray}
Z_0  \!\!\!&=&\!\!\! \prod_{i = 1}^N \mbox{Tr}_i \exp( -\beta \hat{\cal H}_i) = \left[\sum_{n=-S}^S \exp \left(\beta n \sqrt{\gamma_z^2 + \gamma_x^2} \right) \right],  \nonumber \\
G_0  \!\!\!&=&\!\!\! -k_{\rm B} T \ln {\cal Z}_0 = - N k_{\rm B} T \ln \left[\sum_{n=-S}^S \cosh \left(\beta n \sqrt{\gamma_z^2 + \gamma_x^2} \right) \right],  \nonumber \\
m_z \!\!\!&=&\!\!\! -\frac{1}{N} \frac{\partial G_0}{\gamma_z} = \frac{\gamma_z}{\sqrt{\gamma_z^2 + \gamma_x^2}} 
\frac{\displaystyle\sum_{n=-S}^S n \sinh \left(\beta n \sqrt{\gamma_z^2 + \gamma_x^2} \right)}{\displaystyle\sum_{n=-S}^S \cosh \left(\beta n \sqrt{\gamma_z^2 + \gamma_x^2} \right)}, \nonumber \\
m_x \!\!\!&=&\!\!\! -\frac{1}{N} \frac{\partial G_0}{\gamma_x} = \frac{\gamma_x}{\sqrt{\gamma_z^2 + \gamma_x^2}} 
\frac{\displaystyle\sum_{n=-S}^S n \sinh \left(\beta n \sqrt{\gamma_z^2 + \gamma_x^2} \right)}{\displaystyle \sum_{n=-S}^S \cosh \left(\beta n \sqrt{\gamma_z^2 + \gamma_x^2} \right)}.
\label{pfgmtim}
\end{eqnarray}

At this stage, it is easy to compute the canonical ensemble average for a difference between the true and trial Hamiltonians within the simplified model system given by Eq. (\ref{thtim}):
\begin{equation}
\left \langle {\cal H} - {\cal H}_0 \right \rangle_{0} = - \frac{Nq}{2} J m_z^2 - N \gamma_z m_z - N (\Omega - \gamma_x) m_x,
\label{mvtim}
\end{equation} 
which in combination with the trial Gibbs free energy $G_0$ given by Eq. (\ref{pfgmtim}) provides the following expression for the variational Gibbs free energy $\Phi$: 
\begin{eqnarray}
\Phi = \! {-} N k_{\rm B} T \ln \! \left[\sum_{n=-S}^S \! \cosh \left(\beta n \sqrt{\gamma_z^2 + \gamma_x^2} \right) \! \right] \! {-} \frac{Nq}{2} J m_z^2 {-} N \gamma_z m_z {-} N (\Omega {-} \gamma_x) m_x.
\label{vptim}
\end{eqnarray} 
The minimalization of the variational Gibbs free energy (\ref{vptim}) with respect to both variational parameters $\gamma_z$ and $\gamma_x$ affords two equations:
\begin{eqnarray}
\frac{\partial \Phi}{\partial \gamma_z} \!\!\!&=&\!\!\! 0 \quad \Longleftrightarrow \quad 
\left(q J m_z - \gamma_z \right) \frac{\partial m_z}{\partial \gamma_z} + \left(\Omega - \gamma_x \right) \frac{\partial m_x}{\partial \gamma_z} = 0, \nonumber \\
\frac{\partial \Phi}{\partial \gamma_x} \!\!\!&=&\!\!\! 0 \quad \Longleftrightarrow \quad 
\left(q J m_z - \gamma_z \right) \frac{\partial m_z}{\partial \gamma_x} + \left(\Omega - \gamma_x \right) \frac{\partial m_x}{\partial \gamma_x} = 0, 
\label{ectim}
\end{eqnarray} 
which are simultaneously satisfied for the following choice of the variational parameters:
\begin{eqnarray}
\gamma_z =  q J m_z \quad \mbox{and} \quad \gamma_x =  \Omega.                        
\label{vpetim}
\end{eqnarray}
The mean-field estimate of the variational Gibbs free energy can be now obtained upon substitution of the variational parameters given by the extremum condition (\ref{vpetim}) into Eq. (\ref{vptim}):
\begin{eqnarray}
G = - N k_{\rm B} T \ln \left\{\sum_{n=-S}^S \cosh \left[\beta n \sqrt{(q J m_z)^2 + \Omega^2} \right] \right\} + \frac{Nq}{2} J m_z^2,
\label{vgfetim}
\end{eqnarray}
whereas the longitudinal and transverse magnetizations follow from the relations:
\begin{eqnarray}
m_z \!\!\!&=&\!\!\! \frac{q J m_z}{\sqrt{(q J m_z)^2 + \Omega^2}} 
\frac{\displaystyle\sum_{n=-S}^S n \sinh \left[\beta n \sqrt{(q J m_z)^2 + \Omega^2} \right]}{\displaystyle\sum_{n=-S}^S \cosh \left[\beta n \sqrt{(q J m_z)^2 + \Omega^2} \right]}, 
\nonumber \\
m_x \!\!\!&=&\!\!\! \frac{\Omega}{\sqrt{(q J m_z)^2 + \Omega^2}} 
\frac{\displaystyle\sum_{n=-S}^S n \sinh \left[\beta n \sqrt{(q J m_z)^2 + \Omega^2} \right]}{\displaystyle \sum_{n=-S}^S \cosh \left[\beta n \sqrt{(q J m_z)^2 + \Omega^2} \right]}.
\label{mxztim}
\end{eqnarray}

\begin{figure}[tb]
\begin{center}
\hspace{-1.0cm}
\includegraphics[clip,width=7.5cm]{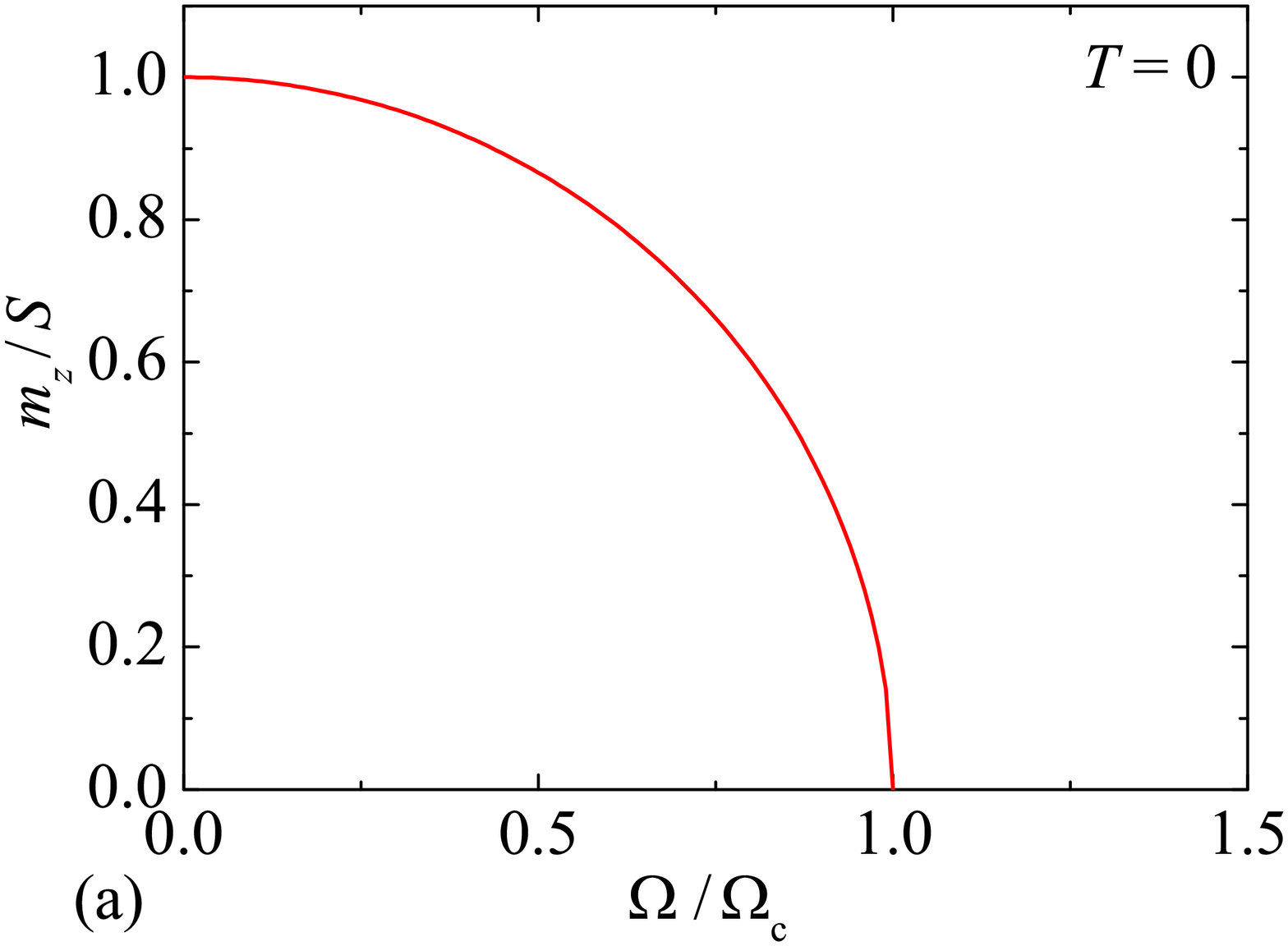}
\hspace{-1.0cm}
\includegraphics[clip,width=7.5cm]{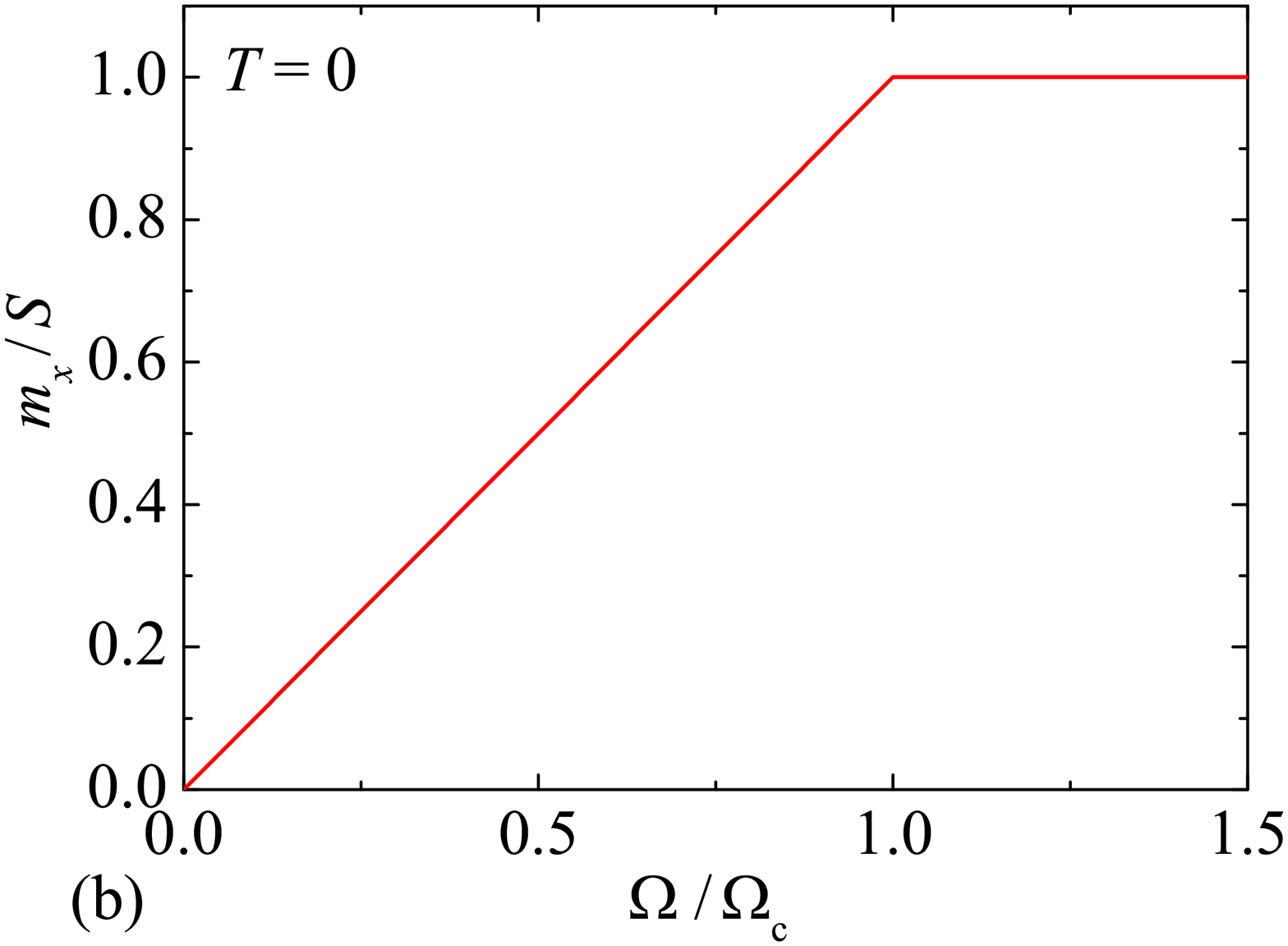}
\hspace{-1.5cm}
\vspace{-0.25cm}
\caption{The zero-temperature magnetization curve of the spin-$S$ Ising model in a transverse magnetic field: (a) longitudinal magnetization; (b) transverse magnetization. Both magnetizations are normalized with respect to the saturation value and the transverse field is scaled with respect to its critical value $\Omega_{\rm c} = q J S$.}
\label{timgs}
\end{center}
\end{figure}

At first, let us examine in detail the ground state of the spin-$S$ Ising model in a transverse magnetic field. It can be readily proved from Eqs. (\ref{vgfetim}) and (\ref{mxztim}) that the longitudinal and transverse magnetizations undergo a dramatic qualitative change at the critical field $\Omega_{\rm c} = q J S$:
\begin{eqnarray}
m_z = \Bigl \{ \begin{array}{l}
                   S \Bigl[1 - \left(\frac{\Omega}{\Omega_{\rm c}}\right)^{\!2}\Bigr]^{\frac{1}{2}},  \\
                   0, 
               \end{array} \Bigr. \qquad
               m_x = \Bigl \{ \begin{array}{l}
                   S \frac{\Omega}{\Omega_{\rm c}},  \qquad  \mbox{for} \quad \Omega < \Omega_{\rm c}, \\
                   S, \hspace{1.15cm}  \mbox{for} \quad \Omega \geq \Omega_{\rm c}.
               \end{array} \Bigr.                  
\label{gsod}
\end{eqnarray}
It is quite evident from Eq. (\ref{gsod}) that the ground state exhibits a peculiar ferromagnetic order with a non-zero longitudinal magnetization $m_z \neq 0$ whenever the transverse field is smaller than the critical value $\Omega < \Omega_{\rm c}$. However, the spontaneous longitudinal magnetization gradually diminishes with an increase in the transverse field until it completely disappears at the critical field $\Omega_{\rm c}$, while the transverse magnetization steadily (linearly) increases with the rising transverse field until it reaches the saturated value at the critical field $\Omega_{\rm c}$ (see Fig. \ref{timgs}). The spin-$S$ Ising model additionally exhibits a disordered paramagnetic ground state with the zero longitudinal magnetization $m_z = 0$ and the fully saturated transverse magnetization $m_x = S$ for the  stronger transverse fields $\Omega \geq \Omega_{\rm c}$. A continuous disappearance of the spontaneous longitudinal magnetization in a vicinity of the critical field $\Omega_{\rm c}$ evidently bears a close resemblance with a continuous phase transition of the spin-1/2 Ising model in a zero magnetic field invoked by the rising temperature [cf. Eq. (\ref{gsod}) with Eq. (\ref{ce12})]. It is worthwhile to remark, however, that the breakdown of the spontaneous ferromagnetic long-range order of the transverse Ising model is driven by quantum fluctuations invoked by the transverse field and not by thermal fluctuations. The continuous order-disorder transition emerging in the spin-$S$ Ising model due to the transverse magnetic field at absolute zero temperature is therefore referred to as a {\it quantum phase transition}. 

\begin{figure}[tb]
\begin{center}
\hspace{-1.0cm}
\includegraphics[clip,width=7.5cm]{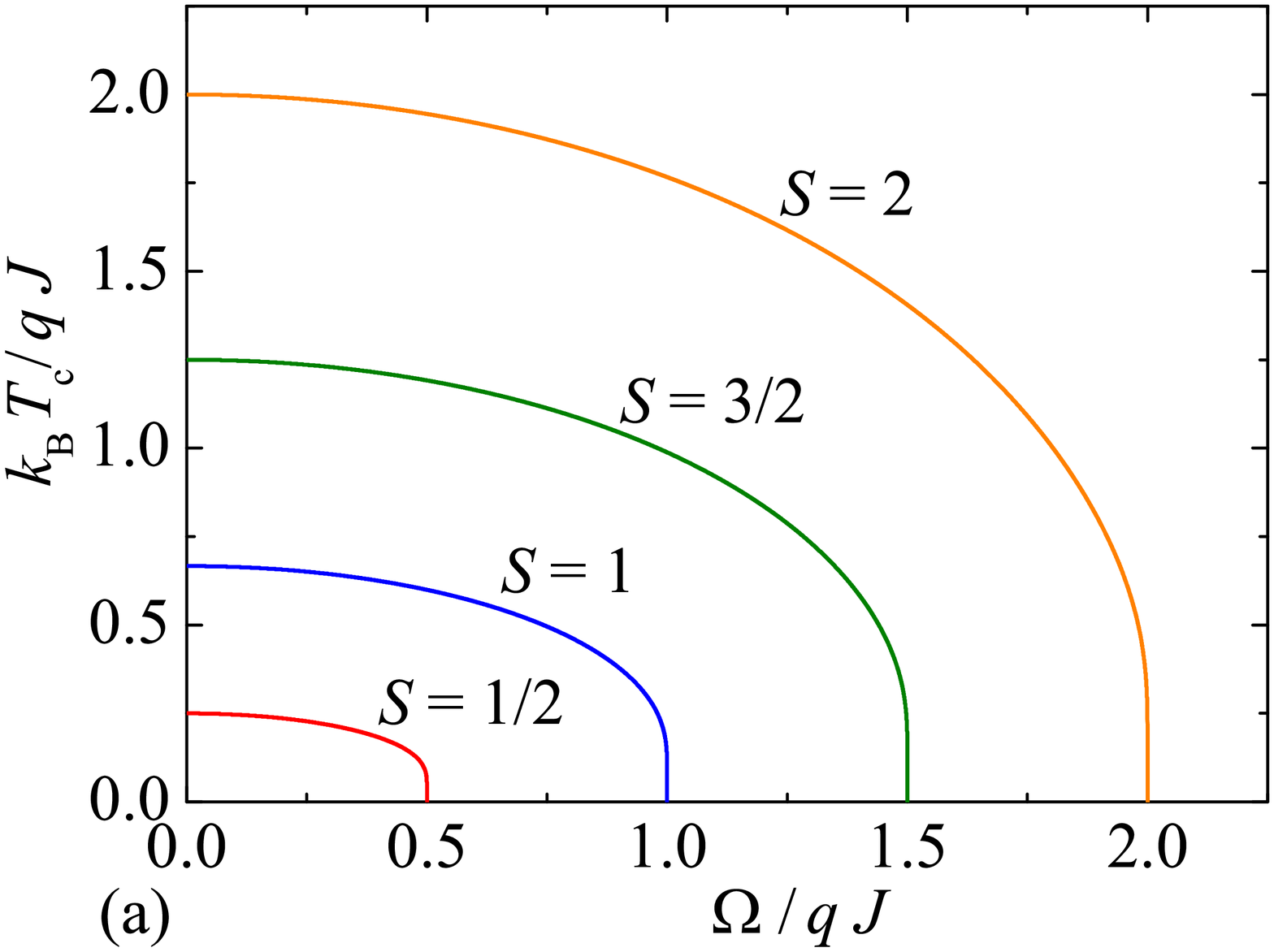}
\hspace{-1.0cm}
\includegraphics[clip,width=7.5cm]{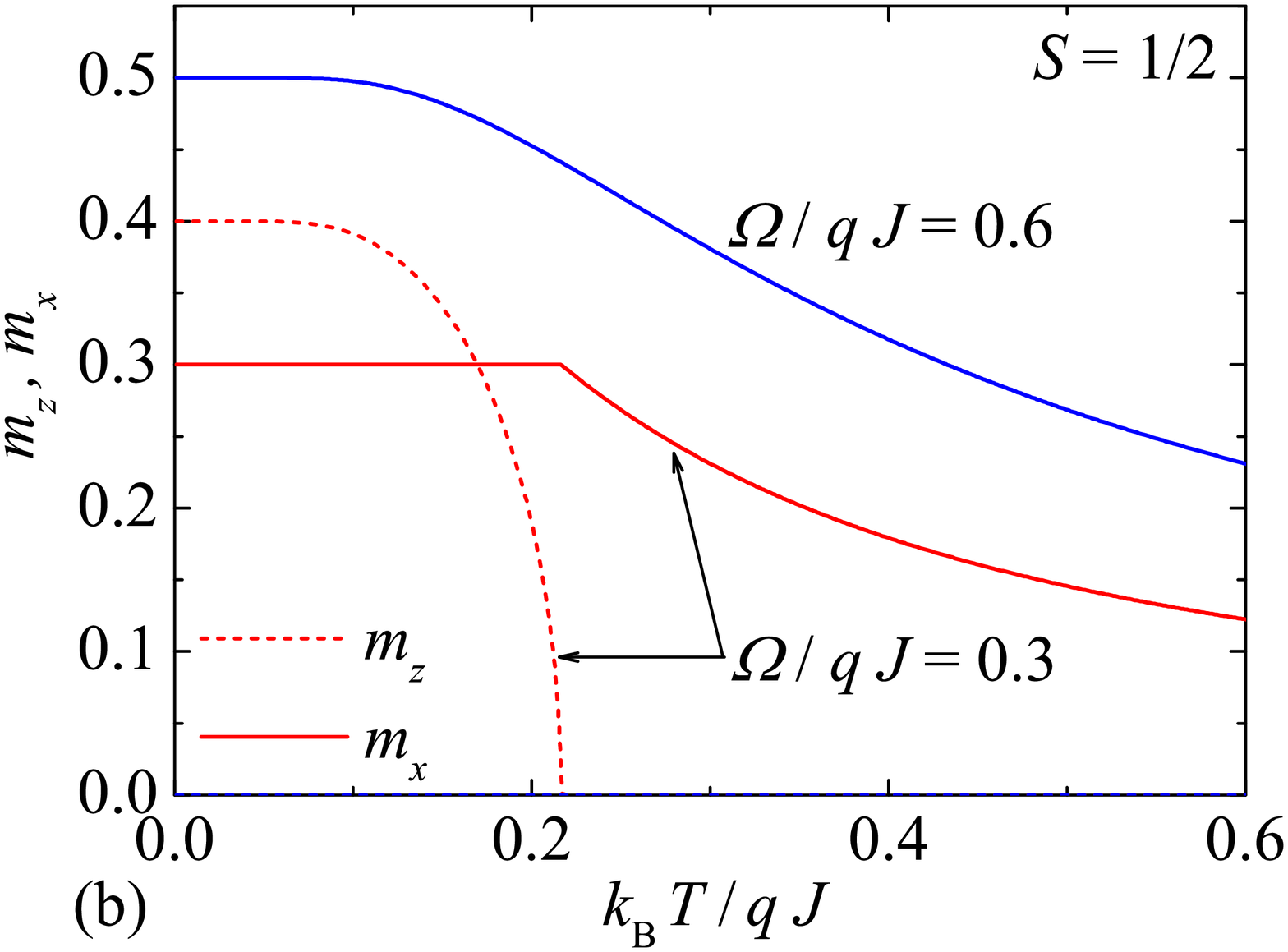}
\hspace{-1.5cm}
\vspace{-0.25cm}
\caption{(a) The critical temperature of the spin-$S$ Ising model as a function of the transverse magnetic field for several values of the quantum spin number $S$; (b) Typical temperature dependences of the longitudinal magnetization (broken line) and transverse magnetization (solid lines) of the spin-$1/2$ Ising model for two different transverse fields selected below and above the critical field $\Omega_{\rm c}/q J = 1/2$.}
\label{timcc}
\end{center}
\end{figure}

Of course, it might be also of particular interest to investigate a precise nature of the order-disorder phase transition between the ferromagnetic and paramagnetic states of the transverse Ising model at finite temperatures. The vanishing longitudinal magnetization at a critical point of the continuous ferromagnetic-to-paramagnetic phase transition allows us to derive from Eq. (\ref{mxztim}) the following critical condition:  
\begin{eqnarray}
\frac{\displaystyle\sum_{n=-S}^S n \sinh \left(\beta_{\rm c} n \Omega \right)}{\displaystyle\sum_{n=-S}^S \cosh \left(\beta_{\rm c} n \Omega \right)} = \frac{\Omega}{q J} \, ,
\qquad \qquad \qquad \left(\beta_{\rm c} = \frac{1}{k_{\rm B} T_{\rm c}}\right).
\label{cctim}
\end{eqnarray}
The critical temperature of the spin-$S$ Ising model computed according to Eq. (\ref{cctim}) is plotted in Fig. \ref{timcc}(a) against a relative strength of the transverse magnetic field for several values of the quantum spin number $S$. Apparently, the transverse field generally causes a progressive reduction of the critical temperature until it entirely tends to zero at the critical field $\Omega_{\rm c} = q J S$ and this generic feature remains valid irrespective of the quantum spin number $S$. The gradual decline of the critical temperature due to the transverse magnetic field can be attributed to reinforcing quantum fluctuations, which support thermal fluctuations in destroying the spontaneous ferromagnetic long-range order. The highest critical temperature of the spin-$S$ Ising model can be accordingly found in the limit of zero transverse magnetic field:
\begin{eqnarray}
\displaystyle \lim_{\Omega \to 0} \frac{k_{\rm B} T_{\rm c}}{q J} = \frac{S(S + 1)}{3},
\label{tctim}
\end{eqnarray}
which can be derived from Eq. (\ref{mxztim}) in the asymptotic limit of vanishing transverse field $\Omega$. 

The longitudinal and transverse magnetizations of the spin-$1/2$ Ising model are depicted in Fig. \ref{timcc}(b) as a function of temperature for two different transverse magnetic fields selected below and above the critical field $\Omega_{\rm c}/q J = 1/2$. In the former case $\Omega < \Omega_{\rm c}$, the longitudinal magnetization already undergoes at zero temperature a quantum reduction of magnetization due to the transverse magnetic field, whereas this reduced initial value is further progressively suppressed by thermal fluctuations upon increasing temperature until the longitudinal magnetization completely vanishes at the critical temperature. By contrast, the transverse magnetization keeps constant unless the critical temperature is reached and then it shows monotonous decline with increasing the temperature. In the latter case $\Omega > \Omega_{\rm c}$, the longitudinal magnetization is zero regardless of the temperature, while the transverse magnetization shows the monotonous decrease upon raising temperature. Finally, it is worthy to notice that the spin-$S$ Ising model in a transverse magnetic field cannot exhibit discontinuous phase transitions provided that the uniaxial single-ion anisotropy is set to zero.

\subsection{Historical survey and future perspectives}

Although it is beyond the scope of the present review to provide a comprehensive list on all previous mean-field calculations, let us conclude this section with a brief summary of a few selected achievements acquired by making use of the mean-field theory. A versatility and conceptual simplicity are two apparent advantages of the mean-field approximation, which can be rather easily adapted to complex lattice-statistical spin models that are fully intractable or difficult of approach within more sophisticated analytical methods. Among other matters, the mean-field theory has played an important role for a deeper understanding of more intricate phase transitions and critical phenomena. For instance, the early mean-field results for the spin-$S$ Blume-Capel \cite{blu66,cap66,cap67,tag70,osa80,wan91,pla93,bak94} and Blume-Emery-Griffiths \cite{blu71,siv72,muk74,kri75,hos91,bar91} models predicted a variety of multicritical phenomena including tricritical points, isolated critical points, critical end points, bicritical points, triple points and tetracritical points in addition to the usual critical points. It is worthy to notice, moreover, that an intriguing multicritical behavior have been later found using the mean-field calculations also in various mixed-spin Ising models with the uniaxial crystal-field anisotropy \cite{kan91,ben97,abu01,bob04,god04,mac04,wei06,bab07,ben09,mia09,wei09,cru13}. All aforementioned mean-field results were derived for the Ising-like spin systems, which represent plausible models for magnetic materials with a regular crystal structure. 

Another wide family of magnetic materials form amorphous magnets with a certain degree of randomness, which makes theoretical modeling of respective magnetic properties particularly intriguing. However, the mean-field theory has been also adapted to the Ising-like models accounting for the site and/or bond disorder \cite{ber71,tho79} as well as the random distribution of magnetic ions within the binary \cite{mej81,upt89,jap93,jas96,jas97}, ternary \cite{bob03,del04,del06,bob07} or quaternary \cite{wei04} alloys. It has been also demonstrated by the mean-field calculations that the effect of structural disorder on some magnetic properties of amorphous materials can be captured by the random magnetic field \cite{sax83,mat85,kau86,seb90,had10,had11} or random crystal field \cite{ben87,boc89,kan90,vie01,ben07,bah07,bah08,ben11}.

Quantum effects introduced into the Ising spin systems either by the transverse magnetic field or the transverse crystal field represent another challenging topic to account for within the framework of the mean-field treatment. The effect of transverse magnetic field on a magnetic behavior of the Ising models has been thoroughly investigated for example in the mixed-spin systems \cite{bar85,kan88,abu02}, ternary alloy \cite{del07} and thin films \cite{wan95,wan00,ane03,kan03,kan04,kan05}. A lot of the mean-field calculations has been carried out for the transverse Ising models, which take into consideration a random character of the transverse magnetic field \cite{dut86,cas91}, the multispin interactions \cite{ane05,ten05}, the longitudinal magnetic field \cite{wei07}, the longitudinal \cite{yan99,zha06} or transverse \cite{sou08} crystal field. 

Although the results obtained within the mean-field theory might be largely affected by a rather crude approximation ignoring all spin-spin correlations, the mean-field theory will certainly remain the primary method for obtaining a rough estimate of a magnetic behavior of more complex lattice-statistical spin models. 
 
\setcounter{equation}{0} \newpage 
\section{Effective-field theory}
\label{eft}

To go beyond the accuracy of the mean-field theory Honmura and Kaneyoshi have suggested a new type of the effective-field theory utilizing the differential operator technique \cite{honm78,honm79}. It is noteworthy that the effective-field theory has been originally applied to the simplest spin-1/2 Ising model, for which the obtained results indeed turned out to be quite superior with respect to the relevant mean-field results. In addition, the simplicity and versatility of the effective-field approach based on the differential operator technique has been recognized soon after pioneering works due to Honmura and Kaneyoshi  \cite{honm78,honm79} when several research groups have applied this relatively simple mathematical technique to diverse Ising-like spin systems of much more complex nature.  Before describing the crucial steps of this interesting method it is worthwhile to remark that several alternative mathematical formulations of the effective-field theory have been later discovered using the integral operator formulation \cite{miel86} or finite-cluster approximation \cite{beny83}, but these alternative mathematical approaches have never gained the popularity and applicability of the original formulation based on the differential operator technique even though they provide results of exactly the same numerical accuracy. 

\subsection{Callen-Suzuki identity}
\label{Callen_identity}
The common starting point for developing various effective-field theories will be a simple equation, which is currently known in the literature as the Callen-Suzuki identity. This identity has  been originally derived by Callen for a spin-1/2 Ising model using Green function technique \cite{call63,suzu65}. Using different mathematical methods this useful relation has been later generalized for the Ising-like models with arbitrary spin values, amorphous systems, site- and bond-diluted models, the systems with random exchange interactions, random crystal and/or external magnetic fields and many others physically interesting systems \cite{balc02}. The quantum version of Callen-Suzuki identity has been at first derived and applied to the transverse Ising model by S\'{a}~Barreto, Fittipaldi a Zeks \cite{saba81}.  

Our first and very important goal in this work is a detailed derivation of the Callen-Suzuki identity within the standard canonical ensemble of statistical mechanics. For this purpose we will consider a quantum system consisting  of $N$ spins that are localized on the sites of an infinite  crystalline lattice with the coordination number  $q$. Next, we will assume that our system is described by the Hamilton operator $\hat{\cal H} $, which in general depends on spin operators  of all lattice sites. Naturally, such a Hamiltonian  also includes various physical parameters like exchange interactions, 
external and internal fields and so on.   

Under these very general assumptions, we at first separate the total Hamiltonian into two parts as follows:  
\begin{eqnarray} 
\label{eq1b1}
           \hat {\cal H} = \hat{\cal H}_i   +\hat {\cal H}^{'},
\end{eqnarray}  
where $\hat{\cal H}_i$ contains all interaction terms related to the $i$th site of the lattice and all remaining terms are of course included in $\hat{\cal H}^{'}$. It is useful noticing here that the Hamiltonians $\hat{\cal H}_i$ and  $\hat{\cal H}^{'}$ do not commute in general, i.e.,
\begin{eqnarray} 
\label{eq1b2}
            \bigl [  \hat{\cal H}_i , \hat{\cal H}^{'} \bigr ]  \neq 0.
\end{eqnarray}  
The canonical expectation value of a physical quantity $\hat {\cal O}_i$  related to the $i$th site of the lattice can be calculated from the well-known relation:
\begin{eqnarray} 
\label{eq1b3}
             \bigl \langle \hat {\cal O}_i \bigr \rangle = 
             \dfrac{1}{Z} \mbox{Tr}\; \hat{\cal O}_i \exp(-\beta \hat{\cal H}) ,
\end{eqnarray}  
where the canonical partition function $Z$ is given by the formula:
\begin{eqnarray} 
\label{eq1b4}
             Z = \mbox{Tr} \exp(-\beta \hat {\cal H}).
\end{eqnarray} 
Here, $\beta = 1/(k_{\rm B} T)$, $T$ is the absolute temperature and $k_{\rm B}$ represents the Boltzmann's constant. The total trace (Tr) in Eqs. \eqref{eq1b3} and \eqref{eq1b4} can be formally expressed as a product of partial traces over the spin operators on individual lattice sites. Taking this fact into account, we can rewrite  \eqref{eq1b3} in the form:
\begin{eqnarray} 
\label{eq1b5}
\nonumber
             \bigl \langle \hat{\cal  O}_i \bigr \rangle &=& \dfrac{1}{Z} \mbox{Tr}^{'}\mbox{Tr}_i\; \hat{\cal O}_i 
             \exp \bigl [-\beta (\hat{\cal H}_i   + \hat{\cal H}^{'} )\bigr],\\ 
\nonumber            
             \\
            &=& \dfrac{1}{Z} \mbox{Tr}^{'}\mbox{Tr}_i\; \hat{\cal O}_i 
            \bigl [ \exp \bigl (-\beta \hat{\cal H}_i \bigr)  \exp \bigl (-\beta \hat{\cal H}^{'} \bigr)  +   \hat\Delta \bigr],
\end{eqnarray}  
where $\mbox{Tr}_i$ denotes the partial trace over the spin operator related to the $i$th lattice site, $\mbox{Tr}^{'}$ represents the trace over all remaining degrees of freedom and the operator 
$\hat \Delta$ is defined by the formula:
\begin{eqnarray} 
\label{eq1b6}
            \hat \Delta = 
             \exp \bigl [-\beta (\hat{\cal H}_i   +\hat {\cal H}^{'} )\bigr] -
            \exp \bigl (-\beta \hat{\cal H}_i \bigr)  \exp \bigl (-\beta \hat{\cal H}^{'} \bigr).
\end{eqnarray}  
For further manipulations it is useful to rewrite \eqref{eq1b5} as follows:
\begin{eqnarray} 
\label{eq1b7}
             \bigl \langle \hat{\cal O}_i \bigr \rangle = \dfrac{1}{Z} \mbox{Tr}^{'}\Bigl 
             [\mbox{Tr}_i\; \hat {\cal O}_i 
             \exp \bigl (-\beta \hat{\cal H}_i \bigr ) \Bigr]  \exp \bigl (-\beta \hat {\cal  H}^{'} \bigr) +
            \dfrac{1}{Z} \mbox{Tr}\; \bigl (\hat {\cal O}_i \hat {\Delta} \bigr).
\end{eqnarray}  
If we now insert the expression $\mbox{Tr}_i\;\exp \bigl (-\beta \hat {\cal H}_i \bigr) /\mbox{Tr}_i\;\exp \bigl (-\beta \hat {\cal H}_i \bigr)$ into first term of Eq. \eqref{eq1b7}, then after a 
small manipulation we obtain equation:
\begin{eqnarray} 
\label{eq1b8}
\nonumber 
\bigl \langle \hat{\cal O}_i \bigr \rangle &=& 
\dfrac{1}{Z} \mbox{Tr}^{'}  \biggl \{ \dfrac{ \mbox{Tr}_i\;
	\hat{\cal O}_i \exp \bigl (-\beta \hat{\cal H}_i \bigr)  }			  	        	                                                  
	{\mbox{Tr}_i\;\exp \bigl (-\beta \hat{\cal H}_i \bigr)}   \biggr \}
     \mbox{Tr}_i\;  
    \exp \bigl (-\beta \hat{\cal H}_i \bigr )   \exp \bigl (-\beta \hat{\cal H}^{'} \bigr) \\ \nonumber \\
&+&
\dfrac{1}{Z} \mbox{Tr}\; \bigl (\hat{\cal O}_i \hat\Delta \bigr).
\end{eqnarray}  
Using the fact that expression in curled brackets does not depend on degrees of freedom of the $i$th lattice site one may rearrange Eq. \eqref{eq1b8}  by taking the partial trace $\mbox{Tr}_i$ before this expression and employing $\mbox{Tr}^{'}\mbox{Tr}_i = \mbox{ Tr}$:
\begin{eqnarray} 
\label{eq1b9}
\nonumber 
             \bigl \langle \hat{\cal O}_i \bigr \rangle &=& 
             \dfrac{1}{Z} \mbox{Tr} \biggl \{ \dfrac{ \mbox{Tr}_i\;
             \hat{\cal O}_i \exp \bigl (-\beta \hat{\cal H}_i \bigr)  } 
             {\mbox{Tr}_i\; \exp \bigl (-\beta \hat{\cal H}_i \bigr) }    \biggr \}       
             \exp \bigl (-\beta \hat{\cal H}_i \bigr )   \exp \bigl (-\beta \hat{\cal H}^{'} \bigr) \\\nonumber \\
             &+&
            \dfrac{1}{Z} \mbox{Tr}\; \bigl (\hat{\cal O}_i \hat\Delta \bigr).
\end{eqnarray}  
Next, one may rewrite \eqref{eq1b9}  with the help of Eq. \eqref{eq1b6} into the form:
\begin{eqnarray} 
\label{eq1b10}
\nonumber 
             \bigl \langle \hat{\cal O}_i \bigr \rangle &=& 
             \dfrac{1}{Z} \mbox{Tr} \biggl \{ \dfrac{ \mbox{Tr}_i\;
            \hat{\cal O}_i \exp \bigl (-\beta \hat{\cal H}_i \bigr)  } 
           {\mbox{Tr}_i\; \exp \bigl (-\beta \hat{\cal H}_i \bigr) }   \biggr \}        
             \exp \bigl (-\beta \hat {\cal H} \bigr )   \\ \nonumber \\
             &-&
            \dfrac{1}{Z} \mbox{Tr} \biggl \{ \dfrac{ \mbox{Tr}_i\;
           \hat{\cal O}_i \exp \bigl (-\beta \hat{\cal H}_i \bigr)  } 
            {\mbox{Tr}_i\; \exp \bigl (-\beta \hat{\cal H}_i \bigr)  }  
           - \hat{\cal O}_i \biggr \}         
             \hat\Delta,
\end{eqnarray}  
or alternatively:
\begin{eqnarray} 
\label{eq1b11}
\nonumber 
             \bigl \langle \hat{\cal O}_i \bigr \rangle &=& 
          \biggl \langle \dfrac{ \mbox{Tr}_i\;
          \hat{\cal O}_i \exp \bigl (-\beta \hat{\cal H}_i \bigr)  } 
          {\mbox{Tr}_i\; \exp \bigl (-\beta \hat{\cal H}_i \bigr) }   \biggr \rangle       
              \\ \nonumber \\
             &-&
              \biggl \langle \biggl ( \dfrac{ \mbox{Tr}_i\;
              \hat{\cal O}_i \exp \bigl (-\beta \hat{\cal H}_i \bigr)  } 
             {\mbox{Tr}_i\; \exp \bigl (-\beta \hat{\cal H}_i \bigr)  }  - \hat{\cal O}_i \biggr )       
             \hat{\cal D } \biggr \rangle, 
\end{eqnarray}  
where we have introduced the operator $ \hat{\cal D }$ as:
\begin{eqnarray} 
\label{eq1b12}
\nonumber
              \hat{\cal D } &=& 
             \hat \Delta \exp \bigl (\beta \hat{\cal H} \bigr) \\
             &=&            
             \hat  I - \exp \bigl (-\beta \hat{\cal H}_i \bigr)  \exp \bigl (-\beta \hat{\cal H}^{'} \bigr) \exp \bigl (\beta \hat{\cal H}\bigr)   
\end{eqnarray}  
with $\hat I$ standing for the identity operator in the relevant Hilbert space.

It should be emphasized here that Eq. \eqref{eq1b11} cannot be further manipulated exactly, therefore to proceed further with calculations, we now apply the following decoupling procedure:
\begin{eqnarray} 
\label{eq1b13}  
\nonumber       
              &&\biggl \langle \biggl ( \dfrac{ \mbox{Tr}_i\;
              \hat{\cal O}_i \exp \bigl (-\beta \hat{\cal H}_i \bigr)  } 
              {\mbox{Tr}_i\; \exp \bigl (-\beta \hat{\cal H}_i \bigr)  }  
              - \hat{\cal O}_i \biggr )  \hat{\cal D } \biggr \rangle
             \\ \nonumber \\
             \approx 
           && \biggl \langle \biggl ( \dfrac{ \mbox{Tr}_i\;
           	\hat{\cal O}_i \exp \bigl (-\beta \hat{\cal H}_i \bigr)  } 
           {\mbox{Tr}_i\; \exp \bigl (-\beta \hat{\cal H}_i \bigr)  }  
         - \hat{\cal O}_i \biggr )       
             \biggr \rangle  \bigl \langle\hat{\cal D }  \bigr \rangle
\end{eqnarray}  
and express Eq. \eqref{eq1b11} in the form:
\begin{eqnarray} 
\label{eq1b14}  
       \biggl (   \bigl \langle \hat{\cal O}_i \bigr \rangle -      
              \biggl \langle \dfrac{ \mbox{Tr}_i\;
              \hat{\cal O}_i \exp \bigl (-\beta \hat{\cal H}_i \bigr)  } 
            {\mbox{Tr}_i\; \exp \bigl (-\beta \hat{\cal H}_i \bigr) }    \biggr \rangle  \biggr )  
            \Bigl(  \hat I -  \bigl \langle \hat{\cal D }\bigr \rangle \Bigr)   = 0.            
\end{eqnarray}  
Taking into account that  $\bigl \langle \hat{\cal D } \bigr \rangle \neq \hat I$  one finally gets from \eqref{eq1b14} the following simple equation:
\begin{eqnarray} 
\label{eq1b15}  
          \bigl \langle \hat{\cal O}_i \bigr \rangle =      
          \biggl \langle \dfrac{ \mbox{Tr}_i\; \hat{\cal O}_i \exp \bigl (-\beta \hat{\cal H}_i \bigr)  } 
         {\mbox{Tr}_i\; \exp \bigl (-\beta \hat{\cal H}_i \bigr) }    \biggr \rangle,  
\end{eqnarray}  
which represents a generalized form of the original Callen-Suzuki identity. This relation frequently serves as a starting point for thermodynamic descriptions of  localized spin models 
with short-range interactions. Moreover, it is clear that the Callen-Suzuki identity given by Eq.~\eqref{eq1b15} represents only an approximate expression whenever Eq. \eqref{eq1b2} is 
fulfilled due to the decoupling procedure used in \eqref{eq1b13}. On the other hand, the Callen-Suzuki identity \eqref{eq1b15} becomes naturally exact in the particular case of commuting Hamiltonians $\bigl [  \hat{\cal H}_i , \hat{\cal H}^{'} \bigr ]  = 0$ since $\hat \Delta = 0$. 

In order to analyze physical properties of localized spin models it is frequently very useful to evaluate correlation functions among various spin variables. For this purpose, we may use instead of Eq.  \eqref{eq1b15} its extended version, namely,  
\begin{eqnarray} 
\label{eq1b16}  
          \bigl \langle \hat{\cal O}_i  \bigl\{   f_i \bigr\} \bigr \rangle =      
          \biggl \langle \bigl\{   f_i \bigr\}\dfrac{ \mbox{Tr}_i\;
          \hat {\cal O}_i \exp \bigl (-\beta \hat{\cal H}_i \bigr)  } 
           {\mbox{Tr}_i\; \exp \bigl (-\beta \hat{\cal H}_i \bigr) }    \biggr \rangle, 
\end{eqnarray}  
where $\bigl\{  f_i \bigr\}$ represent a function of arbitrary spin operators except those belonging to the $i$th lattice site. 

In all previous calculations we have considered a spin cluster consisting of one central spin only. It is however quite clear that a finite-size spin cluster generalization of the Callen-Suzuki identity can be readily obtained by using the same mathematical procedures as formerly described for the single-spin cluster version. The derivation of the generalized Callen-Suzuki identity for a finite-size spin cluster is straightforward and very similar to that of the single-spin cluster case and we will therefore not repeat it. Hence, let us merely quote the final form of the generalized Callen-Suzuki identity for the $n$-spin central cluster,  which takes the following form:
\begin{eqnarray} 
\label{eq1b17}  
          \bigl \langle \hat{\cal O}_{ij ... n}  \bigl\{   f_{ij ... n} \bigr\} \bigr \rangle =      
          \biggl \langle  \bigl\{   f_{ij ... n} \bigr\}
          \dfrac{ \mbox{Tr}_{ij ... n}\;\hat{\cal O}_{ij ... n} \exp \bigl (-\beta\hat {\cal H}_{ij ... n}  \bigr)  }
                  {\mbox{Tr}_{ij ... n} \; \exp \bigl (-\beta \hat{\cal H}_{ij ... n}  \bigr) }    \biggr \rangle,
\end{eqnarray}  
where the operator $\hat {\cal O}_{ij ... n}$ may contain all spin operators belonging to the $n$-spin central cluster and $ \bigl\{   f_{ij ... n} \bigr\}$ represents a function of arbitrary spin operators except those of the $n$-spin central cluster. The cluster Hamiltonian  $\hat {\cal H}_{ij ... n} $ exactly accounts for all interaction terms within the central cluster and also the terms describing interactions of the central cluster with its nearest-neighboring spins and also the interactions with an external magnetic field.

\subsection{Spin-1/2 Ising model within a single-spin cluster approach}
\label{0S_EFT}
In this part we will study a spin-1/2 Ising model on a crystalline lattice characterized through the coordination number $q$. The Ising model represents a  strongly anisotropic limit of a more general Heisenberg model in which  we  only consider interactions between $z$-components of spins. For that reason one can avoid a strict operator formalism in all subsequent calculations. 

Let us start our analysis with the Hamiltonian of the spin-1/2 Ising model in a longitudinal magnetic field, which can be written in the form:
\begin{eqnarray} 
\label{eq3b1}
             {\cal H} = - \sum_{i,j} J_{ij} \sigma_i \sigma_j - h \sum_i \sigma_i,
\end{eqnarray}  
where $\sigma_i = \pm 1/2$ are Ising spin variables representing two possible spin projections onto the $z$ direction, $h$ denotes value of external magnetic field and $J_{ij}$ stands for the exchange interaction between the spins. Most frequently the exchange interaction $J_{ij}$ is assumed to be non-zero only between nearest-neighbor spins on the lattice, however, an inclusion of the exchange couplings between the next-nearest or next-next-nearest neighbors is also possible within the present formulation. It should be noticed here that the summation in the first term of Eq. \eqref{eq3b1} will be hereafter restricted to all different pairs of nearest-neighbor spins and that one in the second term of Eq. \eqref{eq3b1} is performed over all spins.

In order to employ the Callen-Suzuki identity \eqref{eq1b15} for developing a new type of the effective-field theory it is convenient to separate the total Hamiltonian \eqref{eq3b1} into two parts:  
\begin{eqnarray} 
\label{eq3b2}
             {\cal H} =  {\cal H}_i  +  {\cal H}^{'}, 
\end{eqnarray} 
with
\begin{eqnarray} 
\label{eq3b3}
             {\cal H}_i = -\sigma_i \Bigl (\sum_{j= 1}^q J_{ij} \sigma_j + h\Bigr). 
\end{eqnarray}
It is worth recalling here that $q$ denotes the coordination number (i.e. the number of nearest neighbors) and the Hamiltonian $ {\cal H}_i $ contains all interaction terms of the central spin $\sigma_i$ with its nearest-neighbor spins on a lattice and also with an external magnetic field. All other interaction terms are included in the remaining Hamiltonian ${\cal H}^{'}$. Our main task is now the derivation of closed-form equations for important thermodynamic quantities such as the magnetization, Curie temperature, magnetic susceptibility, internal energy, specific heat and so on.  

It will be exemplified in subsequent calculations that one can obtain relatively simple expressions for all relevant physical quantities of the spin-1/2 Ising model within such an effective-field theory, in which the structure of the lattice is taken into account only by one parameter, namely, the coordination number $q$. This fact is very comfortable and useful from the mathematical point of view, since it enables a unified formulation of the theory for all crystalline structures. On the other hand, this is an undesirable deficiency of the theory from the physical point of view, which  does not enable to distinguish physical properties of the spin-1/2 Ising model on lattices with the same coordination numbers but different geometry. Therefore, the single-spin cluster formulation of the effective-field theory gives identical results for the spin-1/2 Ising model on several lattices (for example, plane triangular and simple cubic lattices, or square and kagom\'e lattices) despite of their different topology.

The original quantum-mechanical formulation of the Callen-Suzuki identity \eqref{eq1b15} can be rewritten for the spin-1/2 Ising model in a longitudinal magnetic field 
in the following simple exact form: 
\begin{eqnarray} 
\label{eq3b4}  
\displaystyle
          \bigl \langle \sigma_i \bigr \rangle =      
          \Biggl \langle \dfrac{ \displaystyle\sum_{\sigma_i =\pm 1/2} \;
          	\sigma_i \exp \bigl (-\beta {\cal H}_i \bigr)  } 
            {\displaystyle \sum_{\sigma_i =\pm 1/2}\; \exp \bigl (-\beta {\cal H}_i \bigr) }    \Biggr \rangle,  
\end{eqnarray}  
where we have at first  set $\hat {\cal O}_i = \sigma_i$  into Eq. \eqref{eq1b15}  and then we have substituted the  summation over spin variable $\sigma_i = \pm 1/2$ instead of  Tr$_i$. The angular brackets naturally denote the standard canonical ensemble averaging. 

Performing the summation over two available states of a central spin $\sigma_i $ one obtains from Eq. \eqref{eq3b4} the following expression for the magnetization $m$:
\begin{eqnarray} 
\label{eq3b5}  
        m \equiv  \bigl \langle \sigma_i \bigr \rangle =      
          \biggl  \langle \frac{1}{2} \tanh\Bigl [\frac{\beta}{2} \Bigl ( \sum_{j= 1}^q J_{ij} \sigma_j + h\Bigr) 
          \Bigr ]  \biggr \rangle. 
\end{eqnarray}  
At this stage, we are faced with a non-trivial problem how to preform  averaging over  spin variables $\sigma_j = \pm 1/2$  entering the hyperbolic tangent. Obviously, one cannot treat this problem exactly, so we are forced to adopt some approximation. The simplest but the least accurate approximation is obtained by a direct averaging the argument of hyperbolic tangent in Eq.  \eqref{eq3b5}. In this way one gets:
\begin{eqnarray} 
\label{eq3b6}  
        m  =   \frac{1}{2} \tanh\Bigl [\frac{\beta}{2} \bigl (qJm + h\bigr) \Bigr], 
\end{eqnarray}  
where we have employed the relation $m = \langle \sigma_i \rangle \; \forall i$ due to translation invariance of the lattice and we have also assumed a unique ferromagnetic exchange interaction $J_{ij} = J > 0 \; \forall i, j$  between the nearest-neighbor spins. It is worthy to note that the previous equation is identical with that one obtained within the standard mean-field theory for the spin-1/2 Ising model in a longitudinal magnetic field (c.f. Eq. \eqref{m12}). One should also notice that the same procedure can be used to obtain the mean-field results also for other more complex Ising-like spin models although the standard mean-field approximation is usually formulated in a somewhat different way (c.f. with the approach elaborated in Section \ref{mft} on the basis of the Gibbs-Bogoliubov inequality). However, the numerical accuracy and physical predictions of the mean-field theory are frequently rather crude, so it is desirable to develop more precise approximation procedures going beyond the standard mean-field theory. 

It has been already mentioned that an improved approximation can be obtained using the effective-field method due to Honmura and Kaneyoshi, which takes advantage of the differential operator technique \cite{honm78,honm79}. Thus, the effective-field theory is in its essence based on the following property of differential operator:  
\begin{eqnarray} 
\label{eq3b7}  
        \exp \bigl ( \alpha \nabla_x \bigr) f(x) =  f (x + \alpha),       
\end{eqnarray}  
where $\alpha$ is an arbitrary parameter and $\nabla_x =\partial / \partial x $ represents the differential operator. 

Using Eq. \eqref{eq3b7} we may rewrite Eq. \eqref{eq3b5} in the form:
\begin{eqnarray} 
\label{eq3b8}  
        m \equiv  \bigl \langle \sigma_i \bigr \rangle =      
          \Bigl  \langle \exp \Bigl [ \Bigl (J \sum_{j= 1}^q  \sigma_j \Bigr) \nabla_x  \Bigr ]  
          \Bigr \rangle  \frac{1}{2}\tanh \Bigl [\frac{\beta}{2}\bigl(x + h \bigr)\Bigr]\Bigr|_{x=0},   
\end{eqnarray}  
or alternatively:
\begin{eqnarray} 
\label{eq3b9}  
        m  =      
          \Bigl  \langle 
          \prod_{j= 1}^q  \exp \bigl ( J   \sigma_j  \nabla_x \bigr ) 
          \Bigr \rangle   \frac{1}{2}\tanh \Bigl [\frac{\beta}{2}\bigl(x + h \bigr)\Bigr]\Bigr|_{x=0},   
\end{eqnarray}  
where we have employed the commutation of all terms in the argument of exponential function. The next step in calculations can be achieved by applying the exact van der Waerden identity \cite{waer41}:  
\begin{eqnarray} 
\label{eq3b10}  
        \exp \bigl (\alpha  \sigma_j \bigr)  =  \cosh\Bigl( \frac{\alpha}{2}\Bigr) + 
                                                                 2\sigma_j\sinh\Bigl( \frac{\alpha}{2}\Bigr)       
\end{eqnarray}  
which is evidently fulfilled for any parameter $\alpha$ and $\sigma_j = \pm 1/2$. With the help of Eq. \eqref{eq3b10} one may rewrite Eq. \eqref{eq3b9} into the form:
\begin{eqnarray} 
\label{eq3b11}  
        m  =      
          \Bigl  \langle 
          \prod_{j= 1}^q  \Bigl [ \cosh \Bigl ( \frac{J\nabla_x}{2} \Bigr ) +  2\sigma_j \sinh \Bigl ( \frac{J\nabla_x}{2}
           \Bigr ) \Bigr]\Bigr \rangle 
         \frac{1}{2}\tanh \Bigl [\frac{\beta}{2}\bigl(x + h \bigr)\Bigr]\Bigr|_{x=0}.  
\end{eqnarray}  
It should be emphasized here that the previous equation is still exact and any kind of approximation made at this stage will be superior with respect to the mean-field theory, since the exact van der Waerden identity \eqref{eq3b10} exactly accounts for the following relations:   
\begin{eqnarray} 
\label{eq3b12}  
                     \sigma_j^{2n+1} = 4^{-n} \sigma_j, \quad \qquad \sigma_j^{2n}  = 4^{-n}. 
\end{eqnarray}  
In order to proceed further in formulation of the single-spin cluster effective-field theory, we may split on the r.h.s of Eq. \eqref{eq3b11} correlations between all nearest-neighbor spins of the central spin in the following way: 
\begin{eqnarray} 
\label{eq3b13a}
\langle \sigma_j \sigma_k \sigma_\ell \ldots \sigma_n\rangle =
\langle \sigma_j \rangle \langle \sigma_k \rangle \langle\sigma_\ell\rangle \ldots \langle\sigma_n\rangle
\end{eqnarray} 
and obtain the relatively simple relation for the magnetization:  
\begin{eqnarray} 
\label{eq3b14}  
        m  =      
           \biggl [ \cosh \Bigl ( \frac{J\nabla_x}{2} \Bigr ) +  2 m\sinh \Bigl ( \frac{J\nabla_x}{2} \Bigr ) \biggr] ^q  
          \frac{1}{2}\tanh \Bigl [\frac{\beta}{2}\bigl(x + h \bigr)\Bigr]\Bigr|_{x=0}.
\end{eqnarray}  
Applying the binomial theorem one can rewrite the previous equation into the form: 
\begin{eqnarray} 
\label{eq3b15}  
        m  = \sum_{k=0}^{q}   {q \choose  k} m^k 2^{k}
        									\Bigl [\cosh \Bigl ( \frac{J\nabla_x}{2} \Bigr )\Bigr ]^{q-k} 
                                            \Bigl [ \sinh\Bigl ( \frac{J\nabla_x}{2} \Bigr )\Bigr ] ^k  
                                            \frac{1}{2}\tanh \Bigl [\frac{\beta}{2}\bigl(x + h \bigr)\Bigr]\Bigr|_{x=0},
\end{eqnarray}
which can be finally simplified to:
\begin{eqnarray} 
\label{eq3b16}  
        m  =  \sum_{k=0}^{q}  {q \choose  k} m^k A^{(q)}_k.                                     
\end{eqnarray}
In above, we have introduced the coefficients:
\begin{eqnarray} 
\label{eq3b17}  
									A^{(q)}_k  = 2^{k}
        							\Bigl [\cosh \Bigl ( \frac{J\nabla_x}{2} \Bigr )\Bigr ]^{q-k} 
                                    \Bigl [\sinh \Bigl ( \frac{J\nabla_x}{2} \Bigr )\Bigr ]^{k} 
                                 \frac{1}{2}\tanh \Bigl [\frac{\beta}{2}\bigl(x + h \bigr)\Bigr]\Bigr|_{x=0}.
\end{eqnarray}
Any numerical calculation, of course, requires the explicit knowledge of all relevant coefficients in previous equation. Let us now explain in detail how to calculate these coefficients.
At first, we use definitions of hyperbolic functions to transform  Eq. \eqref{eq3b17} into form:
\begin{eqnarray}
\label{eq3b18}  
									A^{(q)}_k =
									2^{ k - q } \mbox{e}^{-\frac{q }{2} J\nabla_{x}} 
        						    \Bigl ( \mbox{e}^{J\nabla_{x}} +1 \Bigr )^{q-k}  
        								 \Bigl ( \mbox{e}^{J\nabla_{x}} -1 \Bigr )^{k}                                           
                               \frac{1}{2}\tanh \Bigl [\frac{\beta}{2}\bigl(x + h \bigr)\Bigr]\Bigr|_{x=0}
\end{eqnarray}
and then after the straightforward application of the binomial theorem and Eq. \eqref{eq3b7}  we obtain the final result:
\begin{eqnarray} 
\label{eq3b19}  
\nonumber
  A^{(q)}_k = A^{(q)}_k (t, \tilde{h}) =    2^{ k - q - 1}  \sum_{i=0}^{q-k} \sum_{j=0}^{k} \bigl ( -1 
 \bigr)^j   {q - k \choose  i}{ k \choose   j}\tanh \Bigl[\frac{t}{4}\Bigl( q - 2i - 2j + 2\tilde{h}  \Bigr)\Bigr],\\
\end{eqnarray} 
where we have introduced the dimensionless  parameters $t = \beta J$ and $ \tilde{h}  = h/J$. To complete this subsection one should emphasize that on the basis of Eqs. \eqref{eq3b16} and \eqref{eq3b19} we can numerically study either  magnetization processes at fixed temperatures or alternatively to investigate the temperature variations of magnetization at fixed external magnetic field for an arbitrary lattice defined solely through the coordination number $q$.

\subsubsection{Spontaneous magnetization and critical temperature}
\label{mag_Tc}
The spontaneous magnetization is defined as a magnetic moment per unit volume of a magnetic system in a zero external magnetic field. Within the single-spin cluster effective-field theory one determines the spontaneous magnetization from Eq. \eqref{eq3b16} by setting $\tilde{h}=0$:
\begin{eqnarray} 
\label{eq3b20}  
        m  =  \sum_{k=0}^{q}  {q \choose  k} m^k A^{(q)}_{k}(t),                                            
\end{eqnarray}
where temperature-dependent coefficients  $ A^{(q)}_k (t) = A^{(q)}_k (t, 0)$ are given by:
\begin{eqnarray} 
\label{eq3b21}  
  A^{(q)}_k (t)=   2^{ k - q- 1 }  \sum_{i=0}^{q-k} \sum_{j=0}^{k} \bigl ( -1 
 \bigr)^j   {q - k \choose  i}{ k \choose   j}\tanh \Bigl[\frac{t}{4}\Bigl( q - 2i - 2j   \Bigr)\Bigr].
\end{eqnarray}
One can easily verify that for an arbitrary $q$ and even values of $k$ one has 
\begin{eqnarray} 
\label{eq3b22}  
                           A^{(q)}_{k}(t )= 0.
\end{eqnarray} 
Thus, Eq. \eqref{eq3b20} contains only terms proportional to odd powers of the spontaneous magnetization $m$ simultaneously serving also as the order parameter. Consequently, this polynomial equation has always one trivial root $m = 0$  corresponding to the well-known disordered paramagnetic phase. Although the trivial solution exists at all temperatures, the paramagnetic phase is under normal circumstances  stable only above some characteristic temperature, which is for ferromagnets called as a Curie temperature. The Curie temperature represents a critical temperature, at which the disordered paramagnetic phase and the spontaneously ordered ferromagnetic phase coincide, and the system under investigation undergoes a second-order phase transition. The ferromagnetic phase exhibits a nonzero spontaneous magnetization, which takes its maximum possible value $m = 1/2$ at $T = 0$ and then gradually decreases with the increasing temperature until it completely vanishes at the critical temperature. The regions of stability for the paramagnetic and ferromagnetic phases are obviously
determined by comparing the values of Helmholtz free energy of these phases.

It follows from the general theory of phase transitions and critical phenomena that different quantities of very diverse physical systems may exhibit near the critical point a universal power-law behavior, which can be mathematically described by critical indexes. Thus, various physical systems exhibiting phase transitions can be divided into a few universality classes depending on the spatial dimension, kind of the order parameter and symmetry of the Hamiltonian. Now, let us find the Curie temperature of the ferromagnetic spin-1/2 Ising model on several lattices within single-spin cluster  effective-field theory and afterwards we will investigate the behavior of the spontaneous magnetization near a critical point.  
\begin{table}[!b]
\caption[Comparisom of the Curie temperatures for Ising model with spin 1/2 within different approximations]{A comparison of the Curie 
temperatures for the ferromagnetic spin-1/2 Ising model on several lattices obtained within different approximate and exact theories (see Ref. \cite{kane93r}).
The abbreviation used: MFA  - mean-field approximation, SEFA - single-spin cluster formulation of the effective-field theory, BPA - Bethe-Peierls approximation,  Exact/HTSE - exact results/high-temperature series expansion, lc - linear chain, hc - honeycomb, sq - square, sc - simple cubic, bcc - body centered cubic, fcc - face centered cubic.}
\bigskip
\centering 
\begin{tabular}{ccccc} 
\hline\hline 
     & MFA& SEFA & BPA & Exact/HTSE\\ 
$q$ &$k_{\rm B} T_{\rm c}/J $ & $k_{\rm B} T_{\rm c}/J $ & $k_{\rm B} T_{\rm c}/J $ & $k_{\rm B} T_{\rm c}/J $\\ 
\hline 
2   & 0.500         & 0.000           & 0.000      &  0.000 \, (lc) \\ 
3   & 0.750         & 0.526           & 0.455      &  0.380 \, (hc)\\ 
4   & 1.000         & 0.773           & 0.721      &  0.567 \, (sq)\\
6   & 1.500         & 1.268           & 1.233      &  1.128 \, (sc)\\
8   & 2.000         & 1.765           & 1.738      &  1.588 \, (bcc)\\
12  & 3.000         & 2.761           & 2.743      &  2.449 \, (fcc) \\
\hline 
\end{tabular}
\label{JAEP}
\end{table}
For this purpose, one may take advantage of the aforementioned fact that the spontaneous magnetization (i.e. the order parameter) gradually decreases with temperature until it continuously tends to zero at the Curie temperature. Hence, one can neglect in Eq. \eqref{eq3b20} all terms including higher-order powers of the spontaneous magnetization and retain only terms linear in the spontaneous magnetization $m$, because the spontaneous magnetization is sufficiently small $m\ll 1$ close to a phase transition point. Then, one may evaluate the Curie temperature from the following simple equation using the fact that $m \ne 0$ for $T \nearrow T_c $:
\begin{eqnarray} 
\label{eq3b23}  
        1  =  q  A^{(q)}_{1}(t_{\rm c}  ),                                          
\end{eqnarray}
with $ t_{\rm c} = J/(k_{\rm B} T_{\rm c})$.

The numerical solution of Eq. \eqref{eq3b23} gives Curie temperatures of the ferromagnetic spin-1/2 Ising model on various lattices within the single-spin cluster effective-field theory, which are listed in Tab. \ref{JAEP} along with the critical temperatures obtained for the same model within the framework of other theories. The results obtained within the single-spin cluster effective-field theory  clearly improve those of the standard mean-field theory, but at the same time, they do not overcome the accuracy of the Bethe-Peierls approximation. In the case of 1D Ising model one obtains within the single-spin cluster approach the zero critical temperature in accordance with the exact result. It should be also pointed out that the single-spin cluster effective-field theory does not discern between the spin-1/2 Ising lattices with the same coordination number but different topological structure, for example, the obtained Curie temperature for $q=6$ is equally valid for the ferromagnetic spin-1/2 Ising model on 2D triangular lattice as well as 3D simple cubic lattice. This is of course an unpleasant feature of this approximation, however, it is possible to improve the formulation of effective-field theory in order to eliminate this failure as it will be discussed later. 

Similarly as in the case of mean-field approximation,  we can now investigate the behavior of spontaneous magnetization in the vicinity of the Curie temperature also within the single-spin cluster effective-field theory. To this end, we retain in Eq. \eqref{eq3b20} only terms up to $m^3$  and obtain the following expression:  
\begin{eqnarray} 
\label{eq3b23a}  
          m = \pm \sqrt{6 \dfrac{ 1- q  A^{(q)}_{1}(t)}{ q ( q -1)( q -2)  A^{(q)}_{3}(t)}},                                  
\end{eqnarray}
where 
\begin{eqnarray}
\label{eq3b23c}  
  A^{(q)}_1 (t) =    \dfrac{1}{2^q} \sum_{i=0}^{q-1}  \sum_{j = 0}^{1} (-1)^j {q - 1 \choose  i}
{1 \choose  j} \tanh \Bigl[\frac{t}{4}\Bigl(q -2i -2j  \Bigr)\Bigr].
\end{eqnarray}
\begin{eqnarray}
\label{eq3b23d}  
  A^{(q)}_3 (t) =    \dfrac{1}{2^{2-q}} \sum_{i=0}^{q-1}  \sum_{j = 0}^{3} (-1)^j {q - 1 \choose  i}
{3 \choose  j}  \tanh \Bigl[\frac{t}{4}\Bigl(q -2i -2j  \Bigr)\Bigr].
\end{eqnarray}

\begin{figure}[tb]
\begin{center}
\includegraphics[clip,width=8cm]{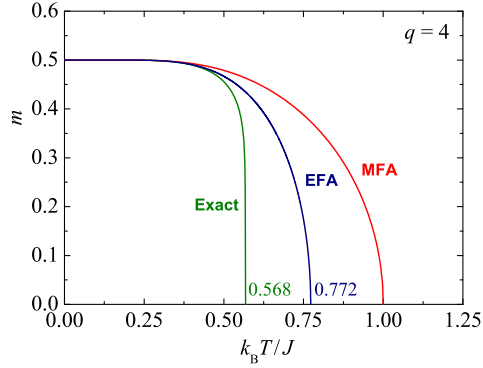}
\vspace{-0.25cm}
\caption{Temperature dependences of the spontaneous magnetization for the spin-1/2 Ising model on a square lattice ($q=4$) within different approximations. The red curve represents the mean-field result, the blue one the single-spin cluster formulation of the effective-field theory and the green one is the exact result according to Yang \cite{yang52}. The values of the Curie temperatures are also indicated.}
\label{mt_q4}
\end{center}
\end{figure}

Expanding the r.h.s of \eqref{eq3b23a} with respect to $ \delta = T_c - T $ one obtains the familiar formula:
\begin{eqnarray} 
\label{eq3b23b}  
          m \approx \mbox{const}\cdot (T_c - T)^{\frac{1}{2}},                                 
\end{eqnarray}
which proves that the critical index of the spontaneous magnetization equals $\beta_m = 1/2$ independently of spatial dimensionality and geometry of a given lattice. This result is identical with the standard mean-field prediction and it is obviously in contradiction with the results obtained from exact or more accurate approximate theories. It could be therefore concluded that the single-spin cluster effective-field theory has the same accuracy as traditional mean-field theory in the critical region and hence, it fails to reproduce correct critical indexes. On the other hand, this theory gives numerically better results than the mean-field theory outside of the critical region as it is quantitatively illustrated in Fig. \ref{mt_q4}, where we have depicted the thermal variations of the spontaneous magnetization obtained within the single-spin cluster effective-field theory, mean-field theory and exact calculations for the particular case of the spin-1/2 Ising model on a square lattice  with the coordination number $q = 4$. 

\subsubsection{Initial magnetic susceptibility}
In magnetic systems, the initial magnetic susceptibility represents an important physical quantity characterizing a response of the spin system with respect to variations of the external magnetic field and it is defined as:
\begin{eqnarray} 
\label{eq3b24a}  
        \chi  =  \lim_{H_{ext}\to 0} \biggl ( \dfrac{\partial M}{\partial H_{ext}} \biggr )_{T},                                             
\end{eqnarray}
where $M  = N g\mu_{\rm B} m $ is the total magnetization of the system, $H_{ext} = h/ (g \mu_{\rm B}) $ denotes the intensity of external magnetic field, $g$ is the Land\'{e} factor and  $\mu_{\rm B}$ represents the Bohr magneton. For our purpose it is more useful to express the previous equation in the form:  
\begin{eqnarray} 
\label{eq3b24}  
      \chi  = \frac{N}{J} g^2 \mu_{\rm B}^2 \lim_{\tilde{h} \to 0} \biggl ( \dfrac{\partial m}{\partial \tilde{h} } \biggr )_{t},                                             
\end{eqnarray}

In order to determine the initial susceptibility of the spin-1/2 Ising model on a lattice with the coordination number $q$ within the single-spin cluster effective-field theory one has at first to  evaluate the expression $ ( \partial m/\partial \tilde{h } )_{t}$ from the equation of state \eqref{eq3b16} and then to take the limit of vanishing external field. For this purpose it is sufficient to perform the Taylor expansion of coefficients on the r.h.s of Eq.~\eqref{eq3b16} and to retain only expressions  up to linear term in $\tilde{h}$ since all other terms will vanish for $\tilde{h } \to 0$. Following this procedure one obtains the expression: 
\begin{eqnarray} 
\label{eq3b26}  
        m  =  \sum_{k=0}^{q}  {q \choose  k} m^k
        									\Bigl[A^{(q)}_k (t) + \frac{t}{2} \tilde{h}  B^{(q)}_k (t) \Bigr ],                                        
\end{eqnarray}
where  $ A^{(q)}_k (t) = A^{(q)}_k (t, 0) $ is given by \eqref{eq3b21} and the coefficients $B^{(q)}_k (t)  $ take the following explicit form:
\begin{eqnarray}
\label{eq3b27}   
  B^{(q)}_k (t) =    2^{ k - q  -1} \sum_{i=0}^{q-k} \sum_{j=0}^{k} \bigl ( -1 \bigr)^j   {q - k \choose  i}{ k \choose   j}
  \mbox{sech}^2\Bigl[\frac{t}{4}\Bigl( q - 2i - 2j   \Bigr)\Bigr].
\end{eqnarray}
It is useful to note for further calculation that the coefficients  $B^{(q)}_k (t) = 0$ for all odd values of $k$. Now, one may differentiate Eq. \eqref{eq3b26} with respect to $\tilde{h}$ in order to get:
\begin{eqnarray} 
\label{eq3b28}  
  \biggl ( \dfrac{\partial m}{\partial \tilde{h}} \biggr )_{t} =
  \dfrac{\frac{t}{2} \displaystyle \sum_{k=0}^{q}  {q \choose  k} m^k  B^{(q)}_k (t)  }{1 - \displaystyle 
   \sum_{k=0}^{q}  {q \choose  k} k m^{k -1}  A^{(q)}_k (t) }.    									
\end{eqnarray}
After inserting Eq. \eqref{eq3b28} into Eq. \eqref{eq3b24} one obtains the following relation for the initial magnetic susceptibility:
\begin{eqnarray} 
\label{eq3b29}  
        \chi  = \dfrac{N g^2 \mu_{\rm B}^2}{2J} 
        \dfrac{ t\displaystyle \sum_{k=0}^{q} {q \choose  k} m^k  B^{(q)}_k (t)  }{1 - \displaystyle \sum_{k=0}^{q}  {q\choose  k} k m^{k -1}  	A^{(q)}_k (t) }.
\end{eqnarray}
In the paramagnetic region ($T>T_{\rm c}$) one obviously has zero spontaneous magnetization ($m = 0$), so the  previous equation can be simplified as follows: 
\begin{eqnarray} 
\label{eq3b30}  
        \chi  =  \dfrac{N g^2 \mu_{\rm B}^2}{2J} 
        \dfrac{  t B^{(q)}_0 (t)  }{1 - q A^{(q)}_1 (t) }, \quad (T > T_{\rm c})
\end{eqnarray}
where $A^{(q)}_1 (t)$ is given by Eq. \eqref{eq3b23c} and the coefficients $B^{(q)}_0(t)$ follow from:
\begin{eqnarray}
\label{eq3b32}   
  B^{(q)}_0(t) =    \dfrac{1}{2^{q+1}} \sum_{i=0}^{q}   {q \choose  i}\mbox{sech}^2\Bigl[\frac{t}{4}\Bigl( q - 2i  \Bigr)\Bigr].
\end{eqnarray}
The result (\ref{eq3b30}) offers a unique opportunity to test our calculations against the experimental observations, since the initial paramagnetic susceptibility obeys at high enough temperatures 
the Curie-Weiss law for many real ferromagnetic materials.

In order to investigate the high-temperature behavior of the initial paramagnetic susceptibility, we take into account the fact that for  $ T \gg T_{\rm c}$ we have  $ t \ll 1  $,  so that  we can expand the hyperbolic function in  Eqs. \eqref{eq3b23c} and \eqref{eq3b32} and retain only  absolute terms and those  linear in $t$. Completing this mathematical procedure one obtains the relations:
\begin{eqnarray}
\label{eq3b33}   
   A^{(q)}_1(t) \approx \frac{t}{4}, \qquad
  B^{(q)}_0(t) \approx \frac{1}{2},
\end{eqnarray}
which enable to rewrite Eq. \eqref{eq3b30} as:
\begin{eqnarray} 
\label{eq3b34}  
        \chi  = \dfrac{ N g^2 \mu_{\rm B}^2}{4J}  
        \dfrac{   t }{1 - \frac{qt}{4} }, 
\end{eqnarray}
or alternatively 
\begin{eqnarray} 
\label{eq3b35}  
        \chi  =  \dfrac{  C }{T - \Theta }.
\end{eqnarray}
Above, $C = N g^2 \mu_{\rm B}^2 /(4k_{\rm B})$ denotes the Curie constant and $\Theta = q J / (4k_{\rm B})$ represents the Weiss constant. The previous equation represents nothing but the famous Curie-Weiss law for ferromagnets and hence, one may conclude that our theory agrees well with experimental findings in high-temperature region.

\subsection{Internal energy, enthalpy and  specific heat}
In general, the internal energy is  in statistical mechanics defined as the expectation value of the Hamiltonian with vanishing external fields. Thus, the internal energy of the Ising model with the unique nearest-neighbor exchange interaction $J$ is given by: 
\begin{eqnarray} 
\label{eq3b36}  
        U  =   \bigl \langle{\cal H}_0 \bigr \rangle  = -  \Bigl \langle J \sum_{ \langle ij \rangle}  \sigma_i \sigma_j \Bigr \rangle .  
\end{eqnarray} 
where the subscript 0 indicates that no external field is included in the Hamiltonian. By taking into account the translation invariance of the lattice one obtains:
\begin{eqnarray} 
\label{eq3b37}  
        U  =   \dfrac{N}{2} \bigl \langle {\cal H}_{i0} \bigr \rangle = - \dfrac{N}{2} \bigl \langle \sigma_i E_i \bigr \rangle, 
\end{eqnarray} 
where  $N$ is the number of lattice sites and  $E_i$ is given by:     
\begin{eqnarray} 
\label{eq3b38}  
        E_i=  J\sum_{j= 1}^q \sigma_j. 
\end{eqnarray} 
It is quite evident that one has to determine the expectation value of $\bigl \langle \sigma_i E_i \bigr \rangle$ on the r.h.s of Eq. \eqref{eq3b7} in order to get the closed-form expression for the internal energy. To this end, one may exploit Eq. \eqref{eq1b16} by setting $\hat {\cal O} = \sigma_i$, $\{f_i\} = E_i$ and $\mbox{Tr}_i = \sum_{\sigma_i = \pm 1/2}$. Then,  one may rewrite Eq. \eqref{eq1b16} using the fact that ${\cal H}_{i0} = - \sigma_i E_i $ into the following form:
\begin{eqnarray} 
\label{eq1b39}  
          \bigl \langle \sigma_i   E_i \bigr \rangle =      
          \biggl \langle    E_i
          \dfrac{  \displaystyle \sum_{\sigma_i =\pm 1/2}  \;\sigma_i \exp \bigl (\beta \sigma_i E_i\bigr)  } 
                   {\displaystyle \sum_{\sigma_i =\pm 1/2}\; \exp \bigl (\beta \sigma_i E_i \bigr) }    \biggr \rangle.
\end{eqnarray}
Performing the summation over $\sigma_i$ and applying Eq. \eqref{eq3b7} we obtain the following relation:
\begin{eqnarray} 
\label{eq3b40}  
       \bigl \langle \sigma_i    E_i  \bigr \rangle   =      
          \Bigl  \langle 
         E_i  \exp \bigl ( E_i \nabla_x     \bigr ) 
          \Bigr \rangle 
          \frac{1}{2}\tanh \Bigl (\frac{\beta}{2} x \Bigr) \biggr|_{x=0},
\end{eqnarray}  
which can be rewritten in the following more elegant form:
\begin{eqnarray} 
\label{eq3b40a}  
       \bigl \langle \sigma_i    E_i  \bigr \rangle   =      
          \biggl   [   \dfrac{\partial}{\partial y}  \Bigl  \langle  \exp \bigl ( E_i y     \bigr ) 
          \Bigr \rangle \biggr ]_{y =\nabla_x }
           \frac{1}{2}\tanh \Bigl (\frac{\beta}{2} x \Bigr) \biggr|_{x=0}.
\end{eqnarray}  
After substituting Eq. \eqref{eq3b40a} to Eq. \eqref{eq3b37} we obtain the formula for the internal energy of the spin-1/2 Ising model:
\begin{eqnarray} 
\label{eq3b41}  
       U = -\dfrac{N}{4}  \biggl \{ \dfrac{\partial}{\partial y}  \Bigl  \langle  
       \prod_{j =1}^q   \Bigl [ \cosh \Bigl (\frac{Jy}{2}  \Bigr ) +  2\sigma_j \sinh \Bigl (\frac{Jy}{2}  \Bigr ) \Bigr ]\Bigr    		  		\rangle  \biggr \}_{y 
       =\nabla_x }\tanh \Bigl (\frac{\beta}{2} x \Bigr) \biggr|_{x=0}.
\end{eqnarray}  
To be consistent with previous calculations we now apply the same decoupling procedure as in Eq. \eqref{eq3b13a} and rewrite previous equation into the form:
\begin{eqnarray} 
\label{eq3b42}  
        U =     -\dfrac{N}{4}  \biggl \{   \dfrac{\partial}{\partial y}  
                           \Bigl [ \cosh \Bigl (\frac{Jy}{2}  \Bigr ) +  2m \sinh \Bigl (\frac{Jy}{2}  \Bigr )  \Bigr] ^q                           
           \biggr \}_{y =\nabla_x }
           \tanh \Bigl (\frac{\beta}{2} x \Bigr) \biggr|_{x=0},
\end{eqnarray}
or alternatively:
\begin{eqnarray} 
\label{eq3b43}  
\nonumber
        - \dfrac{2U}{qNJ}  &=&     
                           \Bigl [ \sinh \Bigl (\frac{J\nabla_x }{2}  \Bigr ) +  2m \cosh \Bigl (\frac{J\nabla_x }{2}  \Bigr )  \Bigr]
                           \Bigl [ \sinh \Bigl (\frac{J\nabla_x }{2}  \Bigr ) +  2m \cosh \Bigl (\frac{J\nabla_x }{2}  \Bigr )  \Bigr] ^{q-1}  \\                         
                       &\times &\frac 12 \tanh \Bigl (\frac{\beta}{2} x \Bigr) \biggr|_{x=0}.
\end{eqnarray}
In order to obtain an expression suitable for numerical calculations we again utilize the binomial theorem to recast Eq.  \eqref{eq3b43} as follows:
\begin{eqnarray} 
\label{eq3b44}  
\nonumber
        - \dfrac{4U}{qNJ}& = &\sum_{k=0}^{q -1}  {q-1 \choose  k} \Bigl(2m \Bigr)^{k+1}
        									\Bigl [\cosh \Bigl (\frac{J\nabla_x }{2}  \Bigr ) \Bigr ]^{q -k-1} 
                                            \Bigl [ \sinh \Bigl (\frac{J\nabla_x }{2}  \Bigr ) \Bigr ] ^{k+1}  
                                            \tanh \Bigl (\frac{\beta}{2} x \Bigr) \biggr|_{x=0} \\ \nonumber 
                                  &+& \sum_{k=0}^{q-1}  {q- 1 \choose  k} \Bigl(2m \Bigr)^{k}
        									\Bigl [\cosh \Bigl (\frac{J\nabla_x }{2}  \Bigr ) \Bigr ]^{q -k } 
                                            \Bigl [ \sinh \Bigl (\frac{J\nabla_x }{2}  \Bigr ) \Bigr ] ^{k}  
                                            \tanh \Bigl (\frac{\beta}{2} x \Bigr) \biggr|_{x=0}.\\ \nonumber \\
\end{eqnarray}
The previous equation  can be naturally expressed in the following abbreviated and simpler form:
\begin{eqnarray} 
\label{eq3b45}  
        - \dfrac{2U}{qNJ}& = &\sum_{k=0}^{q  -1}  {q - 1 \choose  k} \Bigl [ m^k  C^{(q)}_k \bigl (t\bigr)  
        						+ m^{k +1} D^{(q)}_k \bigl (t\bigr) \Bigr ], 
\end{eqnarray}
where we have denoted $t = \beta J$ and we have also introduced the thermal coefficients $C^{(q)}_k (t) $ a $D^{(q)}_k (t) $ as follows:
\begin{eqnarray}
\label{eq3b46}  
C^{(q)}_k (t)  = 2^{k-1}\Bigl [\cosh \Bigl (\frac{J\nabla_x }{2}  \Bigr ) \Bigr ]^{q -k-1} 
                                            \Bigl [ \sinh \Bigl (\frac{J\nabla_x }{2}  \Bigr ) \Bigr ] ^{k+1}  
                                            \tanh \Bigl (\frac{\beta}{2} x \Bigr) \biggr|_{x=0},
\end{eqnarray}
\begin{eqnarray}
\label{eq3b47}  
D^{(q)}_k (t) = 2^{k}\Bigl [\cosh \Bigl (\frac{J\nabla_x }{2}  \Bigr ) \Bigr ]^{q -k } 
                                            \Bigl [ \sinh \Bigl (\frac{J\nabla_x }{2}  \Bigr ) \Bigr ] ^{k}  
                                            \tanh \Bigl (\frac{\beta}{2} x \Bigr) \biggr|_{x=0}.
\end{eqnarray}
After a straightforward application of the binomial theorem and Eq. \eqref{eq3b7} one easily gets these coefficients in the following final form:
\begin{eqnarray} 
\label{eq3b48}  
  C^{(q)}_k (t) =    2^{k-q -1} \sum_{i=0}^{q-k - 1} \sum_{j=0}^{k +1} \bigl ( -1 \bigr)^j   {q - k  - 1\choose  i}{ k + 1\choose   j}\tanh \Bigl[\frac{t}{4}\Bigl( q - 2i - 2j   \Bigr)\Bigr]
\end{eqnarray}
\begin{eqnarray} 
\label{eq3b49}  
  D^{(q)}_k (t) =     2^{k-q}\sum_{i=0}^{q-k} \sum_{j=0}^{k} \bigl ( -1 \bigr)^j   {q - k \choose  i}{ k \choose   j}\tanh \Bigl[\frac{t}{4}\Bigl( q - 2i - 2j  \Bigr)\Bigr].
\end{eqnarray}
For practical applications it is useful to take into account the following relations:
\begin{eqnarray} 
\label{eq3b50}  
                             && C^{(q)}_k (t) =   0, \qquad \mbox{for} \;\; k  \; \mbox{even} \\
\label{eq3b51}                               
                             && D^{(q)}_k (t) =   0, \qquad \mbox{for} \;\; k  \; \mbox{odd}. 
\end{eqnarray}
Altogether, we have obtained the polynomial equation \eqref{eq3b45} with temperature dependent coefficients for the internal energy of the spin-1/2 Ising model on the arbitrary crystalline lattice,  which can be easily numerically solved. In this equation the lattice structure is specified solely by the coordination number $q$, whereas the spontaneous magnetization entering the r.h.s of this equation must be calculated from Eq. \eqref{eq3b20}.

Having obtained the explicit form of the internal energy one can simply calculate other physical quantities using basic thermodynamic relations. At first, the enthalpy $H$ of the spin-1/2 Ising model can be easily obtained from the equation: 
\begin{eqnarray} 
\label{eq3b52}  
                             H = U - Nh m, 
\end{eqnarray}
where the internal energy is given by \eqref{eq3b45}, magnetization by \eqref{eq3b16} and $h$ represents an external magnetic field. Note that the internal energy and enthalpy represent important  thermodynamic potentials, which enable to understand behavior of the system  under different physical conditions. Within the effective-field theories, the significance of these two quantities is reinforced because they also serve as a starting point for a calculation of other thermodynamic quantities such as magnetic specific heat, Helmholtz and/or Gibbs free energy. In fact, the specific heat at constant magnetization or constant external magnetic field are respectively calculated from the relations:
\begin{eqnarray} 
\label{eq3b53}  
                             C_m = \biggl ( \dfrac{\partial U}{\partial T}\biggr )_m
\end{eqnarray}
and
\begin{eqnarray} 
\label{eq3b54}  
                             C_h = \biggl ( \dfrac{\partial H}{\partial T}\biggr )_h.
\end{eqnarray}
Of course, one can in principle obtain the relevant expressions by the differentiation of \eqref{eq3b45} and \eqref{eq3b52}, however, it is easy to see that the final analytical results will be too intricate. For that reason it is in practice  very comfortable to calculate both specific heats using a numerical differentiation of relevant thermal variations of the internal energy or enthalpy. 
 
\subsection{Helmholtz and Gibbs free energy. Entropy}
The Helmholtz and Gibbs free energies play an important role in theoretical investigation of magnetic systems, because they enable to analyze the thermodynamic stability of all possible phases appearing in system under various external physical conditions. Both these quantities are in thermodynamics defined as a Lagrange transform of the internal energy or enthalpy of the system and  these procedures require explicit knowledge of the entropy.  It is well known fact that the phenomenological thermodynamics does not provide a mathematical tool for explicit calculation of the entropy, so that one cannot use the Lagrange transform to determine relevant free energies. 

Let us utilize the framework of statistical mechanics when defining some magnetic system through the Hamiltonian, so that the Helmholtz and Gibbs free energy are formally given by the same 
equations:
\begin{eqnarray} 
\label{eq3b55}  
                             F=  - \dfrac{1}{ \beta} \ln  Z(\beta)
\end{eqnarray}  
and
\begin{eqnarray} 
\label{eq3b56}  
                             G=  - \dfrac{1}{ \beta} \ln  Z(\beta, h)
\end{eqnarray}  
where $\beta = 1/k_{\rm B} T$, $h$ denotes the external magnetic field and the relevant partition functions are respectively given by:
\begin{eqnarray} 
\label{eq3b57}  
                             Z(\beta) = \mbox{Tr e}^{-\beta \hat{\cal H}_0} 
\end{eqnarray}
and
\begin{eqnarray} 
\label{eq3b58}  
                             Z(\beta, h)=  \mbox{Tr e}^{-\beta \hat{\cal H}}. 
\end{eqnarray}
In order to avoid confusion one should notice that in Eqs.  \eqref{eq3b56} and \eqref{eq3b58} the Hamiltonian $\hat {\cal H}$ of a given magnetic system includes also the Zeeman's term associated with the external magnetic field, while the Hamiltonian $\hat {\cal H}_0$ in Eqs.  \eqref{eq3b55} and \eqref{eq3b57} must not depend on the external magnetic field as formally denoted by the subscript zero. Moreover, it should be emphasized here that the direct application of the previous equations is not straightforward, since it requires the calculation of partition function that is usually very hard problem even in some approximate theories.

As far as the effective-field theories based on the differential operator technique are concerned, it is necessary to note that the calculation of free energies represents a crucial problem of the theory, which has not been solved yet. The basic problem lies in fact that one always uses as a starting point within these theories the Callen-Suzuki identities, which do not enable to obtain the closed-form expression  for the partition function. In order to avoid this serious limitation of effective-field theory, we now derive general equations for the Helmholtz and Gibbs free energy allowing to calculate these quantities numerically for a wide class of the Ising-like spin systems. It will be shown below that the proposed numerical method for a calculation of the free energy assumes the knowledge of the internal energy. This is much easier task in comparison with the evaluation of the partition function,  because it is enough to evaluate the nearest-neighbor correlation function 
and then to use the following general relation: 
\begin{eqnarray}   
\label{eq3b59}  
                             U= \langle {\cal H}_0 \rangle_0  = - 
                             \dfrac{N z J}{2} \langle \sigma_i \sigma_{j} \rangle_0,
\end{eqnarray} 
where the subscript 0 means that relevant expectations are calculated with the Hamiltonian ${\cal H}_0$. Of course, one can readily use Eq. \eqref{eq3b44} in the case of a spin-1/2 Ising model.

The starting point for a calculation of Helmholtz free energy is the well-known equation for the internal energy: 
\begin{eqnarray}   
\label{eq3b60}  
                             U(\beta)  = - \dfrac{\partial }{\partial \beta} \ln  Z(\beta),
\end{eqnarray} 
where the partition function is given by Eq. \eqref{eq3b57}. Now, integrating the previous equation over the (inverse) temperature range from 0 to $\beta$ one obtains the relation:
\begin{eqnarray} 
\label{eq3b61}  
                             \int_0^\beta U (\beta') d\beta' =  - \Bigl [ \ln Z(\beta) - \ln Z(0) \Bigr].  
\end{eqnarray}  
The partition function of the spin-$S$ Ising model on a lattice with the total number of spins $N$ can be simply evaluated at infinite (zero inverse) temperature:
\begin{eqnarray} 
\label{eq3b62}  
                            Z(0)  = Tr {\cal I}  = (2S +1)^N
\end{eqnarray}  
since ${\cal I} $ represents in general a $(2S +1)^N \times (2S +1)^N$ identity matrix. Finally, one obtains for the Helmholtz free energy the following formula from Eqs. \eqref{eq3b55} and \eqref{eq3b62}: 
\begin{eqnarray} 
\label{eq3b63}  
                            F = - \frac{N}{\beta}  \ln (2S +1) +  \frac{1}{\beta} \int_0^\beta U (\beta') d\beta'. 
\end{eqnarray}  
Moreover, it is quite clear that one gets very similar expression for the Gibbs free energy of the spin-$S$ Ising model when the same procedure is repeated under a mere replacement of the internal energy through the enthalpy:  
\begin{eqnarray} 
\label{eq3b64}  
                            G = - \frac{N}{\beta}  \ln (2S +1) +  \frac{1}{\beta} \int_0^\beta H (\beta', h) d\beta'. 
\end{eqnarray}  
For the sake of completeness, it is worth recalling that the enthalpy $H$ can be simply obtained from Eq. \eqref{eq3b52}. It should be stressed that Eqs. \eqref{eq3b63} and \eqref{eq3b64} are very general and they are applicable within any statistical-mechanical theory, which enables to calculate the internal energy and/or enthalpy of the system under investigation. Of course, the numerical accuracy of Eqs. \eqref{eq3b63} and \eqref{eq3b64} is basically determined by the applied analytical approach, so that one may obtain only approximate results in some particular cases, whereas in some others one may even obtain the free energy exactly. 

Finally, let us note that the entropy of the system can be very simply obtained from the following formulas: 
\begin{eqnarray} 
\label{eq3b64entrf}  
                            S = - \Bigl (\frac{\partial F}{\partial T}  \Bigr)_m
\end{eqnarray}  
and
\begin{eqnarray} 
\label{eq3b64entrg}  
                            S = - \Bigl (\frac{\partial G}{\partial T}  \Bigr)_h. 
\end{eqnarray} 
Now, performing the differentiation of \eqref{eq3b63} and \eqref{eq3b64} with respect to temperature one obtains the final expressions for entropy:
\begin{eqnarray} 
\label{eq3b64ent1}  
                            S =  Nk_{\rm B}   \ln (2S +1) - k_{\rm B}\int_0^\beta U(\beta', h) d\beta' + k_{\rm B} \beta U(\beta)
\end{eqnarray}  
\begin{eqnarray} 
\label{eq3b64ent2}  
                            S = Nk_{\rm B}  \ln (2S +1) - k_{\rm B}\int_0^\beta H (\beta', h) d\beta' + k_{\rm B} \beta H(\beta).
\end{eqnarray} 
Here, one should emphasize that our Eq. \eqref{eq3b64ent1} is identical with the formula for the entropy obtained by \v{Z}ukovi\v{c} using a different approach \cite{zuki13}.

Let us illustrate the applicability of Eq. \eqref{eq3b64} by calculating the Gibbs free energy  of the spin-1/2 Ising model within the mean-field approach. The enthalpy of the spin-1/2 Ising model is within the mean-field approximation given by: 
\begin{eqnarray} 
\label{eq3b64a}  
      H(\beta, h) = - \frac{NqJ}{2} m^2 - Nmh,
\end{eqnarray}  
whereas the magnetization should be calculated from the self-consistency condition: 
\begin{eqnarray} 
\label{eq3b64b}  
      m(\beta, h) = \frac 12 \tanh\Bigl [ \frac \beta2 \bigl (q J m + h \bigr)\Bigr].
\end{eqnarray}  
Substituting  Eq. \eqref{eq3b64a} into Eq. \eqref{eq3b64} one obtains the expression:
\begin{eqnarray} 
\label{eq3b65}  
                            G = - \frac{N}{\beta}  \ln 2 -  \frac{NqJ}{2\beta} \int_0^\beta m ^2(\beta',h) d\beta' 
                                                                       - Nh \int_0^\beta m (\beta',h) d\beta'.
\end{eqnarray}
It is advisable to change an integration variable before evaluating the integrals on the r.h.s of the previous equation:
\begin{eqnarray} 
\label{eq3b66}  
      G = - \frac{N}{\beta}  \ln 2 -  \frac{NqJ}{2\beta} \int_0^{m} m'^{2} \frac{d\beta'}{dm'}  dm'
                                                                       - \frac{Nh}{\beta} \int_0^{m} m' \frac{d\beta'}{dm'}  dm',
\end{eqnarray}
and subsequently to utilize the integration by parts in combination with the expression:  
\begin{eqnarray} 
\label{eq3b67}  
       \beta  = \frac{1}{(qJm + h)}\ln \Bigl (\frac{1+2m}{1-2m} \Bigr )
\end{eqnarray}
in order to obtain the Gibbs free energy in the following form: 
\begin{eqnarray} 
\label{eq3b68}  
      G (\beta, h) =   \frac{N}{2 \beta} \ln \Bigl(\frac{ 1- 4 m^2}{4} \Bigr)  
                            +\frac{NqJ}{2 } m^2.
\end{eqnarray}  
Finally, substituting the r.h.s of Eq. \eqref{eq3b64b} into argument of logarithmic function one may recast the previous equation into the identical form as Eq. \eqref{g12}:
\begin{eqnarray} 
G = - N k_{\rm B} T \ln 2 - N k_{\rm B} T \ln \cosh \Bigl[\frac{\beta}{2} \left(q J m + h \right) \Bigr] + \frac{Nq}{2} J m^2.
\label{eq3b68a}  
\end{eqnarray} 
This calculation clearly illustrates that the proposed formulas for the Helmholtz and Gibbs free energies can be very powerful in combination with various analytical (even approximate) theories.
These formulas can be readily combined with all exact or approximate theories, which provide the explicit knowledge of the internal energy and/or the enthalpy. In the following, this property will be explicitly illustrated by an exact calculation of the Helmholtz free energy for the spin-1/2 Ising chain. Another pleasant feature of the approach based on Eqs. \eqref{eq3b63} and  \eqref{eq3b64} is that it is also compatible with the Monte Carlo simulations and others computational techniques. Last but not least, it should be pointed out that the expressions for the Helmholtz and Gibbs free energy are not limited to magnetic systems, so the procedure can also be simply applied to more general statistical-mechanical systems. It is of course clear that the approach for obtaining the Helmholtz and Gibbs free energies presented in above is of principal importance especially for those cases, for which the free energies cannot be obtained by means of other methods.

\subsection{Exact results for a spin-1/2 Ising chain}
It has been demonstrated in previous parts that one can develop a simple approximate theory using the differential operator method, which preserves a single-spin kinematic relations exactly and neglects all correlations among nearest-neighbors spins of the central spin of the selected spin cluster. This approach represents a kind of effective-field theory with a better  numerical  accuracy than that one of the standard mean-field theory. Surprisingly, the critical temperature for the spin-1/2 Ising chain with the coordination number $q=2$ obtained from the single-spin cluster effective-field theory coincides with the exact result ($T_{\rm c} =0$). It is worthwhile to recall that the mean-field theory predicts for the spin-1/2 Ising chain the non-zero Curie temperature ($k_{\rm B} T_{\rm c}/J = 2$) and thus, it fails to reproduce exact Curie temperature in one spatial dimension. It is of course very interesting to verify whether the single-spin cluster effective-field theory may reproduce exact results  also for other physical quantities of the spin-1/2 Ising chain. 

The starting point for  our calculation will be the following equation: 
\begin{eqnarray}
\label{eq3b69}  
      \bigl \langle \sigma_i  \bigl\{ f_i \bigr\} \bigr \rangle  =      
          \Bigl  \langle 
           \bigl\{ f_i\bigr\} \Bigl  \langle 
          \prod_{j= \pm 1}  \Bigl [ \cosh \Bigl ( \frac{J\nabla_x}{2} \Bigr ) +  2\sigma_{i+j} \sinh \Bigl ( \frac{J\nabla_x}{2}
           \Bigr ) \Bigr]\Bigr \rangle 
          \frac{1}{2}\tanh \Bigl[\frac{\beta}{2}\bigl(x + h \bigr)\Bigr] \biggr|_{x=0} 
\end{eqnarray}  
which can be easily obtained from the generalized Callen-Suzuki identity \eqref{eq1b16} by repeating the same procedure as described in the subsection \ref{0S_EFT}. In order to obtain exact results for physical quantities of the spin-1/2 Ising chain one may expand the previous equation with respect to the magnetic field $h$ up to linear term:   
\begin{eqnarray}
\label{eq3b69a}  
\nonumber
 \!\!\!\!    \!\!\!\!   \bigl \langle \sigma_i  \bigl\{ f_i \bigr\} \bigr \rangle \!\! &=& \!\!     
       \Bigl  \langle 
       \bigl\{ f_i\bigr\} \Bigl  \langle 
       \prod_{j= \pm 1}  \Bigl [ \cosh \Bigl ( \frac{J\nabla_x}{2} \Bigr ) +  2\sigma_{i+j} \sinh \Bigl ( \frac{J\nabla_x}{2}
        \Bigr ) \Bigr]\Bigr \rangle 
          \frac{1}{2}\tanh \Bigl(\frac{\beta x}{2} \Bigr) \biggr|_{x=0}\\ \nonumber
 \!\!&+& \!\!   h  \Bigl  \langle 
       \bigl\{ f_i\bigr\} \Bigl  \langle 
       \prod_{j= \pm 1}  \Bigl [ \cosh \Bigl ( \frac{J\nabla_x}{2} \Bigr ) +  2\sigma_{i+j} \sinh \Bigl ( \frac{J\nabla_x}{2}
        \Bigr ) \Bigr]\Bigr \rangle 
         \frac \beta 4\mbox{sech}^2  \Bigl(\frac{\beta x}{2} \Bigr) \biggr|_{x=0}.\\             
\end{eqnarray}  
To make a progress with calculations,  one may take advantage of Eq. \eqref{eq3b7} and express the previous relation in a more suitable form for further manipulation:
\begin{eqnarray}
\label{eq3b70}  
        \bigl \langle \sigma_i  \bigl\{ f_i \bigr\} \bigr \rangle   =    
            \bigl \langle \bigl\{ f_i \bigr\} \bigl( 2\sigma_{i-1}  + 2\sigma_{i+1} \bigr) \bigr \rangle  A^{(2)}_1 
       +  h \bigl \langle  \bigl\{ f_i \bigr\} \bigl ( A^{(2)}_0 +  4 \sigma_{i-1}  \sigma_{i+1}   A^{(2)}_2  \bigr)\bigl \rangle, 
\end{eqnarray}    
where $\sigma_{i+1} $ and $\sigma_{i-1} $ denote two nearest-neighbor spins of the central spin $\sigma_{i}$ and the coefficients $A^{(2)}_i $ are respectively given by:
\begin{eqnarray}
\label{eq3b71a}  
   A^{(2)}_0 &=& \frac \beta 4 \cosh^2\Bigl (\frac {J\nabla}{2} \Bigr)  
                     \mbox{sech}^2  \Bigl(\frac{\beta x}{2} \Bigr) \biggr|_{x=0}  = 
                     \frac \beta 8 \Bigl [ \mbox{sech}^2  \Bigl(\frac{\beta J}{2} \Bigr) + 1   \Bigr ],                   
                         \\           
\label{eq3b71b}  
  A^{(2)}_1 &=& \frac{1}{2}\cosh \Bigl (\frac {J\nabla}{2} \Bigr) \sinh \Bigl (\frac {J\nabla}{2} \Bigr)  
   \tanh \Bigl(\frac{\beta x}{2} \Bigr) \biggr|_{x=0} = 
     \frac{1}{4}\tanh \Bigl(\frac{\beta J}{2}  \Bigr),\\
     \label{eq3b71c}  
     A^{(2)}_2 &=& \frac \beta 4 \sinh^2\Bigl (\frac {J\nabla}{2} \Bigr)  
                     \mbox{sech}^2  \Bigl(\frac{\beta x}{2} \Bigr) \biggr|_{x=0}  = 
                     \frac \beta 8 \Bigl [ \mbox{sech}^2  \Bigl(\frac{\beta J}{2} \Bigr) - 1   \Bigr ].     
\end{eqnarray}    
Setting $\bigl\{ f_i \bigr\}  =1$ and $h = 0$ to Eq. \eqref{eq3b70} one obtains the following simple relation for the spontaneous magnetization $m = \bigl \langle \sigma_i \bigr \rangle$ of the spin-1/2 Ising chain:  
\begin{eqnarray}
\label{eq3b71}      
           m =  4 m   A^{(2)}_1.    
\end{eqnarray}    
The critical temperature of the spin-1/2 Ising chain can be accordingly determined form the expression:
\begin{eqnarray}
\label{eq3b72}  
          1 = \tanh \Bigl (\dfrac{J}{ 2 k_{\rm B} T_{\rm c}} \Bigr), 
\end{eqnarray}
which implies only a trivial solution $T_{\rm c} =0$ in agreement with the exact result obtained by means of the transfer-matrix method in Section \ref{exact}.
 
In order to calculate exact results for some other physical quantities, let us set  $\bigl\{ f_i \bigr\} = \sigma_k$, $h = 0$ and substitute Eq. \eqref{eq3b71b} into Eq. \eqref{eq3b70} to obtain:
\begin{eqnarray}
\label{eq3b73}  
        \bigl \langle  \sigma_k  \sigma_i \bigr \rangle   =    
         \frac{1}{2} \tanh \Bigl(\frac{\beta J}{2}  \Bigr)
            \Bigl (\bigl \langle  \sigma_k\sigma_{i-1} \bigr \rangle + 
            \bigl \langle\sigma_k\sigma_{i+1}  \bigr \rangle \Bigr).    
\end{eqnarray}    
Here, $\sigma_k$ represent an arbitrary Ising spin variable situated the $k$th lattice site, which is located at the absolute distance $r = |i-k|$ with respect to the central spin $\sigma_i $ of the considered spin cluster.

At this stage, one should realize that the correlation function $\bigl \langle  \sigma_k  \sigma_i \bigr \rangle $ depends only on the absolute distance between the relevant lattice sites at a fixed temperature. We can therefore introduce the correlation function $g (r)$: 
\begin{eqnarray}
\label{eq3b74}  
        g (r) = g(|i-k|) = \bigl \langle  \sigma_k  \sigma_i \bigr \rangle  
\end{eqnarray}    
and rewrite Eq. \eqref{eq3b73} into the form:
\begin{eqnarray}
\label{eq3b75}  
        2\coth \Bigl(\frac{\beta J}{2}  \Bigr) = \dfrac{g (r +1)}{  g (r)}   +\dfrac{g (r - 1)}{  g (r)}.   
\end{eqnarray}    
Since the l.h.s of this functional equation is independent of $r$, so the the r.h.s must not depend on the distance as well. Thus, the following relations must be satisfied quite generally:
\begin{eqnarray}
\label{eq3b76}  
       \dfrac{g (r +1)}{  g (r)}   = \gamma(\beta J),
\end{eqnarray}  
where the  function $\gamma(\beta J)$ can be simply determined after substitution of Eq. \eqref{eq3b75} into Eq. \eqref{eq3b76}. In this way one obtains a quadratic equation, which unambiguously determines the unknown function  $\gamma$: 
\begin{eqnarray}
\label{eq3b77}  
        \gamma = \tanh \Bigl(\frac{\beta J}{4}  \Bigr).
\end{eqnarray}  
Using the property $g(0) = 1/4$ one easily obtains the well-known exact result for the correlation function of the spin-1/2 Ising chain (see Eq. \eqref{s36} in Section \ref{exact}):
\begin{eqnarray}
\label{eq3b78}  
     \bigl \langle  \sigma_k  \sigma_i \bigr \rangle   = g (r) = \frac{1}{4}\Bigl [\tanh \Bigl(\frac{\beta J}{4}  \Bigr) \Bigr ]^r.
\end{eqnarray}
Our further goal is to illustrate how to obtain exact results for the internal energy, specific heat and initial magnetic susceptibility of the spin-1/2 Ising chain within the single-spin cluster formulation of the effective-field theory.  

The internal energy can be determined from Eq. \eqref{eq3b37} after substituting instead of $E_i$ the following expression:
\begin{eqnarray}
\label{eq3b79}  
        E_i  = J ( \sigma_{i-1}+ \sigma_{i+1}). 
\end{eqnarray}
In this way one gets the relation:
\begin{eqnarray}
\label{eq3b80}  
      U = - \dfrac{ N J}{2} \Bigl ( \bigl \langle \sigma_{i} \sigma_{i-1}\bigr \rangle + \bigl \langle \sigma_{i} \sigma_{i+1}\bigr \rangle \Bigr ),
\end{eqnarray}
which can be rewritten using Eq.  \eqref{eq3b78} as: 
\begin{eqnarray}
\label{eq3b81}  
      U = -  \dfrac{ N J}{4} \tanh \biggl(\frac{\beta J}{4}  \biggr). 
\end{eqnarray}
By differentiating the previous equation with respect to temperature one obtains the specific heat: 
\begin{eqnarray}
\label{eq3b82}  
     C =   \dfrac{NJ^2}{16 k_{\rm B} T^2}   \mbox {sech}^2  \biggl(  \dfrac{J}{4 k_{\rm B} T} \biggr). 
\end{eqnarray}
Next, substituting Eq.  \eqref{eq3b81} into Eq. \eqref{eq3b63} one easily finds the Helmholtz free energy: 
\begin{eqnarray}
\label{eq3b82b}  
     F =  -N  k_{\rm B} T  \ln \biggl [2\cosh\biggl(\dfrac{J}{4 k_{\rm B} T}  \biggr) \biggr],
\end{eqnarray}
by virtue of which the entropy of the spin-1/2 Ising chain is given by:
\begin{eqnarray}
\label{eq3b82a}  
    S = N  k_{\rm B}   \ln \biggl [2\cosh\biggl(\dfrac{J}{4 k_{\rm B} T}  \biggr) \biggr] + 
     \frac{NJ}{4k_{\rm B} T} \tanh\biggl(\frac{J}{4k_{\rm B} T}\biggr).         
\end{eqnarray}
Finally, let us calculate the initial magnetic susceptibility from Eq. \eqref{eq3b24}. One obtains from Eq. \eqref{eq3b70} by setting $\{f_i \} =1$: 
\begin{eqnarray}
\label{eq3b83}  
       m =   4m  A^{(2)}_1 
       +  h  \bigl [ A^{(2)}_0 +  4 \bigl \langle\sigma_{i-1}  \sigma_{i+1} \bigl \rangle  A^{(2)}_2  \bigr],
\end{eqnarray}    
where the two-spin correlation function $\langle\sigma_{i-1}  \sigma_{i+1} \rangle$ is given by (see Eq. \eqref{eq3b78}):
\begin{eqnarray}
\label{eq3b83a}  
     g (2) = \bigl \langle  \sigma_{i-1}  \sigma_{i +1} \bigr \rangle   =  \frac{1}{4}\tanh^2 \biggl(\frac{\beta J}{4}  \biggr).
\end{eqnarray}
Substituting Eqs. \eqref{eq3b71}-\eqref{eq3b73} into Eq. \eqref{eq3b83} and taking the limit of $h \to 0$ one easily derives the following relation:
 \begin{eqnarray}
\label{eq3b83b}  
     \biggl ( \frac{\partial m}{\partial h } \biggr )_\beta = 
     \frac{ \beta \biggl \{ 1 + \mbox{sech}^2 \biggl( \dfrac{\beta J}{2}\biggr)   + 
     \tanh^2 \biggl ( \dfrac{\beta J}{4}\biggr)\biggl[ \mbox{sech}^2 \biggl( \dfrac{\beta J}{2}\biggr) - 1 \biggr]  
     \biggr \} }{ 8 \biggl [ 1 - \tanh^2 \biggl( \dfrac{\beta J}{2}\biggr) \biggr ]},
\end{eqnarray}
which can be further simplified to:
\begin{eqnarray}
\label{eq3b84}  
     \chi(T) =    \dfrac{ N g^2 \mu_{\rm B}^2} {k_{\rm B} T}\exp\biggl(  \frac{ J } {2 k_{\rm B} T} \biggr ). 
\end{eqnarray}
To summarize, the single-spin cluster formulation of the effective-field theory gives exact results for all thermodynamic quantities of the spin-1/2 Ising chain.

\subsection{Spin-1/2 Ising model in a transverse field within a single-spin cluster approach}
\label{tim12efa}
The transverse Ising model is one of the simplest possible quantum models describing  cooperative phenomena in the solid-state physics. The model  includes the effect of  external magnetic field applied both along the $z$  and $x$ directions. It has been originally introduced by Kirkwood in 1933, who has clearly demonstrated the existence of long-range order and phase transitions in the system with nonzero perpendicular component of the external magnetic field  \cite{kirk33}.
 
Later on, the transverse Ising model has been applied to investigate physical properties in various physical systems, such as hydrogen bonded ferro- and antiferroelectrics \cite{dege63,blin72}, ferromagnets with induced magnetic moments or cooperative Jahn-Teller systems \cite{birg72}. The most natural usage of this model is, of course, its application in description of various magnetic systems in an external transverse  magnetic field (see for example \cite{stin73,kane87tim,kane88tim1}). The transverse Ising model has been theoretically investigated by means of the standard mean-field theory \cite{stin73}, pair approximation \cite{yqma92}, high-temperature series expansions \cite{oitm81}, transfer-matrix method and Monte Carlo simulations \cite{suzu85} and rarely it was also treated exactly \cite{pfeu70}.

Our main aim in this part is to explain how the single-spin cluster formulation of the effective-field theory based on the differential operator technique can be extended to the spin-1/2  Ising model in a transverse magnetic field. For this purpose, we will study the model defined by the Hamiltonian: 
\begin{eqnarray} 
\label{eq5b1}
             {\cal \hat H} = - J \sum_{ \langle ij \rangle} \hat S_i^z \hat S_j^z - h \sum_{i=1}^N \hat S_i^z  - \Omega \sum_{i=1}^N \hat S_i^x,
\end{eqnarray} 
where the constant $J$ denotes the nearest-neighbor exchange interaction, $h$ and $\Omega$ represent the longitudinal and transverse components of external magnetic field along the $z$ and $x$ direction, respectively. The spin-1/2 operators  $\hat S_i^x$ and $\hat S_i^z$  ($i = 1, \dots , N $) take the following standard matrix form: 
\begin{eqnarray} 
\label{eq5b2}
 \hat S_i^x= \dfrac{1}{2}\begin{pmatrix} 
                                    0 & 1\\
			                       1 & 0 
                           \end{pmatrix}_i, \qquad
  \hat S_i^z=   \dfrac{1}{2}\begin{pmatrix} 
                                         1 & 0\\
			   							0 & -1 
                          \end{pmatrix}_i.
\end{eqnarray}
Similarly as in previous calculations, we again separate the total Hamiltonian  \eqref{eq5b1} into two parts: 
\begin{eqnarray} 
\label{eq5b3}
           \hat  {\cal H} = \hat{\cal H}_i   +  \hat {\cal H}^{'},
\end{eqnarray}  
whereas the single-spin cluster Hamiltonian $\hat{\cal H}_i $ is given by:
\begin{eqnarray} 
\label{eq5b4}
         \hat{\cal H}_i   =  -\hat  S_i^z \Bigl (J \sum_{j=1}^q \hat S_j^z + h \Bigr ) 
                                 -   \Omega \hat  S_i^x. 
\end{eqnarray}  
The  single-spin cluster Hamiltonian \eqref{eq5b4} includes all interactions of the central $i$th spin with its nearest neighbors and external magnetic field, while the Hamiltonian $ \hat {\cal H}^{'}$ accounts for all remaining interactions in the system under investigation.

The starting point for the formulation of the single-spin cluster effective-field theory will be again the Callen-Suzuki identity  \eqref{eq1b15}, which after setting $ \hat{ \cal O}_i = \hat  S_i^{\alpha} $ takes the form:
\begin{eqnarray} 
\label{eq5b5}  
          \bigl \langle \hat  S_i^\alpha \bigr \rangle =      
          \biggl \langle \dfrac{ \mbox{Tr}_i\;\hat  S_i^\alpha  \exp \bigl (-\beta \hat {\cal H}_i \bigr)  } 
                                                         {\mbox{Tr}_i\; \exp \bigl (-\beta \hat {\cal H}_i \bigr) }    \biggr \rangle,   
\end{eqnarray}  
where $\alpha = x, z$ and $ \hat{\cal H}_i $ is given by equation \eqref{eq5b4}. It is wortwhile to recall here that the previous identity is only approximate expression due to a non-commutative nature of the Hamiltonians $ \hat {\cal  H}_i $ and $ \hat {\cal H}^{'}$, i.e. $[\hat {\cal  H}_i, \hat {\cal H}^{'}] \neq 0$. Despite of this fact, such a theory will be in several regards still superior to the standard mean-field theory.

The matrix representation of the single-spin cluster Hamiltonian $\hat{\cal H}_i $ takes the following explicit form:
\begin{eqnarray} 
\label{eq5b6}
  \hat{\cal H}_i = -\dfrac{1}{2}\begin{pmatrix} 
                                    E_i  + h&\Omega\\
			                       \Omega   & -E_i-  h 
                           \end{pmatrix}_i,
\end{eqnarray}
where we have introduced the new parameter:
\begin{eqnarray} 
\label{eq5b7}
E_i = J \sum_{j=1}^q  \hat S_j^z.
\end{eqnarray} 
In order to proceed further with a calculation, one has to evaluate  the  exponential function in \eqref{eq5b5}. The most elegant way to perform this calculation is  the diagonalization of Hamiltonian \eqref{eq5b4} by applying the following unitary rotation transformation: 
\begin{eqnarray} 
\label{eq5b8}
					\hat  S_i^z &=& \hat  S_i^{z'} \cos \varphi_i - \hat  S_i^{x'} \sin \varphi_i, \\
\label{eq5b9}								
				 \hat  S_i^x &= &\hat  S_i^{z'} \sin \varphi_i  + \hat  S_i^{x'} \cos \varphi_i,  
\end{eqnarray} 
where
\begin{eqnarray} 
\label{eq5b10}
						\cos \varphi_i = \dfrac{E_i + h}{\sqrt{\Omega^2    + \bigl (E_i + h \bigr) ^2}}, \quad
						\sin \varphi_i = \dfrac{\Omega}{\sqrt{\Omega^2    + \bigl (E_i + h \bigr) ^2}}.						
\end{eqnarray} 
Using the relations \eqref{eq5b8} and \eqref{eq5b9}, we respectively express  the  operators $ \hat S_i^z$,  $\hat S_i^x $ and $ \hat{\cal H}_i $ in the following matrix form:
\begin{eqnarray} 
\label{eq5b12}
  \hat S_i^z= -\dfrac{1}{2}
 \begin{pmatrix} 
\dfrac{E_i + h}{\sqrt{\Omega^2   +  \bigl (E_i + h \bigr) ^2}} &  - \dfrac{\Omega}{\sqrt{\Omega^2   +  \bigl (E_i + h \bigr) ^2}}\\
 - \dfrac{\Omega}{\sqrt{\Omega^2   +  \bigl (E_i + h \bigr) ^2}}   &  - \dfrac{E_i +h}{\sqrt{\Omega^2   +  \bigl (E_i + h \bigr) ^2}}
\end{pmatrix}_i,
\end{eqnarray}
\begin{eqnarray} 
\label{eq5b13}
   \hat S_i^x= -\dfrac{1}{2}
 \begin{pmatrix} 
\dfrac{\Omega}{\sqrt{\Omega^2   +  \bigl (E_i + h \bigr) ^2}} &   \dfrac{E_i + h}{\sqrt{\Omega^2   +  \bigl (E_i + h \bigr) ^2}}\\
  \dfrac{E_i + h}{\sqrt{\Omega^2   +  \bigl (E_i + h \bigr) ^2}}   &  - \dfrac{\Omega}{\sqrt{\Omega^2   +  \bigl (E_i + h \bigr) ^2}}
\end{pmatrix}_i
\end{eqnarray}
and 
\begin{eqnarray} 
\label{eq5b11}
  \hat{\cal H}_i = -\dfrac{1}{2}
  							\begin{pmatrix} 
                                    \sqrt{\Omega^2   +  \bigl (E_i + h \bigr) ^2} & 0 \\
			                       0   & -\sqrt{\Omega^2   + \bigl (E_i + h \bigr) ^2}
                           \end{pmatrix}_i,
\end{eqnarray}
After substituting Eqs. \eqref{eq5b12} and \eqref{eq5b11} into Eq. \eqref{eq5b5}, we simply evaluate relevant traces and obtain the following expression for the longitudinal  component of magnetization:
\begin{eqnarray} 
\label{eq5b14}  
        m_z \equiv  \bigl \langle \hat S_i^z \bigr \rangle =      
          \Biggl  \langle \dfrac{1}{2}\dfrac{E_i +h}{\sqrt{\Omega^2   + \bigl (E_i + h \bigr) ^2}}
          \tanh \biggl [\dfrac{\beta}{2} \sqrt{\Omega^2   +  \bigl (E_i + h \bigr) ^2}  \biggr]  \Biggr \rangle.  
\end{eqnarray}  
The same procedures can be routinely repeated in order to find the following   relation  for the transverse component of magnetization: 
\begin{eqnarray} 
\label{eq5b15}  
        m_x \equiv  \bigl \langle \hat S_i^x \bigr \rangle =      
          \Biggl  \langle \dfrac{1}{2}\dfrac{\Omega}{\sqrt{\Omega^2   + \bigl (E_i + h \bigr) ^2}}
          \tanh \biggl [ \dfrac{\beta}{2} \sqrt{\Omega^2   +  \bigl (E_i + h \bigr) ^2}\biggr ]  \Biggr \rangle.   
\end{eqnarray}  
Before proceeding further, one can notice that the least accurate approximation one gets by simple averaging the quantity $E_i$ in Eqs. \eqref{eq5b14} and \eqref{eq5b15}. In this way one obtains the results:
\begin{eqnarray} 
\label{eq5b16}  
        m_z  =      
           \dfrac{1}{2}\dfrac{qJm_z + h }{\sqrt{\Omega^2   + (q J m_z + h) ^2}}
          \tanh \biggl [\dfrac{\beta}{2} \sqrt{\Omega^2   +  \bigl (q J m_z + h \bigr) ^2}  \biggr]   
\end{eqnarray}  
and
\begin{eqnarray} 
\label{eq5b17}  
        m_x  =      
           \dfrac{1}{2}\dfrac{\Omega }{\sqrt{\Omega^2   + ( qJ  m_z + h) ^2}}
          \tanh \biggl [\dfrac{\beta}{2} \sqrt{\Omega^2   +  \bigl (q J m_z + h \bigr) ^2} \biggr ],   
\end{eqnarray}  
which are identical with the expressions (\ref{mxztim}) obtained within the mean-field theory presented in the subsection \ref{mfa12t} for the particular case with $h = 0$ and $S = 1/2$.

In order to obtain more accurate results, we now develop the effective-field theory using the exact van der Waerden identity \eqref{eq3b7} and differential operator technique \eqref{eq3b10}.
In doing so, we rewrite Eqs.  \eqref{eq5b14} and \eqref{eq5b15} in the form:
\begin{eqnarray} 
\label{eq5b14a}  
        m_{z}  =      
          \Bigl  \langle 
          \prod_{j= 1}^{q}  \Bigl [\cosh \Bigl (\dfrac{J}{2} \nabla_x \Bigr )  +  2\hat S_j^z  \sinh \Bigl (\dfrac{J}{2} \nabla_x \Bigr )  \Bigr]   
          \Bigr \rangle 
          f(x) \Bigr|_{x=0}.  
\end{eqnarray}  
\begin{eqnarray} 
\label{eq5b15a}  
        m_x =      
          \Bigl  \langle 
          \prod_{j= 1}^{q}  \Bigl [ \cosh \Bigl (\dfrac{J}{2} \nabla_x \Bigr )  +  2\hat S_j^z  \sinh \Bigl (\dfrac{J}{2} \nabla_x \Bigr )  \Bigr]   
          \Bigr \rangle 
          g(x) \Bigr|_{x=0}, 
\end{eqnarray}  
where we have introduced the functions $f(x)$ and $g(x)$ as: 
\begin{eqnarray} 
\label{eq5b21}  
           f \bigl(x \bigr)  = \dfrac{x+h}{2\sqrt{\Omega^2   + (x+h) ^2}} 
           \tanh \biggl (\dfrac{\beta}{2}\sqrt{\Omega^2   +  (x+h) ^2}\biggr),
\end{eqnarray}  
and
\begin{eqnarray} 
\label{eq5b22}  
           g \bigl(x\bigr)  = \dfrac{\Omega}{2\sqrt{\Omega^2   + (x+h) ^2}} 
           \tanh \biggl (\dfrac{\beta}{2}\sqrt{\Omega^2   +  (x+h) ^2}\biggr).
\end{eqnarray}  
Neglecting all multispin correlations on the r.h.s. of \eqref{eq5b14a} and \eqref{eq5b15a},  we obtain the single-spin cluster effective-field theory for the spin-1/2 transverse Ising model. Within this approximation the longitudinal and transverse components of magnetization are respectively given by:    
\begin{eqnarray} 
\label{eq5b19}  
        m_z  =      
           \Bigl [ \cosh \Bigl (\dfrac{J}{2} \nabla_x \Bigr ) +  2 m_z\sinh \Bigl ( \dfrac{J}{2}\nabla_x \Bigr ) \Bigr] ^q  
           f \bigl(x\bigr) \Bigr|_{x=0}, 
\end{eqnarray}  
\begin{eqnarray} 
\label{eq5b20}  
        m_x  =      
           \Bigl [ \cosh \Bigl (\dfrac{J}{2} \nabla_x \Bigr ) +  2 m_z\sinh \Bigl ( \dfrac{J}{2}\nabla_x \Bigr ) \Bigr] ^q
           g \bigl(x\bigr) \Bigr|_{x=0}.
\end{eqnarray}  
In order to complete the calculation,  we follow the same steps as in section~\ref{0S_EFT} and get the following expressions for the longitudinal and transverse magnetizations $m_z$ and $m_x$:
\begin{eqnarray} 
\label{eq5b23}  
        m_z  = \sum_{k=0}^{q}   {q \choose  k} m^k_z A^{(q)}_k ,
\end{eqnarray}
\begin{eqnarray} 
\label{eq5b24}  
        m_x = \sum_{k=0}^{q}   {q \choose  k} m^k_z B^{(q)}_k ,
\end{eqnarray}
As one can see, these equations have formally the same form as Eq. \eqref{eq3b10}, however, the coefficients $A^{(q)}_k$  a $B^{(q)}_k$ are now defined by the relations:
\begin{eqnarray} 
\label{eq5b25}  
        A^{(q)}_k =
        				2^k\Bigl [\cosh \Bigl ( \dfrac{J}{2} \nabla_x \Bigr )\Bigr ]^{q-k} 
                                     \Bigl [ \sinh \Bigl ( \dfrac{J}{2}\nabla_x \Bigr )\Bigr ] ^k  
                                           f\bigl (x\bigr) \Bigr|_{x=0}
\end{eqnarray}
and
\begin{eqnarray} 
\label{eq5b26}  
        B^{(q)}_k =
        				2^k\Bigl [\cosh \bigl ( \dfrac{J}{2}\nabla_x \bigr )\Bigr]^{q-k} 
                                    \Bigl [ \sinh \bigl ( \dfrac{J}{2}\nabla_x \bigr )\Bigr ] ^k  
                                           g\bigl (x\bigr) \Bigr|_{x=0},
\end{eqnarray}
\begin{figure}[tb]
\begin{center}
\hspace{-1.0cm}
\includegraphics[clip,width=7.5cm]{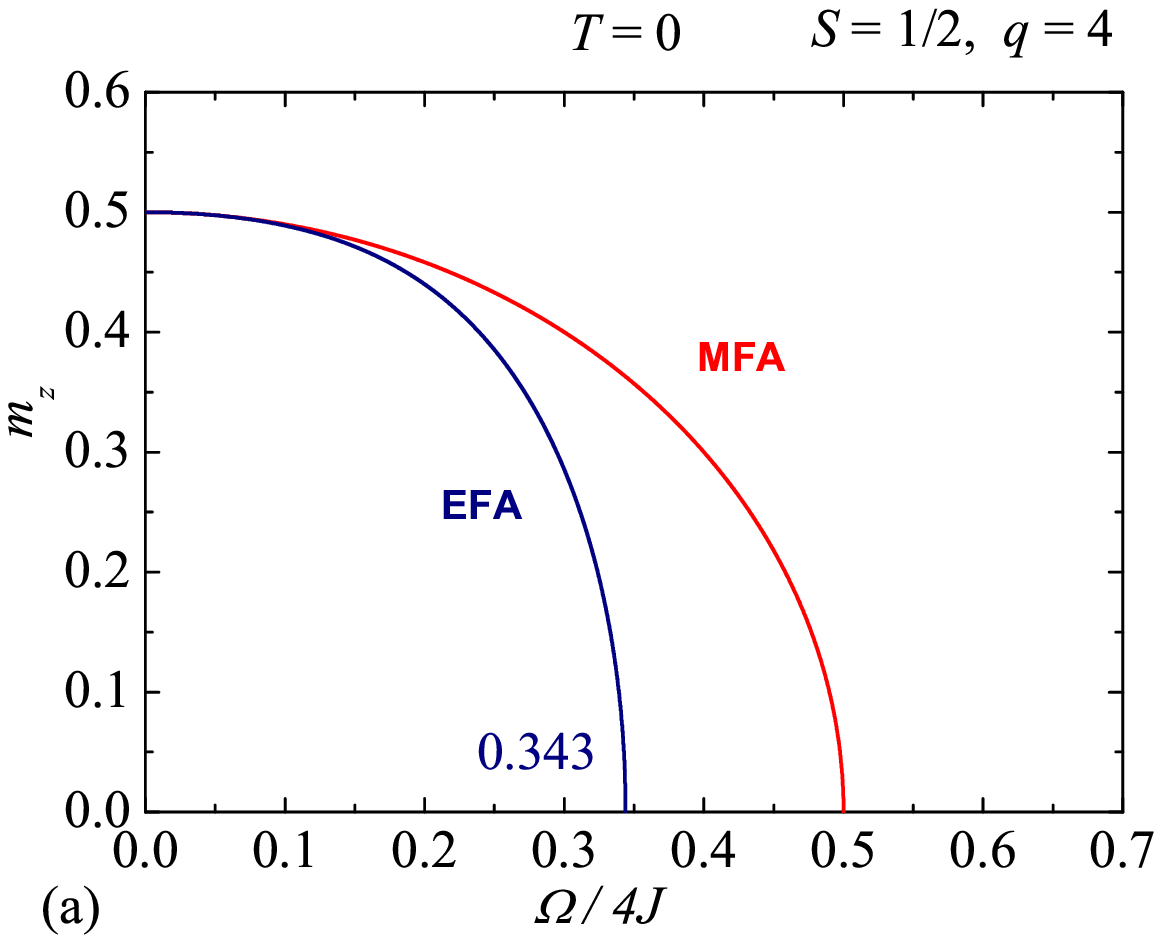}
\hspace{-1.0cm}
\includegraphics[clip,width=7.5cm]{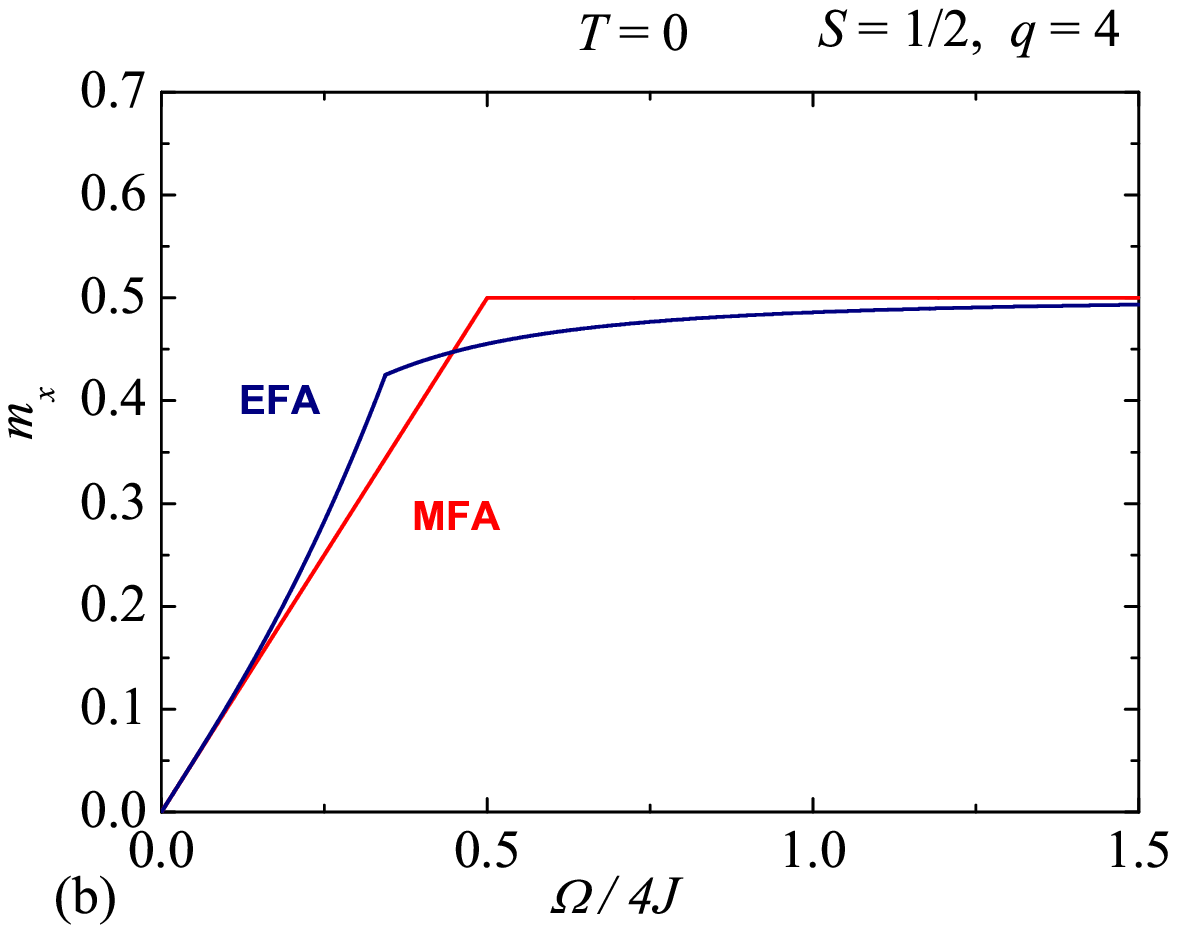}
\hspace{-1.5cm}
\vspace{-0.25cm}
\caption{The zero-temperature magnetization curve of the spin-$1/2$ Ising model in a transverse magnetic field: (a) longitudinal magnetization; (b) transverse magnetization. The red and blue curves  represent the results obtained within the mean-field approximation (MFA) and the single-spin cluster formulation of the effective-field approximation (EFA), respectively. }
\label{timgsefa}
\end{center}
\end{figure}    
where the functions $ f(x)$ a $g(x)$ are given by \eqref{eq5b21} a \eqref{eq5b22}.

At first, let us verify  the accuracy of the present theory at zero temperature $T=0$. For this purpose,  we have 
depicted in Fig. \ref{timgsefa}  the transverse-field dependences of the longitudinal and transverse magnetizations for the spin-1/2 Ising model on a 
square lattice within the mean-field and single-spin cluster effective-field approximations. As one can see, the effective-field theory 
remarkably improves the standard mean-field theory. In particular, the region of the existence of ferromagnetically ordered  
phase is significantly reduced, since  the critical values of the transverse field are respectively given by  
$\Omega_{\rm c}^{EFA}/4J = 0.343$ and  $\Omega_{\rm c}^{MFA}/4J = 0.5$. One can also notice that the effective-field 
approximation predicts more realistic and accurate result (see Fig. \ref{timgsefa}(b)) when the transverse magnetization does not reach the saturation value right at the critical field. 

\begin{figure}[tb]
\begin{center}
\hspace{-1.0cm}
\includegraphics[clip,width=7.5cm]{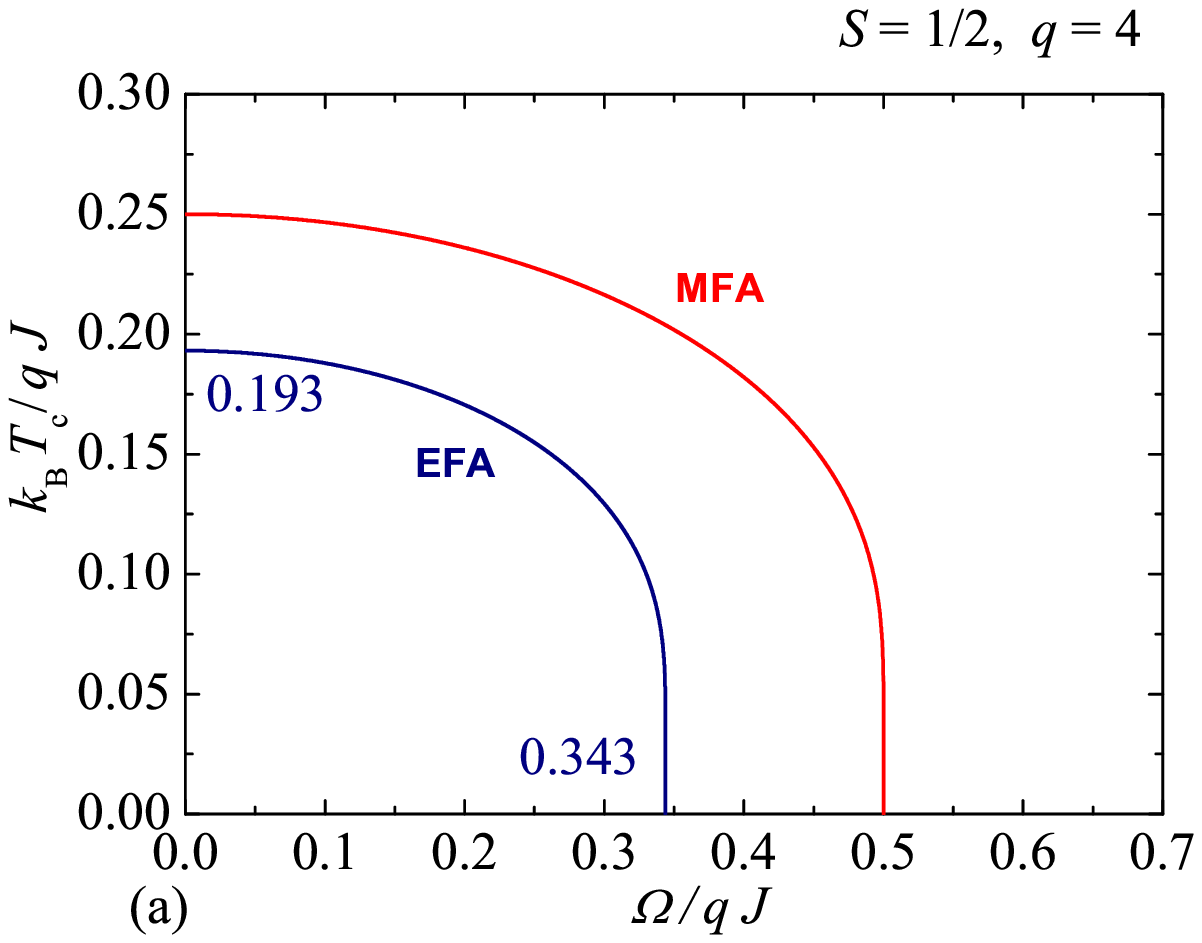}
\hspace{-1.0cm}
\includegraphics[clip,width=7.5cm]{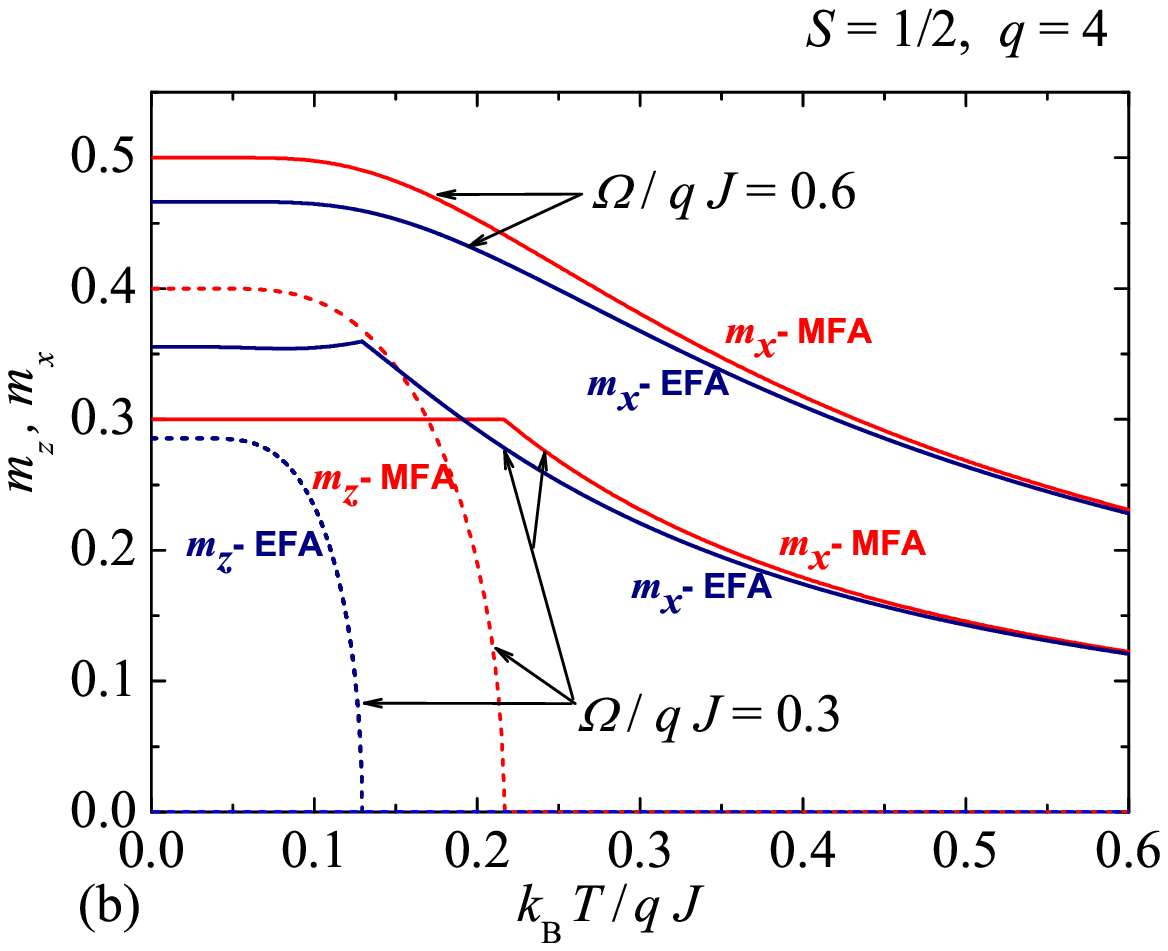}
\hspace{-1.5cm}
\vspace{-0.25cm}
\caption{(a) The critical temperature of the spin-$1/2$ Ising model on a square lattice ($q=4$) as a function of the reduced transverse magnetic field. The red curve (labeled MFA) represents the result of the mean-field approximation and the blue one (labeled EFA) is the result of the single-spin cluster approach within the effective-field approximation; 
(b) Typical temperature dependences of the longitudinal magnetization (broken line) and transverse magnetization (solid lines) of the spin-$1/2$ Ising model for two different transverse fields selected below and above the critical fields. Different colors have the same meaning as in case (a).}
\label{timefa_tc}
\end{center}
\end{figure}

Now, let us investigate the numerical accuracy of the single-spin cluster formulation of the effective-field theory also at finite temperatures.  
In order to illustrate finite-temperature predictions we have shown in Fig. \ref{timefa_tc} the phase diagram in the 
$\Omega - T_{\rm c}$ space and also temperature variations of the longitudinal and transverse magnetizations.
The blue curves in the figures represent the results of the  effective-field approximation (EFA)  and the red ones those of the mean-field approximation (MFA).
As one can see, the effective-field approach is superior to the mean-field calculation and provides a notable numerical improvement  and physically more realistic dependences of relevant quantities than the standard mean-field theory. 

The effective-field theory has been of course also applied to the system with higher spin values despite of growing 
complexity of mathematical formulation \cite{kane93bc1}. In this work  we will show extension of simplified version of this 
theory to the system with arbitrary spin in the next section.  One should again recall that the effective-field theory
based on the differential operator technique has been successfully  applied to study many other transverse Ising models
including the amorphous systems, magnetically diluted systems, the systems with random exchange interactions
and/or random single-ion anisotropy, semi infinite system, thin magnetic films, magnetic  multilayers an so on (see the subsequent part \ref{histmis} and references cited therein).   

The relevant discussion of these complex systems  is clearly beyond the scope of this paper and it requires a separate work.
Nevertheless, one should emphasize here that  applications of the effective-field theory with the differential operator
technique in all cases have significantly improved the standard mean-field results for the same systems. Moreover,   in 
many cases the effective-field theories also rule out non-physical predictions of  simpler approaches such as mean-field or Oguchi approximation \cite{smart66} .

\subsection{Spin-1 Blume-Capel model within a single-spin cluster approach}
\label{BC_1}
In this part we will show how the differential operator technique can be adapted to the Ising-like systems with higher spin value. For this purpose, we will choose the well-known spin-1 Blume-Capel model in a longitudinal magnetic field described by the Hamiltonian: 
\begin{eqnarray} 
\label{eq6b1}
             {\cal H} = - J \sum_{\langle ij \rangle} S_i  S_j  - D \sum_{i=1}^N S_i ^2 - h \sum_{i=1}^N S_i,
\end{eqnarray}  
where $S_i = 0, \pm 1$, the coupling constant $J$ describes the exchange interaction between the nearest-neighbor spin pairs, the parameter $D$ represents the single-ion anisotropy and $h$ is the external magnetic field.

It immediately follows from Eq. \eqref{eq6b1} that all spins will prefer at $T=0, \; h = 0$ two magnetic states  $S_i =\pm 1$ for strong positive values of $D$, while they should reside the non-magnetic state $S_i = 0$ for strong negative values of $D$. Consequently, the spin-1  Blume-Capel model exhibits a zero-temperature phase transition between the ferromagnetically ordered and disordered paramagnetic phases. Indeed, a simple comparison of the relevant ground-state energies indicates that such a discontinuous phase transition takes place at $D_{\rm c}/J = - q/2$.  

Let us separate the Hamiltonian \eqref{eq6b1} into two parts in order to develop the effective-field theory:
\begin{eqnarray} 
\label{eq6b1a}
             {\cal H} =  {\cal H}_i  +  {\cal H}^{'}. 
\end{eqnarray} 
The single-spin cluster Hamiltonian, which includes all interaction terms depending on the central $i$th spin of the cluster, is given by: 
\begin{eqnarray} 
\label{eq6b1b}
             {\cal H}_i = -S_i \bigl (E_i+ h\bigr)   - D  S_i ^2,  
\end{eqnarray}
whereas the parameter $E_i$ is used to denote the internal exchange field coming from the nearest-neighbor coupling of the central $i$th spin with all its nearest neighbors:
\begin{eqnarray} 
\label{eq6b2}  
        E_i =  J \sum_{j= 1}^q S_j.
\end{eqnarray}

Similarly as for the spin-1/2 Ising model, one may also obtain the Callen-Suzuki identities for the quantities  $\bigl \langle S_i  \bigl\{   f_i \bigr\}\bigr \rangle$ and
$  \bigl \langle S_i ^2 \bigl\{   f_i \bigr\}\bigr \rangle $ within the spin-1 Blume-Capel model:
\begin{eqnarray} 
\label{eq6b2a}  
\displaystyle
             \bigl \langle S_i  \bigl\{   f_i \bigr\}\bigr \rangle =      
          \Biggl \langle    \bigl\{   f_i \bigr\} \dfrac{ \displaystyle\sum_{S_i =0, \pm 1} \;
          	S_i \exp \bigl (-\beta {\cal H}_i \bigr)  } 
            {\displaystyle \sum_{S_i = 0,\pm 1}\; \exp \bigl (-\beta {\cal H}_i \bigr) }    \Biggr \rangle
\end{eqnarray}  
and 
\begin{eqnarray} 
\label{eq6b2b}  
\displaystyle
         \bigl \langle S_i ^2 \bigl\{   f_i \bigr\}\bigr \rangle =      
          \Biggl \langle   \bigl\{   f_i \bigr\} \dfrac{ \displaystyle\sum_{S_i =0, \pm 1} \;
          	S_i^2 \exp \bigl (-\beta {\cal H}_i \bigr)  } 
            {\displaystyle \sum_{S_i = 0,\pm 1}\; \exp \bigl (-\beta {\cal H}_i \bigr) }    \Biggr \rangle, 
\end{eqnarray}  
where $\bigl\{   f_i \bigr\}$ stands for any function of arbitrary spin with exception of $S_i$.

Now, performing the summation over the central spin one gets exact relations:
\begin{eqnarray} 
\label{eq6b3}  
       \bigl \langle S_i  \bigl\{   f_i \bigr\}\bigr \rangle =              
          \Biggl \langle \bigl\{   f_i \bigr\}\dfrac{ 2 \sinh\bigl [ \beta (E_i + h)  \bigr] }
                  {  2 \cosh\bigl [ \beta( E_i + h)  \bigr]+ \mbox{e}^{-\beta D} \ }    \Biggr \rangle
\end{eqnarray} 
and
\begin{eqnarray} 
\label{eq6b4}  
       \bigl \langle S_i ^2 \bigl\{   f_i \bigr\}\bigr \rangle =               
          \Biggl \langle \bigl\{   f_i \bigr\} \dfrac{ 2 \cosh\bigl [ \beta (E_i + h)  \bigr] }
                  {  2 \cosh\bigl [ \beta( E_i + h)  \bigr]+ \mbox{e}^{-\beta D} \ }    \Biggr \rangle,
\end{eqnarray} 
where we are again faced with the problem of averaging the r.h.s of these equations. If one sets $\bigl\{   f_i \bigr\}=1$ and applies the simplest possible way of averaging, i.e. the averaging of arguments of all functions, then one obtains the following simple expressions for the magnetization $m =\bigl \langle S_i \bigr \rangle$ and quadrupolar moment $q_D =\bigl \langle S_i^2 \bigr \rangle$: 
\begin{eqnarray} 
\label{eq6b3a}  
       m =                
          \dfrac{ 2 \sinh\bigl [ \beta (qJm + h)  \bigr] }
                  {  2 \cosh\bigl [ \beta( qJm + h)  \bigr]+ \mbox{e}^{-\beta D} \ },
\end{eqnarray} 
\begin{eqnarray} 
\label{eq6b4a}  
       q _D =                
         \dfrac{ 2 \cosh\bigl [ \beta (qJm + h)  \bigr] }
                  {  2 \cosh\bigl [ \beta( qJm+ h)  \bigr]+ \mbox{e}^{-\beta D} \ }.   
\end{eqnarray} 
These relations are of course nothing but the expected  mean-field results derived alternatively in the subsection~\ref{mfa1l} 
using the Gibbs-Bogoliubov inequality (c.f. with Eq. \eqref{mq1}).

However, the main goal of this part is to show how the differential operator technique can be employed to obtain more accurate results within the effective-field theory, which will provide superior results with respect the mean-field and Oguchi approximations \cite{smart66}. As a first step towards  developing such a theory for the model under investigation, we set $\bigl\{   f_i \bigr\}=1$ and then apply Eq. \eqref{eq3b7} to rewrite Eqs. \eqref{eq6b3} and \eqref{eq6b4} into form:
\begin{eqnarray} 
\label{eq6b5}  
          m =  \Bigl \langle  \exp\Bigl (E_i\nabla_x \Bigr) \Bigl \rangle f(x) \biggl |_{x = 0}  
\end{eqnarray} 
\begin{eqnarray} 
\label{eq6b6}  
         q_D =   \Bigl \langle \exp\Bigl (E_i\nabla_x \Bigr) \Bigl \rangle  g(x) \biggl |_{x = 0},  
\end{eqnarray} 
where for the sake of simplicity we have introduced the functions:
\begin{eqnarray} 
\label{eq6b10}  
          f(x) =  \dfrac{ 2 \sinh\bigl [ \beta (x + h)  \bigr] }
                  {  2 \sinh\bigl [ \beta( x + h)  \bigr]+ \mbox{e}^{-\beta D}  }
\end{eqnarray} 
and 
\begin{eqnarray} 
\label{eq6b11}  
               g(x) =  \dfrac{ 2 \cosh\bigl [ \beta (x + h)  \bigr] }
                  {  2 \sinh\bigl [ \beta( x + h)  \bigr]+ \mbox{e}^{-\beta D}  }. 
\end{eqnarray} 
In order to proceed further with a calculation, we have to use the generalized van der Waerden identity for the spin 1, which takes the following explicit form:
\begin{eqnarray} 
\label{eq6b7}  
                 \exp\bigl({\alpha S_j}\bigr ) =  S_j ^2 \cosh(\alpha)    +    S_j \sinh(\alpha) + 1 - S_j ^2, \qquad 
                 S_j = 0, \pm 1.
\end{eqnarray} 
The application of this identity  enables one  to recast Eqs. \eqref{eq6b5} and \eqref{eq6b6} into the following exact form: 
\begin{eqnarray}
\label{eq6b8}  
  m  =      
 \Bigl  \langle 
 \prod_{j= 1}^{q}  \Bigl [  S_j ^2\cosh \bigl ( J\nabla_x \bigr ) + S_j \sinh\bigl ( J\nabla_x \bigr ) + 1 - S_j^2  \Bigr]   \Bigr \rangle f \bigl(x \bigr) \Bigr|_{x=0}  
\end{eqnarray}  
\begin{eqnarray}
\label{eq6b9}  
     q_D  =  \Bigl  \langle 
          \prod_{j= 1}^{q}  \Bigl [  S_j ^2\cosh \bigl ( J\nabla_x \bigr ) + S_j \sinh\bigl ( J\nabla_x \bigr ) + 1 - S_j^2  \Bigr]   \Bigr \rangle g \bigl(x \bigr) \Bigr|_{x=0}.   
\end{eqnarray}  
It is clear that after expanding of the r.h.s. of \eqref{eq6b8} and \eqref{eq6b9} there appear various multispin correlations that cannot be further treated exactly and  must be approximated. 
In fact, the simplest rational approximation is based on the following  splitting scheme of  multispin  correlations: 
\begin{eqnarray} 
\label{eq6b12}  
               \bigl \langle S_j S_k^2 \dots  S_\ell \bigr \rangle \approx 
               \bigl \langle S_j \bigr \rangle\bigl \langle S_k ^2 \bigr \rangle\dots  \bigl \langle S_\ell \bigr \rangle.
\end{eqnarray} 
The resulting theory is again called as a single-spin cluster effective-field theory of the spin-1 Blume-Capel model, whereas arbitrary complex correlations among different spins can be expressed as powers of magnetization and quadrupolar moment within this theory. This splitting procedure is mathematically equivalent to the approximation $\langle \prod_{j= 1}^{q} [\dots]\rangle  \approx \langle[\dots]\rangle ^q$ on the r.h.s  in  Eqs. \eqref{eq6b8} and \eqref{eq6b9},  which enables to  obtain  a closed system of two polynomial equations:
\begin{eqnarray}
\label{eq6b13}  
 m  =  \Bigl [q_D\cosh \bigl ( J\nabla_x \bigr ) + m \sinh\bigl ( J\nabla_x \bigr ) + 1 - q_D \Bigr]^{q}  
              f \bigl(x \bigr) \Bigr|_{x=0}  
\end{eqnarray}  
\begin{eqnarray}
\label{eq6b14}  
   q_D  =   \Bigl [q_D\cosh \bigl ( J\nabla_x \bigr ) + m \sinh\bigl ( J\nabla_x \bigr ) + 1 - q_D \Bigr]^{q}  
              g \bigl(x \bigr) \Bigr|_{x=0}.    
\end{eqnarray}  
As one can see, the application of differential operator technique to the spin-1 Blume-Capel model leads to more complex equations than that obtained for the spin-1/2 Ising model. Moreover,  one  can also notice here that a unified mathematical approach for all crystalline structures based on the application of binomial expansion would be in this case much more involved, thus we will not continue to develop our calculations along this direction. Nonetheless, in the following subsection,  we will develop some alternative mathematical procedures that enable the derivation of relatively simple closed-form effective-field equations for the phase boundaries and tricritical point of the spin-1 Blume Capel model on arbitrary crystal lattice.

\subsubsection{Continuous phase transitions and tricritical point}
\label{Curie_temp_BC1}
It has been already noted earlier that the spin-1 Blume-Capel model can exist either in the ferromagnetically ordered or disordered paramagnetic phase at $T = 0$ and $h = 0$ depending on the value of  the single-ion anisotropy $D$. It is therefore natural to expect that these phases may also exists at finite temperatures under condition of zero external field. The finite-temperature phase transition between the ferromagnetic ordered and disordered paramagnetic phase can be either continuous or discontinuous as exemplified by our previous mean-field calculations. For that reason, the calculation of the phase boundary requires separate mathematical procedures for each particular case.

At first, let us derive explicit equations for the continuous phase transitions in the  system under investigation. For this purpose, we use the fact that the spontaneous magnetization  can take arbitrarily small values when the temperature approaches the Curie point from below. Consequently, in the neighborhood of the phase boundary it is always possible to express the magnetization in the form:
\begin{eqnarray}
\label{eq6b15}  
  m = am + b m^3 ,
\end{eqnarray}   
where 
\begin{eqnarray}
\label{eq6b16}  
    a  =    q \sinh\bigl ( J\nabla_x \bigr )   \Bigl [ q_0\cosh \bigl ( J\nabla_x \bigr ) +  1 - q_0 \Bigr]^{ q -1 }   
              f \bigl(x \bigr) \Bigr|_{x=0},   
\end{eqnarray}  
and 
\begin{eqnarray}
\label{eq6b25}  
\nonumber 
 b   &=&        
           \dfrac{q !}{2! (q -2)!} q_2 \bigl [ q_0\cosh \bigl ( J\nabla_x \bigr ) +  1 - q_0 \bigr]^
           {q-2}  
           \\ \nonumber 
          &\times&    \sinh \bigl ( J\nabla_x \bigr ) \bigl [ \cosh \bigl ( J\nabla_x \bigr ) -1   \bigr]
              f\bigl(x \bigr) \Bigr|_{x=0} \\ 
        &+&   \dfrac{q !}{3! (q -3)!}  \bigl [ q_0\cosh \bigl ( J\nabla_x \bigr ) +  1 - q_0 \bigr]^{q-3}   
          \sinh^3 \bigl ( J\nabla_x \bigr )
              f\bigl(x \bigr) \Bigr|_{x=0}.  
\end{eqnarray}         
The parameter  $q_0 \equiv q_{D}(m=0) $ is the solution of the equation:
\begin{eqnarray}
\label{eq6b17}  
    q_0  =     \Bigl [ q_0\cosh \bigl ( J\nabla_x \bigr ) +  1 - q_0 \Bigr]^{q} 
              g\bigl(x \bigr) \Bigr|_{x=0}.  
\end{eqnarray} 
The  derivation of the equations above  requires to analyze  the behavior of the system near the critical boundary.  For this purpose we utilize  the fact that the spontaneous magnetization continuously decreases near a critical point, so that one can  express the quadrupolar moment as:   
\begin{eqnarray}
\label{eq6b21}  
  								q_D = q_0+ q_2 m^2  
\end{eqnarray}
where  $q_0 \equiv q_{D0}$ is given by \eqref{eq6b17}. If we now substitute this expression into Eq.  \eqref{eq6b14} and retain on the r.h.s. only the terms  up to $m^2$, then, we obtain the coefficient $q_2$ in the form:
\begin{eqnarray}
\label{eq6b22}  
  						q_2 = \dfrac{f}{1-e},  
\end{eqnarray} 
with
\begin{eqnarray}
\label{eq6b23}  
   e  =      
          q\bigl [ q_0\cosh \bigl ( J\nabla_x \bigr ) +  1 - q_0 \bigr]^{q -1}   
           \bigl [ \cosh \bigl ( J\nabla_x \bigr ) -1 \bigr ] 
              g\bigl(x \bigr) \Bigr|_{x=0},   
\end{eqnarray} 
\begin{eqnarray}
\label{eq6b24}  
   f  =      
           \dfrac{q !}{2! (q -2)!} \bigl [ q_0\cosh \bigl ( J\nabla_x \bigr ) +  1 - q_0 \bigr]^{q -2}   
           \sinh^2 \bigl ( J\nabla_x \bigr ) 
              g\bigl(x \bigr) \Bigr|_{x=0}.   
\end{eqnarray} 
By putting  Eq. \eqref{eq6b21} into Eq.   \eqref{eq6b13} and retaining  on the r.h.s. only the terms up to $m^3$, one confirms that the coefficient $a$ is given by Eq.   \eqref{eq6b16}.

Now, one easily finds from Eq. \eqref{eq6b15} that the spontaneous magnetization is governed by the following expression in a vicinity of critical point:
\begin{eqnarray}
\label{eq6b19a}  
m = \left \{  
\begin{array}{ll} 
\sqrt{\dfrac{1 - a}{b}} & \, {\rm if} \, \, \, T < T_{\rm c}, \\ 
\qquad  \: \: 0 & \, \quad {\rm if} \, \, \, T \geq T_{\rm c}. 
\end{array} \right.
\end{eqnarray} 
Moreover, one can also verify that $a>1$ in the whole parameter space and hence, the inequality $b<0$ must be satisfied to ensure the existence of real continuous solutions of  Eq. \eqref{eq6b19a} near the critical temperature. Consequently, the second-order critical boundary (i.e. Curie temperatures) can be numerically calculated from the relations:
\begin{eqnarray}
\label{eq6b18}  
  								a = 1, \qquad   b <0
\end{eqnarray} 
It is also clear that if the parameter $b$ becomes positive in a certain region of the parameter space, then, the spontaneous magnetization must jump from nonzero to zero values, what indicates the first-order phase transition.  This behavior takes place below the tricritical point, which is a special critical point on a phase boundary that separates the first-order phase transitions from the second-order ones. The tricritical point can be calculated within the present formalism from the relations:
\begin{eqnarray}
\label{eq6b20}  
  								a = 1, \qquad   b  = 0.
\end{eqnarray}

\subsubsection{Discontinuous phase transitions}
It was already mentioned in the previous part that the spin-1 Blume-Capel model may exhibit the discontinuous (first-order) phase transition between the ferromagnetically ordered phase and the disordered paramagnetic phase in a relatively narrow parameter region around $D/J \approx -q/2$.  It is also well known that the best way to locate the curve of first-order phase transitions is 
based on the comparison of the Helmholtz free energies of the relevant phases. These phases coexist at all points along the first-order phase transition line, so that within the effective-field theory  this boundary can be mathematically found from the condition:
\begin{eqnarray}
\label{eq6b20a}  
  								F_{ferro}(\beta_{\rm c}) = F_{para}(\beta_{\rm c}),
\end{eqnarray}
where  $\beta_{\rm c} = 1/k_{\rm B}T_{\rm c}$, $F_{ferro}$ and $F_{para}$ are respectively given by (see Eq. \eqref{eq3b63}): 
\begin{eqnarray} 
\label{eq6b20b}  
                            F_{ferro}(\beta_{\rm c})  = - \frac{N}{\beta_c}  \ln (2S +1) + 
                             \frac{1}{\beta_{\rm c}} \int_0^{\beta_{\rm c}} U_{ferro} (\beta') d\beta'. 
\end{eqnarray}  
and
 \begin{eqnarray} 
\label{eq6b20c}  
                            F_{para}(\beta_{\rm c})  = - \frac{N}{\beta_{\rm c}}  \ln (2S +1) + 
                             \frac{1}{\beta_{\rm c}} \int_0^{\beta_{\rm c}} U_{para} (\beta') d\beta'. 
\end{eqnarray}  
The internal energies $U_{ferro}$ and $U_{para}$ are easily evaluated within the effective-field theories based on the differential operator technique.

\subsection{Exact van der Waerden identities for higher-spin Ising models}
In the previous section we have  used the differential operator technique to formulate the effective-field theory for the spin-1 Blume-Capel model. We have found that this theory is superior to the standard mean-field theory, although it is mathematically a bit more complex than that one of the spin-1/2 Ising model. These mathematical complications arise out from the generalized van der Waerden identity and it is therefore of principal importance to investigate the precise form of this identity for higher spin values $S \geq 1$.

It is well known that the following general expansion is valid for arbitrary spin value $S$: 
\begin{eqnarray}
\label{eq6b26}  
  					 \exp\bigl({\alpha S_j}\bigr ) =\sum_{k = 0}^{2S}  A_k(\alpha) S_j^k,
\end{eqnarray} 
where $\alpha$ is an arbitrary parameter,  $S_j= -S, -S+1, ...., S-1, S$ represents $2S+1$ projections of a spin towards the $z$ axis and $A_k(\alpha)$ are unknown expansion coefficients depending on a spin magnitude. Now, one gets the explicit form of van der Waerden identity  by setting all available values of spin into Eq. \eqref{eq6b26} and finding all relevant coefficients $A_k$ from the resulting set of equations. In this way, one of course obtains for the spin 1/2 the exact spin identity given by Eq.  \eqref{eq3b10} and for the spin 1 the exact spin identity given by Eq. \eqref{eq6b7}. 

Following this approach one may also obtain the explicit form of van der Waerden identities for larger spin values like for instance for $S = 3/2$:
\begin{eqnarray}
\label{eq6b27}  
\nonumber
  	\exp\bigl({\alpha S_j}\bigr ) =A_0 + A_1 S_j  + A_2 S_j^2 + A_3 S_j^3, 
\end{eqnarray} 
with the expansion coefficients:
\begin{eqnarray}
\label{eq6b28}  
  		&& A_0(\alpha) = \dfrac{1}{8}\biggl [ 9 \cosh\Bigl ( \dfrac{\alpha}{2}\Bigr) 
  														  - \cosh\Bigl ( \dfrac{3\alpha}{2}\Bigr) \biggr ] \\
\label{eq6b29}  
  		&& A_1(\alpha) = \dfrac{1}{12}\biggl [ 27 \sinh\Bigl ( \dfrac{\alpha}{2}\Bigr) 
  														  - \sinh\Bigl ( \dfrac{3\alpha}{2}\Bigr) \biggr ] \\  													
\label{eq6b30}  
  		&& A_2(\alpha) = \dfrac{1}{2}\biggl [  \cosh\Bigl ( \dfrac{3\alpha}{2}\Bigr) 
  														  - \cosh\Bigl ( \dfrac{\alpha}{2}\Bigr) \biggr ] \\
\label{eq6b31}  
  		&& A_3(\alpha) = \dfrac{1}{3}\biggl [  \sinh\Bigl ( \dfrac{3\alpha}{2}\Bigr) 
  		 														  - 3\sinh\Bigl ( \dfrac{\alpha}{2}\Bigr) \biggr ].  		 									
 \end{eqnarray} 
or for $S = 2$:
\begin{eqnarray}
\label{eq6b32}  
\nonumber
 \exp\bigl({\alpha S_j}\bigr ) =A_0 + A_1 S_j  + A_2 S_j^2 + A_3 S_j^3 + A_4 S_j^4, 
\end{eqnarray} 
with the expansion coefficients:
\begin{eqnarray}
\label{eq6b33}  
			&& 	A_0(\alpha)  = 1\\
\label{eq6b34}
  			&& A_1(\alpha) = \dfrac{1}{6}\Bigl [ 8 \sinh( \alpha )  - \sinh ( 2\alpha) \Bigr ] \\
\label{eq6b35}  
  			&& A_2(\alpha) = \dfrac{1}{12}\Bigl [ 16 \cosh( \alpha )  - \cosh ( 2\alpha) - 15 \Bigr ] \\ 									\label{eq6b36}  
			&& 	A_3(\alpha) = \dfrac{1}{6}\Bigl [ 8 \sinh( 2\alpha )  - 2\sinh ( \alpha) \Bigr ] \\
\label{eq6b37}  
  			&& A_4(\alpha) = \dfrac{1}{12}\Bigl [  \cosh( 2\alpha )  - 4\cosh ( \alpha) + 3\Bigr ].							 
\end{eqnarray}  
It is clear from the aforementioned examples that the complexity of van der Waerden identity rapidly increases with increasing of the spin value. Moreover, one can also see that  a consistent application of the differential operator technique requires for an arbitrary value of $S$ to account for $2S$ expectations $\langle S_j ^n\rangle , \; n = 1, \ldots, 2S $, that naturally bring
additional complexity into formulation of the effective-field theory.  Nonetheless, the effective-field theory based on the differential operator technique  has been successfully applied  
to investigate many higher-spin  systems under very diverse physical conditions.   

To complete our discussion, one should notice that the problems mentioned above can be avoided using some clever approximations for the van der Waerden identity. One possible approach  to such a formulation of the effective-field theory, which is also based on the differential operator technique, will be the subject matter of the subsequent part.

\subsection{Approximate van der Waerden identities for higher-spin Ising models}
The increasing mathematical complexity of the effective-field formulation makes analysis of many physical systems with higher spins too much intricate and time-consuming. For that reason it would be extremely useful somehow to simplify and/or approximate the exact van der Waerden identities for higher spin values. Of course, one is limited by the requirement of obtaining  numerical results that will be accurate enough for a wide class of complex spin systems when looking for such a simplification.

A possible way how to avoid annoying complexity of effective-field theory for the higher-spin problems has been found by Kaneyoshi, Tucker and Ja\v{s}\v{c}ur \cite{kane93bc1} by proposing a mathematical procedure simplifying the exact van der Waerden identity. This approach is based on approximating some terms in Taylor series of the exponential function entering the general expansion \eqref{eq6b26}. Following the reference \cite{kane93bc1}, we at first  formally separate the Taylor series of $\exp(\alpha S_i)$ into even and odd powers to obtain:   
\begin{eqnarray}
\label{eq6b37a}  
  						 \exp\bigl({\alpha S_i}\bigr ) = \sum_{n=0}^{\infty} \frac{\alpha^{2n}}{(2n)!} S_i^{2n}
  						                                          + \sum_{n=0}^{\infty} \frac{\alpha^{2n+1}}{(2n+1)!} S_i^{2n+1}.
\end{eqnarray}
Now,  after introducing the following approximations  $S_i^{2n} \approx \bigl \langle S_i^2 \bigr \rangle^ n$, $S_i^{2n +1} \approx S_i \bigl \langle S_i^2 \bigr \rangle ^n$ and defining $\eta^2 =\bigl \langle S_i ^2 \bigr \rangle$ one may rewrite the previous equation to the form:
\begin{eqnarray}
\label{eq6b38}  
  						 \exp\bigl({\alpha S_i}\bigr ) = \cosh\bigl ( \alpha  \eta \bigr) +\dfrac{S_i}{\eta} 
  						  \sinh\bigl (\alpha   \eta \bigr),
\end{eqnarray} 
which represents the generalized but approximate van der Waerden identity. One should notice here that in the special case of $ S= 1/2$ we have  $\eta =\sqrt{\bigl \langle S_i ^2 \bigr \rangle }= 
1/2$, thus, Eq. \eqref{eq6b38} reduces to the exact relation \eqref{eq3b10}. On the other hand, the relation \eqref{eq6b38} is always approximate for  $\forall S \geq 1$, nonetheless providing  numerical results that are accurate enough in comparison with those obtained using the exact van der Waerden identity. 

It is apparent from Eq. \eqref{eq6b38} that the complexity of effective-field theory for higher-spin Ising models will be significantly reduced if we use the approximate instead of the exact van der Waerden identity \eqref{eq6b26}. Now, let us illustrate the applications of approximate van der Waerden identity in developing of the single-spin cluster effective-field theory for the spin-$S$ Blume-Capel model with an arbitrary spin value $S \geq 1$.
 
The Hamiltonian of the spin-$S$ Blume-Capel model takes formally the same form as that one given by Eq. \eqref{eq6b1}:
\begin{eqnarray} 
\label{eq6b39}
             {\cal  H} = - J \sum_{\langle ij \rangle} S_i  S_j  - D \sum_{i=1}^N S_i ^2 - h \sum_{i=1}^N  S_i,
\end{eqnarray}  
For an arbitrary but fixed value of spin, the spin variables $S_i$ are allowed to take $(2S +1)$ possible values ranging from $-S, -S+1$ up to $S$.  The meaning of other parameters is the same as described earlier for the Hamiltonian \eqref{eq6b1}. One should emphasize here that the generalization of the spin-$S$ Blume-Capel model for arbitrary spin values does not bring any significant modification in deriving a single-spin cluster effective-field theory. Therefore,  starting from the generalized Hamiltonian \eqref{eq6b39} and repeating the same steps as in the subsection  
\ref{BC_1} one simply arrives at the following exact Callen-Suzuki identities: 
\begin{eqnarray} 
\label{eq6b38a}  
      \bigl \langle S_i  \bigl\{   f_i \bigr\} \bigr \rangle =  \bigl \langle \bigl\{   f_i \bigr\} F_S(E_i) \bigr \rangle,
\end{eqnarray} 
\begin{eqnarray} 
\label{eq6b38b}  
        \bigl \langle S_i^2  \bigl\{   f_i \bigr\} \bigr \rangle = \bigl \langle \bigl\{   f_i \bigr\}G_S(E_i) \bigr \rangle,
\end{eqnarray} 
where $E_i = J \sum _{j=1}^q S_j$ and the functions $F_S(x)$ and $G_S(x)$ are defined as: 
\begin{eqnarray}
\label{eq6b40}  
  						F_S(x) =  \dfrac{\sum\limits_{k =-S}^{S}  k \exp\bigl(\beta D k^2\bigr) \sinh\bigl[ \beta k(x +h)\bigr ]}
  						{\sum\limits_{k =-S}^{S}   \exp\bigl(\beta D k^2\bigr)\cosh\bigl[ \beta k(x +h)\bigr ]},
\end{eqnarray} 
\begin{eqnarray}
\label{eq6b41}  
  						G_S(x) = \dfrac{\sum\limits_{k =-S}^{S}  k^2 \exp\bigl(\beta D k^2\bigr) \cosh\bigl[ \beta k(x +h)\bigr ]}
  						{\sum\limits_{k =-S}^{S}   \exp\bigl(\beta D k^2\bigr)\cosh\bigl[ \beta k(x +h)\bigr ]}.
\end{eqnarray}  
Similarly as in the previous text, the symbol  $\bigl\{   f_i \bigr\}$  represents an arbitrary functions of spin variables with exception of those involving the central spin $S_i$.

Now, one may rewrite the relations \eqref{eq6b38a} and  \eqref{eq6b38b} using Eqs. \eqref{eq3b10}, \eqref{eq6b38} by setting $\bigl\{   f_i \bigr\} = 1$:
\begin{eqnarray}
\label{eq6b42}  
  				 m  =  \bigl \langle S_i \bigr \rangle     =
          					\Bigl  \langle 
          						\prod_{j= 1}^q   \Bigl [ \cosh \bigl ( J \eta \nabla_x \bigr ) + 
        						\dfrac{S_j}{\eta}          \sinh \bigl ( J \eta \nabla_x \bigr )  \Bigr]  
                              			 \Bigr \rangle F_S \bigl(x \bigr) \Bigr|_{x=0},  
\end{eqnarray}  
\begin{eqnarray}
\label{eq6b43}  
      			\eta^2 =    \bigl \langle S_i^2 \bigr \rangle     =
          					\Bigl  \langle 
          						\prod_{j= 1}^q     \Bigl [ \cosh \bigl ( J \eta \nabla_x \bigr ) + 
        						\dfrac{S_j}{\eta}          \sinh \bigl ( J \eta \nabla_x \bigr )  \Bigr]
                              			 \Bigr \rangle G_S \bigl(x \bigr) \Bigr|_{x=0}.
\end{eqnarray}  
Here, one should notice that Eqs. \eqref{eq6b42} and \eqref{eq6b43} are not exact unlike of those ones obtained using the exact van der Waerden identity (c.f. Eqs.  \eqref{eq6b13} and \eqref{eq6b14}), nevertheless, the resulting loss of the numerical accuracy is negligible.

In order to obtain the single-spin cluster effective-field theory based on the approximate van der Waerden identity we again use the approximation scheme \eqref{eq6b12} to split the multispin correlations on the  r.h.s. of  Eqs. \eqref{eq6b42} and \eqref{eq6b43}. In this way one obtains the following coupled equations for the spontaneous magnetization $m$ and the quadrupolar moment 
$q_D = \eta^2$:
\begin{eqnarray}
\label{eq6b44}  
  				 m  =      
          					   \Bigl [ \cosh \bigl ( J \eta\nabla_x \bigr ) + 
        						\dfrac{m}{\eta}          \sinh \bigl ( J \eta \nabla_x \bigr )  \Bigr]^{q}  
                              			 F_S \bigl(x \bigr) \Bigr|_{x=0},  
\end{eqnarray}  
\begin{eqnarray}
\label{eq6b45}  
  				\eta^2 =      
          					   \Bigl [ \cosh \bigl ( J \eta \nabla_x \bigr ) + 
        						\dfrac{m}{\eta}          \sinh \bigl ( J \eta \nabla_x \bigr )  \Bigr]^{q}  
                              			 G_S \bigl(x \bigr) \Bigr|_{x=0}.  
\end{eqnarray}  
It should be emphasized that changing the spin value in the Hamiltonian of the system under investigation modifies only the functions $F_S(x)$ and $G_S(x)$, while Eqs. \eqref{eq6b44} and \eqref{eq6b45} remain untouched, which is the principal advantage of this approach. Moreover, these equations possess only the mathematical complexity of the simplest spin-1 Blume-Capel model independently of a spin value. This huge simplification  is very clear manifestation of the power of approximate van der 
Waerden identity.  

Owing to the aforementioned simplicity, one can again apply the binomial theorem and rewrite  previous relations as: 
\begin{eqnarray} 
\label{eq6b46}  
        m  =  \sum_{k=0}^{ q}  { q \choose  k} \dfrac{m^k}{\eta^k} 
        									A^{( q)}_k,                                               
\end{eqnarray}
\begin{eqnarray} 
\label{eq6b47}  
        q  =  \eta^2 = \sum_{\ell=0}^{ q}  { q \choose  \ell} \dfrac{m^\ell}{\eta^\ell} 
        									B^{( q)}_\ell,                                              
\end{eqnarray} 
where using the same mathematical procedures as in section \ref{0S_EFT} we obtain the corresponding coefficients in the form:
\begin{eqnarray} 
\label{eq6b48}  
\nonumber 
  A^{( q)}_k =A^{(q)}_k (\beta J, \eta, h) =    \dfrac{1}{2^q} \sum_{i=0}^{q-k} \sum_{j=0}^{k} \bigl ( -1 \bigr)^j   
  {q - k \choose  i}{ k \choose   j}F_S \bigl[\eta\beta  J\bigl(q -2i -2j + \tilde{h}  \bigr)\bigr]\\
\end{eqnarray} 
and
\begin{eqnarray}
\label{eq6b49}  
\nonumber 
 B^{( q)}_\ell = B^{(q)}_\ell (\beta J, \eta, h) =    \dfrac{1}{2^q} \sum_{i=0}^{q-\ell} \sum_{j=0}^{\ell} \bigl ( -1 
 \bigr)^j   {q - \ell \choose  i}{ \ell \choose   j}G_S \bigl[\eta\beta  J\bigl(q -2i -2j + \tilde{h}  \bigr)\bigr].\\
\end{eqnarray} 

Thus, we have obtained a closed systems of equations that are valid for arbitrary crystalline lattice with the coordination number $q$ and arbitrary spin values. It is useful to emphasize here that the mathematical complexity of these equations is of nearly the same level as those of the simple spin-1/2 Ising model (c.f. Eqs. \eqref{eq3b14} and  \eqref{eq3b15}). This is an extremely pleasant consequence of the approximate van der Waerden identity, the application of which is the keystone in developing a powerful effective-field theories for various physical systems. 

The previous part \ref{Curie_temp_BC1} was devoted to a comprehensive analysis of phase boundaries of the spin-1 Blume Capel model within the single-spin cluster formulation of the effective-field theory. Since the physical arguments remain the same also for phase boundaries of the spin-$S$ Blume-Capel model, we will not repeat them again here. Instead, let us directly proceed to a derivation of relevant phase boundaries and tricritical point for the more general spin-$S$ Blume-Capel model with an arbitrary spin value. If we set $h=0$ in Eqs. \eqref{eq6b40} and \eqref{eq6b41} and make relevant expansions of  Eqs. \eqref{eq6b44} and \eqref{eq6b45} for small values of the magnetization, then, one obtains formally the same equations for the continuous phase transitions and tricritical point as in subsection \ref{Curie_temp_BC1}, i.e., 
\begin{eqnarray}
\label{eq6b50}  
  								a= 1, \qquad   b <0 
\end{eqnarray} 
and
\begin{eqnarray}
\label{eq6b51}  
  								a= 1, \qquad   b = 0. 
\end{eqnarray} 
However, the coefficients $a$ and $b$ are now respectively given by: 
\begin{eqnarray}
\label{eq6b52}  
    a  =       
        \dfrac{ q}{\eta_0}  \sinh\bigl ( J\eta_0 \nabla_x \bigr )   
        \Bigl [ \cosh \bigl ( J\eta_0 \nabla_x \bigr )  \Bigr] ^{ q -1}  
             F_S\bigl(x \bigr) \Bigr|_{  x=0},   
\end{eqnarray}  
and
\begin{eqnarray}
\label{eq6b53}  
\nonumber  
\!\!\!\! b   &=&        
            \dfrac{1}{\eta_0^3} \dfrac{q !}{3! (q-3)!}\bigl [ \cosh \bigl ( J \eta_0 \nabla_x \bigr )  \bigr]^
           {q-3}  \sinh^3 \bigl ( J \eta_0 \nabla_x \bigr ) F_S\bigl(x \bigr) \Bigr|_{ x = 0} 
           \\ \nonumber   \\ \nonumber 
          &-&  \dfrac{q_1}{\eta_0^3} \dfrac{q}{2} \bigl [ \cosh \bigl ( J \eta_0 \nabla_x \bigr )  \bigr]^
           {q -1}  \sinh \bigl ( J \eta_0 \nabla_x \bigr ) F_S\bigl(x \bigr) \Bigr|_{ x = 0}  \\ \nonumber \\ \nonumber
           &+& \dfrac{q_1}{\eta_0^2}  \beta J \dfrac{q(q -1)}{2}
           \bigl [ \cosh \bigl ( J \eta_0 \nabla_x \bigr )   \bigr]^{q}  \sinh^2 \bigl ( J \eta_0 \nabla_x \bigr ) 
           \tilde {F_S}\bigl(x \bigr) \Bigr|_{ x = 0} \\ \nonumber \\ 
        &+&   \dfrac{q_1}{\eta_0^2} \beta J  \bigl [ \cosh \bigl ( J \eta_0 \nabla_x \bigr )  \bigr]^
           {q}  \tilde {F_S}\bigl(x \bigr) \Bigr|_{ x=0},
\end{eqnarray} 
where the quantity $\eta_0$ can be evaluated from the relation: 
\begin{eqnarray}
\label{eq6b54}  
  				 q_0 = \eta_0^2 =       
          					   \Bigl [ \cosh \bigl ( J \eta_0 \nabla_x \bigr )  \Bigr]^{q}  
                              			 \ G_S \bigl(x \bigr) \Bigr|_{ x=0}.  
\end{eqnarray}  
The  quantity $q_1$ can be explicitly obtained substituting the expansion  $\eta^2 = \eta_0^2 + q_1 m^2$ into Eq.
\eqref{eq6b45} and performing once more the Taylor expansion and retaining only the terms up to $m^2$. 
Completing this calculations one obtains  $q_1$ in the form:
\begin{eqnarray}
\label{eq6b55}  
  				 q_1 = \dfrac{e}{1-f},
\end{eqnarray}  
where the coefficients $e$ a $f$ are respectively given by the relations:
\begin{eqnarray}
\label{eq6b56}  
    e  =       
          \dfrac{1}{\eta_0^2} \dfrac{q !}{2! (q -2)!} \sinh^2\bigl ( J\eta_0 \nabla_x \bigr )   
        \Bigl [ \cosh \bigl ( J\eta_0 \nabla_x \bigr )  \Bigr] ^{ q -2}  
             G_S\bigl(x \bigr) \Bigr|_{ x=0},   
\end{eqnarray}  
and 
\begin{eqnarray}
\label{eq6b57}  
    f =       
          \dfrac{1}{\eta_0} \dfrac{q }{2} \beta J\sinh\bigl ( J\eta_0 \nabla_x \bigr )   
        \Bigl [ \cosh \bigl ( J\eta_0 \nabla_x \bigr )  \Bigr] ^{ q -1}  
            \tilde {G_S}\bigl(x \bigr) \Bigr|_{ x=0},  
\end{eqnarray} 
while the function $\tilde {F_S}$ a $\tilde {G_S}$ are defined as:
\begin{eqnarray}
\label{eq6b58}  
    \tilde {F_S} (x) = \nabla_x F_S(x), \qquad \tilde {G_S} (x) = \nabla_x G_S(x) .
\end{eqnarray}
\begin{table}[tb]
\centering 
\begin{tabular}{lcccc} 
\hline\hline 
                             & MFA           &  MFA       &   SEFA    & SEFA    \\ 
$q$                &$k_{\rm B} T_{\rm t}/J$ & $D_{\rm t}/J $  & $k_{\rm B} T_{\rm t}/J$ & $D_{\rm t}/J $ \\ 
\hline \hline 
3                 & 1.0000         & -1.3863           &     0.7137      & -1.4412    \\
4                & 1.3333          & -1.8484           &    1.0884      & -1.8848    \\
6                & 2.0000          & -2.7726           &    1.7382      & -2.8068    \\
\hline 
\end{tabular}
\caption{The locus of the tricritical points for the spin-1  Blume-Capel on lattices with different values of the coordination number $q$ within the mean-field (MFA) and the single-spin cluster effective-field (SEFA) approximation (see Ref. \cite{kane93tim}).}
\label{spin1}
\end{table}
\begin{table}[!b]
\caption{The same as in Table \ref{spin1}, but for the spin-2 Blume-Capel model (see Ref. \cite{kane93tim}).}
\bigskip
\centering 
\begin{tabular}{lcccc} 
\hline\hline 
                             & MFA           &  MFA       &   SEFA    & SEFA    \\ 
$q$                &$k_{\rm B} T_{\rm t}/J$ & $D_{\rm t}/J $  & $k_{\rm B} T_{\rm t}/J$ & $D_{\rm t}/J $ \\ 
\hline \hline 
3                & 1.5229         & -1.4931            & 0.8340      & -1.4955    \\
4                & 2.0306          & -1.9911           & 1.6456      & -1.9940  \\
6                & 3.0459          & -2.9866           & 2.6436      & -2.9892    \\
\hline 
\end{tabular}
\label{spin2}
\end{table}
The numerical accuracy of the single-spin cluster effective-field theory is illustrated in Tables \ref{spin1} and \ref{spin2}, where we have listed the coordinates of tricritical points for the spin-1 and spin-2 Blume-Capel model on several crystal lattices. These results again confirm notable improvement over the mean-field estimate for the tricritical point and moreover, the significant improvement concerns also other physical quantities of the spin-$S$ Blume-Capel systems when
they are treated within the effective-field theories.  

To conclude, the single-spin cluster effective-field theory has successfully been applied to investigate diverse Ising spin systems of rather complex nature.  Among the most important applications
one must point out the work of Kaneyoshi et al. \cite{kane93tim}, in which the authors have utilized the approximate effective-field theory to calculate the phase diagrams, magnetization, internal energy and specific heat of the spin-$S$ transverse Ising model on arbitrary crystalline lattice. Another very important application represents the extension  of the method to the two-spin cluster approximation, which has been done by Jur\v{c}i\v{s}in et al. \cite{jurc96bc}. Numerical results obtained in  all these applications have confirmed  very good numerical accuracy of the method
and revealed a huge application potential to further complex systems.

\subsection{Spin-$S$ Ising model in a transverse field within a single-spin cluster approach}
Now, let us briefly sketch the application of the effective-field theory developed in the previous part to the transverse  Ising model with an arbitrary spin.
The spin-$S$ Ising model in a longitudinal as well as transverse magnetic field is described by the Hamiltonian:
\begin{eqnarray} 
\label{eq7b1}
             {\cal \hat H} = - J \sum_{\langle ij \rangle} \hat S_i^z \hat S_j^z  - h \sum_{i=1}^N \hat S_i^z -  \Omega \sum_{i=1}^N \hat S_i^x,
\end{eqnarray} 
For  each spin value,  the  operators  $\hat S_i^x$ and $\hat  S_i^z$  ($i = 1, \dots , N $) can be expressed  as $(2S +1) \times (2S +1)$  symmetric matrices with the following elements:
\begin{eqnarray} 
\label{eq7b2}
           &&\bigl \langle   \sigma | \hat {S}_i^x  | \sigma'  \bigr \rangle =
           \bigl \langle   \sigma' | \hat {S}_i^x  | \sigma  \bigr \rangle =
           \dfrac{1}{2}\delta_{\sigma', \sigma - 1}\sqrt{( S + \sigma)(S - \sigma + 1)}, \\
         &&  \bigl \langle   \sigma | \hat {S}_i^z  | \sigma'  \bigr \rangle =
           \bigl \langle   \sigma' | \hat {S}_i^z | \sigma  \bigr \rangle = \sigma \delta_{\sigma, \sigma' },
\end{eqnarray}
where $\delta$ is the usual Kronecker delta symbol, $\sigma$ and $ \sigma'$ take $ (2S +1) $ possible values, namely,  
$-S, -S +1, \dots,  S -1,  S  $. The other parameters in the Hamiltonian  have the same  meaning as those in Eq. \eqref{eq5b1}.
The partial Hamiltonian ${\cal \hat H}_i $ entering the Callen-Suzuki identity takes formally the form as that one in section 
\ref{tim12efa}, i.e.
\begin{eqnarray} 
\label{eq7b3}
         \hat{\cal H}_i   =  -\hat  S_i^z \Bigl (J \sum_{j=1}^{q}  \hat S_j^z + h \Bigr ) 
                                 -   \Omega \hat  S_i^x.
\end{eqnarray}  
Applying the rotation transformation \eqref{eq5b8} one may rewrite the Hamiltonian  $\hat{\cal H}_i$  into the following diagonal form:
\begin{eqnarray} 
\label{eq7b3b}
         \hat{\cal H}_i   =  -\hat  S_i^{z'} \sqrt{\Omega^2 + E_i^2 }, \quad \mbox{with} \;  E_i = J \sum_{j=1}^q  S_j^{z'}.
\end{eqnarray}  
Following the same steps as in previous sections, we obtain three approximate Callen-Suzuki identities:  
\begin{eqnarray} 
\label{eq7b4}  
      \bigl \langle \hat S_i^x  \bigl\{   f_i \bigr\} \bigr \rangle =  \bigl \langle \bigl\{   f_i \bigr\} F^x_S(E_i) \bigr \rangle,
\end{eqnarray} 
\begin{eqnarray} 
\label{eq7b5}  
      \bigl \langle \hat S_i^z  \bigl\{   f_i \bigr\} \bigr \rangle =  \bigl \langle \bigl\{   f_i \bigr\} F^z_S(E_i) \bigr \rangle,
\end{eqnarray} 
and
\begin{eqnarray} 
\label{eq7b6}  
      \bigl \langle (\hat S_i^z)^2  \bigl\{   f_i \bigr\} \bigr \rangle =  \bigl \langle \bigl\{   f_i \bigr\} G^z_S(E_i) \bigr \rangle.
\end{eqnarray} 
The functions $F^x_S(x)$ and $F^z_S(x)$ can be expressed as:
\begin{eqnarray}
\label{eq7b7}  
  						F^x_S(x) = \dfrac{\Omega}{\sqrt{\Omega^2 + (x + h)^2}} \dfrac{\sum\limits_{k =-S}^{S}  
  						k  \sinh\bigl[ \beta k\sqrt{\Omega^2 + (x + h)^2}\bigr ]}
  						{\sum\limits_{k =-S}^{S}   \cosh\bigl[ \beta k\sqrt{\Omega^2 + (x + h)^2}\bigr ]}
\end{eqnarray} 
and
\begin{eqnarray}
\label{eq7b8}  
						F^z_S(x) = \dfrac{x + h}{\sqrt{\Omega^2 + (x + h)^2}} \dfrac{\sum\limits_{k =-S}^{S}  
  						k  \sinh\bigl[ \beta k\sqrt{\Omega^2 + (x + h)^2}\bigr ]}
  						{\sum\limits_{k =-S}^{S}  \cosh\bigl[ \beta k\sqrt{\Omega^2 + (x + h)^2}\bigr]}.  						
\end{eqnarray}  
On the other hand, the function  $G^z_S(x) $ must  be calculated for each particular value of spin separately and for instance
for $S = 1$ it is  given by
 \begin{eqnarray}
\label{eq7b9}  
  				G^z_1(x) = \dfrac{\Omega^2 + \bigl[2\Omega^2 + (x + h)^2 \bigr ]\cosh\bigl[ \beta \sqrt{\Omega^2 + (x + h)^2}\bigr ]}
  				{\bigl( \Omega^2 + (x + h)^2\bigr ) \bigl( 1 + \cosh\bigl[ \beta \sqrt{\Omega^2 + (x + h)^2}\bigr ] \bigr)}.	
\end{eqnarray} 
After putting ${f_i} = 1$ to the approximate Callen-Suzuki identities \eqref{eq7b4}-\eqref{eq7b6} and utilizing the approximate van der Waerden identity \eqref{eq6b38} and decoupling approximation \eqref{eq6b12} one obtains from Eqs. \eqref{eq7b7}-\eqref{eq7b9} three inter-connected relations for the magnetizations and quadrupolar moment $m_x$, $m_z$ and $\eta^2 $, which can be written in the following compact form:
\begin{eqnarray} 
\label{eq7b9b}
  							\begin{pmatrix} 
                                    m_x\\
			                        m_z \\
			                        \eta^2  
                           \end{pmatrix} =                  
                                                  \Bigl [ \cosh \bigl ( J \eta\nabla_x \bigr ) + 
        		           \dfrac{m}{\eta}          \sinh \bigl ( J \eta \nabla_x \bigr )  \Bigr]^{q}      
                           \begin{pmatrix} 
                                    F_S^x \bigl(x \bigr) \\
			                        F_S^z \bigl(x \bigr) \\
			                       G_S^z \bigl(x \bigr)
                           \end{pmatrix}_{x=0} .
\end{eqnarray}
\begin{figure}[tb]
\begin{center}
\includegraphics[clip,width=8.0cm]{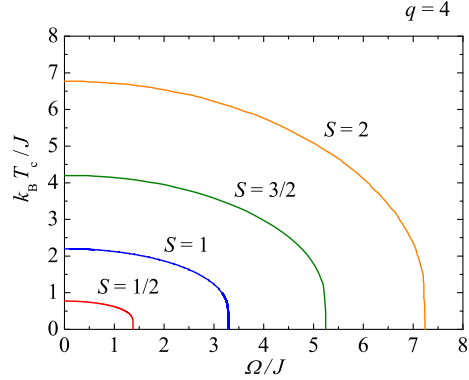}
\vspace{-0.7cm}
\caption{Critical temperature as a function of the transverse magnetic field for the spin-$S$ Ising model on a square lattice ($q=4$).}
\label{Tc_S}
\end{center}
\end{figure}    
At this stage, it is clear that all relevant physical quantities can be calculated using the procedures described in previous section, so that we will not repeat details here. Instead of reproducing these calculations, we have shown in Fig. \ref{Tc_S} the finite-temperature  phase diagrams for the transverse Ising model with spin 1/2, 1, 3/2 and 2 on the square lattice. The results indicate that the behavior of the systems with different spins is very similar, however, the results indicate that both the values of Curie temperature as well as the critical value of transverse field increase with the increasing spin of the system.   The accuracy of the effective-field results is satisfactory as it is shown in Table \ref{Tc_Soit}, in which we compare the estimations of $\Omega_{\rm c}$ obtained for coordination numbers within different theoretical methods. 

\begin{table}[tb]
\centering 
\begin{tabular}{cccccccc} 
\hline
\multicolumn{8}{c}{$\Omega_{\rm c}/J$}\\
\hline
                  &  MFA                  & PA                   &  SEFA           & MFA          & PA             & SEFA           & SE                  \\ 
$q$            &  $S = 3/2$         & $S = 3/2$       &  $S = 3/2$    & $S = 2$     & $S = 2$     & $S = 2$       & $S = 2$     \\ 
\hline \hline 
3                & 4.50                   & 3.00                & 3.74            & 6.00          & 4.00           & 5.21           & \\
4                & 6.00                   & 4.50                & 5.25            & 8.00          & 6.00           & 7.23           & \\
6                & 9.00                   & 7.50                & 8.26            & 12.00        & 10.00         & 11.25         & 11.60\\
8                & 12.00                 & 10.50              & 11.26          & 16.00        & 14.00         & 15.25         & 15.60\\
12              & 18.00                 & 16.50              & 17.26          & 24.00        & 22.00         & 23.25         & 23.35\\
\hline 
\end{tabular}
\caption{Critical values of the transverse field $\Omega_{\rm c}/J$  for the spin-3/2 and spin-2 Ising models on various crystal lattices obtained within the mean-field approximation (MFA),  the pair approximation (PA) \cite{yqma92}, single-spin cluster approximation (SEFA), and series expansion (SE) \cite{oitm81}. }
\label{Tc_Soit}
\end{table}
\subsection{Spin-1/2 Ising model within a two-spin cluster approach}  
\label{two-spinclustereft}

In the previous sections we have shown how the effective-field theory based on the differential operator technique
can be developed for various Ising-like spin systems. The common feature of all these applications is the usage of the 
single-spin Callen-Suzuki identity, which  always serves as a starting point for calculation of relevant physical quantities.    On the 
other hand, it is evident from Eq. \eqref{eq1b17} that   the Callen-Suzuki identity can be very simply extended for more than one 
spin, which opens a possibility to develop more accurate and complex effective-field theories. 

The first generalization along this direction has been done by Bob\'{a}k and Ja\v{s}\v{c}ur \cite{boby86}, who
have proposed to use as a starting point the Callen-Suzuki identity for  two nearest-neighbor spins. Using the
differential operator technique the authors have subsequently developed a more accurate theory, which is usually called in the literature 
as two-spin cluster effective-field theory.  This generalized effective-field theory exactly accounts for all 
single-site kinematic relations and additionally all interactions of a central pair of spins are also treated exactly.  
Except of further improvement of numerical accuracy the generalized effective-field theory brings also further crucial 
benefits in comparison with the single-spin cluster counterpart of the effective-field theory.  The  most remarkable improvement represents  the 
possibility to take into account different topology of lattices with the same coordination number.  In fact, the two-spin 
cluster effective-field theory is able to discern for example between the magnetic behavior of the Ising model on square and kagom\'e lattices 
with the same coordination number $q=4$, or that one of the Ising model on triangular and simple cubic lattices with the coordination number $q=6$ \cite{tomc87}. Moreover, the two-spin cluster effective-field theory in combination with the single-spin version serves as a basis for developing the effective-field renormalization-group theory, which works well in the 
critical region of various Ising models  \cite{liya88}.

Now, let us formulate the two-spin cluster approximation for the simplest possible case of the spin-1/2 Ising model defined by the Hamiltonian  \eqref{eq3b1}. Similarly as in the section \ref{0S_EFT}, we at first separate total Hamiltonian into two parts: 
\begin{eqnarray} 
\label{eq4b1}
             {\cal H} =  {\cal H}_{ij}  +  {\cal H}^{'} , 
\end{eqnarray} 
where two-spin cluster Hamiltonian
\begin{eqnarray} 
\label{eq4b2}
             {\cal H}_{ij} = - J \sigma_i \sigma_j -\sigma_i \Bigl (J \sum_{k= 1}^{q-1} \sigma_k + h\Bigr) 
                -\sigma_j \Bigl (J \sum_{\ell = 1}^{q-1} \sigma_\ell + h\Bigr) 
\end{eqnarray}
accounts for all interaction terms of the central two-spin cluster with its nearest neighbors and external magnetic field. All remaining interaction terms of the total Hamiltonian are included in the Hamiltonian ${\cal H}^{'}$.

Now, we substitute  ${\cal \hat O}_{ij ... n}  = (\sigma_i + \sigma_j )/2 $ and $ \{   f_{ij ... n} \} =1 $ into Eq.  \eqref{eq1b17} and  obtain the generalized Callen-Suzuki identity in the form:
\begin{eqnarray} 
\label{eq4b3}  
     m \equiv    \bigl \langle \frac12 (\sigma_i + \sigma_j) \bigr \rangle =                
          \biggl \langle \dfrac{\displaystyle{ \sum_{ \sigma_i, \sigma_j   =\pm \frac{1}{2}} }  \;
                                 \frac12 \bigl  (\sigma_i + \sigma_j \bigr )\exp \bigl (-\beta {\cal H}_{ij}  \bigr)  }
                  { \displaystyle{ \sum_{ \sigma_i, \sigma_j   =\pm \frac{1}{2}} }  \;
                                 \exp \bigl (-\beta {\cal H}_{ij}  \bigr) }    \biggr \rangle.
\end{eqnarray} 
Let us denote the local fields in Eq. \eqref{eq4b2} as: 
\begin{eqnarray} 
\label{eq4b4}
             E_i =  J \sum_{k= 1}^{q -1}  \sigma_k  \quad \mbox{and} \quad     
             E_j =  J \sum_{\ell= 1}^{q-1}  \sigma_\ell.   
\end{eqnarray}
After performing the summation over the central two-spin cluster in Eq. \eqref{eq4b3} one gets the following explicit form of the two-spin Callen-Suzuki identity: 
\begin{eqnarray} 
\nonumber
       m \equiv     \bigl \langle \frac12 (\sigma_i + \sigma_j) \bigr \rangle =                
          \Biggl \langle \dfrac{  \sinh\bigl [ \frac\beta2 (E_i + E_j)  +  \beta h \bigr] }
                  { 2 \cosh\bigl [ \frac\beta2 (E_i + E_j  )+  \beta h \bigr] + 2\mbox{e}^{-\frac{\beta J}{2}} 
                       \cosh\bigl [ \frac\beta2 (E_i - E_j   )  \bigr] }  \Biggr \rangle . \\ 
\label{eq4b5}  
\end{eqnarray} 
As we have already noted, this exact equation can be used as a starting point for developing various approximate 
theories for the spin-1/2 Ising model. 

Similarly as in the case of the single-spin Callen-Suzuki identity (see \eqref{eq3b5}) by averaging arguments of hyperbolic functions in 
Eq. \eqref{eq4b5} one obtains the following simple equation for the reduced magnetization:
\begin{eqnarray} 
\label{eq4b6}  
          m =                
           \dfrac{  \sinh\bigl [ \beta J (q - 1)m +  \beta h \bigr] }
                  {  \cosh\bigl [ \beta J (q - 1)m +  \beta h  \bigr] + \mbox{e}^{-\frac{\beta J}{2}}} , 
\end{eqnarray} 
which is identical with that obtained within the standard Oguchi approximation \cite{smart66}. The Oguchi approximation is numerically more accurate than the standard mean-field theory, nevertheless, it still fails to discern between the magnetic properties of the Ising model on lattices with the same coordination numbers but different geometry. 

In order to evaluate the average on the r.h.s of Eq. \eqref{eq4b5} more accurately than within the Oguchi approximation, we may utilize the following relation:
\begin{eqnarray} 
\label{eq4b7}  
          \mbox{e}^{\lambda_x \nabla_x + \lambda_y \nabla_y}f(x, y)  =  f(x + \lambda_x,  y + \lambda_y ), 
\end{eqnarray} 
where $\nabla_x = \partial/\partial x$,  $ \nabla_y = \partial/\partial y$ and $\lambda_x$, $\lambda_y$ are arbitrary parameters.
Using the relation \eqref{eq4b7} we may rewrite Eq. \eqref{eq4b5} into the form:
\begin{eqnarray} 
\label{eq4b8}  
       m =   \dfrac{1}{2}\bigl \langle \sigma_i + \sigma_j \bigr \rangle =                
          \Bigl \langle \mbox{e}^{  (E_i +E_j)\nabla_x +   (E_i - E_j)\nabla_y}  \Bigr \rangle F(x, y)
          \Bigl|_{x= 0, y=0}.        
\end{eqnarray} 
where for the sake of simplicity we have  defined the function $F(x,y)$ as:
\begin{eqnarray} 
\label{eq4b9}  
            F(x, y) =
          \dfrac{ \sinh\bigl [ \frac\beta2 (x  +   2h) \bigr] }
                  {  2\cosh\bigl [ \frac\beta2 (x  +  2h)  \bigr] + 2\mbox{e}^{-\frac{\beta J}{2}} \cosh\bigl ( \frac\beta2 y \bigr)}.
\end{eqnarray} 
In order to proceed further,  we at first  rearrange \eqref{eq4b8} in the form:
\begin{eqnarray} 
\label{eq4b10}  
   m =   \dfrac{1}{2}    \bigl \langle \sigma_i + \sigma_j \bigr \rangle =                
          \Bigl \langle \mbox{e}^{  E_i (\nabla_x +\nabla_y )} \mbox{e}^{  E_j (\nabla_x - \nabla_y )}  \Bigr \rangle F(x, y)
          \Bigl|_{x= 0, y=0},          
\end{eqnarray}
and then we apply the exact van der Waerden identity \eqref{eq3b10} to obtain the exact equation:
\begin{eqnarray} 
\label{eq4b11}  
\nonumber 
        m \!\!\!   &=&   \!\!\!   
          \Bigl  \langle 
          \displaystyle \prod_{k= 1}^{q -1}
         \Bigl \{ \cosh \Bigl [ \frac J2 (\nabla_x + \nabla_y) \Bigr ] +  2\sigma_k
          \sinh \Bigl [  \frac J2 (\nabla_x + \nabla_y) \Bigr ] \Bigr \} \\ \nonumber
          &\times  & \displaystyle \prod_{\ell= 1}^{q -1} 
          \Bigl \{ \cosh \Bigl [ \frac J2 (\nabla_x - \nabla_y) \Bigr ] +  2\sigma_\ell
          \sinh \Bigl [  \frac J2 (\nabla_x - \nabla_y) \Bigr ] \Bigr \} \Bigr \rangle  \nonumber \\
           &\times & F(x, y)\Bigl|_{x= 0, y=0}.              
\end{eqnarray} 
It is quite evident that after  expanding the r.h.s of previous equation there again appear various multispin correlations that cannot be further treated exactly and must be split into simpler expressions. If we use the decoupling scheme \eqref{eq3b13a}, then, we will of course obtain the aforementioned two-spin cluster effective-field theory.  

Applying such an approximation, one gets the equation for magnetization in the form:
\begin{eqnarray} 
\label{eq4b12}  
\nonumber 
        m  &=&               
         \Bigl \{ \cosh \bigl [ \frac J2 (\nabla_x + \nabla_y) \bigr ] + 2m
          \sinh \bigl [ \frac J2 (\nabla_x + \nabla_y) \bigr ] \Bigr \}^{q-1} \\ \nonumber
         &\times& \Bigl \{ \cosh \bigl [ \frac J2 (\nabla_x - \nabla_y) \bigr ] +  2m 
          \sinh \bigl [ \frac J2 (\nabla_x - \nabla_y) \bigr ] \Bigr \}^{q-1}    \nonumber \\
          & \times & F(x, y)\Bigl|_{x= 0, y=0},              
\end{eqnarray} 
which can be rewritten using the binomial theorem as:
\begin{eqnarray} 
\label{eq4b13}  
\nonumber
       m & = &\sum_{k=0}^{q -1}  {q-1 \choose  k} 2^k m^k
        							\Bigl \{ \cosh \Bigl [ \frac J2 (\nabla_x + \nabla_y) \Bigr ] \Bigr\}^{q-k -1} 
                                    \Bigl \{ \sinh \Bigl [ \frac J2 (\nabla_x + \nabla_y) \Bigr ] \Bigr\}^{k}
                                            \\ \nonumber 
                                  &\times & 
   \sum_{\ell=0}^{q -1}  {q-1 \choose  \ell} 2^\ell m^\ell
        							\Bigl \{ \cosh \Bigl [ \frac J2 (\nabla_x - \nabla_y) \Bigr ] \Bigr\}^{q-\ell -1} 
                                    \Bigl \{ \sinh \Bigl [ \frac J2 (\nabla_x - \nabla_y) \Bigr ] \Bigr\}^{\ell}                               
                                  \\ \nonumber \\                              
                                 &\times & F (x, y)\Bigl|_{x= 0, y=0}.
\end{eqnarray}
If one introduces the coefficients  $A_{k\ell}^{(q)}$  as follows:
\begin{eqnarray} 
\label{eq4b15}  
\nonumber
       A_{k\ell}^{(q)} = A_{k\ell}^{(q)}(\beta J, h)  &=& 	\Bigl \{ \cosh \Bigl [  \frac J2 (\nabla_x + \nabla_y) \Bigr ] \Bigr\}^{q-k-1} 
                                    \Bigl \{ \sinh \Bigl [  \frac J2 (\nabla_x + \nabla_y) \Bigr ] \Bigr\}^{k} 
\\ \nonumber \\    \nonumber     					
        		&\times &\Bigl \{ \cosh \Bigl [  \frac J2 (\nabla_x - \nabla_y) \Bigr ] \Bigr\}^{q-\ell - 1} 
                                    \Bigl \{ \sinh \Bigl [  \frac J2 (\nabla_x - \nabla_y) \Bigr ] \Bigr\}^{\ell}                               
                                  \\ \nonumber \\                              
                                 &\times & F (x, y)\Bigl|_{x= 0, y=0}, 
\end{eqnarray}
then, the equation \eqref{eq4b13} can be expressed in the following abbreviated  form:
\begin{eqnarray} 
\label{eq4b14}  
       m & = &\sum_{k=0}^{q-1} \sum_{\ell=0}^{q-1} {q-1 \choose  k}     {q-1 \choose  \ell} 2^{k +\ell} m^{k + \ell} A_{k\ell}^{(q)}.
\end{eqnarray}
Similarly as in the case of the single-spin cluster effective-field theory, we can now straightforwardly evaluate the coefficients $A_{k\ell}^{(q)}$ using the binomial theorem and Eq. \eqref{eq4b7}. In this way one obtains the following expression for $h = 0$:
\begin{eqnarray}
\nonumber
  A_{k\ell}^{(q)}   & = & \sum_{i=0}^{q - k  -1} \sum_{j =0}^{k}  \sum_{n=0}^{q - \ell  -1} \sum_{p =0}^{\ell} 
  \Bigl ( -1 \Bigr)^{j +p}   {q-k-1 \choose  i}     {q- \ell -1 \choose  n}  {q \choose  j} {q\choose  p}  \\ \nonumber \\ \nonumber
   & \times &  \dfrac{ \sinh\bigl [ \frac{\beta J}{2}(q-1-i-j-n-p) \bigr] }
                  {  \cosh\bigl [ \frac{\beta J}{2}(q-1-i-j-n-p) \bigr] + \mbox{e}^{-\frac{\beta J}{2}} 
                  \cosh\bigl [\frac{\beta J}{2}(n+p -i-j) \bigr ]}.  \\ \nonumber \\
                  \label{eq4b16}
\end{eqnarray}
The relations \eqref{eq4b14} and \eqref{eq4b16} completely determine the equation of state of the spin-1/2 Ising model within the two-spin cluster effective-field theory and solving them numerically one can investigate the critical behavior of the model as well as the thermal dependencies of the magnetization. Despite of increasing complexity of Eqs. \eqref{eq4b14} and \eqref{eq4b16}, the numerical calculations  do not bring significant complications in comparison with the ones following from the single-spin cluster effective-field theory (c.f. \eqref{eq3b16} and \eqref{eq3b19}).

\begin{figure}[t]
\begin{center}
\vspace{-0.2cm}
\includegraphics[scale=0.4]{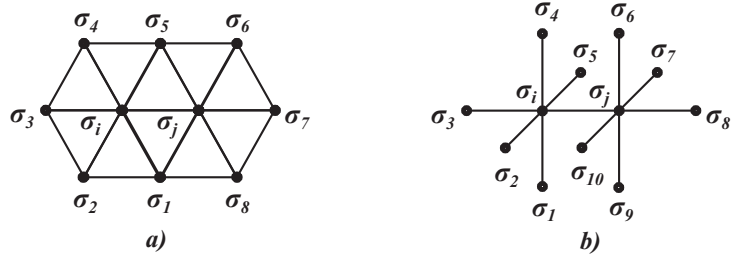}
\vspace{-0.5cm}
\end{center}
\caption{The central two-spin cluster composed of the Ising spins $\sigma_i$, $\sigma_j$ and its nearest neighbor for: a) plane triangular lattice ($q=6$), b) simple cubic lattice ($q=6$).}
\label{daepvilo}
\end{figure}

Now, let us continue with the discussion of some new topological aspects of the two-spin cluster effective-field theory. As one can see, the structure of the lattice is again taken into account in this formulation by coordination number $q$, however, one must be now very careful when applying the decoupling scheme. Looking more closely on the Eq. \eqref{eq4b14} , one finds that this equation is valid for the square and simple cubic lattice in the case of $q=4$ and $q=6$,  respectively. This statement follows directly from the fact that in both cases all nearest neighbors of the 
$i$th central spin differ from those of the $j$th central spin. On the other hand,  when applying the two-spin cluster effective-field  theory, for instance, to the kagom\'e or triangular lattice, we have to take into account that  some of the spins are the common nearest neighbors for both central spins $\sigma_i $  and $\sigma_j $.  In order to clarify this situation, we have depicted in Fig. \ref{daepvilo}  two-spin clusters including its nearest neighbors for the case of triangular and simple cubic lattice with $q=6$.  It is seen from this figure that the central sites of the cluster ($i,j$) have two common nearest neighbors (denoted as $\sigma_1$  and  $\sigma_5$) in the case of a triangular lattice, while no common nearest-neighbors exist in the 
case of the cubic lattice.

Taking into account this fact, one obtains for the spin-1/2 Ising model on a triangular lattice from  Eq. \eqref{eq4b11} the following expression:
\begin{eqnarray} 
\label{eq4b17}  
\nonumber 
        m  &=&               
         \Bigl \{ \cosh \Bigl [ \frac J2 \bigl (\nabla_x + \nabla_y \bigr) \Bigr ] + 2m
          \sinh \Bigl [ \frac J2 \bigl (\nabla_x + \nabla_y \bigr ) \Bigr ] \Bigr \}^3 \\ \nonumber
         &\times& \Bigl \{ \cosh \Bigl [ \frac J2 \bigl (\nabla_x - \nabla_y \bigr) \Bigr ] +  2m 
          \sinh \Bigl [ \frac J2 \bigl(\nabla_x - \nabla_y\bigr) \Bigr ] \Bigr \}^3    \nonumber \\
         &\times& \Bigl \{ \cosh \bigl ( J\nabla_x  \bigr ) +  2m 
          \sinh \bigl ( J \nabla_x \bigr ) \Bigr \}^2    \nonumber \\  
          & \times & F(x, y)\Bigl|_{x= 0, y=0}.              
\end{eqnarray} 
On the other hand, one gets for the spin-1/2 Ising model on a simple cubic lattice from Eq. \eqref{eq1b12} the relation: 
\begin{eqnarray} 
\label{eq4b18}  
\nonumber 
        m  &=&               
         \Bigl \{ \cosh \Bigl [ \frac J2 \bigl (\nabla_x + \nabla_y \bigr) \Bigr ] + 2m
          \sinh \Bigl [ \frac J2 \bigl (\nabla_x + \nabla_y \bigr ) \Bigr ] \Bigr \}^5 \\ \nonumber
         &\times& \Bigl \{ \cosh \Bigl [ \frac J2 \bigl (\nabla_x - \nabla_y \bigr) \Bigr ] +  2m 
          \sinh \Bigl [ \frac J2 \bigl(\nabla_x - \nabla_y\bigr) \Bigr ] \Bigr \}^5    \nonumber \\
          & \times & F(x, y)\Bigl|_{x= 0, y=0}.              
\end{eqnarray} 
Since previous two equations are different, then one can conclude that  unlike other approximate theories,  the two-spin cluster  effective-field theory is really able to take into account the topology of lattice and  to distinguish lattices with the same coordination number but different crystalline structure.
In order to illustrate this fact numerically we have listed in Table \ref{TCDAEP} numerical values of the critical temperatures for the spin-1/2 Ising model on lattices with different topological structure.

\begin{table}[t]
\caption[Comparisom of the Curie temperatures for  Ising model with spin 1/2 within different approximations]{A comparison of the Curie temperatures for the spin-1/2 Ising model on different lattices within several approximate and exact methods (see Refs. \cite{kane93r,jurc96bc}). The abbrevations used: MFA - mean-field approximation, OA  - Oguchi approximation, BPA - Bethe-Peierls approximation, SEFA - single-spin cluster effective-field approximation, TEFA - two-spin cluster effective-field approximation, Exact/HTSE - exact results/high-temperature series expansion.}
\medskip
\centering 
\begin{tabular}{lcccccc} 
\hline\hline 
                             & MFA           &  OA       &   BPA    & SEFA     & TEFA & Exact/HTSE\\ 
 lattice                 &$k_{\rm B} T_{\rm c}/J$ & $k_{\rm B} T_{\rm c}/J $  & $k_{\rm B} T_{\rm c}/J$ & $k_{\rm B} T_{\rm c}/J $ & $k_{\rm B} T_{\rm c}/J $ & $k_{\rm B} T_{\rm c}/J $\\ 
\hline \hline 
linear                   & 0.500          & 0.391           & 0.000      & 0.000      &  0.000  & 0.000 \\ 
honeycomb          & 0.750          & 0.677           & 0.455      & 0.526      &  0.497 & 0.380 \\
square                 & 1.000          & 0.944           & 0.721      & 0.773      &  0.756 & 0.567 \\
kagom\'e            & 1.000           & 0.944           & 0.721      & 0.773     &  0.731 & 0.536 \\
triangular            & 1.500          & 1.462            & 1.233      & 1.268     &  1.238 & 0.910 \\
simple cubic        & 1.500          & 1.462           & 1.233       & 1.268     &  1.260 & 1.128 \\
\hline 
\end{tabular}
\label{TCDAEP}
\vspace*{-0.4cm}
\end{table}
Finally, let us conclude this part by emphasizing that other physical quantities, for example, the internal energy, initial magnetic susceptibility, specific heat and many others, can be calculated within the two-spin cluster effective-field theory by reproducing   the same mathematical procedures as those described in the section \ref{0S_EFT}. In obtaining the aforementioned quantities there are no additional tricks to be explained, however, the calculations are rather tedious and time-consuming, therefore  we will not present all this stuff  in this work. Of course, the interested reader can find all details in the relevant references.   

\subsection{Historical survey and future perspectives}
\label{histmis}

Over the last few decades, the effective-field theory formulated on the basis of differential operator technique has become quite popular what can be convincingly evidenced by a rather extensive list of existing literature dealing with magnetic properties of various Ising-like spin systems obtained within this approximate theory. From the theoretical and experimental point of view, the most important applications of the effective-field theory based on differential operator can be divided into following groups:

A) Studies of  crystalline magnetic systems:
\begin{enumerate}
\item Ising \cite{honm78, honm79}, \cite{kane82} -\cite{balc07}, Blume-Capel \cite{kane90bc} -\cite{yuks09bc},  Blume-Emery-Griffiths models  with arbitrary spins 
\cite{kane87beg} -\cite{kane93beg} and  mixed-spin  Ising models \cite{kane88mix} -\cite{canp07mix} including decorated structures \cite{kane96dec1} -\cite{kane97dec1}.
\item  Semi-infinite Ising models and surface magnetism \cite{tamu83sf} -\cite{jasc93sf}, thin magnetic films \cite{wiat86film} -\cite{akta14film2}, magnetic bilayers and multilayers \cite{kane93lay1} -\cite{jasc97lay4}.
\item Transverse Ising  models \cite{kane88tim2} -\cite{yuks10tim}.
\end{enumerate}

B) Studies of  disordered magnetic systems:
\begin{enumerate}
\item  Ising,  Blume-Capel,  Blume-Emery-Griffiths models and mixed-spin Ising models,   including the  semi-infinite  models, thin magnetic films, magnetic bilayers and multilayers in a random longitudinal and/or transverse
external magnetic field \cite{kane86rf} -\cite{yuks12rf2}.
\item Amorphous spin systems \cite{kane80am} -\cite{kane92am}, Ising models with a random crystal-field or random exchange interactions \cite{tamu81ran} -\cite{zuki15ran}, 
site- or bond diluted magnetic systems \cite{kane80dil} -\cite{kane15dil}.
\item Magnetic binary and ternary alloys \cite{honm82al} -\cite{vata15al}.
\end{enumerate}

Having sketched the historical and recent development of the effective-field theory, it is now clear that one cannot even very briefly review so many papers in so diverse research areas. However, it  should be noticed here that in all aforementioned applications of the effective-field theory based on the differential operator technique has provided the results that are both numerically and physically superior with respect to the standard mean-field theory. This statement concerns with the behavior of all physical quantities outside of the critical region. On the other hand, the effective-field theory usually significantly improves also determination of critical boundaries, while it does not overcome the standard mean-field approaches in determining the values of critical exponents. Nevertheless, this fault of the effective-field theory can removed by developing a phenomenological effective-field renormalization group theory combining the results of the effective-field theories for  central clusters of different sizes (see Ref. \cite{liya88}).  

After almost four decades from the beginning of this story, it is very pleasant to find out that the effective-field theories based on differential operator technique are still alive when being actively engaged in solving new lattice-statistical spin systems of current research interest. Among the most important recent developments and applications of the effective-field theories based on the differential operator technique one could mention recent works by Polat, Akinci, Yuksel and others \cite{yuks09bc,canp07mix,akta14film2,yuks10tim,akin11rf,yuks12rf1,yuks12rf2,akin11dil,vata12dil,vata15al}.  
Thus, despite of huge number of applications done until now, the potential of the differential operator method seems to be far from being exhausted and we surely may expect many new interesting applications of this theoretical approach in a near future.

\setcounter{equation}{0} \newpage
\section{Exact results}
\label{exact}

The present section will be devoted to the exact solutions of the Ising models. It is worthy to recall that the Ising models are exactly tractable in one and two dimensions only, so our attention will be henceforth focused just on 1D Ising chains and 2D Ising lattices. In particular, we will examine a magnetic behavior of the spin-1/2 Ising models, the spin-1 Blume-Capel models and the mixed-spin Ising models either in a presence or absence of the magnetic field. 

\subsection{One-dimensional Ising models}

The main advantage of 1D Ising chains lies in that they can be exactly treated with the help of the transfer-matrix method even in a presence of the external magnetic field \cite{kram41,baxt82}. It should be nevertheless pointed out that 1D Ising models cannot exhibit a non-trivial criticality at finite temperatures in contrast with their 2D counterparts.    

\subsubsection{Spin-1/2 Ising chain in a longitudinal field}

Let us consider first the spin-1/2 Ising chain in a longitudinal magnetic field, which can be defined through the Hamiltonian:
\begin{eqnarray}
{\cal H} = - J \sum_{i=1}^{N} \sigma_i \sigma_{i+1} - h \sum_{i=1}^{N} \sigma_i.
\label{s1}	
\end{eqnarray}
Here, $\sigma_i = \pm 1/2$ denotes the Ising spin placed at the $i$-th lattice point, the parameter $J$ labels the nearest-neighbor Ising coupling and the parameter $h$ relates to the magnetostatic Zeeman's energy. For simplicity, we will further assume a periodic boundary condition imposed by the constraint $\sigma_{N+1} \equiv \sigma_1$, because the overall magnetic behavior does not depend in the thermodynamic limit $N \to \infty$ on a particular choice of the boundary condition. It is worth mentioning, however, that the periodic boundary condition $\sigma_{N+1} \equiv \sigma_1$ ensures a translational invariance of the Ising chain defined by the Hamiltonian (\ref{s1}), which considerably simplifies further treatment. The exact solution of the spin-1/2 Ising chain in a magnetic field can be most easily found by the \textit{transfer-matrix method} originally introduced by Kramers and Wannier \cite{kram41}. This rather powerful technique is of particular importance, since it formulates the exact solution in a relatively tractable matrix form that can be straightforwardly adapted to more complicated 1D and 2D Ising spin systems.  

First, it is very advisable to rewrite the total Hamiltonian (\ref{s1}) into the most symmetric form: 
\begin{eqnarray}
{\cal H} = \sum_{i=1}^{N} {\cal H}_i, \qquad {\cal H}_i = - J \sigma_i \sigma_{i+1} - \frac{h}{2} \left(\sigma_i + \sigma_{i+1} \right),
\label{s2}	
\end{eqnarray}
which can be used for a simple factorization of the partition function into a product of several terms each involving just two adjacent Ising spins: 
\begin{eqnarray}
Z = \sum_{\{\sigma_i \}} \prod_{i=1}^{N} \exp(- \beta {\cal H}_i) = \sum_{\{\sigma_i \}} \prod_{i=1}^{N} 
\exp\left[\beta J \sigma_i \sigma_{i+1} + \frac{\beta h}{2} \left(\sigma_i + \sigma_{i+1} \right) \right]. 
\label{s3}	
\end{eqnarray}
Next, let us formally substitute each factor in the product (\ref{s3}) by the function $T (\sigma_i, \sigma_{i+1})$ depending just on the two nearest-neighboring spins $\sigma_i$ and $\sigma_{i+1}$: 
\begin{eqnarray}
Z = \sum_{\sigma_1 = \pm \frac{1}{2}} \sum_{\sigma_2 = \pm \frac{1}{2}} \! \! \cdots \! \! \sum_{\sigma_N = \pm \frac{1}{2}}   
T (\sigma_1, \sigma_2) T (\sigma_2, \sigma_3) \ldots T (\sigma_i, \sigma_{i+1}) \ldots T (\sigma_N, \sigma_1), 
\label{s4}	
\end{eqnarray}
whereas the definition of the function $T (\sigma_i, \sigma_{i+1})$ is obvious:
\begin{eqnarray}
T (\sigma_i, \sigma_{i+1}) = \exp\left[\beta J \sigma_i \sigma_{i+1} + \frac{\beta h}{2} \left(\sigma_i + \sigma_{i+1} \right)\right]. 
\label{s5}	
\end{eqnarray}
It should be mentioned that the function (\ref{s5}) is not the only possible choice for $T (\sigma_i, \sigma_{i+1})$, it can be multiplied for instance by any factor $\exp[a (\sigma_i - \sigma_{i+1})]$ ($a$ is arbitrary constant) without loosing a validity of the overall product (\ref{s3}). However, this choice is the only one that preserves a complete symmetry with respect to the interchange: 
\begin{eqnarray}
T (\sigma_i, \sigma_{i+1}) = T (\sigma_{i+1}, \sigma_i), 
\label{s6}	
\end{eqnarray}
which is in accord with a translational symmetry ensured by the periodic boundary condition. 

Now, let us perform a small calculation that reveals an essence of the expression (\ref{s5}). It is worthy to recall that the Ising spin $\sigma_2$, for example, enters in Eq.~(\ref{s4}) just to 
two side-by-side standing expressions $T (\sigma_1, \sigma_2)$ and $T (\sigma_2, \sigma_3)$ and hence, one may perform the summation over two available states of the Ising spin $\sigma_2$  independently of other expressions appearing within this product. After summing over two available states of the Ising spin $\sigma_2$ one consequently obtains the following Boltzmann's weight:
\begin{eqnarray}
\sum_{\sigma_2 = \pm \frac{1}{2}} \! \! \! \! \! \! \!\!\! && \!\!\! \! \! \! \! \! \! T (\sigma_1, \sigma_2) 
T (\sigma_2, \sigma_3) = \sum_{\sigma_2 = \pm \frac{1}{2}} \! \! \exp \left[\beta J \sigma_2 (\sigma_1 + \sigma_3) + 
                                                      \frac{\beta h}{2} (\sigma_1 + 2 \sigma_2 + \sigma_3)\right]  \label{s7}	 \\ \!\!\!\!\!\! &=&  \!\!\!\!\!\!
\exp\!\left[\frac{\beta h}{2} (\sigma_1 + \sigma_3)\right] \! \left\{ \exp\!\left[\frac{\beta J}{2}(\sigma_1 + \sigma_3) + \frac{\beta h}{2}\right]  
                       \!+ \exp\!\left[-\frac{\beta J}{2}(\sigma_1 + \sigma_3) - \frac{\beta h}{2}\right] \! \right \}\!. \nonumber
\end{eqnarray}
It is worthwhile to remark that one arrives to the same result when the expression $T (\sigma_i, \sigma_{i+1})$ is alternatively viewed as a two-by-two matrix with appropriately chosen matrix elements:
\begin{eqnarray}
T (\sigma_i, \sigma_{i+1}) \! = \! 
\left(\!
\begin{array}{cc} 
   T(+,+)     &     T(+,-)    \\
   T(-,+)     &     T(-,-)    \\
\end{array}
                 \! \right) \! = \! 
\left(\!\!
\begin{array}{cc} 
   \exp\!\left(\frac{\beta J}{4} + \frac{\beta h}{2}\right)      &    \exp\!\left(-\frac{\beta J}{4}\right)             \\
         \exp\!\left(-\frac{\beta J}{4}\right)         &    \exp\!\left(\frac{\beta J}{4} - \frac{\beta h}{2}\right)    \\
\end{array}
                \!\!  \right)\!, 
\label{s8}	
\end{eqnarray}
which are related to four possible spin configurations of two adjacent Ising spins $\sigma_i$ and $\sigma_{i+1}$. The matrix element $T(+,-)$ marks for instance the Boltzmann's factor to be obtained from Eq.~(\ref{s5}) by considering the particular spin configuration $\sigma_i = +1/2$ and $\sigma_{i+1} = -1/2$. Each row in the matrix $T (\sigma_i, \sigma_{i+1})$ thus accounts for the spin states of the former spin $\sigma_i$ (in other words, by changing a row one changes a spin state of the former spin $\sigma_i$), while each column determines the spin states of the latter spin $\sigma_{i+1}$ (by changing a column one changes a spin state of the latter spin $\sigma_{i+1}$). It should be noted that the spin variable $\sigma_2$ enters in the first expression $T (\sigma_1, \sigma_2)$ as the latter spin, while in the second expression $T (\sigma_2, \sigma_3)$ it acts as the former spin. Taking into account the spin states of the spin $\sigma_2$ changes the first expression $T (\sigma_1, \sigma_2)$ to the one-by-two (row) matrix and the second expression $T (\sigma_2, \sigma_3)$ to the two-by-one (column) matrix. As a result, the summation over the spin configurations of the Ising spin $\sigma_2$ turns out to be equivalent with a matrix multiplication between the row matrix $T (\sigma_1, \sigma_2)$ and the column matrix $T (\sigma_2, \sigma_3)$: 
\begin{eqnarray}
 \sum_{\sigma_2 = \pm \frac{1}{2}}  \! \! \! \! \! \! \!\!\! && \! \! \! \! \!  \! \!\!\!
 T (\sigma_1, \sigma_2) T (\sigma_2, \sigma_3) = \nonumber \\ 
 \! \! \!\!\! &&  \!\!\! \! \! \left(
  \begin{array}{cc} 
    \exp \Bigl [\frac{\beta J}{2} \sigma_1 + \frac{\beta h}{2}(\sigma_1 + \frac{1}{2}) \Bigr]  &  
    \exp \Bigl [-\frac{\beta J}{2} \sigma_1 + \frac{\beta h}{2} (\sigma_1 - \frac{1}{2}) \Bigr] \\              
  \end{array}
 \right) 
 \nonumber \\  \! \! \!\!\! &&  \!\!\! \! \!
\left(
  \begin{array}{c} 
   \exp \Bigl [\frac{\beta J}{2} \sigma_3 + \frac{\beta h}{2} (\sigma_3 + \frac{1}{2})  \Bigr]   \\
   \exp \Bigl [-\frac{\beta J}{2} \sigma_3 + \frac{\beta h}{2} (\sigma_3 - \frac{1}{2}) \Bigr]  \\              
  \end{array}
\right) \nonumber	\\
  \! \! \!\!\! &=&  \! \! \!\!\! 
  \exp \Bigl[\frac{\beta h}{2}(\sigma_1 + \sigma_3) \Bigr] \!\left \{ \exp\!\left[\frac{\beta J}{2}(\sigma_1 + \sigma_3) 
  + \frac{\beta h}{2}\right] +  \exp\!\left[-\frac{\beta J}{2}(\sigma_1 + \sigma_3) - \frac{\beta h}{2}\right]\! \right \}\!. \nonumber \\
\label{s9}
\end{eqnarray}
As a matter of fact, the expression (\ref{s9}) obtained from the matrix product is indeed completely identical with the result (\ref{s7}), which has been previously acquired by a straightforward summation over the spin states of the spin variable $\sigma_2$. In addition, the final expression obtained after the respective matrix multiplication might be considered as some two-by-two matrix $T^2 (\sigma_1, \sigma_3)$, the elements of which depend on the Ising spins $\sigma_1$ and $\sigma_3$ according to the definition (\ref{s9}): 
\begin{eqnarray}
\sum_{\sigma_2 = \pm \frac{1}{2}}  T (\sigma_1, \sigma_2) T (\sigma_2, \sigma_3) = T^2 (\sigma_1, \sigma_3).
\label{s10}	
\end{eqnarray}  
The summations over the spin states of a set of the Ising spins $\{ \sigma_2, \sigma_3, \ldots, \sigma_N \}$ can in turn be regarded as successive matrix multiplications yielding:                   \begin{eqnarray}
Z  \! \! \! &=&  \! \! \! \sum_{\sigma_1 = \pm \frac{1}{2}} \sum_{\sigma_2 = \pm \frac{1}{2}} \! \! \ldots \! \! 
\sum_{\sigma_N = \pm \frac{1}{2}} T (\sigma_1, \sigma_2) T (\sigma_2, \sigma_3) \ldots T (\sigma_i, \sigma_{i+1}) \ldots T (\sigma_N, \sigma_1) \nonumber \\
\! \! \! &=&  \! \! \! \sum_{\sigma_1 = \pm \frac{1}{2}} \sum_{\sigma_3 = \pm \frac{1}{2}} \! \! \ldots \! \! 
\sum_{\sigma_N = \pm \frac{1}{2}} T^2 (\sigma_1, \sigma_3) \ldots T (\sigma_i, \sigma_{i+1}) \ldots 
T (\sigma_N, \sigma_1) \nonumber \\
 \! \! \! &=&  \! \! \! 
 \sum_{\sigma_1 = \pm \frac{1}{2}} T^N (\sigma_1, \sigma_1) = T^N (+,+) + T^N (-,-) = \mbox{Tr} \, T^N.
\label{s11}	
\end{eqnarray}                    
At each step of this procedure, the matrix multiplication by $T$ corresponds to the summation over the spin configurations of one more Ising spin. That is why the matrix $T$ is referred to as the transfer matrix, namely, it transfers the dependence from the one spin to its neighboring spin. Besides, it can be also easily understood from Eq.~(\ref{s11}) that the last summation, which is carried out over the spin states of the first Ising spin $\sigma_1$, is equivalent to taking a trace of the matrix $T^{N}$, i.e. to performing a summation over elements of the matrix $T^{N}$ on its main diagonal. The problem of finding the exact solution for the spin-1/2 Ising chain in a longitudinal magnetic field thus reduces to the calculation of a trace of so far unknown matrix $T^{N}$ when assuming the periodic boundary condition.  

At this stage, it is very convenient to use an invariance of the trace with respect to a cyclic permutation of the product in order to calculate a trace of the matrix $T^{N}$ (the product of three square matrices $A$, $B$ and $C$ has for instance the same trace upon following cyclic permutations Tr$(ABC)$=Tr$(BCA)$=Tr$(CAB)$). With this in mind, the unitary matrix $U$ and its inverse $U^{-1}$ ($U^{-1}U=UU^{-1}=1$) can be employed to convert the transfer matrix $T$ into a diagonal form: 
\begin{eqnarray}
U^{-1} T U = \Lambda = \left(
\begin{array}{cc} 
   \lambda_{+}     &          0          \\
       0           &      \lambda_{-}    \\
\end{array}
                  \right).
\label{s12}	
\end{eqnarray}   
After inserting their products into the relation (\ref{s11}) one obtains:
\begin{eqnarray}
Z  \! \! \! &=& \! \! \! 
\mbox{Tr} \, T^N  = \mbox{Tr}(T T \ldots T) = \mbox{Tr}(UU^{-1}TUU^{-1}TU \ldots U^{-1}T) \nonumber \\
          \! \! \! &=& \! \! \!            \mbox{Tr}(U^{-1}TUU^{-1}TU \ldots U^{-1}TU) =  
          \mbox{Tr}(\Lambda \Lambda \ldots \Lambda)=\mbox{Tr} \, \Lambda^N.
\label{s13}	
\end{eqnarray}                                                                 
The above result is nothing but a similarity invariance, which means that $T^{N}$ and $\Lambda^{N}$ matrices have the same trace (the trace does not depend on a particular choice of basis). The matrix $\Lambda^{N}$ is $N$th power of the diagonal matrix (\ref{s12}) and hence, its diagonal elements are simply $N$th power of diagonal elements of the matrix $\Lambda$: 
\begin{eqnarray}
\Lambda^N = \left(
\begin{array}{cc} 
   \lambda_{+}^N     &          0            \\
       0             &      \lambda_{-}^N    \\
\end{array}
                  \right).
\label{s14}	
\end{eqnarray}                              
It is therefore sufficient to find elements of the diagonal matrix $\Lambda$ in order to calculate the desired trace $\mbox{Tr} \, T^{N} = \mbox{Tr} \, \Lambda^{N} = \lambda_{+}^N + \lambda_{-}^N $. 
This eigenvalue problem can be attacked with the help of unitary transformation $T U = U \Lambda$, which can be written in the matrix representation: 
\begin{eqnarray}
                   \left(
\begin{array}{cc} 
   T(+,+)     &          T(+,-)            \\
   T(-,+)     &          T(-,-)            \\
\end{array}
                  \right) 
                  \left(
\begin{array}{cc} 
   a_{+}     &          a_{-}            \\
   b_{+}     &          b_{-}            \\
\end{array}
                  \right) 
                  =
                  \left(
\begin{array}{cc} 
    a_{+}     &          a_{-}            \\
    b_{+}     &          b_{-}            \\
\end{array}
                  \right) 
                  \left(
\begin{array}{cc} 
   \lambda_{+}     &          0            \\
       0             &      \lambda_{-}    \\
\end{array}
                  \right)  
\label{s15}	
\end{eqnarray} 
and is equivalent to solving the so-called characteristic (secular) equation:                            
\begin{eqnarray}
T V_{\pm} = \lambda_{\pm} V_{\pm} \Longleftrightarrow  
                  \left(
\begin{array}{cc} 
   T(+,+)     &          T(+,-)            \\
   T(-,+)     &          T(-,-)            \\
\end{array}
                  \right) 
                  \left(
\begin{array}{cc} 
   a_{\pm}                \\
   b_{\pm}                \\
\end{array}
                  \right) 
                  = 
                  \lambda_{\pm}
                  \left(
\begin{array}{cc} 
   a_{\pm}               \\
   b_{\pm}               \\          
\end{array}
                  \right).                  
\label{s16}	
\end{eqnarray}                                                                                          
Here, $V_{\pm}$ denote eigenvectors of the transfer matrix $T$ and $\lambda_{\pm}$ are their corresponding eigenvalues. Both secular equations are in fact a homogeneous system of linear 
equations, which has a non-trivial solution if and only if its determinant equals zero: 
\begin{eqnarray}
|T - \lambda_i I| = 0. \qquad (i \in \{ \pm \})
\label{s17}	
\end{eqnarray}
Here, $I$ is used for labeling the unit matrix of appropriate size. In this way, the eigenvalues of the transfer matrix $T$ can be calculated without knowing explicit form of the corresponding eigenvectors $V_{\pm}$. As a matter of fact, both eigenvalues can readily be calculated by solving the determinant:
\begin{eqnarray}
\left|
\!
\begin{array}{cc} 
   \exp\!\left(\frac{\beta J}{4} + \frac{\beta h}{2}\right) - \lambda_i      &    \exp\!\left(-\frac{\beta J}{4}\right)             \\
         \exp\!\left(-\frac{\beta J}{4}\right)         &    \exp\!\left(\frac{\beta J}{4} - \frac{\beta h}{2}\right) - \lambda_i    \\
\end{array}
                \!                    \right| = 0,
\label{s18}	
\end{eqnarray}
which gives after straightforward calculation two eigenvalues:             
\begin{eqnarray}
\lambda_{\pm} = \exp\!\left(\frac{\beta J}{4}\right) 
\left[\cosh\!\left(\frac{\beta h}{2}\right) \pm \sqrt{\sinh^2\!\left(\frac{\beta h}{2}\right) + \exp(- \beta J)} \right].
\label{s19}	
\end{eqnarray} 
The final expression for the partition function of the spin-1/2 Ising chain in a presence of the longitudinal magnetic field is subsequently given by:
\begin{eqnarray}
Z = \mbox{Tr} \, T^{N} = \mbox{Tr} \, \Lambda^{N} = \lambda_{+}^N + \lambda_{-}^N. 
\label{s20}	
\end{eqnarray}
The equation~(\ref{s20}) represents a central result of our calculation, since the whole thermodynamics is afterwards accessible from the partition function. In the thermodynamic limit $N \to \infty$, the final expression for the Gibbs free energy per one spin considerably simplifies by realizing that the ratio between two transfer-matrix eigenvalues is always less than unity ($\lambda_{-}/\lambda_{+}<1$):
\begin{eqnarray}
G  \! \! \! &=&  \! \! \! - k_{\rm B} T \frac{1}{N} \lim_{N \to \infty} \ln Z 
         = - k_{\rm B} T \frac{1}{N} \lim_{N \to \infty} \ln (\lambda_{+}^N + \lambda_{-}^N) 
         \nonumber \\ 
          \! \! \! &=&  \! \! \! - k_{\rm B} T \frac{1}{N} \lim_{N \to \infty} 
         \ln \left \{ \lambda_{+}^N \left[1 + \left(\frac{\lambda_{-}}{\lambda_{+}} \right)^N \right]  
         \right \} = - k_{\rm B} T \ln \lambda_{+}.  
\label{s21}	
\end{eqnarray} 
According to Eq.~(\ref{s21}), the Gibbs free energy (and also all other important thermodynamic quantities) depend exclusively on the greatest eigenvalue $\lambda_{+}$ of the transfer matrix. It is noteworthy that this is even true for any other interacting many-particle system, which can be treated within the transfer-matrix method. Thus, the problem of finding the exact solution 
within the transfer-matrix approach is essentially equivalent to the problem of finding the greatest eigenvalue of the relevant transfer matrix. 
  
Now, let us perform a more comprehensive analysis of the obtained results. It can be readily understood from Eq.~(\ref{s21}) that the Gibbs free energy is a smooth analytic function of the temperature for all non-zero temperatures and thus, the exact solution admits no phase transition. The absence of phase transition in 1D Ising chain is consistent with the fact that this model cannot sustain a spontaneous magnetization at any finite temperature. To clarify this issue in detail, let us derive the exact expression for the magnetization as a function of the temperature and magnetic field. One possible route how to obtain the magnetization is to differentiate the Gibbs free energy with respect to the external magnetic field: 
\begin{eqnarray}
m = \langle \sigma_i \rangle = - \frac{\partial G}{\partial h} =  \frac{1}{2} \frac{\sinh\!\left(\frac{\beta h}{2}\right)}{\sqrt{\sinh^2\!\left(\frac{\beta h}{2}\right) + \exp(-\beta J)}}.  
\label{s24}	
\end{eqnarray} 
It is easy to prove that the magnetization equals zero ($m=0$) on assumption that the zero magnetic field ($h=0$) is set to Eq.~(\ref{s24}) at any finite (non-zero) temperature. This result gives  a rigorous proof for the absence of a spontaneous long-range ordering (spontaneous magnetization) at any finite temperature. There is nevertheless still possibility that the absolute zero temperature ($T=0$~K) by itself is a critical point at which the spontaneous long-range order might appear. Under the assumption of the ferromagnetic interaction ($J>0$), it can be readily verified that the spontaneous magnetization takes its saturation values in the zero-temperature limit:
\begin{eqnarray}
\lim_{h \to 0} \lim_{T \to 0^{+}} m (T, h) = \pm \frac{1}{2}.  
\label{s25}	
\end{eqnarray}   
and the conjecture about the critical point at absolute zero temperature is indeed verified. 

To provide a deeper insight into this rather intricate critical behavior, it is necessary to study the pair correlation function between two spins and its dependence on a lattice spacing 
between them. For this purpose, we will need an explicit expression for both the eigenvectors $V_{\pm}$ determining the matrices of unitary transformation. According to Eq.~(\ref{s16}), 
the coefficients that determine both eigenvectors $V_{\pm}$ must obey following conditions:
\begin{eqnarray}
a_{\pm} =  b_{\pm} \frac{T(+,-)}{\lambda_{\pm} - T(+,+)} \quad \mbox{and} \quad 
a_{\pm} =  b_{\pm} \frac{\lambda_{\pm} - T(-,-)}{T(-,+)}, 
\label{s26}	
\end{eqnarray} 
which can be symmetrized upon their multiplication to:
\begin{eqnarray}
a_{\pm}^2 = b_{\pm}^2 \frac{\lambda_{\pm} - T(-,-)}{\lambda_{\pm} - T(+,+)}.
\label{s27}	
\end{eqnarray}  
The relationship between these coefficients can be rewritten in a more useful form by substituting the eigenvalues (\ref{s19}) and the elements of transfer matrix (\ref{s8}) into Eq.~(\ref{s27}):
\begin{eqnarray}
a_{\pm}^2 = b_{\pm}^2 \frac{\sinh\!\left(\frac{\beta h}{2}\right) \pm \sqrt{\sinh^2\!\left(\frac{\beta h}{2}\right) + \exp(-\beta J)}}
                           {-\sinh\!\left(\frac{\beta h}{2}\right) \pm \sqrt{\sinh^2\!\left(\frac{\beta h}{2}\right) + \exp(-\beta J)}}.
\label{s28}	
\end{eqnarray}
For further convenience, let us introduce a new variable that is connected to the temperature $T$, the exchange constant $J$ and the magnetic field $h$ by means of the relation 
$\mbox{cotg} 2 \phi = \exp(\beta J/2)$ $\times\sinh(\beta h/2)$, which will significantly simplify further steps in the calculation. Within the new notation, the coefficients $a_{\pm}$ 
and $b_{\pm}$ satisfy the relation: 
\begin{eqnarray}
a_{\pm}^2 = -b_{\pm}^2 \frac{\cos 2 \phi \pm 1}{\cos 2 \phi \mp 1},
\label{s29a}	
\end{eqnarray}
which together with the normalization condition $a_{\pm}^2 + b_{\pm}^2 = 1$ unambiguously determines both the eigenvectors $V_{\pm}$. After straightforward calculation, it is possible 
to derive explicit expressions for the coefficients $a_{\pm}$ and $b_{\pm}$:
\begin{eqnarray}
a_{+} = \cos \phi, \quad b_{+} = \sin \phi, \qquad \mbox{and} \qquad
a_{-} = -\sin \phi, \quad b_{-} = \cos \phi.  
\label{s29} 
\end{eqnarray} 
With the help of the coefficients (\ref{s29}), the explicit expression of the unitary matrix $U$ and its inverse matrix $U^{-1}$ can be found following the standard algebraic procedures: 
\begin{eqnarray}
U = \left( \begin{array}{cc} 
     \cos \phi  &  -\sin \phi  \\
     \sin \phi  &   \cos \phi  \\
\end{array} \right) \qquad \mbox{and} \qquad
U^{-1} = \left( \begin{array}{cc} 
    \cos \phi  &  \sin \phi  \\
   -\sin \phi  &  \cos \phi  \\
\end{array} \right).
\label{s30} 
\end{eqnarray} 
Now, the pairwise correlation $\langle \sigma_i \sigma_j \rangle$ between the two spins that reside some general $i$th and $j$th position within the closed Ising chain can be immediately calculated. The statistical definition of the canonical ensemble average allows us to calculate the correlation $\langle \sigma_i \sigma_j \rangle$ in terms of transfer matrices:
\begin{eqnarray}
\langle \sigma_i \sigma_j \rangle \! \! \! &=& \! \! \! \frac{1}{Z} \sum_{\{ \sigma_i \}} \sigma_i \sigma_j \exp(- \beta {\cal H}) = \frac{1}{Z} \sum_{\{ \sigma_i \}} \Bigl[ T (\sigma_1, \sigma_2) T (\sigma_2, \sigma_3) \ldots \\ \nonumber \! \! \! && \! \! \! T (\sigma_{i-1}, \sigma_{i}) \sigma_i T (\sigma_i, \sigma_{i+1}) \ldots T (\sigma_{j-1}, \sigma_{j}) \sigma_j T (\sigma_j, \sigma_{j+1}) \ldots T (\sigma_N, \sigma_1) \Bigr],
\label{s31} 
\end{eqnarray} 
where $j>i$ is assumed without loss of the generality. If the product of the unitary and its inverse matrix is inserted to each term of the matrix product (\ref{s31}), then, the two-spin correlation (\ref{s31}) can be rearranged to a more appropriate form using the cyclic permutation of matrices behind the relevant trace: 
\begin{eqnarray}
\langle \sigma_i \sigma_j \rangle = \frac{1}{Z} \mbox{Tr} \left[S T^{j-i} S T^{N+i-j} \right]  
= \frac{1}{Z}  \mbox{Tr} \left[U^{-1}SU \Lambda^{j-i} U^{-1}SU \Lambda^{N+i-j} \right].
\label{s32} 
\end{eqnarray} 
In above, the diagonal matrix $S$ with elements $S(\sigma_i, \sigma_j) = \sigma_i \delta_{ij}$ ($\delta_{ij}$ is Kronecker symbol) should account for possible spin states of $\sigma_i$ and $\sigma_j$ spins. An elementary calculation gives from Eq.~(\ref{s30}):
\begin{eqnarray}
U^{-1}SU = \frac{1}{2} \left( \begin{array}{cc} 
     \cos 2 \phi  &  -\sin 2 \phi  \\
    -\sin 2 \phi  &  -\cos 2 \phi  \\
\end{array} \right).
\label{s33} 
\end{eqnarray}
The matrix (\ref{s33}) together with the explicit expression (\ref{s12}) of the matrix $\Lambda$ yields after straightforward but bit tedious modification the following result for the two-spin correlation: 
\begin{eqnarray}
\langle \sigma_i \sigma_{i + r} \rangle = \frac{1}{4}
\frac{\lambda_{+}^{N} \cos^2 2 \phi  + \lambda_{+}^{N-r} \lambda_{-}^{r} \sin^2 2 \phi + 
\lambda_{+}^{r} \lambda_{-}^{N-r} \sin^2 2 \phi +\lambda_{-}^{N} \cos^2 2 \phi}
     {\lambda_{+}^{N} + \lambda_{-}^{N}},
\label{s34} 
\end{eqnarray}  
where $r = i - j$ is used to measure a distance in between the $i$th and $j$th spins. In the thermodynamic limit, the derived result (\ref{s34}) for the pair correlation largely simplifies to:
\begin{eqnarray}
\langle \sigma_i \sigma_{i + r} \rangle = \frac{1}{4} \left[
\cos^2 2 \phi  + \left( \frac{\lambda_{-}}{\lambda_{+}} \right)^r \sin^2 2 \phi \right].
\label{s35} 
\end{eqnarray}  
It should be mentioned that the expression (\ref{s35}) represents the most general result for the two-spin correlation, which depends on a distance $r$ between spins, the ratio between 
the smaller and the larger eigenvalue of the transfer matrix and the two expressions directly connected to the relation $\mbox{cotg} 2 \phi = \exp(\beta J/2) \sinh(\beta h/2)$ via terms:
\begin{eqnarray}
\cos 2 \phi = \frac{\sinh\!\left(\frac{\beta h}{2}\right)}{\sqrt{\sinh^2\!\left(\frac{\beta h}{2}\right) + \exp(-\beta J)}}, \quad 
\sin 2 \phi = \frac{\exp\!\left(-\frac{\beta J}{2}\right)}{\sqrt{\sinh^2\!\left(\frac{\beta h}{2}\right) + \exp(-\beta J)}}.
\label{s35a} 
\end{eqnarray}  
In an absence of the longitudinal magnetic field ($h=0$), the expression (\ref{s35}) for the two-spin correlation further reduces, because of the validity of $\cos 2 \phi = 0$ and $\sin 2 \phi = 1$, 
to the famous result:
\begin{eqnarray}
\langle \sigma_i \sigma_{i + r} \rangle = \frac{1}{4} \left( \frac{\lambda_{-}}{\lambda_{+}} \right)^r = \frac{1}{4} \left[\tanh \left(\frac{\beta J}{4}\right)\right]^r.
\label{s36} 
\end{eqnarray}  
It can be easily understood from Eq.~(\ref{s36}) that the two-spin correlation is determined by $r$th power of the ratio between the smaller and larger eigenvalue of the transfer matrix. 
With regard to this, the correlation between two spins exhibits a peculiar \textit{power-law decay} with the distance $r$ in between them. In other words, the correlation between distant spins is 
simply $r$th power of the correlation between the nearest-neighbor spins. It should be realized that the expression $\tanh (\beta J/4)$ is at any finite temperature always less than unity and thus, the two-spin correlation gradually tends to zero with increasing the lattice spacing $r$ between two spins. One may easily prove the following asymptotic behavior of the pair correlation 
in the limiting case of infinite distance $r \to \infty$:
\begin{eqnarray}
\lim_{r \to \infty} \langle \sigma_i \sigma_{i + r} \rangle = 0 \qquad \mbox{for any} \quad T \neq 0.
\label{s37} 
\end{eqnarray}  
The expression (\ref{s37}) furnishes proof for an absence of the spontaneous long-range order at any finite temperature, which may consequently emerge only at zero temperature. One actually finds different asymptotic behavior of the pair correlation in the zero-temperature limit: 
\begin{eqnarray}
\lim_{r \to \infty} \lim_{T \to 0} \langle \sigma_i \sigma_{i + r} \rangle = \frac{1}{4},
\label{s38} 
\end{eqnarray} 
which serves in evidence of a perfect correlation between distant spins and an existence of the spontaneous long-range order. The zero temperature can be accordingly considered as a very special critical point, at which an onset of spontaneous ordering appears.

\subsubsection{Spin-1 Blume-Capel chain in a longitudinal field}

Next, we will demonstrate versatility of the transfer-matrix approach by considering the Hamiltonian of the spin-1 Blume-Capel chain in a longitudinal magnetic field:
\begin{eqnarray}
{\cal H} = - J \sum_{i=1}^{N} S_i S_{i+1} - D \sum_{i=1}^{N} S_i^2 - h \sum_{i=1}^{N} S_i,
\label{beg}
\end{eqnarray}
which includes at the $i$th lattice point the three-valued Ising spin variable $S_i = \pm 1, 0$. The interaction terms in the Hamiltonian (\ref{beg}) have apparent meaning: $J$ is the bilinear nearest-neighbor interaction, $D$ is the single-ion anisotropy, $h$ is the longitudinal magnetic field and the cyclic boundary condition $S_{N+1} \equiv S_1$ is imposed for simplicity.

The exact solution for the partition function of the spin-1 Blume-Capel chain closely follows the procedure thoroughly described for the spin-1/2 Ising chain and thus, let us merely quote a few most important steps of this rigorous calculation. The Hamiltonian (\ref{beg}) can be at first rewritten into the symmetric form:
\begin{eqnarray}
{\cal H} = \sum_{i = 1}^N {\cal H}_i, \qquad {\cal H}_i = - J S_i S_{i+1} - D (S_i^2 + S_{i+1}^2) - \frac{h}{2} (S_i + S_{i+1})
\label{begi}
\end{eqnarray}
and afterwards, the partition function of the spin-1 Blume-Capel chain can be factorized to the following product:
\begin{eqnarray}
Z = \sum_{S_1 = \pm 1,0} \sum_{S_2 = \pm 1,0} \! \! \cdots \! \! \sum_{S_N = \pm 1,0} \prod_{i=1}^N  T (S_i, S_{i+1}),
\label{pfbeg}
\end{eqnarray}
where the expression $T (S_i, S_{i+1})$ is defined as:
\begin{eqnarray}
\nonumber
T (S_i, S_{i+1}) = \exp(-\beta {\cal H}_i) = \exp \left[\beta J S_i S_{i+1} + \frac{\beta D}{2} \left(S_i^2 + S_{i+1}^2 \right) + \frac{\beta h}{2} \left(S_i + S_{i+1} \right)\right]\!\!.\\
\label{tmbc}
\end{eqnarray}
Apparently, the relevant expression (\ref{tmbc}) can be identified as the three-by-three transfer matrix:
\begin{eqnarray}
T (S_i, S_{i+1}) = \left( \begin{array}{ccc}
T (1,1) & T (1,0) & T (1,-1) \\
T (0,1) & T (0,0) & T (0,-1) \\
T (-1,1) & T (-1,0) & T (-1,-1)
\end{array}
\right),
\label{tmbcmf}
\end{eqnarray}
whose rows (columns) determine the spin state of the former (latter) spin. It is quite evident that a sequential summation over the spin states in the derived formula (\ref{pfbeg}) for the partition function of the spin-1 Blume-Capel chain will correspond to a multiplication of the relevant transfer matrices. Owing to this fact, the partition function can easily be calculated from the eigenvalues $\lambda_i$ of the transfer matrix (\ref{tmbc})-(\ref{tmbcmf}) according to:
\begin{eqnarray}
Z = \sum_{S_1 = \pm 1, 0} T^N (S_1, S_1) = \mbox{Tr} \, T^N  = \sum_{i=1}^3 \lambda_i^N.
\label{pfbege}
\end{eqnarray}
For completeness, let us quote the final expressions for three eigenvalues of the transfer matrix defined by Eqs.~(\ref{tmbc})-(\ref{tmbcmf}):
\begin{eqnarray}
\lambda_i = r + 2 \, {\rm sgn} (q) \, \sqrt{p} \cos \left[\phi + (i-1) \frac{2\pi}{3} \right],
\label{evtm}
\end{eqnarray}
where
\begin{eqnarray}
{\rm sgn} (q) = \left \{\begin{array}{rl}
 -1, & ~q < 0
\\[3pt]
  1, & ~q \geq 0
\end{array}\right. , \nonumber
\end{eqnarray}
and the parameters $r$, $p$ and $q$ are defined as follows:
\begin{eqnarray}
r &=& \frac{1}{3} \left[1 + 2 \exp(\beta J + \beta D) \cosh \left(\beta h \right) \right], \nonumber \\
p &=& \frac{1}{4} (r-1)^2 + \frac{1}{3} \exp(2 \beta J + 2 \beta D) \sinh^2 \left(\beta h \right) \nonumber \\
&& + \frac{1}{3} \exp(-2 \beta J + 2 \beta D) + \frac{2}{3} \exp(\beta D) \cosh \left(\beta h \right), \nonumber \\
q &=& r^3 - \exp(\beta J + 2 \beta D) + \exp(-\beta J + 2 \beta D) \nonumber \\
&& + r \exp(\beta D) \cosh \left(\beta h \right) + (1 - r) \exp(2 \beta D) \sinh (2 \beta J) \nonumber \\
&& - r \exp(\beta J + \beta D) \cosh (\beta h),  \nonumber \\
\phi &=& \frac{1}{3} \arctan \left(\frac{\sqrt{p^3 - q^2}}{q} \right).
\label{pqr}
\end{eqnarray}
In the thermodynamic limit $N \to \infty$, the partition function as well as the associated Gibbs free energy per site of the spin-1 Blume-Capel chain is simply given by the largest transfer-matrix eigenvalue $\lambda_{\rm max} = {\rm max} \{ \lambda_1, \lambda_2, \lambda_3 \}$: 
\begin{eqnarray}
G = - k_{\rm B} T \lim_{N \to \infty} \frac{1}{N} \ln Z = - k_{\rm B} T \ln \lambda_{\rm max}.
\label{frebege}
\end{eqnarray}
In an absence of the external field ($h = 0$), the analytic formula for the largest transfer-matrix eigenvalue reduces to much simpler form:
\begin{eqnarray}
\lambda_{\rm max} = \frac{1}{2} \left\{ 1 + 2 \exp(\beta D) \cosh(\beta J) + \sqrt{\left[1 - 2 \exp(\beta D) \cosh(\beta J) \right]^2 + 8 \exp(\beta D)} \right\}\!. \nonumber \\
\label{evtml}
\end{eqnarray}
Obviously, the largest transfer-matrix eigenvalue (\ref{evtml}) and also any of its derivative is relatively simple analytic function of temperature and so is the associated Gibbs free energy (and any of its derivatives). This result furnishes a rigorous proof about a non-existence of finite-temperature phase transition in the spin-1 Blume-Capel chain quite similarly as we have previously evidenced (in somewhat more detail) for the spin-1/2 Ising chain.  

\subsubsection{Mixed-spin Ising chain in a longitudinal field}

We will consider the mixed spin-1/2 and spin-1 Ising chain as the last example of one-dimensional spin model, which is exactly tractable through the general scheme offered by the transfer-matrix method. The mixed spin-1/2 and spin-1 Ising chain accounting for the uniaxial single-ion aniso\-tropy and the longitudinal magnetic field can be defined by means of the Hamiltonian:
\begin{equation}
{\cal H} = - J \sum_{i=1}^{N} S_i (\sigma_i + \sigma_{i+1}) - D \sum_{j=1}^{N} S_j^2 - h_A \sum_{i=1}^{N} \sigma_i - h_B \sum_{j=1}^{N} S_j,
\label{hamix}
\end{equation}
where $\sigma_i = \pm 1/2$ and $S_j = \pm 1, 0$ are the spin-1/2 and spin-1 Ising variables located at the $i$th and $j$th lattice point, respectively. The first summation accounts for the Ising interaction $J$ between the nearest-neighbor spin-1/2 and spin-1 particles, the second summation takes into consideration the uniaxial single-ion anisotropy $D$ acting on the spin-1 particles, while the third and fourth summations account for the Zeeman's energy of the spin-1/2 and spin-1 particles in longitudinal magnetic fields $h_A$ and $h_B$, respectively. Note furthermore that the primitive unit cell of the mixed-spin Ising chain defined through the Hamiltonian (\ref{hamix}) contains two different spins (one spin-1/2 particle and one spin-1 particle) regularly alternating along the chain axis, whereas the periodic boundary condition is imposed by the constraint $\sigma_{N+1} \equiv \sigma_{1}$.

Now, let us decompose the total Hamiltonian (\ref{hamix}) into the symmetric parts:
\begin{equation}
{\cal H} = \sum_{i=1}^N {\cal H}_i, \quad {\cal H}_i = - J S_i (\sigma_i + \sigma_{i+1}) - D (S_i^2 + S_{i+1}^2) - \frac{h_A}{2} (\sigma_i + \sigma_{i+1}) - h_B S_i,
\label{hamixi}
\end{equation}
which allows us to write the partition function of the mixed spin-1/2 and spin-1 Ising chain in the factorized form:
\begin{eqnarray}
Z  \!\!\!&=&\!\!\! \sum_{\sigma_1 = \pm \frac{1}{2}} \sum_{\sigma_2 = \pm \frac{1}{2}} \! \! \cdots \! \! \sum_{\sigma_N = \pm \frac{1}{2}} 
\prod_{i=1}^N \sum_{S_i = \pm 1,0} \exp(-\beta {\cal H}_i) \nonumber \\ 
\!\!\!&=&\!\!\! \sum_{\sigma_1 = \pm \frac{1}{2}} \sum_{\sigma_2 = \pm \frac{1}{2}} \! \! \cdots \! \! \sum_{\sigma_N = \pm \frac{1}{2}} \prod_{i=1}^N  T (\sigma_i, \sigma_{i+1}).
\label{pfmixic}
\end{eqnarray}
In above, we took advantage of the fact that spin degrees of freedom of the spin-1 particles can be accounted for independently of each other and before performing a summation over spin degrees of freedom of the spin-1/2 particles. Consequently, the straightforward summation over the spin states of the spin-1 variable leads to the effective Boltzmann's weight with a physical meaning of two-by-two transfer matrix:  
\begin{eqnarray}
T (\sigma_i, \sigma_{i+1}) \!\!\!&=&\!\!\! \sum_{S_i = \pm 1,0} \exp(-\beta {\cal H}_i) \nonumber \\
\!\!\!&=&\!\!\! \exp \left[\frac{\beta h_A}{2} (\sigma_i + \sigma_{i+1})\right] \left \{1 + 2 \exp(\beta D) \cosh [\beta J (\sigma_i + \sigma_{i+1}) ] \right\}.
\label{tmixic}
\end{eqnarray}
Owing to the left-right symmetry, the transfer matrix (\ref{tmixic}) includes just three different matrix elements: 
\begin{eqnarray}
T (\sigma_i, \sigma_{i+1}) = 
\left(\!
\begin{array}{cc} 
   T(+,+)     &     T(+,-)    \\
   T(-,+)     &     T(-,-)    \\
\end{array}
                 \! \right)  =
\left(\!
\begin{array}{cc} 
   V_1     &     V_3    \\
   V_3     &     V_2    \\
\end{array}
                 \! \right),
\label{tmixice}
\end{eqnarray}
whose explicit form is given by:
\begin{eqnarray}
V_1 \!\!\!&=&\!\!\! T (+,+) = \exp \left(\frac{\beta h_A}{2} \right) \left [1 + 2 \exp(\beta D) \cosh (\beta J) \right], \nonumber \\
V_2 \!\!\!&=&\!\!\! T (-,-) = \exp \left(-\frac{\beta h_A}{2} \right) \left [1 + 2 \exp(\beta D) \cosh (\beta J) \right], \nonumber \\
V_3 \!\!\!&=&\!\!\! T (+,-) = T (-,+) = 1 + 2 \exp(\beta D).
\label{tmixicev}
\end{eqnarray}
Now, let us proceed in Eq. (\ref{pfmixic}) to subsequent summation over spin degrees of freedom of the spin-1/2 particles, which enables one to express the partition function in terms of two eigenvalues of the transfer matrix (\ref{tmixice}) when taking advantage the transfer-matrix formalism: 
\begin{eqnarray}
Z  = \sum_{\sigma_1 = \pm \frac{1}{2}} \sum_{\sigma_2 = \pm \frac{1}{2}} \! \!\! \cdots \! \! \! \sum_{\sigma_N = \pm \frac{1}{2}} \prod_{i=1}^N  T (\sigma_i, \sigma_{i+1}) 
= \sum_{\sigma_1 = \pm \frac{1}{2}} \!\! T^N(\sigma_1, \sigma_1) = \mbox{Tr} \, T^N = \lambda_{+}^N + \lambda_{-}^N. \nonumber \\
\label{pfmixicr}
\end{eqnarray}
Two eigenvalues $\lambda_{\pm}$ entering into the partition function (\ref{pfmixicr}) can in turn be calculated by solving the eigenvalue problem for the transfer matrix (\ref{tmixice}):
\begin{eqnarray}
\lambda_{\pm} = \frac{1}{2} \left[V_1 + V_2 \pm \sqrt{(V_1 - V_2)^2 + 4 V_3^2} \right].
\label{pfmixev}
\end{eqnarray}
In the thermodynamic limit $N \to \infty$, the Gibbs free energy per elementary unit can be expressed solely in terms of the largest transfer-matrix eigenvalue: 
\begin{eqnarray}
G  = - k_{\rm B} T \frac{1}{N} \lim_{N \to \infty} \ln Z = - k_{\rm B} T \ln \lambda_{+},  
\label{gmix}	
\end{eqnarray} 
which repeatedly points to an absence of the phase transition as neither the exact expression (\ref{gmix}) for the Gibbs free energy nor any of its temperature derivative cannot exhibit singularity at finite temperature.

\subsection{Two-dimensional Ising models}

2D Ising model in a zero magnetic field is still an exactly soluble model within the framework of the transfer-matrix method, but this relatively simple lattice-statistical model may additionally display non-trivial phase transitions and critical phenomena at finite temperature in contrast to its 1D counterpart. It should be nevertheless mentioned that it took almost two decades of intense efforts since the pioneering exact solution of 1D Ising model \cite{isin25} until the exact closed-form  solution was found for 2D Ising model by Onsager \cite{onsa44}. It is worthwhile to remark, moreover, that Ising conjectured in his seminal work \cite{isin25} an absence of phase transitions at non-zero temperature for any (even higher-dimensional) Ising model. This erroneous result was firstly questioned by Peierls \cite{peie36,grif64} who argued that 2D Ising model must necessarily exhibit a phase transition towards the spontaneously ordered phase at sufficiently low but non-zero temperature. Kramers and Wannier were the first who have exactly confirmed Peierls' conjecture by making the fundamental discovery of a duality in low- and high-temperature expansions of the partition function \cite{kram41}. In addition, the self-dual property of the Ising model on a square lattice has enabled them to exactly locate a critical temperature of the order-disorder phase transition. The subsequent Onsager's exact solution was an additional breakthrough by virtue of which 2D Ising model became the most famous exactly solved model in statistical mechanics with a non-trivial phase transition at non-zero temperature. Up to the present time, phase transitions and associated critical phenomena belong to the most attracting issues to deal with in 2D Ising models. In this part we will provide exact results for several 2D Ising models available after modest calculation.

\subsubsection{Dual lattice and dual transformation}
Let us consider the spin-1/2 Ising model on arbitrary 2D lattice in an absence of the external magnetic field, which is given by the following simple Hamiltonian:
\begin{eqnarray}
{\cal H} = - J \sum_{\langle i j \rangle}^{N_{\rm B}} \sigma_i \sigma_j.
\label{2d1}	
\end{eqnarray}
Above, $\sigma_i = \pm 1/2$ represents the Ising spin located at the $i$th lattice point and the summation is carried out over all nearest-neighbor spin pairs on some 2D lattice. Assuming that 
$N$ is a total number of lattice sites and $q$ is its coordination number (number of nearest neighbors), then, there is in total $N_{\rm B}=Nq/2$ pairs of nearest-neighbor spins as along as boundary effects are neglected (i.e. in thermodynamic limit). Each line (bond), which connects two adjacent spins on a lattice, can be thus regarded as a schematic representation of the exchange interaction $J$ between the nearest-neighbor spins. As usual, the central issue is to calculate the canonical partition function:
\begin{eqnarray}
Z = \sum_{\{ \sigma_i \}} \exp(- \beta {\cal H}), 
\label{2d2}	
\end{eqnarray}
where the suffix $\{ \sigma_i \}$ denotes a summation over all possible configurations of a full set of the Ising spins situated on sites of a given lattice. It is worth noticing that there is in total $2^N$ distinct spin configurations, whereas the summation over vast number of spin configurations represents the most important computational difficulty as the number of spins $N$ tends to infinity in the thermodynamic limit. To overcome this problem, let us rewrite at first the Hamiltonian (\ref{2d1}) to the following form:
\begin{eqnarray}
{\cal H} = - \frac{N_{\rm B} J}{4} + J \sum_{\langle i j \rangle}^{N_{\rm B}} \left(\frac{1}{4} - \sigma_i \sigma_j \right).
\label{2d3}	
\end{eqnarray}
Since the canonical ensemble average of Hamiltonian readily represents the internal energy (${\cal U} = \langle {\cal H} \rangle$), it is easy to find the following physical interpretation 
of the Hamiltonian (\ref{2d3}). Each couple of unlike oriented adjacent spins contributes to the sum that appears on the right-hand-side of Eq.~(\ref{2d3}) by the energy gain $J/2$, while each 
couple of equally aligned adjacent spins does not contribute to this sum at all. In this respect, the internal energy acquires its minimum value ${\cal U}=- N_{\rm B} J/4$ just for two fully aligned spin configurations with all spins pointing either 'up' or 'down' when assuming the ferromagnetic coupling constant $J>0$. By substituting the Hamiltonian (\ref{2d3}) into Eq.~(\ref{2d2}), one gets the following expression for the partition function: 
\begin{eqnarray}
Z = \exp \left(\frac{\beta J N_{\rm B}}{4} \right) \sum_{\{ \sigma_i \}} \exp \left(- n  \frac{\beta J}{2} \right), 
\label{2d4}	
\end{eqnarray}  
where $n$ denotes the total number of unequally aligned adjacent spin pairs within each spin configuration. Evidently, the summation on the right-hand-side of Eq.~(\ref{2d4}) may in principle contain just different powers of the expression $\exp(- \beta J/2)$. In the limit of zero temperature ($T \to 0$), the expression $\exp(-\beta J/2)$ tends to zero and hence, the expansion into a power series 
$\sum_n a_n \exp(- n \beta J/2)$ gives very valuable estimate of the partition function in the limit of sufficiently low temperatures and is therefore termed as the \textit{low-temperature series expansion}.

It is possible to find a simple geometric interpretation of each term emerging in the aforementioned power series by introducing the idea of a \textit{dual lattice}. For this purpose, let us introduce a basic terminology of the \textit{graph theory}, where each site of a lattice is called as a \textit{vertex} and each bond (line) connecting the nearest-neighbor sites (vertices) is called an \textit{edge}. An interior of elementary polygon delimited by edges is denoted as a \textit{face}, while an ensemble of all vertices and edges is called a \textit{full lattice graph}. A dual lattice graph can be constructed as follows. Vertices of a dual lattice are simply obtained by situating them into a middle of each face of the original lattice. The vertices situated at adjacent faces, i.e. the faces sharing a common edge on the original lattice, then represent the nearest-neighbor vertices of the dual lattice. Edges of the dual lattice are obtained by connecting all couples of adjacent vertices of the dual lattice.  
\begin{figure}[t]
\begin{center}
\includegraphics[clip,width=0.95\textwidth]{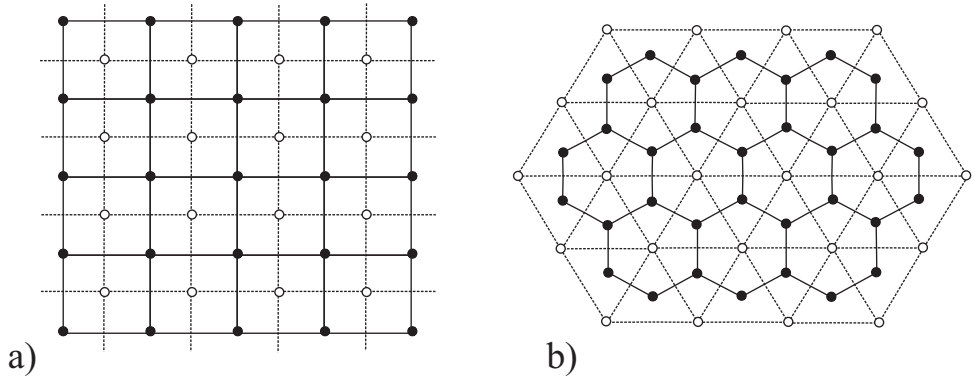}
\end{center}
\vspace{-0.4cm}
\caption{\small Two pairs of dual lattices: a) self-dual square lattices; b) mutually dual hexagonal and triangular lattices. Solid lines and solid circles label edges and vertices of original 
lattices, while broken lines and empty circles stand for edges and vertices of dual lattices, respectively.}
\label{fig:dual}
\end{figure}

\begin{figure}[!t]
\begin{center}
\includegraphics[clip,width=4cm]{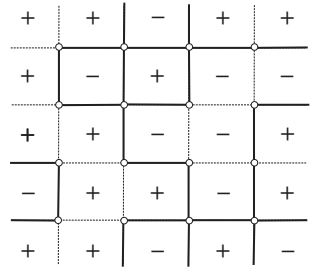}
\end{center}
\vspace{-0.4cm}
\caption{\small A particular spin configuration on a square lattice, whose vertices and edges are not drawn for clarity. The plus (minus) sign at the $i$th lattice point corresponds to the spin 
state $\sigma_i = +1/2$ ($\sigma_i = -1/2$). The system of solid and broken lines unambiguously determines the corresponding polygon line graph at the dual square lattice (for details see the text).}
\label{fig:con}
\end{figure}

For illustration, Fig.~\ref{fig:dual} shows two original lattices -- square and hexagonal -- and their corresponding dual lattices. The bonds of original lattices are displayed by solid lines, while the bonds of their dual lattices are depicted by broken lines. As one can see, the square lattice is a self-dual, i.e. the dual lattice to a square lattice is again a square lattice. Contrary to this, the triangular lattice is a dual lattice to the hexagonal lattice and vice versa. Remembering that $N$ and $N_{\rm B}$ is the total number of vertices and edges of the original lattice, respectively, it is quite clear that $N_{\rm B}$ is at the same time the total number of edges of the dual lattice as well (each bond of the dual lattice intersects one and just one bond of the original lattice). If $N_{\rm D}$ denotes the total number of vertices of the dual lattice, $N_{\rm D}$ is in turn also the total number of faces of the original lattice (each face of the original lattice involves one and just one vertex of the dual lattice). Euler's relation for planar graphs\footnote{The planar graph is a graph, which can be embedded in the plane so that no edges intersect.} then relates the total number of sites of the original and dual lattices with the total number of edges:\footnote{The right-hand-side of Eq.~(\ref{2d5}) is in fact either $N_{\rm B} + 2$ or $N_{\rm B} + 1$ for any planar graph, however, in the thermodynamic limit one can use $N_{\rm B}$ instead of them.} 
\begin{eqnarray}
N + N_{\rm D} = N_{\rm B}.
\label{2d5}	
\end{eqnarray}  

Suppose now some random arrangement of 'up' and 'down' spins on the original lattice, i.e. some particular spin configuration as displayed in Fig.~\ref{fig:con} for some fragment of the whole original lattice (the bonds connecting adjacent spins on the original lattice are not drawn for clarity). If some pair of adjacent spins is not equally aligned, then, let us draw a solid line on a respective edge of a dual lattice situated in between two unequally oriented spins. On the contrary, broken lines will be just drawn on edges of a dual lattice separating equally aligned pairs of adjacent spins. The set of vertices connected on the dual lattice by solid lines then creates a configurational graph, which is a subgraph of the full dual lattice graph (the sets of vertices and edges of a configurational graph are subsets of vertex set and edge set of a full dual lattice graph). The most fundamental property of the configurational graph is that a mutual interchange of any two adjacent spins will not cause any change in the configurational graph, or it will cause an even number of changes. 
The total number of solid and broken lines incident at each vertex of the dual lattice must be therefore either even or zero. This implies that solid (broken) lines of each configurational graph should consists of closed polygons. So, the configurational graph  must form a certain kind of polygon line graph. Another important property of the polygon line graph is that the reversal of all spins does not change the configurational graph. It means that two different spin configurations (one is being obtained from the other by reversing all spins) correspond just to one configurational polygon line graph due to invariance $\sigma_i \to - \sigma_i$ ($\forall i$). The partition function of the planar Ising model can be therefore expressed as: 
\begin{eqnarray}
Z = 2 \exp \left(\frac{\beta J N_{\rm B}}{4} \right) \sum_{{\rm p.g.}} \exp\left(- n \frac{\beta J}{2} \right), 
\label{2d6}	
\end{eqnarray}  
where the summation is now performed over all possible polygon subgraphs on a dual lattice, $n$ denotes the total number of solid lines within each polygon subgraph 
and the factor 2 comes from the two-to-one mapping between spin and polygon configurations.

Another interesting property of the partition function of the spin-1/2 Ising model directly follows from its definition (\ref{2d2}). By adopting the exact van der Waerden identity \cite{waer41}:
\begin{eqnarray}
\exp(\beta J \sigma_i \sigma_j) \!\!\!&=&\!\!\! \cosh \left(\frac{\beta J}{4} \right) + 4 \sigma_i \sigma_j \sinh \left( \frac{\beta J}{4} \right) \nonumber \\
                                \!\!\!&=&\!\!\! \cosh \left(\frac{\beta J}{4} \right) \left[1 + 4 \sigma_i \sigma_j \tanh \left( \frac{\beta J}{4} \right) \right] 
\label{2d7}	
\end{eqnarray}  
and substituting it into Eq.~(\ref{2d2}) one obtains:
\begin{eqnarray}
Z = \left[\cosh \left(\frac{\beta J}{4} \right)\right]^{N_{\rm B}} \sum_{\{ \sigma_i \}} \prod_{\langle ij \rangle} 
           \left[1 + 4 \sigma_i \sigma_j \tanh \left( \frac{\beta J}{4} \right) \right]. 
\label{2d8}	
\end{eqnarray} 
The product on the right-hand-side of Eq.~(\ref{2d8}) involves in total $N_{\rm B}$ terms (one for each bond), which give after formal multiplication a sum of in total $2^{N_{\rm B}}$ terms. However, 
many terms eventually vanish after a summation over all available spin configurations $\sum_{\{ \sigma_i \}}$ is explicitly carried out. For instance, it can be readily proved that all linear terms of the type $4 \sigma_i \sigma_j \tanh(\beta J/4)$ will finally disappear when summing over spin states of the spin $\sigma_i$ or $\sigma_j$. The condition, which ensures that the relevant term makes a non-zero contribution to the partition function, can be simply guessed from the validity of the trivial spin identity $\sigma_i^2 = 1/4$. According to this, all non-zero terms must necessarily consist of spin variables, which enter into these expressions either even number of times or do not enter into these terms at all. The simplest non-vanishing term for the Ising square lattice apparently is: 
\begin{eqnarray}
4^4 (\sigma_i \sigma_j) (\sigma_j \sigma_k) (\sigma_k \sigma_l) (\sigma_l \sigma_i) \tanh^4 \left( \frac{\beta J}{4} \right)
= 4^4 \sigma_i^2 \sigma_j^2 \sigma_k^2 \sigma_l^2 \tanh^4 \left( \frac{\beta J}{4} \right) = \tanh^4 \left( \frac{\beta J}{4} \right), 
\nonumber
\end{eqnarray} 
which is constituted by the product of four nearest-neighbor interactions (edges) forming an elementary square face as the simplest closed polygon on this lattice. The summation over spin configurations of four spins included in this term consequently gives the factor $2^4 \tanh^4(\beta J/4)$, while the summation over spin states of all other spins yields an additional factor $2^{N-4}$, so that the Boltzmann's factor $2^{N} \tanh^4(\beta J/4)$ is finally obtained as the contribution to the partition function stemming from a single square line graph. It is therefore not difficult to construct the following geometric interpretation of the non-vanishing terms: edges corresponding to interactions present in these terms must create closed polygons so that either no lines 
or even number of lines meet at each vertex of the lattice. From this point of view, polygon line graphs very similar to those described by the low-temperature series expansion 
give a non-zero contribution to the partition function. With regard to this, the expression (\ref{2d8}) for the partition function can be replaced by:
\begin{eqnarray}
Z = 2^N \left[\cosh \left(\frac{\beta J}{4} \right)\right]^{N_{\rm B}} \sum_{{\rm p.g.}} \left[\tanh \left( \frac{\beta J}{4} \right) \right]^n, 
\label{2d10}	
\end{eqnarray}  
where $n$ denotes the total number of full lines constituting the particular polygon line graph. It is quite apparent that the summation on the right-hand-side of Eq.~(\ref{2d10}) contains 
just different powers of the expression $\tanh(\beta J/4)$. In the limit of high temperatures ($T \to \infty$), the expression $\tanh(\beta J/4)$ tends to zero and thus, the expansion into a power series $\sum_n a_n [ \tanh (\beta J/4) ]^n$ gives very valuable estimate of the partition function in the limit of sufficiently high temperatures termed as the \textit{high-temperature series expansion}.

There is an interesting correspondence between summations to emerge in Eqs.~(\ref{2d6}) and (\ref{2d10}), since both of them are performed over certain sets of polygon line graphs. The most
essential difference between them consists in the fact that the summation in Eq.~(\ref{2d6}) is carried out over polygon line graphs on the dual lattice, while the summation in Eq.~(\ref{2d10}) 
is performed over polygon line graphs directly on the original lattice. It should be stressed, however, that both the final expressions (\ref{2d4}) and (\ref{2d10}) for the partition function are exact 
when the relevant series is performed up to an infinite order and therefore, they must basically give the same result for the partition function. In the thermodynamic limit $N \to \infty$, the 
summations in Eqs. (\ref{2d6}) and (\ref{2d10}) yield the same partition function provided that: 
\begin{eqnarray}
\exp\left(- \frac{\beta_{\rm D} J}{2} \right) \! \! \! &=& \! \! \! \tanh \left( \frac{\beta J}{4} \right), \label{2d11a}	\\
\frac{Z(N_{\rm D}, \beta_{\rm D} J)}{\exp\left(\frac{\beta_{\rm D} J N_{\rm B}}{4} \right)} 
\! \! \! &=& \! \! \!  \frac{Z(N, \beta J)}{2^{N} \left[\cosh\left( \frac{\beta J}{4} \right)\right]^{N_{\rm B}}}, 
\label{2d11b}	
\end{eqnarray} 
where we have introduced some reciprocal temperature $\beta_{\rm D} = 1/(k_{\rm B} T_{\rm D})$ for the dual lattice and the factor $2$ was omitted from the denominator on the left-hand-side 
of Eq.~(\ref{2d11b}) due to its negligible effect in the thermodynamic limit. The connection between two mutually dual lattices (\ref{2d11a}) and (\ref{2d11b}) can also be inverted because 
of a symmetry in the duality (each from a couple of the mutually dual lattices is dual one to another):
\begin{eqnarray}
\exp\left(- \frac{\beta J}{2} \right) \! \! \! &=& \! \! \! \tanh \left( \frac{\beta_{\rm D} J}{4} \right), \label{2d12a}	\\
\frac{Z(N, \beta J)}{\exp\left(\frac{\beta J N_{\rm B}}{4} \right)} 
\! \! \! &=& \! \! \!  \frac{Z(N_{\rm D}, \beta_{\rm D} J)}{2^{N_{\rm D}} \left[\cosh\left( \frac{\beta_{\rm D} J}{4} \right)\right]^{N_{\rm B}}}. 
\label{2d12b}	
\end{eqnarray} 
With the help of Eqs.~(\ref{2d11a}) and (\ref{2d11b}) [or equivalently Eqs.~(\ref{2d12a}) and (\ref{2d12b})], it is also possible to write this so-called \textit{dual transformation} 
even in a symmetric form as would be expected from the symmetrical nature of the duality. The relation (\ref{2d11a}) gives after straightforward calculation: 
\begin{eqnarray}
\sinh\left(\frac{\beta J}{2} \right) \sinh\left(\frac{\beta_{\rm D} J}{2} \right) = 1,
\label{2d13}	
\end{eqnarray} 
while the expression (\ref{2d11b}) can be modified by regarding the equality of partition functions, Eqs.~(\ref{2d11a}), (\ref{2d13}) and Euler's relation (\ref{2d5}) to:
\begin{eqnarray}
2^N \! \! \left[\cosh\left(\! \frac{\beta J}{4} \right)\right]^{N_{\rm B}} \!\!\! \exp\!\left(\! - \frac{\beta_{\rm D} J}{4} N_{\rm B} \! \right) \! \! \! &=& \! \! \!  
2^N \left[\cosh\left( \frac{\beta J}{4} \right)\right]^{N_{\rm B}} \left[\tanh\left( \frac{\beta J}{4} \right)\right]^{\frac{N_{\rm B}}{2}} \nonumber \\
\! \! \! &=& \! \! \!  
2^N \left[\sinh\left( \frac{\beta J}{4} \right)\right]^{\frac{N_{\rm B}}{2}} \left[\cosh\left( \frac{\beta J}{4} \right)\right]^{\frac{N_{\rm B}}{2}} \nonumber \\
\! \! \! &=& \! \! \!  2^{N - \frac{N_{\rm B}}{2}} \left[\sinh\left( \frac{\beta J}{2} \right)\right]^{\frac{N_{\rm B}}{2}}  \nonumber \\
\! \! \! &=& \! \! \! 2^{\frac{N - N_{\rm D}}{2}} \left[\sinh\left( \frac{\beta J}{2} \right)\right]^{\frac{N + N_{\rm D}}{2}} \nonumber \\
\! \! \! &=& \! \! \!  \frac{\left[2 \sinh\left(\frac{\beta J}{2} \right)\right]^{\frac{N}{2}}}
                            {\left[2 \sinh\left( \frac{\beta_{\rm D} J}{2} \right)\right]^{\frac{N_{\rm D}}{2}}}.
\label{2d14}	
\end{eqnarray} 
By combining Eq.~(\ref{2d14}) with the relation (\ref{2d11b}), one readily gains another symmetric relationship between the partition functions expressed in terms of the high-temperature series expansion on the original lattice and the low-temperature series expansion on its dual lattice:
\begin{eqnarray}
\frac{Z (N, \beta J)}{\left[2 \sinh\left(\frac{\beta J}{2} \right)\right]^{\frac{N}{2}}} = 
\frac{Z (N_{\rm D}, \beta_{\rm D} J)}{\left[2 \sinh\left(\frac{\beta_{\rm D} J}{2} \right)\right]^{\frac{N_{\rm D}}{2}}}. 
\label{2d15}	
\end{eqnarray} 
The existence of an equivalence between the low- and high-temperature series expansions reflects the fundamental property of the partition function of the spin-1/2 Ising model, namely, its symmetry with respect to the low and high temperatures. This symmetry means, among other matters, that the partition function at some lower temperature can always be mapped on the equivalent partition function at certain higher temperature. This mapping is called as the \textit{dual transformation} and the dual lattices are actually connected one to another by means of the dual transformation. In this respect, the dual lattices are topological representations of the dual transformation and consequently, one says that the dual transformation has a character of the topological transformation. 

The mathematical formulation of the dual transformation connecting effective temperatures of the original and its dual lattice is represented (independently of the lattice topology) either 
by the couple of equivalent Eqs.~(\ref{2d11a}) and (\ref{2d12a}), or, respectively, by a single symmetrized relation (\ref{2d13}). The latter relationship is especially useful for a better understanding of the symmetry of the partition function with respect to low and high temperatures. It is sufficient to realize that the argument of the function $\sinh(\beta J/2)$ must unavoidably decrease when the argument of the other function $\sinh(\beta_{\rm D} J/2)$ increases in order to maintain their constant product as required by the dual transformation (\ref{2d13}). This means 
that the partition function at some lower temperature, which is obtained for instance from the low-temperature expansion on the dual lattice, is equivalent to the partition function at a certain higher temperature obtained from the high-temperature expansion on the original lattice. It is noteworthy that the dual transformation markedly simplifies the exact enumeration of thermodynamic quantities on different lattices, since it permits to obtain the exact solution of some quantity on arbitrary lattice merely from the corresponding exact result for this quantity on its dual lattice. For instance, the critical point is always accompanied by some singularity in the thermodynamic quantities and this non-analyticity is always reflected in the partition function as well. The mapping relation (\ref{2d15}) clearly shows that the partition function of the original lattice exhibits a non-analyticity if and only if the partition function of the dual lattice also has a similar non-analyticity at some corresponding temperature satisfying the duality relation (\ref{2d13}). Besides, the expression (\ref{2d15}) allows to calculate the partition function of the one of mutually dual lattices merely by substituting the corresponding exact result for the partition function of its dual lattice.

It is worthwhile to remark that the square lattice has an extraordinary position among all 2D lattices because of its self-duality. The self-dual property together with the symmetry of the partition function with respect to the low and high temperatures is just enough for determining the critical temperature and other thermodynamic quantities precisely at a critical point. Under the assumption 
of a single critical point, the same lattice topology ensures that critical parameters must be equal one to another on both mutually dual square lattices. According to the dual transformation (\ref{2d13}), the critical temperature of the spin-1/2 Ising model on the square lattice must obey the condition: 
\begin{eqnarray}
\sinh^2\left(\frac{\beta_{\rm c} J}{2} \right) = 1, 
\label{2d16}	
\end{eqnarray} 
which is consistent with this value of the critical temperature: 
\begin{eqnarray}
\frac{k_{\rm B} T_{\rm c}}{|J|} = \frac{1}{2 \ln(1 + \sqrt{2})} \approx 0.567296.
\label{2d17}	
\end{eqnarray} 
It should be pointed out that the critical condition (\ref{2d16}) in fact gives two different solutions for the critical temperature, which are of equal magnitude but of opposite sign. The relevant solutions obviously correspond to the spin-1/2 Ising model with the ferromagnetic $J>0$ or antiferromagnetic $J<0$ coupling constant. Note furthermore that the temperature symmetry of the partition function does not suffice for determining critical parameters of the spin-1/2 Ising model on other 2D lattices (like honeycomb or triangular), which are not self-dual. 

\subsubsection{Star-Triangle Transformation}
The dual transformation (\ref{2d13}) maps the partition function of the spin-1/2 Ising model on a honeycomb lattice into that one of the spin-1/2 Ising model on a triangular lattice (or vice versa) and thus, it does not establish the symmetry relationship between the low- and high-temperature partition function on the same lattice. However, it is worth noticing that such a relation can be established even for other (not self-dual) lattices by combining the dual transformation (\ref{2d13}) with some another transformation. The star-triangle transformation invented originally by Onsager \cite{onsa44} establishes this useful mapping relationship for example for a couple of dual lattices like hexagonal--triangular or kagom\'e--diced. 

\begin{figure}[t]
\begin{center}
\includegraphics[clip,width=12cm]{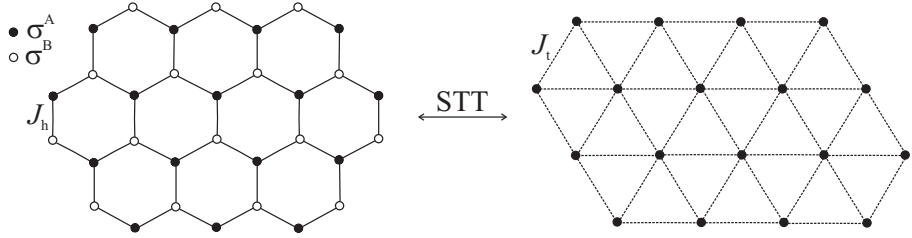}
\end{center}
\vspace{-0.4cm}
\caption{\small The spin-1/2 Ising model on a hexagonal lattice and its relation to the equivalent spin-1/2 Ising model on a triangular lattice. The equivalence between both models can be established by applying the star-triangle transformation (STT) to a half of spins of hexagonal lattice.}
\label{fig:stt}
\end{figure}

Let us consider the spin-1/2 Ising model on the hexagonal lattice, which can be further divided into two equivalent interpenetrating triangular sublattices A and B schematically represented in Fig.~\ref{fig:stt} by full and empty circles, respectively. The division is made in a such way that all nearest neighbors of a site 
from one sublattice (say A) belong to the other sublattice (B) and vice versa. Owing to this fact, one may perform a summation over spin degrees of freedom of spins from the same sublattice independently of each other, because there is no direct interaction between the spins from the same sublattice. In this respect, it is very appropriate to write the total Hamiltonian as a sum of site Hamiltonians:
\begin{eqnarray}
{\cal H} = \sum_{i \in {\rm B}}^{N} {\cal H}_i, 
\label{2d18}	
\end{eqnarray} 
where each site Hamiltonian ${\cal H}_i$ involves all the interaction terms associated with the $i$th spin from the sublattice B:
\begin{eqnarray}
{\cal H}_i = -J_{\rm h} \sigma_i^{\rm B} (\sigma_{j}^{\rm A} + \sigma_{k}^{\rm A} + \sigma_{l}^{\rm A}). 
\label{2d19}	
\end{eqnarray} 
In above, the upper index of spin variable specifies the sublattice to which the relevant spin belongs. By the use of Eqs.~(\ref{2d18}) and (\ref{2d19}), the partition function of the spin-1/2 Ising model on the hexagonal lattice can be partially factorized to the form:
\begin{eqnarray}
Z_{\rm h} = \sum_{\{ \sigma_j^{\rm A} \}} \prod_{i = 1}^N \sum_{\sigma_i^{\rm B} = \pm \frac{1}{2}} 
\exp[\beta J_{\rm h} \sigma_i^{\rm B} (\sigma_{j}^{\rm A} + \sigma_{k}^{\rm A} + \sigma_{l}^{\rm A})], 
\label{2d20}	
\end{eqnarray}  
where the former summation is carried out over all available spin configurations on the sublattice A, the product runs over all sites of the sublattice B and the latter summation accounts for the spin states of one particular spin from the sublattice B. In the consequence of that, it is adequate to consider just one particular spin from the sublattice B and to sum up over spin degrees of freedom of this spin. Each spin from the sublattice B merely interacts with its three nearest-neighbor spins from the sublattice A and hence, this summation gives the Boltzmann's factor:
\begin{eqnarray}
\sum_{\sigma_i^{\rm B} = \pm \frac{1}{2}} \exp[\beta J_{\rm h} \sigma_i^{\rm B} 
(\sigma_{j}^{\rm A} + \sigma_{k}^{\rm A} + \sigma_{l}^{\rm A})] \! \! \! &=& \! \! \! 
2 \cosh \left[\frac{\beta J_{\rm h}}{2} (\sigma_{j}^{\rm A} + \sigma_{k}^{\rm A} + \sigma_{l}^{\rm A})\right] 
\nonumber \\
\! \! \! &=& \! \! \! A \exp[\beta J_{\rm t} (\sigma_{j}^{\rm A} \sigma_{k}^{\rm A} 
          + \sigma_{k}^{\rm A} \sigma_{l}^{\rm A} + \sigma_{l}^{\rm A} \sigma_{j}^{\rm A})],
\label{2d21}	
\end{eqnarray}
which can be substituted by the equivalent expression provided by the \textit{star-triangle transformation}. The physical meaning of the transformation (\ref{2d21}) lies in removing all the interactions associated with a central spin $\sigma_{i}^{\rm B}$ of the \textit{star} and replacing them by new effective interactions between three outer spins $\sigma_{j}^{\rm A}$, $\sigma_{k}^{\rm A}$ and $\sigma_{l}^{\rm A}$ forming the equilateral \textit{triangle}. It is noteworthy that the star-triangle transformation is actually a set of eight equations, which can be obtained from the 
mapping relation (\ref{2d21}) by considering all eight spin configurations available to three outer spins from the sublattice A. However, the consideration of all available spin configurations leads just to two independent equations, which unambiguously determine two yet unknown mapping parameters $A$ and $J_{\rm t}$: 
\begin{eqnarray}
A \! \! \! &=& \! \! \! 2 \left[\cosh\left(\frac{3 \beta J_{\rm h}}{4}\right)\right]^{\frac{1}{4}} 
                          \left[\cosh\left(\frac{\beta J_{\rm h}}{4}\right)\right]^{\frac{3}{4}}, \\ \label{2d22a}	
\beta J_{\rm t}  \! \! \! &=& \! \! \! \ln \left[\frac{\cosh\left(\frac{3 \beta J_{\rm h}}{4}\right)}{\cosh\left(\frac{\beta J_{\rm h}}{4}\right)}\right].
\label{2d22b}	
\end{eqnarray} 
The backward substitution of the transformation (\ref{2d21}) into the partition function (\ref{2d20}), which is equivalent to performing the star-triangle transformation for all spins from 
the sublattice B, yields an exact mapping relationship between the partition functions of the spin-1/2 Ising model on the hexagonal and triangular lattices:
\begin{eqnarray}
Z_{\rm h} (2N, \beta J_{\rm h}) = A^N Z_{\rm t} (N, \beta J_{\rm t}).
\label{2d23}	
\end{eqnarray} 
Notice that the effective couplings (temperatures) of the equivalent spin-1/2 Ising model on the hexagonal and triangular lattices are related through the star-triangle mapping relationship (\ref{2d22b}). As a result, the relation (\ref{2d22b}) established by the use of the star-triangle transformation connects the partition functions of the hexagonal and triangular lattices at two different effective temperatures in a quite similar way as it does the relation (\ref{2d13}) provided by the dual transformation. The most crucial difference consists in a profound essence of both mapping relations; the dual transformation is evidently of the topological origin, whereas the star-triangle mapping is the algebraic transformation in its character.   

At this stage, let us combine the dual and star-triangle transformations in order to bring insight
into the criticality of the spin-1/2 Ising model on the hexagonal and triangular lattices. By 
employing a set of trivial identities\footnote{$\cosh(a \pm b) = \cosh a \cosh b \pm \sinh a \sinh b$,\\ \hspace*{0.44cm} $\cosh 2a = \cosh^2 a + \sinh^2 a$,\\ \hspace*{0.44cm} $\sinh 2a = 2 \sinh a \cosh a$, \\ \hspace*{0.44cm} $\cosh^2 a - \sinh^2 a = 1$,\\ \hspace*{0.44cm} $\cosh^2 a = \frac{1 + \cosh2a}{2}$,\\ \hspace*{0.44cm} $\sinh^2 a = \frac{-1 + \cosh 2a}{2}$.}, the star-triangle transformation (\ref{2d22b}) should be firstly rewritten as follows:  
\begin{eqnarray}
\exp(\beta J_{\rm t})  \! \! \! &=& \! \! \! 
\frac{\cosh\left(\frac{3 \beta J_{\rm h}}{4}\right)}{\cosh\left(\frac{\beta J_{\rm h}}{4}\right)} = 
\frac{\cosh\left(\frac{\beta J_{\rm h}}{2}\right) \cosh\left(\frac{\beta J_{\rm h}}{4}\right)+ 
      \sinh\left(\frac{\beta J_{\rm h}}{2}\right) \sinh\left(\frac{\beta J_{\rm h}}{4}\right)}
     {\cosh\left(\frac{\beta J_{\rm h}}{4}\right)} \nonumber \\
     \! \! \! &=& \! \! \! \cosh\left(\frac{\beta J_{\rm h}}{2}\right) + \tanh\left(\frac{\beta J_{\rm h}}{4}\right) \sinh\left(\frac{\beta J_{\rm h}}{2}\right) \nonumber \\
     \! \! \! &=& \! \! \! \cosh\left(\frac{\beta J_{\rm h}}{2}\right) + 2 \sinh^2 \left(\frac{\beta J_{\rm h}}{4}\right) \nonumber \\
     \! \! \! &=& \! \! \! 2 \cosh\left(\frac{\beta J_{\rm h}}{2}\right) - 1.
\label{2d24}	
\end{eqnarray}
Furthermore, it is appropriate to combine Eq.~(\ref{2d24}) with one of possible representations of the dual transformation (\ref{2d11a}):
\begin{eqnarray}
\exp\left(- \frac{\beta' J_{\rm t}}{2}\right) = \tanh\left(\frac{\beta' J_{\rm h}}{4}\right)
\label{2d25}	
\end{eqnarray}
with the aim to eliminate the effective temperature of the one of lattices with equivalent partition functions. For instance, the procedure that eliminates from Eqs.~(\ref{2d24}) and (\ref{2d25}) the effective temperature of the triangular lattice allows to obtain the symmetrized relationship, which connects the partition function of the spin-1/2 Ising model on the hexagonal lattice at two different temperatures: 
\begin{eqnarray} 
\left[\coth \left(\frac{\beta' J_{\rm h}}{4}\right)\right]^2 \! \! \! &=& \! \! \!  2 \cosh\left(\frac{\beta J_{\rm h}}{2}\right) - 1 \nonumber \\
\frac{\cosh\left(\frac{\beta' J_{\rm h}}{2}\right) + 1}{\cosh\left(\frac{\beta' J_{\rm h}}{2}\right) - 1} \! \! \! &=& \! \! \!  
2 \cosh\left(\frac{\beta J_{\rm h}}{2}\right) - 1 \nonumber \\
\cosh\left(\frac{\beta' J_{\rm h}}{2}\right) + 1 \! \! \! &=& \! \! \!  
\left[2 \cosh\left(\frac{\beta J_{\rm h}}{2}\right) - 1\right] \left[\cosh\left(\frac{\beta' J_{\rm h}}{2}\right) - 1\right] \nonumber \\
\cosh\left(\frac{\beta J_{\rm h}}{2}\right) \cosh\left(\frac{\beta' J_{\rm h}}{2}\right) \! \! \! &-& \! \! \! 
\cosh\left(\frac{\beta J_{\rm h}}{2}\right) - \cosh\left(\frac{\beta' J_{\rm h}}{2}\right) = 0 \nonumber \\ 
 \left[\cosh\left(\frac{\beta J_{\rm h}}{2}\right) \right. \! \! \! &-& \! \! \! 1 \biggr] \left[\cosh\left(\frac{\beta' J_{\rm h}}{2}\right) - 1 \right] = 1.  
\label{2d26}	
\end{eqnarray}
From the physical point of view, the relationship (\ref{2d26}) establishes analogous temperature symmetry of the partition function of the spin-1/2 Ising model on the hexagonal lattice as the dual transformation does for the spin-1/2 Ising model on the self-dual square lattice through the relation (\ref{2d13}). If there exists just a unique critical point, both temperatures connected via the mapping relation (\ref{2d26}) must necessarily meet at a critical point due to the same reasons as given for the spin-1/2 Ising model on a self-dual square lattice. So, the critical temperature of the spin-1/2 Ising model on the hexagonal lattice should obey the condition:
\begin{eqnarray}
\left[\cosh\left(\frac{\beta_{\rm c} J_{\rm h}}{2}\right) - 1\right]^2 = 1,  
\label{2d27}	
\end{eqnarray}
which is consistent with the following double value of the critical temperature: 
\begin{eqnarray}
\frac{k_{\rm B} T_{\rm c}}{|J_{\rm h}|} = \frac{1}{2 \ln(2 + \sqrt{3})} \approx 0.379663.
\label{2d28}	
\end{eqnarray}  
Apparently, the positive (negative) solution corresponds to the spin-1/2 Ising model on the hexagonal lattice with the ferromagnetic (antiferromagnetic) coupling constant $J_{\rm h}>0$ ($J_{\rm h}<0$).

The same procedure can be repeated in order to obtain the critical parameters of the spin-1/2 Ising model on a triangular lattice. An elimination of the effective temperature of spin-1/2 Ising model on the hexagonal lattice from Eqs.~(\ref{2d24}) and (\ref{2d25}) yields the following symmetric mapping: 
\begin{eqnarray}
[\exp(\beta J_{\rm t}) - 1] [\exp(\beta' J_{\rm t}) - 1] = 4,  
\label{2d29}	
\end{eqnarray}
which relates the partition function of the spin-1/2 Ising model on the triangular lattice at two different temperatures. The equation (\ref{2d29}) unambiguously determines the critical condition for the spin-1/2 Ising model on the triangular lattice:
\begin{eqnarray}
[\exp(\beta_{\rm c} J_{\rm t}) - 1]^2 = 4,  
\label{2d30}	
\end{eqnarray}
which affords the following unique value of the critical temperature:
\begin{eqnarray}
\frac{k_{\rm B} T_{\rm c}}{J_{\rm t}} = \frac{1}{\ln 3} \approx 0.910239.
\label{2d31}	
\end{eqnarray}
It is noteworthy that the critical temperature (\ref{2d31}) solely corresponds to the spin-1/2 Ising model on the triangular lattice with the ferromagnetic interaction $J_{\rm t} > 0$, because the analogous model with the antiferromagnetic coupling $J_{\rm t} < 0$ cannot exhibit a spontaneous antiferromagnetism at finite temperature due to a geometric spin frustration inherent to the non-bipartite character of a triangular lattice. It should be also pointed out that the critical temperature of the spin-1/2 Ising model on the triangular lattice (\ref{2d31}) can be alternatively  found by substituting the critical condition of the spin-1/2 Ising model on the hexagonal lattice (\ref{2d27}) to the modified star-triangle transformation (\ref{2d24}). In this way one recovers the same critical temperature as given by Eq. (\ref{2d31}).

\subsubsection{Decoration-Iteration Transformation}

\begin{figure}[t]
\begin{center}
\includegraphics[clip,width=13cm]{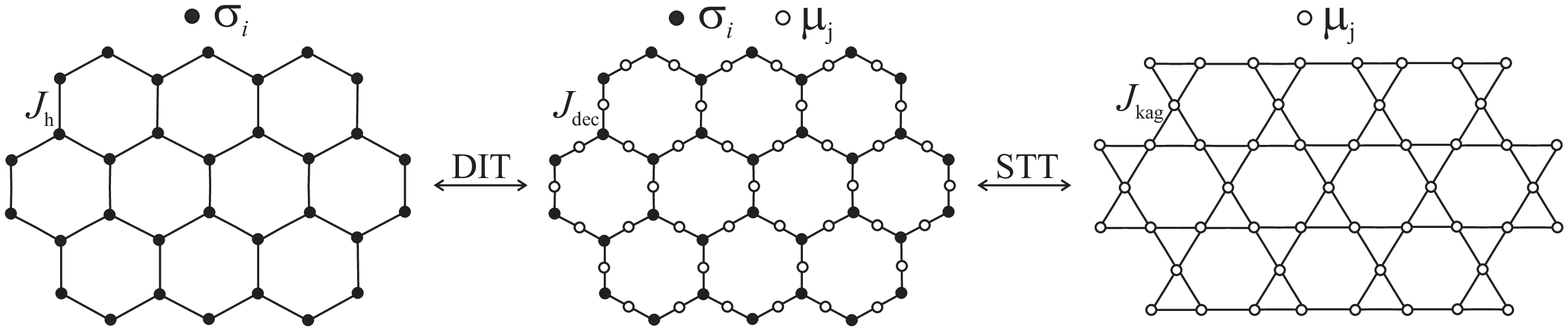}
\end{center}
\vspace{-0.3cm}
\caption{\small The spin-1/2 Ising model on the hexagonal lattice (left), decorated hexagonal lattice (middle) and kagom\'e lattice (right). The equivalence between all three lattice models can be established by employing the decoration-iteration (DIT) and star-triangle (STT) transformations, respectively.}
\label{fig:dit}
\end{figure}

Another important mapping transformation, which is purely of algebraic character, represents the decoration-iteration transformation that was invented by Syozi in an attempt to exactly solve the spin-1/2 Ising model on the kagom\'e lattice \cite{syoz51}. In principle, the approach based on the decoration-iteration transformation enables to obtain an exact solution of the spin-1/2 Ising model on an arbitrary \textit{bond-decorated lattice} from the exact solution of the spin-1/2 Ising model on the corresponding undecorated lattice. The term bond-decorated lattice marks such a lattice, which can be obtained from a simple original lattice (like square, hexagonal or triangular) by placing an additional spin (or eventually a finite cluster of spins) on the bonds of this simpler lattice. The primary sites of the original lattice are then called as  \textit{nodal} sites, while the secondary sites arising out from the decoration procedure are called as \textit{decorating} sites. For illustration, Fig.~\ref{fig:dit} shows the planar topology of the hexagonal lattice, the simply decorated hexagonal lattice and the kagom\'e lattice together with the mapping relations, which can be established between them by employing algebraic transformations. 

Let us consider initially the spin-1/2 Ising model on the decorated hexagonal lattice, which is shown in the middle of Fig.~\ref{fig:dit} and is defined through the Hamiltonian:
\begin{eqnarray}
{\cal H}_{\rm d} = - J_{\rm d} \sum_{\langle ij \rangle}^{3N} \sigma_i \mu_j. 
\label{2d32}	
\end{eqnarray} 
Above, $\sigma_i = \pm 1/2$ and $\mu_j = \pm 1/2$ label Ising spin variables placed at the nodal and decorating sites of the decorated hexagonal lattice, respectively, the summation runs over all nearest-neighbor spin pairs of the lattice and the total number of nodal sites equals $N$. It is convenient to rewrite the total Hamiltonian (\ref{2d32}) as a sum of bond Hamiltonians:
\begin{eqnarray}
{\cal H}_{\rm d} = \sum_{j=1}^{3N/2} {\cal H}_j, 
\label{2d33}	
\end{eqnarray} 
where each bond Hamiltonian ${\cal H}_j$ involves all the interaction terms that include one particular decorating spin $\mu_j$: 
\begin{eqnarray}
{\cal H}_j = - J_{\rm d} \mu_j (\sigma_{k} + \sigma_{l}). 
\label{2d34}	
\end{eqnarray} 
By the use of Eqs.~(\ref{2d33}) and (\ref{2d34}), the partition function of the spin-1/2 Ising model on the decorated hexagonal lattice can be partially factorized to the form:
\begin{eqnarray}
Z_{\rm d} = \sum_{\{ \sigma_i \}} \prod_{j = 1}^{3N/2} \sum_{\mu_j = \pm \frac{1}{2}} \exp[\beta J_{\rm d} \mu_j (\sigma_{k} + \sigma_{l})]. 
\label{2d35}	
\end{eqnarray}  
In the above expression, the former summation is carried out over all available configurations of the nodal spins, the product runs over all decorating sites and the latter summation accounts for spin states of one decorating spin from the $j$th bond of the decorated hexagonal lattice. According to Eq.~(\ref{2d35}), the summation over spin degrees of freedom of the decorating spins can be performed independently of each other and this summation gives the effective Boltzmann's factor, which can be successively replaced by the equivalent expression provided by the \textit{decoration-iteration transformation}:
\begin{eqnarray}
\sum_{\mu_j = \pm \frac{1}{2}} \exp[\beta J_{\rm d} \mu_j (\sigma_{k} + \sigma_{l})] = 2 \cosh \left[\frac{\beta J_{\rm d}}{2} (\sigma_{k} + \sigma_{l})\right] 
= B \exp(\beta J_{\rm h} \sigma_{k} \sigma_{l}).
\label{2d36}	
\end{eqnarray}
The physical meaning of the transformation (\ref{2d36}) consists in removing both interactions associated with the decorating spin $\mu_{j}$ and substituting them through a single effective interaction $J_{\rm h}$ between two nodal spins $\sigma_{k}$ and $\sigma_{l}$, which are being its two nearest neighbors. It should be emphasized that the decoration-iteration transformation (\ref{2d36}) has to satisfy the self-consistency condition, i.e. it must hold for all available spin states of two nodal spins $\sigma_{k}$ and $\sigma_{l}$. In this respect, the mapping relation (\ref{2d36}) is in fact a set of four equations, which can be explicitly obtained by considering all possible spin configurations available to the two nodal spins. It can be readily proved that a substitution of four possible spin configurations yields from the formula (\ref{2d36}) merely two independent equations, which determine the mapping parameters $B$ and $\beta J_{\rm h}$: 
\begin{eqnarray}
B \! \! \! &=& \! \! \! 2 \sqrt{\cosh\left(\frac{\beta J_{\rm d}}{2}\right)}, \\ \label{2d37a}	
\beta J_{\rm h}  \! \! \! &=& \! \! \! 2 \ln \left[\cosh\left(\frac{\beta J_{\rm d}}{2}\right) \right].
\label{2d37b}	
\end{eqnarray} 
By applying the decoration-iteration transformation to all the decorating spins, i.e. substituting the mapping transformation (\ref{2d36}) into the partition function (\ref{2d35}), one acquires an exact mapping correspondence between the partition functions of the spin-1/2 Ising model on the decorated hexagonal lattice and simple hexagonal lattice:
\begin{eqnarray}
Z_{\rm d} (5N/2, \beta J_{\rm d}) = B^{\frac{3N}{2}} Z_{\rm h} (N, \beta J_{\rm h}),
\label{2d38}	
\end{eqnarray} 
whose effective temperatures are connected by means of the mapping relation (\ref{2d37b}). It is quite obvious from Eq.~(\ref{2d38}) that the spin-1/2 Ising model on the decorated hexagonal lattice becomes critical if and only if its corresponding model on the hexagonal lattice becomes critical as well. With regard to this, it is sufficient to substitute the exact critical temperature of 
the hexagonal lattice (\ref{2d28}) into Eq.~(\ref{2d37b}) in order to locate the exact critical temperature of the spin-1/2 Ising model on the decorated hexagonal lattice:
\begin{eqnarray}
\frac{k_{\rm B} T_{\rm c}}{|J_{\rm d}|} = \frac{1}{2 \ln(2 + \sqrt{3} + \sqrt{6 + 4 \sqrt{3}})} \approx 0.251048.
\label{2d39}	
\end{eqnarray} 
It is worthwhile to remark that the decoration-iteration transformation (\ref{2d36}) is restricted neither by the lattice topology nor by its spatial dimensionality and thus, it can be utilized for obtaining the exact results for the spin-1/2 Ising model on arbitrary simply decorated lattice if the exact solution is known for the spin-1/2 Ising model on the corresponding undecorated lattice.

Now, it is possible to make another vital observation. The total Hamiltonian (\ref{2d32}) of the spin-1/2 Ising model on the decorated hexagonal lattice can be also formally written
as a sum of site Hamiltonians:
\begin{eqnarray}
{\cal H}_{\rm d} = \sum_{i=1}^{N} {\cal H}_i, 
\label{2d40}	
\end{eqnarray} 
whereas each site Hamiltonian involves all the interaction terms of the one particular nodal spin: 
\begin{eqnarray}
{\cal H}_i = - J_{\rm d} \sigma_i (\mu_{j} + \mu_{k} + \mu_{l}). 
\label{2d41}	
\end{eqnarray} 
Substituting Eqs.~(\ref{2d40}) and (\ref{2d41}) into a statistical definition of the partition function allows us to partially factorize the partition function of the spin-1/2 Ising model on the decorated hexagonal lattice and write it in the form:
\begin{eqnarray}
Z_{\rm dec} = \sum_{\{ \mu_j \}} \prod_{i = 1}^{N} \sum_{\sigma_i = \pm \frac{1}{2}} \exp[\beta J_{\rm d} \sigma_i (\mu_{j} + \mu_{k} + \mu_{l})]. 
\label{2d42}	
\end{eqnarray}  
Above, the former summation runs over all available configurations of the decorating spins, the product runs over all nodal sites and the latter summation is carried out over spin states 
of one particular nodal spin from the decorated hexagonal lattice. The structure of the partition function (\ref{2d42}) immediately justifies an application of the familiar star-triangle mapping transformation:
\begin{eqnarray}
\sum_{\sigma_i = \pm \frac{1}{2}} \exp[\beta J_{\rm d} \sigma_i (\mu_{j} + \mu_{k} + \mu_{l})] \! \! \! &=& \! \! \! 2 \cosh\left[\frac{\beta J_{\rm d}}{2} (\mu_{j} + \mu_{k} + \mu_{l})\right] \nonumber \\
\! \! \! &=& \! \! \! C \exp[\beta J_{\rm k} (\mu_{j} \mu_{k} + \mu_{k} \mu_{l} + \mu_{l} \mu_{j})],
\label{2d43}	
\end{eqnarray}
which satisfies the self-consistency condition on assumption that:
\begin{eqnarray}
C \! \! \! &=& \! \! \! 2 \left[\cosh\left(\frac{3 \beta J_{\rm d}}{4}\right) \right]^{\frac{1}{4}} 
                          \left[\cosh\left(\frac{\beta J_{\rm d}}{4}\right) \right]^{\frac{3}{4}}, \\ \label{2d44a}	
\beta J_{\rm k}  \! \! \! &=& \! \! \! \ln \left[\frac{\cosh\left(\frac{3 \beta J_{\rm d}}{4}\right)}{\cosh\left(\frac{\beta J_{\rm d}}{4}\right)}\right]
= \ln \left[2 \cosh\left(\frac{\beta J_{\rm d}}{2}\right) - 1 \right].
\label{2d44b}	
\end{eqnarray} 
The star-triangle transformation maps the spin-1/2 Ising model on the decorated hexagonal lattice into the spin-1/2 Ising model on the kagom\'e lattice once it is performed for all nodal spins. 
As a matter of fact, one easily derives the following exact mapping correspondence between the partition functions of both these models by a direct substitution of the star-triangle mapping transformation (\ref{2d43}) into the relation (\ref{2d42}): 
\begin{eqnarray}
Z_{\rm d} (5N/2, \beta J_{\rm d}) = C^{N} Z_{\rm k} (3N/2, \beta J_{\rm k}).
\label{2d45}	
\end{eqnarray}

At this stage, it is possible to combine the decoration-iteration mapping with the star-triangle transformation in order to express the partition function of the spin-1/2 Ising model on the kagom\'e lattice in terms of the corresponding partition function of the spin-1/2 Ising model on the hexagonal lattice. The mapping relations (\ref{2d38}) and (\ref{2d45}) provide this useful connection between 
the partition functions of the spin-1/2 Ising model on the hexagonal and kagom\'e lattices:
\begin{eqnarray}
Z_{\rm k} (3N/2, \beta J_{\rm k}) 
\! \! \! &=& \! \! \! 
\left(\frac{B^{\frac{3}{2}}}{C} \right)^N Z_{\rm h} (N, \beta J_{\rm h}) \nonumber \\
\! \! \! &=& \! \! \!  2^N \left\{ \frac{\exp\left(\frac{3\beta J_{\rm h}}{2}\right) \left[\exp\left(\frac{\beta J_{\rm h}}{2}\right) + 1\right]}
                  {\left[2 \exp\left(\beta J_{\rm h}\right) +  \exp\left(\frac{\beta J_{\rm h}}{2}\right) - 1\right]^3} \right \}^{\frac14}
                  Z_{\rm h} (N, \beta J_{\rm h}),
\label{2d46}	
\end{eqnarray}
whereas Eqs.~(\ref{2d37b}) and (\ref{2d44b}) relate the effective temperatures of the spin-1/2 Ising model on the hexagonal and kagom\'e lattices with equivalent partition functions: 
\begin{eqnarray}
\beta J_{\rm k} = \ln \left[ 2 \exp\left(\frac{\beta J_{\rm h}}{2}\right) - 1 \right].
\label{2d47}	
\end{eqnarray}
Substituting the exact critical temperature of the spin-1/2 Ising model on the hexagonal lattice (\ref{2d28}) to the mapping relation (\ref{2d47}) one simply gets the exact critical temperature
of the spin-1/2 Ising model on the kagom\'e lattice:
\begin{eqnarray}
\frac{k_{\rm B} T_{\rm c}}{J_{\rm k}} = \frac{1}{\ln (3 + 2 \sqrt{3})} \approx 0.535830.
\label{2d48}	
\end{eqnarray}
It is quite evident that the exact critical temperature (\ref{2d48}) of the spin-1/2 Ising model on the kagom\'e lattice is slightly lower than that (\ref{2d17}) of the spin-1/2 Ising model on the square lattice with the same coordination number.

\subsubsection{Critical points of spin-1/2 Ising archimedean lattices}

\begin{figure}[t]
\vspace{0cm}
\begin{center}
\includegraphics[clip,width=11.7cm]{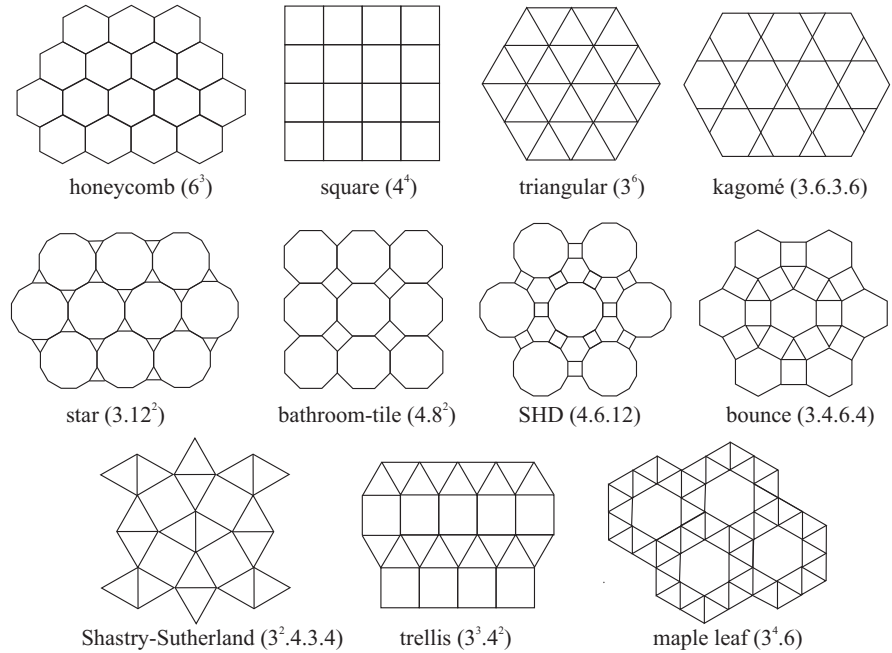}
\end{center}
\vspace{-0.5cm}
\caption{\small A schematic illustration of all archimedean (semi-regular) lattices, which are the only vertex-transitive 2D lattices being composed solely of regular polygons. The first row contains besides three regular (hexagonal, square and triangular) lattices also a kagom\'e lattice, which is the only semi-regular lattice that is edge-transitive quite similarly as a triad of regular lattices. The numbers in round brackets denote the number of sides of regular polygons enclosing each vertex within a given vertex-transitive lattice graphs according to the Gr\"unbaum-Sheppard notation \cite{grun87}.}
\label{fig:arch}
\vspace*{-0.3cm}
\end{figure}

In the previous parts  we have derived exact critical temperatures of the ferromagnetic spin-1/2 Ising model on all three regular planar lattices - honeycomb, square and triangular, which entirely cover the plane by a periodic repetition of the same regular polygon - hexagon, square and triangle, respectively. In addition, we have also exactly calculated the critical point of the ferromagnetic spin-1/2 Ising model on the semi-regular kagom\'e lattice, which consists of two kinds of periodically alternating regular polygons - hexagons and triangles (see Fig. \ref{fig:arch}). This 2D lattice is particularly interesting from the academic point of view, because it represents the only semi-regular (archimedean) tiling that is both vertex-transitive as well as edge-transitive quite similarly as a triad of the three aforementioned regular lattices (c.f. the lattices in the upper row of Fig. \ref{fig:arch} with the semi-regular lattices from other rows). To date, the exact critical temperatures corresponding to the order-disorder phase transition of the ferromagnetic spin-1/2 Ising model are known for all archimedean lattices and they are compiled in Tab. \ref{tabtc} according to Ref. \cite{code10} and references cited therein. The term archimedean lattice denotes such a uniform tiling of the plane by regular polygons, which is vertex-transitive but is not necessarily edge-transitive. There are in total 11 archimedean (semi-regular) lattices, which are schematically illustrated in Fig. \ref{fig:arch}. Due to vertex-transitivity each archimedean lattice can be unambiguously given by a set of numbers determining the number of sides of regular polygons enclosing a given vertex, whereas by convention one starts from a regular polygon with smallest number of sides and proceeds either clockwise or counterclockwise according to the Gr\"unbaum-Sheppard notation (for example 3.6.3.6 for the kagom\'e lattice) \cite{grun87}.

\begin{table}[t]
\begin{center}
\vspace{-0.3cm}
\caption{\small Exact critical temperatures of the ferromagnetic spin-1/2 Ising model on archimedean lattices. The second column determines the coordination number $q$ and the third column specifies a sequence of the regular polygons around each vertex according to the Gr\"unbaum-Sheppard notation \cite{grun87}. The abbreviations SHD and SSL are used for the square-hexagon-dodecagon lattice and the Shastry-Sutherland lattice, respectively (see Fig. \ref{fig:arch}).} 
\vspace*{3mm}
\begin{tabular}{|c|c|c|c|c|} 
\hline  
lattice & $q$ & notation & $\beta_{\rm c} J$ & $k_{\rm B} T_{\rm c}/J$ \\ \hline
honeycomb & 3 & $6^3$ & $2 \ln (2 + \sqrt{3})$ & 0.3797... \\ \hline 
square & 4 & $4^4$ & $2 \ln (1 + \sqrt{2})$ & 0.5673... \\ \hline 
triangular & 6 & $3^6$ & $\ln (3)$ & 0.9102... \\ \hline 
kagom\'e & 4 & $3.6.3.6$ & $\ln (3 + 2 \sqrt{3})$ & 0.5358... \\ \hline 
star & 3 & $3.12^2$ & $ 2 \ln \Bigl[\frac{1}{2} (3 + \sqrt{3} + \sqrt{12 + 10 \sqrt{3}})\Bigr]$ & 0.3079... \\ \hline 
bathroom-tile & 3 & $4.8^2$ & $2 \ln \Bigl[\frac{1}{2} (2 + \sqrt{2} + \sqrt{10  + 8 \sqrt{2}}) \Bigr]$ & 0.3597... \\ \hline 
SHD & 3 & $4.6.12$ & $2 \ln \Bigl[\frac{\sqrt{2} + \sqrt{5  + 3 \sqrt{3} - \sqrt{44 + 26 \sqrt{3}}}}{\sqrt{2} - \sqrt{5  + 3 \sqrt{3} - \sqrt{44 + 26 \sqrt{3}}}} \Bigr]$ & 0.3475... \\ \hline 
bounce & 4 & $3.4.6.4$ & $\ln (3 + 2 \sqrt{3})$ & 0.5358... \\ \hline
SSL & 5 & $3^2.4.3.4$ & root of higher-order polynomial & 0.7316... \\ \hline
trellis & 5 & $3^3.4^2$ & $ \ln (4) $ & 0.7213... \\ \hline
maple leaf & 5 & $3^4.6$ & root of higher-order polynomial & 0.6965... \\ \hline
\end{tabular} 
\label{tabtc}
\end{center}
\vspace{-0.25cm}
\end{table} 

It can be understood from Curie temperatures listed in Tab. \ref{tabtc} that the greater the coordination number of the planar lattice is, the higher is the critical temperature of its order-disorder transition. Accordingly, the cooperativity of spontaneous ferromagnetic ordering seems to be very closely connected to such a topological feature as being the lattice coordination number. However, the coordination number by itself does not entirely determine the critical temperature as it can be clearly seen from a comparison of critical temperatures of the spin-1/2 Ising model on archimedean lattices with the same coordination number $q$ (e.g. hexagonal, star, bathroom-tile, square-hexagon-dodecagon lattices with $q=3$; square, kagom\'e and bounce lattices with $q=4$; or Shastry-Sutherland, trellis and maple-leaf lattices with $q=5$). It turns out that semi-regular lattices being composed of more kinds of regular polygons generally have lower critical temperature than its regular counterpart with the same coordination number. Thus, it might be concluded that the cooperative nature of a spontaneous long-range order is strongly influenced also by other topological features of 2D lattices such as for instance an inhomogeinety of regular polygons within the semi-regular tiling, which apparently affords a greater versatility for low-energy excitations facilitating a breakdown of spontaneous  ordering. It should be emphasized that such an information cannot be elucidated from relatively rough approximate approaches such as the mean-field approximation or the single-spin cluster formulation of the effective-field theory, which predict the same critical temperature for the spin-1/2 Ising model on all lattices with the same coordination number regardless of the lattice geometry. From this perspective, exact solutions of the spin-1/2 Ising model on 2D lattices are indispensable for providing in-depth understanding of critical phenomena closely related to a cooperative nature of spontaneous long-range ordering. In spite of a substantial progress achieved in this research field, there are still a few puzzling features such as the identical Curie temperatures of the ferromagnetic spin-1/2 Ising model on kagom\'e and bounce lattices (see Tab. \ref{tabtc}) that persist comprehensive understanding.
     
\subsubsection{Transfer-Matrix Approach}

Although the dual, star-triangle and decoration-iteration mapping transformations afford several rigorous results for the spin-1/2 Ising model on the planar lattices, they do not solve this problem completely as they afford exact results precisely at a critical point only. In this regard, sophisticated mathematical methods are needed in order to obtain a complete closed-form exact solution of the spin-1/2 Ising model on some 2D lattice in a whole temperature range. Here, we will show some crucial points of the transfer-matrix approach when applying it to the spin-1/2 Ising model on a square lattice. It should be nevertheless mentioned that the procedure to be explained in what follows can be used to obtain the exact solution of the spin-1/2 Ising model on an arbitrary planar lattice without crossing bonds in the qualitatively same manner as explained below for the special case of square lattice. 
  
\begin{figure}[t]
\vspace{-1cm}
\begin{center}
\includegraphics[clip,width=8cm]{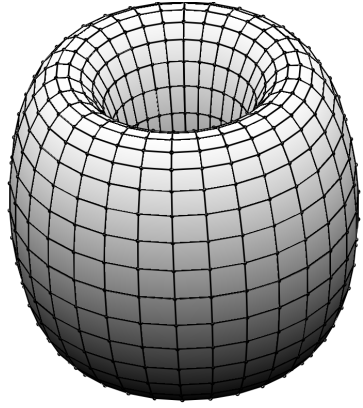}
\end{center}
\vspace{-5.4cm}
\caption{\small The square lattice wrapped on a torus ensures periodic (cyclic) boundary conditions in both horizontal as well as vertical directions.}
\label{fig:torus}
\end{figure}

Let us consider the spin-1/2 Ising model on the square lattice, which consists of $r$ rows and $c$ columns and is defined through the Hamiltonian: 
\begin{eqnarray}
{\cal H} = - J \left( \sum_{i=1}^{r} \sum_{j=1}^{c}  \sigma_{i,j} \sigma_{i+1,j} + 
                      \sum_{i=1}^{r}   \sum_{j=1}^{c} \sigma_{i,j} \sigma_{i,j+1} \right),
\label{2d49}	
\end{eqnarray}
where $\sigma_{i,j} = \pm 1/2$ labels the Ising spin situated at the $i$th row and $j$th column of a square lattice. Apparently, the total Hamiltonian (\ref{2d49}) is constituted by two kinds of summations. The former summation is carried out over all interactions between the nearest-neighbor spins from adjacent rows of the square lattice, while the latter one takes into account all interactions between the nearest-neighbor spins from adjacent columns. As usual, the periodic boundary conditions simplifying further treatment are imposed by the constraints: 
\begin{eqnarray}
\sigma_{r+1,j} \! \! \! &=& \! \! \! \sigma_{1,j}, \qquad {\mbox {where}} \quad
{\mbox{$j$ = 1, 2, \ldots, $c$}}; \label{2d50a}	\\
\sigma_{i,c+1} \! \! \! &=& \! \! \! \sigma_{i,1}, \qquad {\mbox {where}} \quad
{\mbox{$i$ = 1, 2, \ldots, $r$}}. \label{2d50b}	
\end{eqnarray} 
It is noteworthy that the periodicity of the square lattice, which is ensured by the conditions (\ref{2d50a}) and  (\ref{2d50b}), is equivalent to wrapping the square lattice on a torus in such 
a way that the $c$th column is coupled to the first column and the $r$th row is coupled to the 
first row (Fig.~\ref{fig:torus}). 
This is the statistical-mechanical definition of the spin-1/2 Ising model on the square lattice with periodic boundary conditions in both horizontal and vertical directions, which do not affect the behavior of thermodynamic quantities in the thermodynamic limit $r \to \infty$ and $c \to \infty$. It should be noted, however, that the accurate evaluation of the partition function of any 2D Ising model is an extremely difficult task, since it demands a considerable knowledge of sophisticated mathematics required for understanding of the exact solution in its full details. Therefore, we will draw just the main points of this cumbersome derivation.

First, let us denote the overall spin configuration of the $j$th column of spins by:
\begin{eqnarray}
\mu_{j} = \{\sigma_{1,j}, \sigma_{2,j}, \ldots, \sigma_{r,j}\}, 
\label{2d51}	
\end{eqnarray} 
which includes in total $2^{r}$ possible spin configurations. The variable $\mu_{j}$ can be viewed as one 'column' macrospin with $2^{r}$ possible microstates. The total Hamiltonian (\ref{2d49}) can be then written as a sum of two terms, namely, the interaction energy of individual columns and the interaction energy between adjacent columns. The interaction energy pertain to the $j$th column of spins is given by:
\begin{eqnarray}
{\cal H}_1 (\mu_{j}) = -J \sum_{i=1}^{r} \sigma_{i,j} \sigma_{i+1,j}, 
\label{2d52}	
\end{eqnarray} 
while the interaction energy between the $j$th and $(j+1)$st column reads:
\begin{eqnarray}
{\cal H}_2 (\mu_{j}, \mu_{j+1}) = -J \sum_{i=1}^{r} \sigma_{i,j} \sigma_{i,j+1}. 
\label{2d53}	
\end{eqnarray} 
With all this in mind, the total Hamiltonian (\ref{2d49}) can be rewritten as a sum of both these contributions: 
\begin{eqnarray}
{\cal H} = \sum_{j=1}^{c} \left[ {\cal H}_1 (\mu_{j}) + {\cal H}_2 (\mu_{j}, \mu_{j+1}) \right], 
\label{2d54}	
\end{eqnarray}
or it can be alternatively rewritten into the following most symmetric form: 
\begin{eqnarray}
{\cal H} = \sum_{j=1}^{c} \left\{ \frac{1}{2} \left[{\cal H}_1 (\mu_{j}) + {\cal H}_1 (\mu_{j+1})\right] + {\cal H}_2 (\mu_{j}, \mu_{j+1}) \right\}. 
\label{2d55}	
\end{eqnarray}
Furthermore, it is very appropriate to substitute Eq.~(\ref{2d55}) to a statistical definition of the partition function in order to gain the relation:
\begin{eqnarray}
Z \! \! \! &=& \! \! \! \sum_{\mu_1} \sum_{\mu_2} \cdots \sum_{\mu_c} 
\exp \left( \sum_{j=1}^{c} \left \{ - \frac{\beta}{2} \left[ {\cal H}_1 (\mu_{j}) 
+ {\cal H}_1 (\mu_{j+1}) \right] - \beta {\cal H}_2 (\mu_{j}, \mu_{j+1}) \right \} \right) 
\nonumber \\ \! \! \! &=& \! \! \! \sum_{\mu_1} \sum_{\mu_2} \ldots \sum_{\mu_c} 
\prod_{j=1}^{c} \exp \left\{ - \frac{\beta}{2} \left[{\cal H}_1 (\mu_{j}) + {\cal H}_1 (\mu_{j+1}) \right] - \beta {\cal H}_2 (\mu_{j}, \mu_{j+1}) \right\}. 
\label{2d56}	
\end{eqnarray}
Here, the symbol $\sum_{\mu_j}$ marks a summation over all available configurations of spins belonging to the $j$th column. The structure of the relation (\ref{2d56}) already implies that the partition function of the spin-1/2 Ising model on the square lattice can be formally calculated following the standard procedure based on the transfer-matrix method after performing a subsequent summation over individual macrospins:
\begin{eqnarray}
Z = \sum_{\mu_1} \sum_{\mu_2} \cdots \sum_{\mu_c} T (\mu_1, \mu_2) T (\mu_2, \mu_3) \cdots T (\mu_c, \mu_1) = \sum_{\mu_1} T^c (\mu_1, \mu_1) = \mbox{Tr} (T^c), 
\label{2d57}	
\end{eqnarray}
whereas the relevant transfer matrix $T (\mu_j, \mu_{j+1})$ depends on two nearest-neighbor 'column' macrospins $\mu_j$ and $\mu_{j+1}$:
\begin{eqnarray}
T (\mu_j, \mu_{j+1}) \! \! \! &=& \! \! \! 
\exp \left\{ - \frac{\beta}{2} \left[{\cal H}_1 (\mu_{j}) + {\cal H}_1 (\mu_{j+1}) \right]
- \beta {\cal H}_2 (\mu_{j}, \mu_{j+1}) \right\}
\label{2d58}	 \\ \! \! \! &=& \! \! \!
\exp \left[ \frac{\beta J}{2} \left(\sum_{i=1}^{r} \sigma_{i,j} \sigma_{i+1,j} 
+ \sum_{i=1}^{r} \sigma_{i,j+1} \sigma_{i+1,j+1} \right)  
+ \beta J \sum_{i=1}^{r} \sigma_{i,j} \sigma_{i,j+1} \right]. \nonumber 
\end{eqnarray}
The final result (\ref{2d57}) indicates that the partition function of the spin-1/2 Ising model on the square lattice can be calculated as a trace of the matrix $T^c$, which is eventually equal to a sum over all eigenvalues of the $2^r \times 2^r$ transfer matrix $T (\mu_j, \mu_{j+1})$ raised to the $c$th power
\begin{eqnarray}
Z = \mbox{Tr} (T^c) = \sum_{j = 1}^{2^r} \lambda_j^c. 
\label{2d59}	
\end{eqnarray}
In the thermodynamic limit, the Helmholtz free energy normalized per one spin can be largely simplified by sorting the eigenvalues of the transfer matrix $T (\mu_j, \mu_{j+1})$ in descending order $\lambda_1 > \lambda_2 \geq \ldots \geq \lambda_{2^r}$, actually, 
\begin{eqnarray}
F \! \! \! &=& \! \! \!
 - k_{\rm B} T \lim_{r \to \infty} \lim_{c \to \infty} \frac{1}{rc} \ln Z 
= - k_{\rm B} T \lim_{r \to \infty} \lim_{c \to \infty} \frac{1}{rc} \ln 
\left(\lambda_1^c + \lambda_2^c + \ldots + \lambda_{2^r}^c \right) 
\nonumber \\ \! \! \! &=& \! \! \! 
- k_{\rm B} T \lim_{r \to \infty} \frac{1}{r} \ln \lambda_1
- k_{\rm B} T \lim_{r \to \infty} \lim_{c \to \infty} \frac{1}{rc} 
\ln \left[1 + \sum_{k=1}^{2^r} \left( \frac{\lambda_k}{\lambda_1} \right)^c \right]
\nonumber \\ \! \! \! &=& \! \! \! 
- k_{\rm B} T \lim_{r \to \infty} \frac{1}{r} \ln \lambda_1.
\label{2d60}	
\end{eqnarray}
In this way, the finding of the exact solution of the spin-1/2 Ising model on the square lattice essentially reduces to the finding of the largest eigenvalue of the transfer matrix $T (\mu_j, \mu_{j+1})$ defined by Eq. (\ref{2d58}). It should be stressed, nevertheless, that the finding of the largest eigenvalue of the $2^r \times 2^r$ transfer matrix for the spin-1/2 Ising model on the square lattice by letting $r$ tend to infinity is highly non-trivial task unlike the finding of the largest eigenvalue of the simple $2 \times 2$ transfer matrix corresponding to the spin-1/2 Ising chain. This is an origin of major difficulties to be present by solving any planar Ising model.  

\begin{figure}[t]
\begin{center}
\hspace{-1.0cm}
\includegraphics[clip,width=7.5cm]{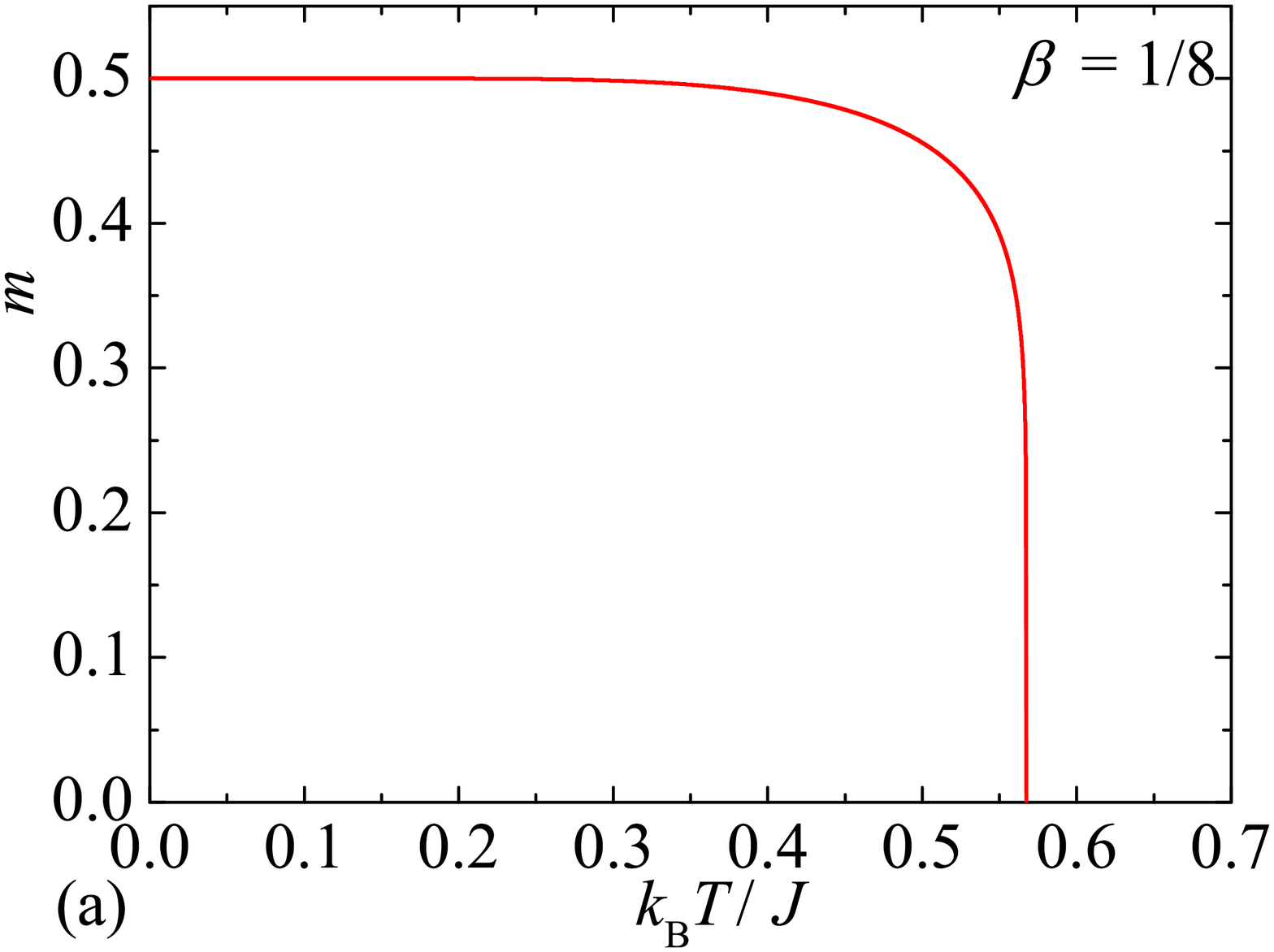}
\hspace{-1.0cm}
\includegraphics[clip,width=7.5cm]{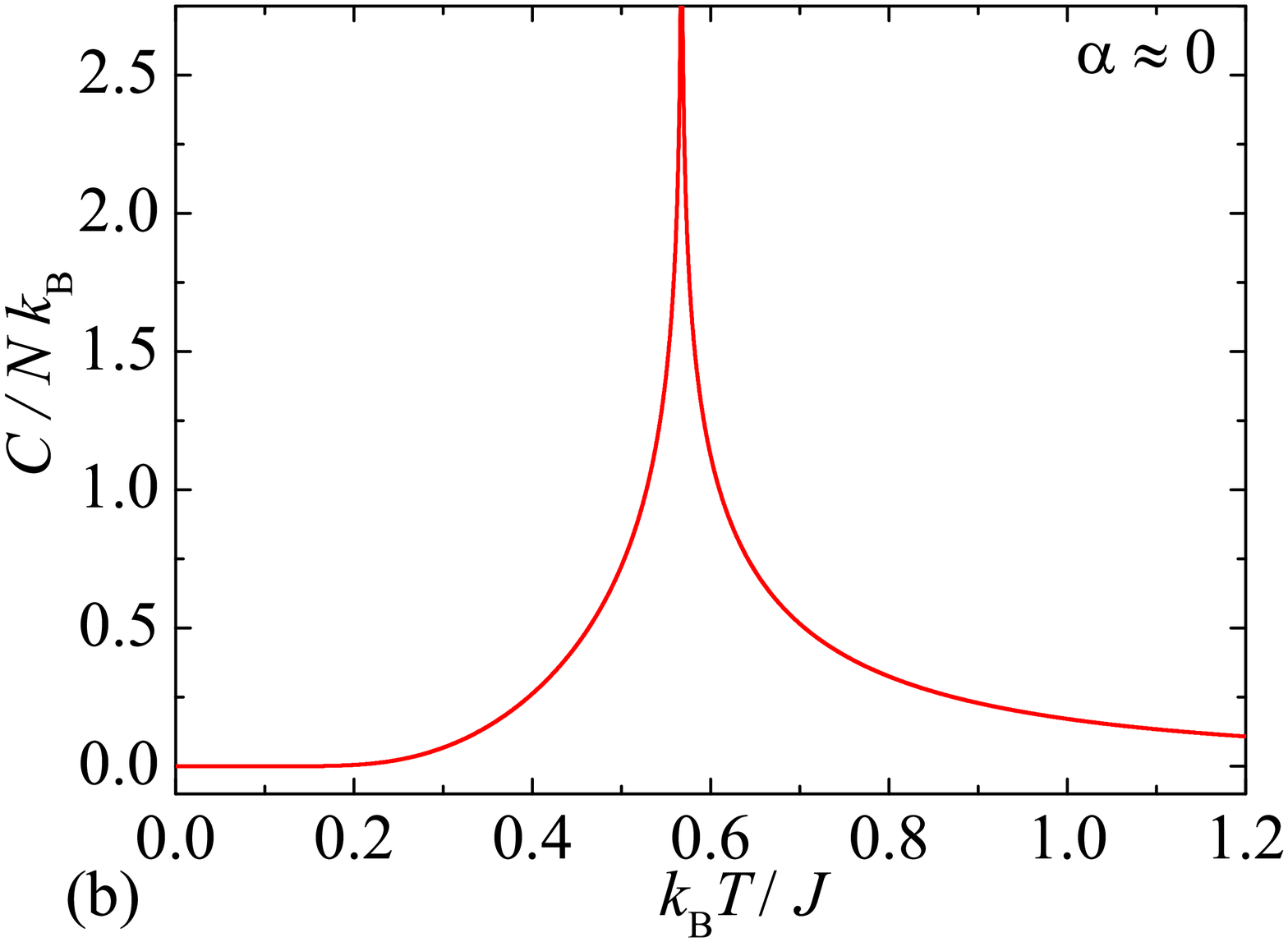}
\hspace{-1.5cm}
\vspace{-0.25cm}
\caption{(a) The spontaneous magnetization of the spin-1/2 Ising model on a square lattice as a function of the reduced temperature; (b) Temperature dependence of the specific heat of the spin-1/2 Ising model on a square lattice. The logarithmic divergence is detected at the critical temperature.}
\label{fig:square}
\end{center}
\end{figure}

The largest eigenvalue of the transfer matrix, which corresponds to the spin-1/2 Ising model on the square lattice, was firstly derived in the highly celebrated Onsager's paper by the use of Lie algebra and group representation \cite{onsa44}. It is worthy to notice that the original Onsager's work as well as its subsequent simplifications are still too intricate to be thoroughly described within this brief review. Therefore, we will henceforth restrict ourselves merely to stating the final result of this rather cumbersome derivation. The largest eigenvalue of the transfer matrix (\ref{2d58}) is:
\begin{eqnarray}
\lambda_1 = \left[ 2 \sinh \left( \frac{\beta J}{2} \right) \right]^{\frac{r}{2}} 
            \exp \left[ \frac{1}{2} (\alpha_1 + \alpha_3 + \ldots + \alpha_{2r-1}) \right],
\label{2d61}	
\end{eqnarray}
where $\alpha_k$ is defined by
\begin{eqnarray}
\cosh(\alpha_k) = \coth\left( \frac{\beta J}{2} \right) \cosh\left( \frac{\beta J}{2} \right) - \cos \left(\frac{\pi k}{r} \right).
\label{2d62}	
\end{eqnarray}
By adopting Eq.~(\ref{2d61}), the Helmholtz free energy (\ref{2d60}) per spin can be modified to: 
\begin{eqnarray}
F = - \frac{k_{\rm B} T}{2} \ln \left[2 \sinh\left( \frac{\beta J}{2} \right) \right] 
- k_{\rm B} T \lim_{r \to \infty} \frac{1}{2r} \sum_{k=1}^{r} \alpha_{2k-1}.
\label{2d63}	
\end{eqnarray}
In the thermodynamic limit ($r \to \infty$), the sum appearing in Eq.~(\ref{2d63}) can be substituted by the integral so that: 
\begin{eqnarray}
\nonumber
F = - \frac{k_{\rm B} T}{2} \ln \left[2 \sinh\left( \frac{\beta J}{2} \right) \right] 
- \frac{k_{\rm B} T}{2 \pi} \! \int\limits_{0}^{\pi} \! \mbox{arccosh} 
\left[\coth\left( \frac{\beta J}{2} \right) \cosh\left( \frac{\beta J}{2} \right) - \cos \theta \right] {\rm d} \theta. \\
\label{2d64}	
\end{eqnarray}
The expression (\ref{2d64}) can be further simplified by the use of identity:
\begin{eqnarray}
\mbox{arccosh} |x| = \frac{1}{\pi} \int\limits_{0}^{\pi} \ln [2(x - \cos \phi)] {\rm d} \phi,
\label{2d65}	
\end{eqnarray}
which consecutively yields from Eq.~(\ref{2d64}) the formula:
\begin{eqnarray}
F = \! \! \! &-& \! \! \! \frac{k_{\rm B} T}{2} \ln \left[2 \sinh\left( \frac{\beta J}{2} \right) \right] 
- \frac{k_{\rm B} T}{2} \ln 2 \nonumber \\ \! \! \! &-& \! \! \! 
\frac{k_{\rm B} T}{2 \pi^2} \int\limits_{0}^{\pi} \int\limits_{0}^{\pi} \ln \left[\coth\left( \frac{\beta J}{2} \right) 
\cosh\left( \frac{\beta J}{2} \right) - \cos \theta - \cos \phi \right] {\rm d} \theta {\rm d} \phi.
\label{2d66}	
\end{eqnarray}
The formula (\ref{2d66}) can finally be symmetrized to:
\begin{eqnarray}
F = \! \! \! &-& \! \! \! k_{\rm B} T \ln 2 \nonumber \\ \! \! \! &-& \! \! \! 
\frac{k_{\rm B} T}{2 \pi^2} \int\limits_{0}^{\pi} \int\limits_{0}^{\pi} 
\ln \left \{ \cosh^2\left( \frac{\beta J}{2} \right) - \sinh\left( \frac{\beta J}{2} \right) \left[\cos \theta + \cos \phi \right] \right \} 
{\rm d} \theta {\rm d} \phi.
\label{2d67}	
\end{eqnarray}
The above equation represents the famous Onsager's result for the Helmholtz free energy of the spin-1/2 Ising model on a square lattice. It should be remarked that exact results for other thermodynamic quantities like entropy or specific heat can be obtained from Eq.~(\ref{2d67}) by making use of standard relations from thermodynamics and statistical physics. By contrast, the derivation of exact expression for the spontaneous magnetization of the ferromagnetic spin-1/2 Ising model on a square lattice is much more complex, because one cannot straightforwardly differentiate the Gibbs free energy with respect to the magnetic field and subsequently take asymptotic limit of vanishing magnetic field as the exact expression for the free energy in a non-zero magnetic field is not known. In this regard, some indirect method must be employed for calculating of the spontaneous magnetization, whereas the overall derivation is regrettably complex in spite of surprisingly simple form of the final expression \cite{yang52}. For simplicity, let us therefore merely quote the final exact result for the spontaneous magnetization of the ferromagnetic spin-1/2 Ising model on a square lattice obtained by Yang \cite{yang52}: 
\begin{eqnarray}
m = \left \{  
\begin{array}{ll} 
\displaystyle \frac{1}{2} \left[1 - \displaystyle \frac{1}{\sinh^4 \left( \frac{\beta J}{2} \right)}\right]^{\frac{1}{8}} & \, {\rm if} \, \, \, 
T < T_{\rm c}, \\ \qquad \quad  \: \: 0 & \, {\rm if} \, \, \, T \geq T_{\rm c}, 
\end{array} \right.
\label{2d68}	
\end{eqnarray}
which simultaneously represents the relevant order parameter inherent to the continuous order-disorder phase transition. It can be easily understood from Eq.~(\ref{2d68}) that the spontaneous magnetization is zero precisely at and above the critical temperature (\ref{2d17}), while it monotonically increases by decreasing temperature until its saturation value is reached at the absolute zero temperature. In addition, the spontaneous magnetization of the ferromagnetic spin-1/2 Ising model on a square lattice disappears at a critical temperature according to the power law:
\begin{eqnarray}
m \cong \left(1 - T/T_{\rm c} \right)^{\frac{1}{8}}, 
\label{2dmagtc}	
\end{eqnarray}
which implies the magnitude of the critical exponent $\beta_m = 1/8$ for the spontaneous magnetization as temperature tends to its critical value (\ref{2d17}) from below. It should be stressed that the same conclusion would be reached for a disappearance of the spontaneous magnetization of the ferromagnetic spin-1/2 Ising model on any 2D lattice. Fig.~\ref{fig:square}(a) therefore only illustrates temperature variations of the spontaneous magnetization of the ferromagnetic spin-1/2 Ising model on a square lattice. The non-zero spontaneous magnetization below the critical temperature means that there is an excess of spins at low enough temperatures, which are spontaneously pointing into the same spatial direction.
In other words, the spins have a tendency to align themselves in the same spatial direction and if we randomly choose one spin from the lattice, then, it is more likely that the spin is pointing in the direction of the spontaneous magnetization rather than in the opposite direction. It is noteworthy that this peculiar long-range correlation occurs spontaneously without being enforced by some external field of force (such as the magnetic field) and notwithstanding of extremely short-range character of the considered interaction potential. 

Interestingly, the specific heat of the spin-1/2 Ising model on a square lattice displays even a more striking temperature dependence with a marked singularity at the critical temperature (\ref{2d17}) as depicted in Fig.~\ref{fig:square}(b). It can be proved that the specific heat diverges logarithmically according to the asymptotic law:
\begin{eqnarray}
C/N k_{\rm B} \cong - \mbox{\it const.} \ln | T - T_{\rm c} |, 
\label{2dshtc}	
\end{eqnarray}
as long as temperature approaches its critical value (\ref{2d17}) either from below or from above. Although the positive constant of proportionality in the asymptotic law (\ref{2dshtc}) varies with the lattice geometry, the logarithmic divergence of specific heat nearby a critical point represents a generic feature of the ferromagnetic spin-1/2 Ising model on arbitrary 2D lattice. Thus, it could be concluded that the ferromagnetic spin-1/2 Ising model on all 2D lattices exhibits strongly universal critical behavior, which can be all characterized by the same asymptotic laws and the same set of critical indices governing the dependences of all important magnetic and thermodynamic response functions (e.g. magnetization, susceptibility or specific heat). The aforedescribed critical behavior is apparently in a sharp contrast with theoretical predictions of approximate methods such as the mean-field or effective-field theories, because the less accurate methods do not take into account long-range correlations that become relevant in a vicinity of the critical temperature where the correlation length diverges. 

\subsubsection{Mixed-spin Ising model on decorated planar lattices} 

\begin{figure}
\begin{center}
\includegraphics[clip,width=0.6\textwidth]{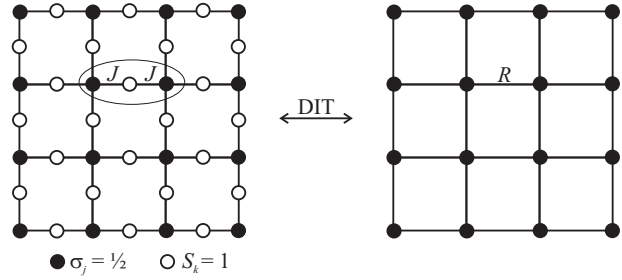}
\end{center}
\vspace{-0.4cm}
\caption{The mixed spin-1/2 and spin-$1$ Ising model on a decorated square lattice (figure on the left) and its exact mapping relationship to the corresponding spin-1/2 Ising model on a simple (undecorated) square lattice (figure on the right) established by means of the decoration-iteration transformation (DIT). The full (empty) circles denote lattice positions of the spin-1/2 (spin-1) Ising spins and the ellipse demarcates the interaction terms belonging to the $k$th bond Hamiltonian.}
\label{figdecsq}
\end{figure}

Next, let us consider the mixed spin-1/2 and spin-1 Ising model on singly decorated planar lattices \cite{jasc98,dakh98} as schematically illustrated in Fig. \ref{figdecsq} on the particular example of a decorated square lattice with the coordination number $q=4$. The magnetic structure of the investigated model system constitute the Ising spins $\sigma=1/2$ placed on nodal sites of some planar lattice and the Ising spins $S=1$ situated on each its bond. The total Hamiltonian of the mixed spin-1/2 and spin-1 Ising model on the decorated planar lattices then reads:
\begin{equation}
{\cal H} = -J\sum_{\langle kj \rangle}^{Nq} S_{k}\sigma_{j} - D \sum_{k=1}^{Nq/2} S_{k}^{2},
\label{deceq1}
\end{equation}
where $\sigma_{j} = \pm 1/2$ and $S_{k} = \pm 1, 0$, the relevant subscript specifies the lattice position, the coupling constant $J$ represents the exchange interaction between the nearest-neighbor spin-1/2 and spin-$1$ atoms, the parameter $D$ stands for the uniaxial single-ion anisotropy acting on the spin-$1$ atoms only. 

Now, we will turn our attention to main points of the transformation method, which provides a rigorous mapping equivalence between the investigated spin model and its corresponding spin-1/2 Ising model on undecorated planar lattice. First, it is quite useful to rewrite the total Hamiltonian (\ref{deceq1}) as a sum over the bond Hamiltonians, i.e. ${\cal H}=\sum_{k=1}^{Nq/2} {\cal H}_{k}$, where $N$ denotes the total number of the spin-1/2 atoms, $q$ is their coordination number (i.e. the number of nearest neighbors) and the summation runs over all bonds of the decorated planar lattice. Each bond Hamiltonian ${\cal H}_{k}$ thus involves all the interaction terms including the decorating Ising spin $S_k$ (see the spin cluster enclosed by ellipse in Fig. \ref{figdecsq}):
\begin{equation}
{\cal H}_k =-JS_{k}(\sigma _{k1}+\sigma _{k2})-DS_{k}^{2}. 
\label{deceq2}
\end{equation}
The partition function of the model under consideration can be rewritten into the following useful form:
\begin{eqnarray}
Z = \sum_{\{\sigma_{i}\}}\prod_{k = 1}^{Nq/2} \sum_{S_{k} = \pm 1,0} {\exp(-\beta {\cal H}_k)},
\label{deceq3}
\end{eqnarray}
where the symbol $\sum_{S_{k} = \pm 1,0}$ denotes a summation over three available spin states of the decorating Ising spin $S_k$, while the symbol $\sum_{\{\sigma_{i}\}}$ marks a summation over all available spin configurations of the complete set of the spin-1/2 Ising atoms. It is noteworthy that the summation over spin states of the decorating spin-$1$ atoms can be performed independently of each other because of the lack of any direct interaction in between them. After performing the summation over the spin states of the decorating Ising spin $S_k$, one gets the effective Boltzmann's weight that implies a possibility of applying the generalized decoration-iteration transformation:
\begin{eqnarray}
\nonumber
\sum_{S_{k} = \pm 1,0} \!\!\! \exp(-\beta {\cal H}_k) = 1 + 2 \exp(\beta D) \cosh[\beta J (\sigma_{k1} + \sigma_{k2}) ] = A \exp( \beta R \sigma_{k1} \sigma_{k2}).\\
\label{deceq4}
\end{eqnarray}
The physical meaning of the mapping transformation (\ref{deceq4}) lies in removing all the interaction parameters associated with the  decorating spin $S_k$ and replacing them by a new unique effective interaction $R$ between its two nearest-neighbor Ising spins $\sigma_{k1}$ and $\sigma_{k2}$. It is worthwhile to mention that the both mapping parameters $A$ and $R$ are 'self-consistently' given directly by the transformation formula (\ref{deceq4}), which must hold for all four possible spin combinations of two Ising spins $\sigma_{k1}$ and $\sigma_{k2}$ that provide just two independent equations from the mapping transformation (\ref{deceq4}). In the consequence of that, two yet unknown mapping parameters $A$ and $R$ can be unambiguously determined by the following formulas:
\begin{eqnarray}
A = (V_{1}V_{2})^{\frac{1}{2}}, \qquad \beta R = 2\ln\left(\frac{V_{1}}{V_{2}}\right),
\label{deceq5}
\end{eqnarray}
which contain two newly defined functions:
\begin{eqnarray}
V_{1} = 1 + 2 \exp(\beta D) \cosh(\beta J), \qquad
V_{2} = 1 + 2 \exp(\beta D).
\label{deceq6}
\end{eqnarray}

After a straightforward substitution of the mapping transformation (\ref{deceq4}) into the formula (\ref{deceq3}), one easily obtains an exact relation between the partition function of the mixed-spin Ising model on the decorated planar lattice and respectively, the partition function of the spin-1/2 Ising model on the simple (undecorated) planar lattice with the temperature-dependent effective interaction $R$:
\begin{equation}
Z = A^{Nq/2} Z_{\rm IM} (\beta R). 
\label{deceq7}
\end{equation}
The mapping relation (\ref{deceq7}) between both partition functions represents a central result of our calculation, because this relationship can in turn be employed for a rigorous calculation of some important quantities such as sublattice magnetizations and free energy. It is worthwhile to recall that this mapping correspondence is valid independently of the lattice coordination number $q$. 

Exact mapping theorems developed by Barry \textit{et al}. \cite{barry1,barry2,barry3,barry4} in combination with Eq. (\ref{deceq7}) yield a straightforward relation between the spontaneous magnetization of the sublattice A of the mixed-spin Ising model on the decorated planar lattice and respectively, the spontaneous magnetization of the corresponding spin-1/2 Ising model on the undecorated planar lattice:   
\begin{eqnarray}
m_{A} \equiv \frac{1}{2} \left( \langle \sigma_{k1} \rangle + \langle \sigma_{k2} \rangle \right) 
 = \frac{1}{2} \left( \langle \sigma_{k1} \rangle + \langle \sigma_{k2} \rangle \right)_{\rm IM}
\equiv m_{\rm IM} (\beta R).  
\label{deceq8a}
\end{eqnarray} 
Here, the symbols $\langle \cdots \rangle $ and $\langle \cdots \rangle_{{\rm IM}}$ denote the standard canonical ensemble average performed in the mixed-spin Ising model on the decorated planar lattice defined through the Hamiltonian (\ref{deceq1}) and respectively, its corresponding spin-1/2 Ising model on the undecorated planar lattice. According to Eq. (\ref{deceq8a}), the sublattice magnetization $m_{A}$ directly equals to the spontaneous magnetization $m_{\rm IM}$ of the corresponding spin-1/2 Ising model on the simple planar lattice. The other spontaneous sublattice magnetization of the mixed-spin Ising model on the decorated planar lattice can be descended from the generalized Callen-Suzuki identity \cite{call63,suzu65,balc02}, which enables to relate it to the previously derived spontaneous sublattice magnetization $m_{A}$:
\begin{eqnarray}
m_{B} \equiv \left \langle S_{k} \right \rangle = 2 m_{A} \, \frac{2 \exp(\beta D) \sinh(\beta J)}{1 + 2 \exp(\beta D) \cosh(\beta J)}.
\label{deceq9}
\end{eqnarray}  
In this way, the exact calculation of both spontaneous sublattice magnetizations of the mixed-spin Ising model on the decorated planar lattice is completed, because it is now sufficient to substitute the exact expression for the spontaneous magnetization of the corresponding spin-1/2 Ising model on undecorated planar lattice into the formerly derived expressions (\ref{deceq8a}) and (\ref{deceq9}) for the sublattice magnetizations $m_A$ and $m_B$. In doing so, one should bear in mind that the effective coupling $\beta R$ given by Eqs. (\ref{deceq5})-(\ref{deceq6}) must enter into exact expressions for the spontaneous magnetization of the corresponding spin-1/2 Ising model on the undecorated planar lattice. For completeness, it is worthy to note that the exact formulas for the spontaneous magnetization of the spin-1/2 Ising model on various planar lattices can be found in Refs. \cite{domb60,link92} . 

Finally, let us examine in detail the critical behavior of the model under investigation. It can be readily understood from the mapping relation (\ref{deceq7}) between both partition functions that the mixed-spin Ising model on the decorated planar lattice may exhibit a critical point only if the corresponding spin-1/2 Ising model on the undecorated planar lattice is at a critical point as well. This result is taken to mean that the critical temperature of the mixed-spin Ising model on the decorated planar lattice can easily be obtained by comparing the effective nearest-neighbor coupling of the corresponding spin-1/2 Ising model on the simple (undecorated) planar lattice with its critical value $\beta_{c} R$. The critical condition for the the mixed-spin Ising model on the decorated planar lattice then reads:
\begin{eqnarray}
\beta_{\rm c} R = 2\ln \left[\frac{1 + 2 \exp(\beta_{\rm c} D) \cosh(\beta_{\rm c} J)}{1 + 2 \exp(\beta_{\rm c} D)} \right] = \left \{ 
\begin{array}{l}
2 \ln (2 + \sqrt{3}) \,\,\,   (\mbox{honeycomb}), \\
\ln (3 + 2 \sqrt{3}) \,\,\,\, (\mbox{kagome}), \\
2 \ln (1 + \sqrt{2}) \,\,\,   (\mbox{square}), \\
\ln (3) \qquad \quad \,\,\,\, (\mbox{triangular}), 
\end{array} \right.
\label{deceqcc}
\end{eqnarray}
\begin{figure}[tb]
\begin{center}
\hspace{-1.0cm}
\includegraphics[clip,width=7.5cm]{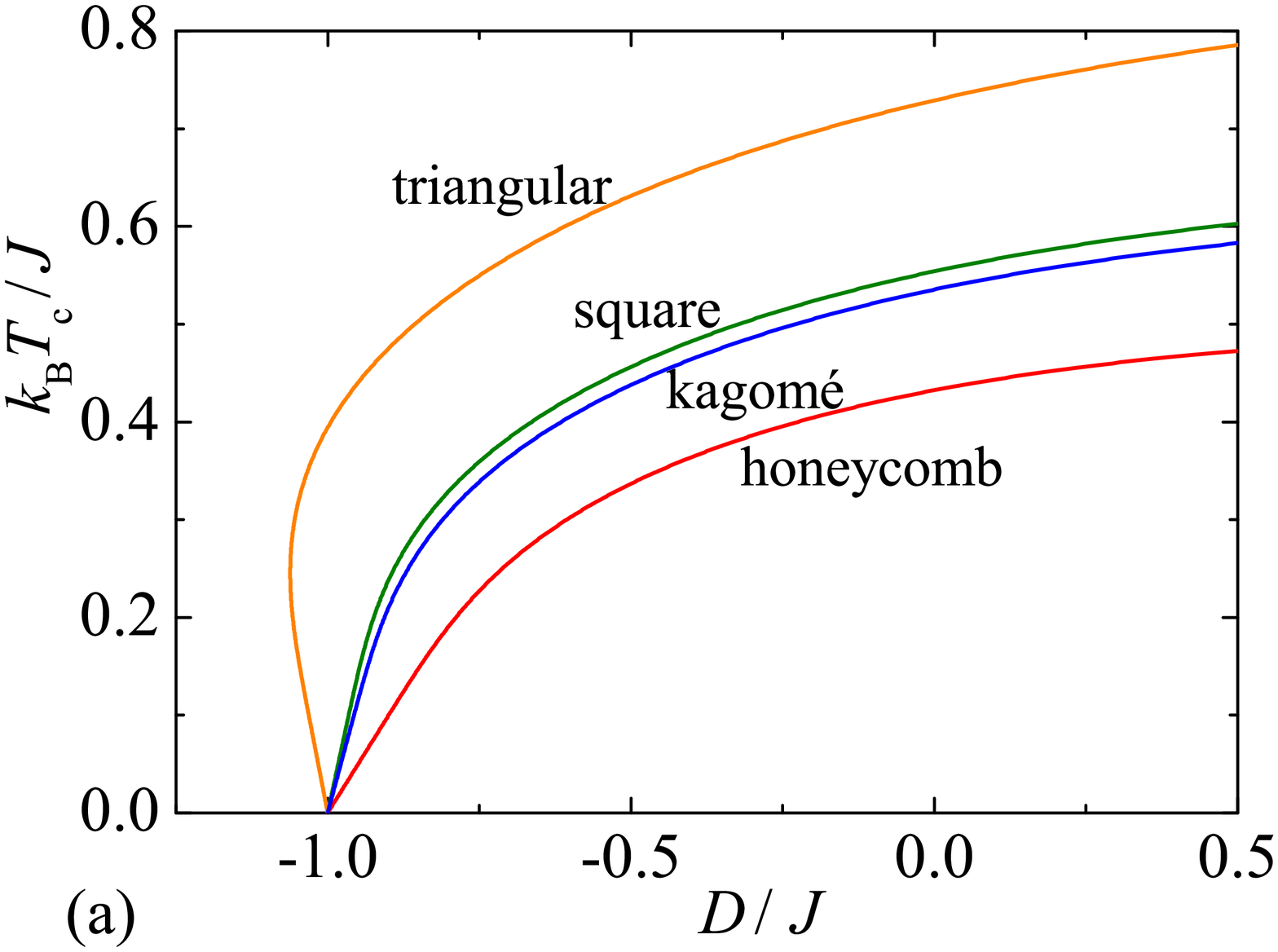}
\hspace{-1.0cm}
\includegraphics[clip,width=7.5cm]{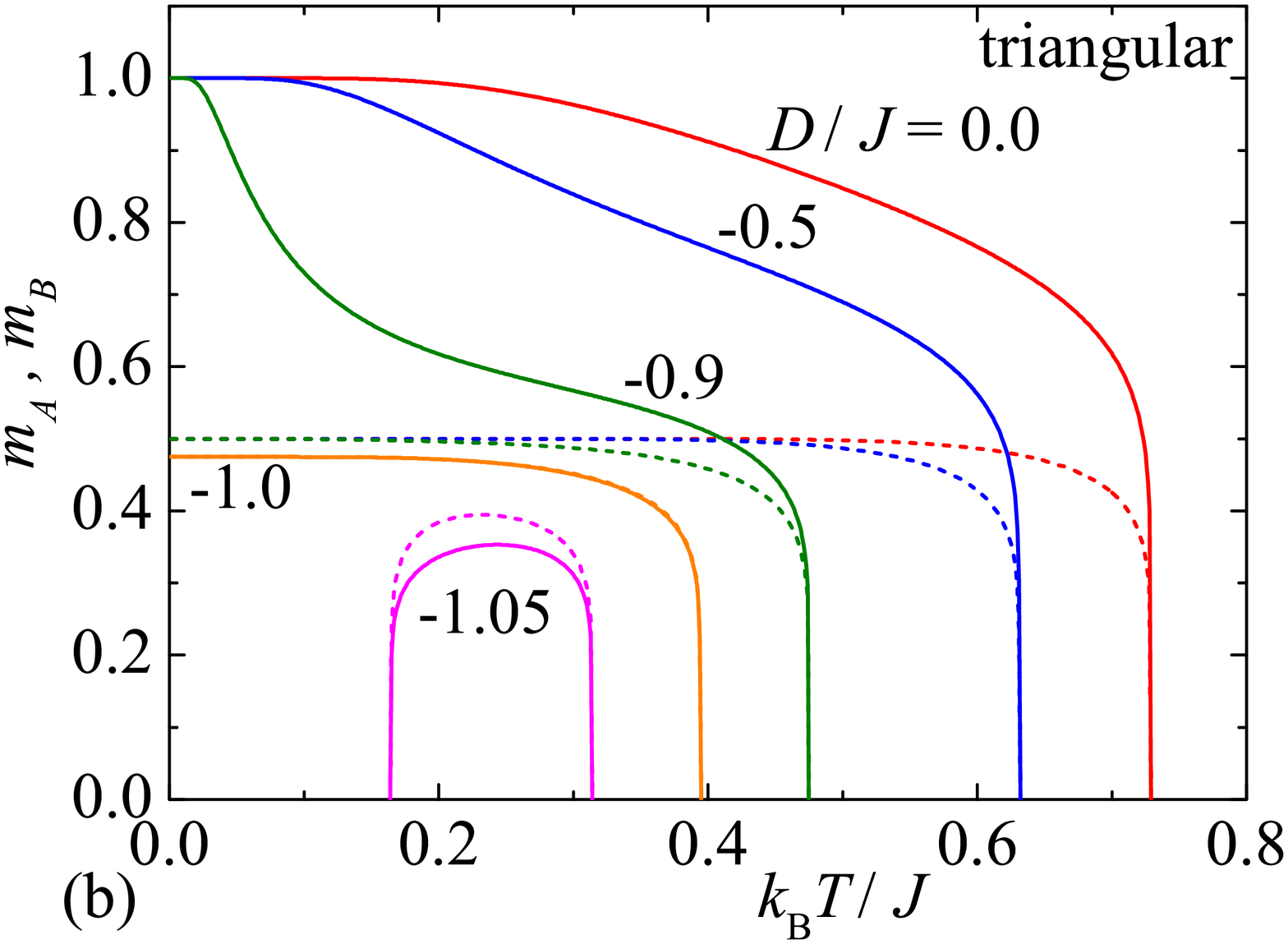}
\hspace{-1.5cm}
\vspace{-0.25cm}
\caption{(a) The critical temperature of the mixed spin-1/2 and spin-1 Ising model on a few different decorated planar lattices as a function of the uniaxial single-ion anisotropy;  (b) Thermal variations of the sublattice magnetizations $m_A$ (broken lines) and $m_B$ (solid lines) of the mixed spin-1/2 and spin-1 Ising model on a decorated triangular lattice for a few different values of the uniaxial single-ion anisotropy.}
\label{figmixdec}
\end{center}
\end{figure}
where $\beta_{\rm c} = 1/(k_{\rm B} T_{\rm c})$ and $T_{\rm c}$ is the critical temperature. Fig. \ref{figmixdec}(a) illustrates typical dependences of the critical temperature on 
the uniaxial single-ion anisotropy for the mixed spin-1/2 and spin-1 Ising model on several decorated planar lattices. As one can see, the same general trends can be detected in the critical temperature versus the uniaxial single-ion anisotropy plot irrespective of the underlying lattice geometry. As a matter of fact, the critical temperature is in general suppressed by the easy-plane (negative) single-ion anisotropy, which energetically favors the non-magnetic spin state $S_k=0$ of the decorating atoms before the magnetic ones $S_k = \pm 1$. If one compares the critical temperatures of the mixed-spin Ising model on decorated kagom\'e and square lattices with the equal lattice coordination number $q=4$, then, one may conclude that the critical temperature of the mixed-spin Ising model on the  decorated kagom\'e lattice is always slightly below that one of the decorated square lattice. Another interesting observation is that the mixed spin-1/2 and spin-1 Ising model on the decorated triangular lattice displays {\it reentrant phase transition} on assumption that the uniaxial single-ion anisotropy is selected slightly below the ground-state boundary between the ferromagnetically ordered and disordered phases $D/J \lesssim -1$. To support this statement, temperature dependences of both spontaneous sublattice magnetizations of the mixed spin-1/2 and spin-1 Ising model on the decorated triangular lattice  are plotted in Fig. \ref{figmixdec}(b) for several values of the uniaxial single-ion anisotropy. Apparently, both spontaneous sublattice magnetizations start from their saturated values just when the uniaxial single-ion anisotropy is greater than the boundary value $D_{\rm b}/J = -1.0$. If the uniaxial single-ion anisotropy is selected sufficiently close but slightly below this boundary value, i.e. $D/J \lesssim -1$, then, the spontaneous magnetizations emerge just at a lower critical temperature and disappear at an upper critical temperature. Finally, it is worth mentioning that the zero-temperature asymptotic limit of both spontaneous sublattice magnetization acquires for the particular case $D_{\rm b}/J = -1.0$ highly non-trivial value $m_A = m_B = \frac{1}{2}(\frac{125}{189})^{1/8} = 0.47482\ldots$.  

\subsubsection{Mixed-spin Ising model on three-coordinated planar lattices}

\begin{figure}[!tbh]
\begin{center}
\includegraphics[clip,width=0.7\textwidth]{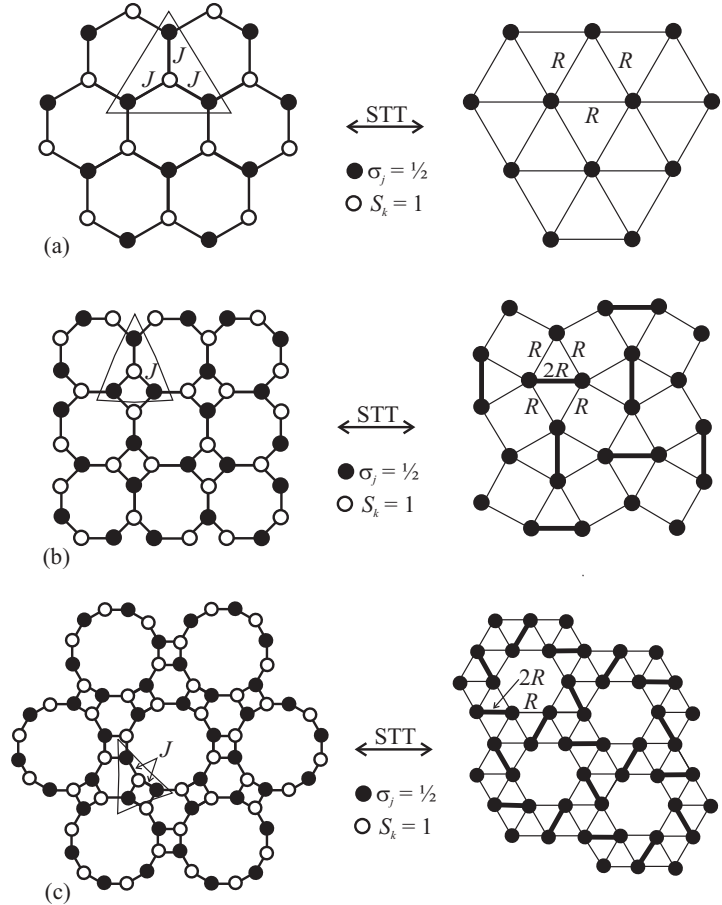}
\end{center}
\vspace{-0.25cm}
\caption{The mixed spin-1/2 and spin-$1$ Ising model on three-coordinated planar lattices (figures on the left) and its exact mapping relationship to the spin-1/2 Ising model on corresponding archimedean lattices (figures on the right) established by means of the star-triangle transformation (STT): (a) honeycomb and triangular lattice; (b) bathroom-tile and Shastry-Sutherland lattice; (c) square-hexagon-dodecagon and maple leaf lattice. The full (empty) circles denote lattice positions of the spin-1/2 (spin-1) atoms and the triangle demarcates the interaction terms belonging to the $k$th site Hamiltonian. Thin (thick) solid lines stand for the effective coupling $R$ ($2R$) of the corresponding spin-1/2 Ising model on the archimedean lattice.}
\label{figmixhon1}
\vspace*{-0.2cm}
\end{figure}

Another exactly soluble model represents the mixed spin-1/2 and spin-1 Ising model on three-coordinated planar lattices such as for instance the honeycomb lattice \cite{gonc85}, bathroom-tile lattice \cite{stre06} or square-hexagon-dodecagon lattice schematically depicted in Fig. \ref{figmixhon1}. To ensure an exact solvability of the model under investigation, we will further suppose that the sites of sublattice A are occupied by the spin-1/2 atoms (full circles) while the sites of sublattice B are occupied by the spin-1 atoms (open circles). The total Hamiltonian of the mixed-spin Ising model on the three-coordinated planar lattices takes the following form:
\begin{equation}
{\cal H} = -J \sum_{\langle kj \rangle}^{3N} S_k \sigma_j - D \sum_{k \in B}^{N} S_k^2. 
\label{mheq1}
\end{equation}
Here, the Ising spin variables $\sigma_j = \pm 1/2$ and $S_k = \pm 1, 0$ are situated at the sites of the sublattice A and B, respectively, $N$ denotes the total number of lattice sites of each sublattice. The first summation in Eq. (\ref{mheq1}) accounts for the Ising interaction $J$ between the nearest-neighbor spin-1/2 and spin-1 atoms, while the second summation involving the parameter $D$ takes into consideration the uniaxial single-ion anisotropy acting on the spin-1 atoms only. 

The model defined through the Hamiltonian (\ref{mheq1}) is exactly tractable within the framework of the generalized star-triangle transformation. First, it is very convenient to write the total Hamiltonian (\ref{mheq1}) as a sum over the site Hamiltonians ${\cal H}_{k}$:
\begin{equation}
{\cal H} = \sum_{k=1}^{N} {\cal H}_{k},
\label{mheq4}
\end{equation}
where each site Hamiltonian ${\cal H}_k$ involves all interaction terms associated with the spin-1 atom residing on the $k$th lattice site of the sublattice $B$:
\begin{equation}
{\cal H}_{k} = - J S_{k} (\sigma_{k1} + \sigma_{k2} + \sigma_{k3}) - D S_{k}^2.
\label{mheq5}
\end{equation}
The partition function of the mixed spin-1/2 and spin-1 Ising model on three-coordinated planar lattices can be partially factorized and subsequently rewritten in the form:
\begin{equation}
Z = \sum_{\{\sigma_j \}} \prod_{k = 1}^{N} \sum_{S_k = \pm 1,0} \exp(- \beta {\cal H}_k).
\label{mheq6}
\end{equation}
In above, the symbol $\sum_{\{\sigma_j \}}$ means a summation over all possible spin configurations of the spin-1/2 atoms from the sublattice A, while the summation symbol $\sum_{S_k = \pm 1, 0}$ accounts for the spin states of $k$th spin from sublattice B. After the latter summation is explicitly carried out one obtains the effective Boltzmann's factor, which can be immediately replaced through the generalized star-triangle transformation:
\begin{eqnarray}
\sum_{S_k = \pm 1,0} \exp(- \beta {\cal H}_k) \! \! \! &=& \! \! \! 1 + 2 \exp(\beta D) \cosh [\beta J (\sigma_{k1} + \sigma_{k2} + \sigma_{k3})] \nonumber \\
\! \! \! &=& \! \! \! A \exp [ \beta R (\sigma_{k1} \sigma_{k2} + \sigma_{k2} \sigma_{k3} + \sigma_{k3} \sigma_{k1})].
\label{mheq9}
\end{eqnarray}
Obviously, the mapping transformation (\ref{mheq9}) replaces the interaction terms of a spin star, i.e. a four-spin cluster consisting of one central spin-1 atom and its three nearest-neighbor spin-1/2 atoms, by the effective interactions of three spin-1/2 atoms forming an equilateral triangle. As usual, both mapping
parameters $A$ and $R$ are 'self-consistently' given by the transformation equation (\ref{mheq9}), which must be valid for any combination of spin states of three spin-1/2 atoms. In a consequence of that one obtains:
\begin{eqnarray}
A = \left(V_{1} V_{2}^3\right)^{\frac{1}{4}}, \qquad \qquad \beta R = \mbox{ln} \left( \frac{V_1}{V_2} \right),
\label{mheq10}
\end{eqnarray}
where we have introduced the functions $V_1$ and $V_2$ defined as follows:
\begin{eqnarray}
V_{1} \! \! \! &=& \! \! \! 1 + 2 \exp(\beta D) \cosh \left(\frac{3}{2} \beta J \right), \nonumber \\
V_{2} \! \! \! &=& \! \! \! 1 + 2 \exp(\beta D) \cosh \left(\frac{1}{2} \beta J \right).
\label{mheq11}
\end{eqnarray}
When the star-triangle mapping transformation (\ref{mheq9}) is applied to all spin-1 atoms from the sublattice B, the mixed-spin Ising model on three-coordinated planar lattices is exactly mapped onto the spin-1/2 Ising model on corresponding archimedean lattices (see Fig. \ref{figmixhon1}):
\begin{equation}
Z = A^{N} Z_{\rm IM} (\beta, R),
\label{mheq12}
\end{equation}
whereas the effective interaction $R$ is given by the 'self-consistency' condition (\ref{mheq10})-(\ref{mheq11}). The only essential difference between the  mapping equivalence of the mixed-spin Ising model on three-coordinated planar lattices and the spin-1/2 Ising model on corresponding archimedean lattices lies in an essence of the effective couplings. While the mixed-spin Ising model on a honeycomb lattice is mapped onto the spin-1/2 Ising model on a triangular lattice with the uniform nearest-neighbor coupling $R$ [Fig. \ref{figmixhon1}(a)], the mixed-spin Ising model on bathroom-tile and square-hexagon-dodecagon lattices is rigorously mapped onto the spin-1/2 Ising model on Shastry-Sutherland and maple leaf lattices with two different effective couplings $R$ and $2R$ schematically illustrated in Fig. \ref{figmixhon1}(b)-(c) by thin and thick solid lines, respectively.  

Exact mapping theorems developed by Barry \textit{et al}. \cite{barry1,barry2,barry3,barry4} in combination with Eq. (\ref{mheq12}) yield a straightforward relation between the spontaneous magnetization of the sublattice A of the mixed-spin Ising model on the three-coordinated planar lattices and respectively, the spontaneous magnetization of the spin-1/2 Ising model on the corresponding archimedean lattice:   
\begin{eqnarray}
m_A \equiv \frac{1}{3} \left( \langle \sigma_{k1} \rangle + \langle \sigma_{k2} \rangle + \langle \sigma_{k3} \rangle \right) 
    = \frac{1}{3} \left( \langle \sigma_{k1} \rangle_{\rm IM} + \langle \sigma_{k2} \rangle_{\rm IM} + \langle \sigma_{k3} \rangle_{\rm IM} \right)    \equiv m_{\rm IM},
\label{mheq13}
\end{eqnarray} 
Here, the symbols $\langle \cdots \rangle $ and $\langle \cdots \rangle_{\rm IM}$ denote the standard canonical ensemble average performed in the mixed-spin Ising model on a three-coordinated planar lattice defined through the Hamiltonian (\ref{mheq1}) and respectively, the spin-1/2 Ising model on a corresponding archimedean lattice. According to Eq. (\ref{mheq13}), the sublattice magnetization $m_{A}$ of the spin-1/2 atoms directly equals to the spontaneous magnetization $m_{\rm IM}$ of the spin-1/2 Ising model on the corresponding archimedean lattice. The other spontaneous sublattice magnetization of the mixed-spin Ising model on the three-coordinated planar lattices can be derived with the help of the generalized Callen-Suzuki identity \cite{call63,suzu65,balc02}, which enables to relate it to the spontaneous sublattice magnetization $m_{A}$ and triplet correlation function $t_A \equiv \langle \sigma_{k1} \sigma_{k2} \sigma_{k3} \rangle_{\rm IM}$:
\begin{eqnarray}
m_{B} \equiv \langle S_k \rangle = \frac{3}{2} m_A \left(\frac{W_1}{V_1} + \frac{W_2}{V_2} \right) + 2 t_A \left(\frac{W_1}{V_1} - 3 \frac{W_2}{V_2}\right).
\label{mheq14}
\end{eqnarray}  
The coefficients $W_1$ and $W_2$ are defined as follows:
\begin{eqnarray}
W_{1} \! \! \! &=& \! \! \! 2 \exp(\beta D) \sinh \left(\frac{3}{2} \beta J \right), \nonumber \\
W_{2} \! \! \! &=& \! \! \! 2 \exp(\beta D) \sinh \left(\frac{1}{2} \beta J \right).
\label{mheq15}
\end{eqnarray}
In this way, the exact calculation of both spontaneous sublattice magnetizations of the mixed-spin Ising model on the three-coordinated planar lattices is formally completed, because it is now sufficient to substitute exact expressions \cite{domb60,link92}  for the spontaneous magnetization and triplet correlation function of the  spin-1/2 Ising model on the corresponding archimedean lattice into the formerly derived expressions (\ref{mheq13}) and (\ref{mheq14}) for the sublattice magnetizations $m_A$ and $m_B$. In doing so, one should bear in mind that the effective coupling $\beta R$ given by Eqs. (\ref{mheq10})-(\ref{mheq11}) must enter into the exact expression for the spontaneous magnetization of the spin-1/2 Ising model on the corresponding archimedean lattice. 

The critical points of the mixed-spin Ising model on three-coordinated planar lattices readily follow from the exact mapping correspondence (\ref{mheq12}) between partition functions. According to this mapping equivalence, the mixed-spin Ising model on a three-coordinated planar lattice may exhibit a critical point only if the spin-1/2 Ising model on a corresponding archimedean lattice is precisely at a critical point. The critical temperature of the mixed-spin Ising model on the three-coordinated planar lattices can be therefore obtained by comparing the effective nearest-neighbor coupling of the spin-1/2 Ising model on the corresponding archimedean lattice with its critical value. The critical temperature of the spin-1/2 Ising model on a triangular lattice with the unique nearest-neighbor interaction $R$ follows from the condition  $\beta_{\rm c} R = \ln 3$ according to Eq. (\ref{2d31}), while the (inverse) critical temperatures of the spin-1/2 Ising model on Shastry-Sutherland and maple leaf lattices with two different nearest-neighbor interactions $R$ and $2R$ follow from \cite{streun}:
\begin{eqnarray}
\beta_{\rm c} R \! \! \! &=& \! \! \! \ln \left[ \frac{8 + 5 \sqrt{2} - \sqrt{10 + 8 \sqrt{2}} + 2 \sqrt{5 + 4 \sqrt{2}}}{2 + \sqrt{2} + \sqrt{10 + 8\sqrt{2}}} \right], \,\, \quad (\mbox{Shastry-Sutherland}) \nonumber \\
\beta_{\rm c} R \! \! \! &=& \! \! \! \ln \left[ \frac{17 + 9 \sqrt{3} - 3 \sqrt{44 + 26 \sqrt{3}}}{\sqrt{44 + 26 \sqrt{3}} - 3 (1 + \sqrt{3})}\right]. \,\, \qquad
\qquad \qquad (\mbox{maple leaf})
\label{tssslshd}
\end{eqnarray}
Comparing the effective nearest-neighbor coupling (\ref{mheq10})-(\ref{mheq11}) with the previously derived critical values one in turn obtains the critical condition for the mixed-spin Ising model on all three considered three-coordinated planar lattices:
\begin{eqnarray}
\nonumber
\frac{1 + 2 \exp(\beta_{\rm c} D) \cosh \left(\frac{3}{2} \beta_{\rm c} J \right)}
        {1 + 2 \exp(\beta_{\rm c} D) \cosh \left(\frac{1}{2} \beta_{\rm c} J \right)} 
        = \left \{ 
\begin{array}{l}
3 \,\,\, \qquad \qquad \qquad \qquad \qquad \qquad \, (\mbox{honeycomb}), \\
\frac{8 + 5 \sqrt{2} - \sqrt{10 + 8 \sqrt{2}} + 2 \sqrt{5 + 4 \sqrt{2}}}{2 + \sqrt{2} + \sqrt{10 + 8\sqrt{2}}} \,\, \quad (\mbox{bathroom-tile}), \\
\frac{17 + 9 \sqrt{3} - 3 \sqrt{44 + 26 \sqrt{3}}}{\sqrt{44 + 26 \sqrt{3}} - 3 (1 + \sqrt{3})} \,\, \quad  (\mbox{square-hexagon-dodecagon}).
\end{array} \right.
\end{eqnarray}
\vspace*{-0.7cm}
\begin{eqnarray}
\label{mhcc}
\end{eqnarray}

\begin{figure}[tb]
\begin{center}
\hspace{-1.0cm}
\includegraphics[clip,width=7.5cm]{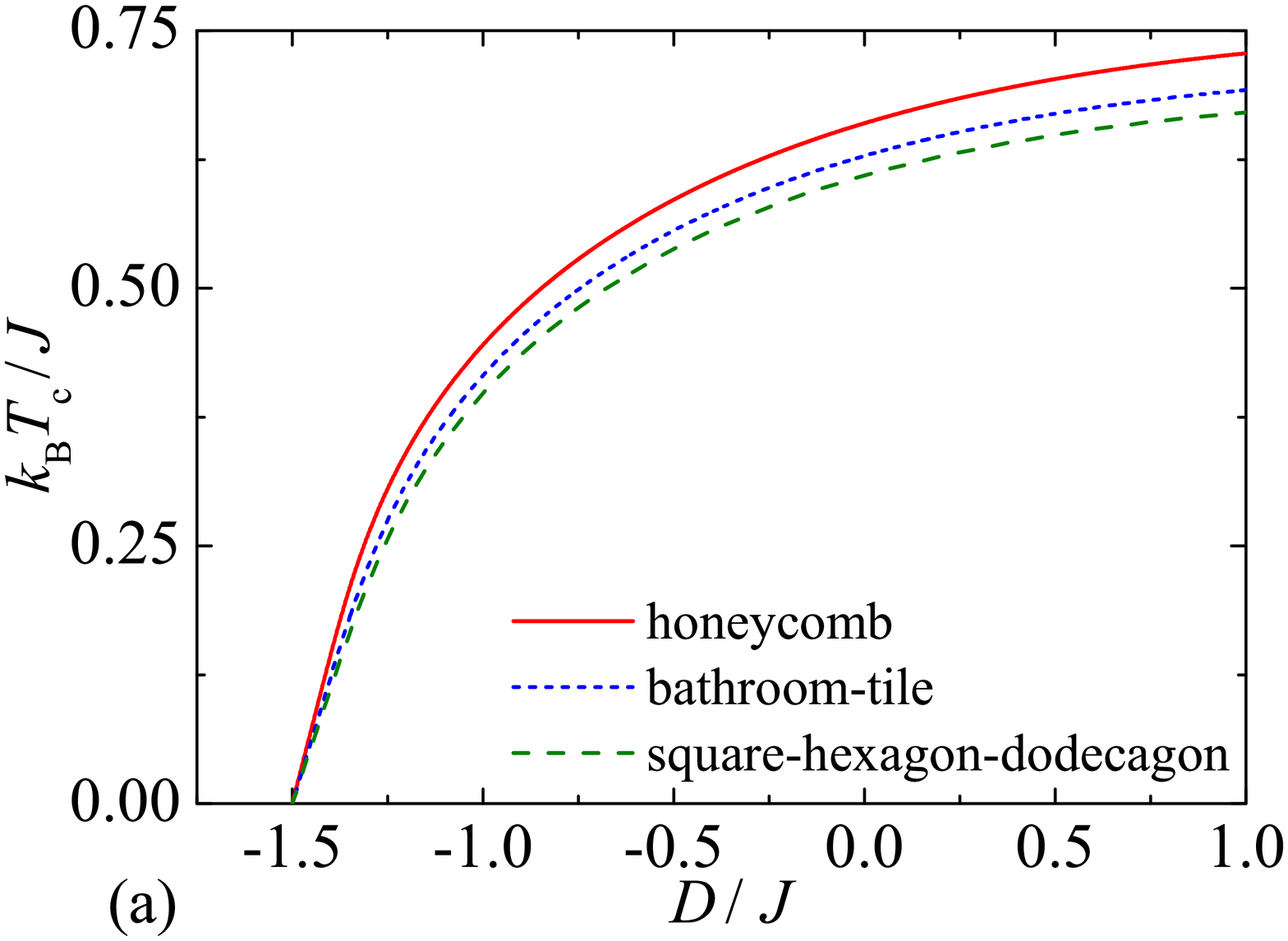}
\hspace{-1.0cm}
\includegraphics[clip,width=7.5cm]{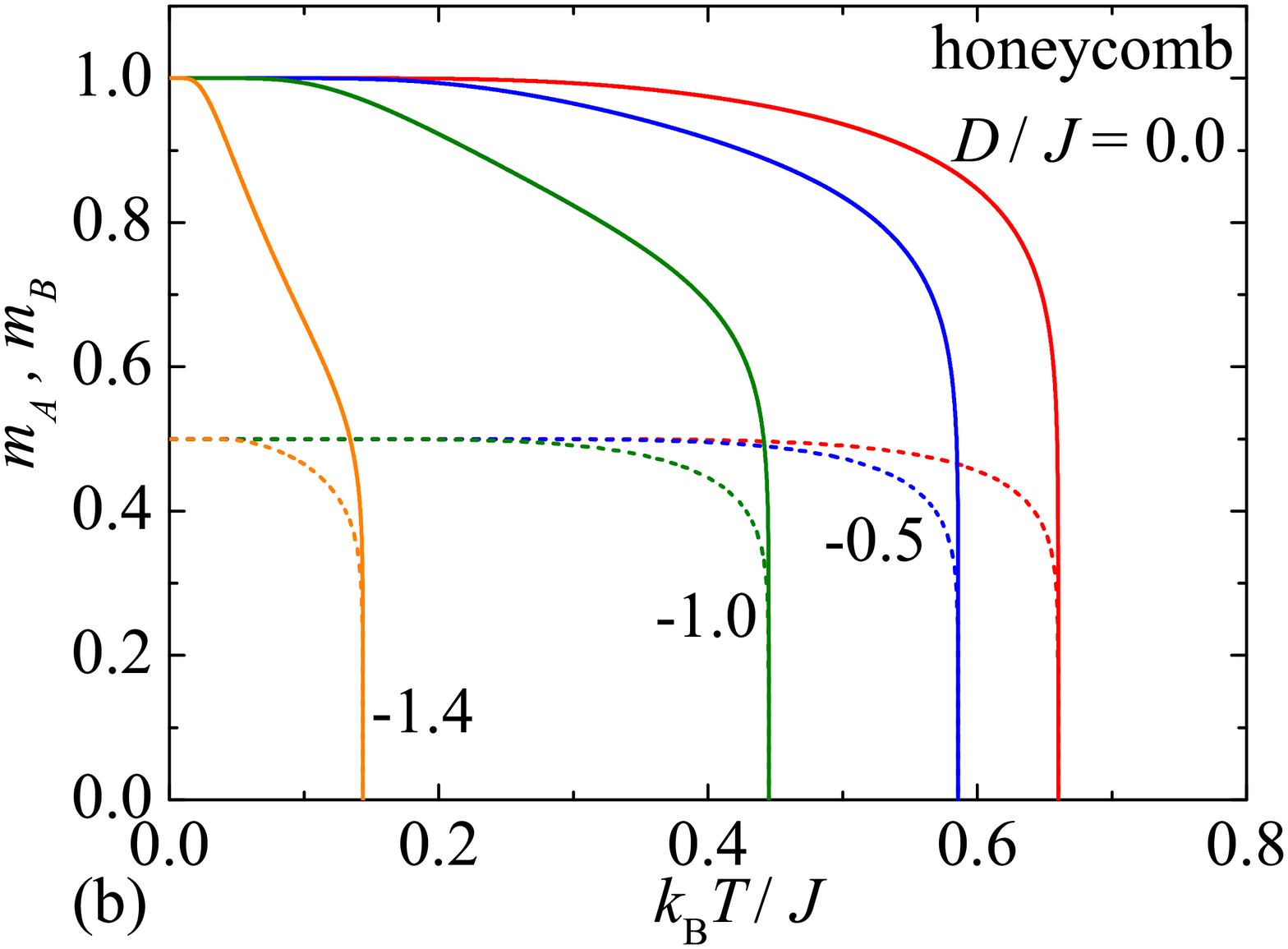}
\hspace{-1.5cm}
\vspace{-0.25cm}
\caption{(a) The critical temperature of the mixed spin-1/2 and spin-1 Ising model on honeycomb, bathroom-tile and square-hexagon-dodecagon lattices as a function of the uniaxial single-ion anisotropy; (b) Thermal variations of the sublattice magnetizations $m_A$ (broken lines) and $m_B$ (solid lines) of the mixed spin-1/2 and spin-1 Ising model on a honeycomb lattice for a few different values of the uniaxial single-ion anisotropy.}
\label{figmixhon}
\end{center}
\end{figure}

The numerical solution of the critical conditions (\ref{mhcc}) is plotted in Fig. \ref{figmixhon}(a) in the form of the critical temperature versus the uniaxial single-ion anisotropy dependence. In agreement with common expectations, the critical temperature of the mixed-spin Ising model on the three-coordinated planar lattices monotonically decreases upon strengthening of the easy-plane (negative) single-ion anisotropy until it completely vanishes at the ground-state boundary $D_{\rm b}/J = -3/2$ between the ferromagnetically ordered phase and the disordered paramagnetic phase. It is quite evident that this generic feature holds irrespective of the lattice topology, whereas the critical temperature of the mixed-spin Ising model on a regular hexagonal lattice slightly exceeds that ones of the same mixed-spin Ising model on semi-regular bathroom-tile and square-hexagon-dodecagon lattices. A gradual decline of the critical temperature with the easy-plane single-ion anisotropy can be explained in terms of energetic favoring of the non-magnetic spin state $S_k=0$ of the spin-1 atoms   from the sublattice B before two magnetic ones $S_k = \pm 1$. It should be remarked that our exact solution gave evidence of only continuous phase transitions in contrast to a more diverse critical behavior including tricritical point and discontinuous phase transitions predicted by approximate methods (c.f. Fig. \ref{figmixhon}(a) with Fig. \ref{figmix}(a)). 

The finite-temperature phase diagram shown in Fig. \ref{figmixhon}(a) for the mixed-spin Ising model on a honeycomb lattice is also corroborated by temperature dependences of both sublattice magnetizations plotted in Fig. \ref{figmixhon}(b). While the spontaneous magnetization of the spin-1/2 atoms from the sublattice A retains the standard Q-type dependence according to N\'eel nomenclature \cite{neel48} regardless of a relative strength of the uniaxial single-ion anisotropy, the spontaneous magnetization of the spin-1 atoms from the sublattice B may display a more interesting R-type dependence owing to predominant thermal excitations from the magnetic state $S_k = 1$ towards the non-magnetic spin state $S_k=0$ on assumption that the uniaxial single-ion anisotropy is sufficiently close but slightly above the ground-state boundary between the ferromagnetically ordered and disordered paramagnetic phases $D/J \gtrsim -3/2$.   

\subsubsection{Selectively diluted mixed-spin Ising model}

To illustrate versatility of the approach based on exact mapping transformations, we will consider in the following the generalized version of the mixed spin-1/2 and spin-1 Ising model on the honeycomb lattice that accounts also for a selective site dilution of the spin-1 atoms on the sublattice B by non-magnetic impurities [Fig. \ref{figmixhon1}(a)]. The sites of the sublattice B will be occupied with the probability $\rho$ by the magnetic spin-1 atom, while they will be either vacant or occupied by the non-magnetic atom with the probability $1-\rho$. The total Hamiltonian of the selectively diluted mixed-spin Ising model on the honeycomb lattice is given by:
\begin{equation}
{\cal H} = -J \sum_{\langle kj \rangle}^{3N} n_k S_k \sigma_j - D \sum_{k \in B}^{N} n_k S_k^2, 
\label{dieq1}
\end{equation}
where the occupation number $n_k = 0,1$ determines whether the $k$th lattice site from the sublattice B is occupied by the magnetic spin-1 atom ($n_k=1$) or the non-magnetic impurity $n_k = 0$. The physical meaning of the other interaction parameters involved in the Hamiltonian (\ref{dieq1}) is the same as described in the previous section. 

The model defined through the Hamiltonian (\ref{dieq1}) can be rigorously solved in the annealed limit by making use of the generalized star-triangle transformation. The total Hamiltonian (\ref{dieq1}) will be at first rewritten as a sum over the site Hamiltonians ${\cal H}_{k}$:
\begin{equation}
{\cal H} = \sum_{k=1}^{N} {\cal H}_{k},
\label{dieq4}
\end{equation}
where each site Hamiltonian $\hat {\cal H}_k$ involves all interaction terms closely connected to the $k$th lattice site of sublattice B:
\begin{equation}
{\cal H}_{k} = - J n_k S_{k} (\sigma_{k1} + \sigma_{k2} + \sigma_{k3}) - D n_k S_{k}^2.
\label{dieq5}
\end{equation}
The grand-canonical partition function of the mixed spin-1/2 and spin-1 Ising model on the honeycomb lattice can be recasted into the following form:
\begin{equation}
\Xi = \sum_{\{\sigma_i \}} \prod_{k = 1}^{N} \sum_{S_k = \pm 1,0} \sum_{n_k = 1,0} \exp(- \beta {\cal H}_k + \beta \mu n_k),
\label{dieq6}
\end{equation}
where $\mu$ is the chemical potential of the magnetic spin-1 atoms and the other symbols have the usual meaning. The effective Boltzmann's weight, which is obtained after carrying out summation over all available microstates of the $k$th site from the sublattice B, can be substituted by the generalized star-triangle transformation:
\begin{eqnarray}
\sum_{S_k = \pm 1,0} \sum_{n_k = 1,0} \exp(- \beta {\cal H}_k + \beta \mu n_k) \! \! \! &=& \! \! \! 
1 + z + 2 z \exp(\beta D) \cosh [\beta J (\sigma_{k1} + \sigma_{k2} + \sigma_{k3})] \nonumber \\
\! \! \! &=& \! \! \! A \exp [ \beta R (\sigma_{k1} \sigma_{k2} + \sigma_{k2} \sigma_{k3} + \sigma_{k3} \sigma_{k1})].
\label{dieq9}
\end{eqnarray}
Here, the parameter $z = \exp(\beta \mu)$ denotes the fugacity of the spin-1 atoms and the mapping parameters $A$, $R$ are given by the "self-consistency" condition of the transformation equation (\ref{dieq9}), which should hold irrespective of spin states of three spin-1/2 atoms involved therein. The yet undetermined mapping parameters $A$ and $R$ are then explicitly given by:
\begin{eqnarray}
A = \left[(1 + zV_{1})(1 + zV_{2})^3\right]^{\frac{1}{4}}, \qquad \qquad \beta R = \mbox{ln} \left( \frac{1 + zV_1}{1 + zV_2} \right),
\label{dieq10}
\end{eqnarray}
where the functions $V_1$ and $V_2$ are defined as follows:
\begin{eqnarray}
V_{1} \! \! \! &=& \! \! \! 1 + 2 \exp(\beta D) \cosh \left(\frac{3}{2} \beta J \right), \nonumber \\
V_{2} \! \! \! &=& \! \! \! 1 + 2 \exp(\beta D) \cosh \left(\frac{1}{2} \beta J \right).
\label{dieq11}
\end{eqnarray}
At this level, the star-triangle mapping transformation (\ref{dieq9}) can be substituted into Eq. (\ref{dieq6}) in order to get the exact mapping equivalence between the grand-canonical partition function of the mixed-spin Ising model on the honeycomb lattice and the canonical partition function of the corresponding spin-1/2 Ising model on a triangular lattice:
\begin{equation}
\Xi = A^{N} Z_{\rm t} (\beta, R),
\label{dieq12}
\end{equation}
whereas the effective interaction $R$ should satisfy the self-consistency condition given by Eqs. (\ref{dieq10})-(\ref{dieq11}). The grand-canonical partition function (\ref{dieq12}) can be straightforwardly used to calculate the grand potential of the selectively diluted mixed spin-1/2 and spin-1 Ising model on the honeycomb lattice from the well-known relation: 
\begin{equation}
\Omega = -k_{\rm B} T \ln \Xi.
\label{dieq13}
\end{equation}

The concentration of the spin-1 atoms can be now readily calculated from the grand potential:
\begin{eqnarray}
\rho = - \frac{1}{N} \frac{\partial \Omega}{\partial \mu} = \frac{z}{N} \frac{\partial \ln \Xi}{\partial z} 
     = \frac{z V_1}{1 + z V_1} \left(\frac{1}{4} + 3 \varepsilon \right) + \frac{3 z V_2}{1 + z V_2} \left(\frac{1}{4} - \varepsilon \right),
\label{dieq14}
\end{eqnarray}
which relates it to the nearest-neighbor pair correlation $\varepsilon = \langle \sigma_i \sigma_j \rangle_{\rm t}$ of the corresponding spin-1/2 Ising model on a triangular lattice \cite{domb60}. After eliminating the fugacity $z$ of the magnetic spin-1 atoms with the help of Eq. (\ref{dieq10}) one obtains from Eq. (\ref{dieq14}) the relation:
\begin{eqnarray}
\rho = \left[1 - \exp(-\beta R) \right]  \frac{V_1}{V_1-V_2} \left(\frac{1}{4} + 3 \varepsilon \right) 
     + 3 \left[\exp(\beta R) - 1 \right] \frac{V_2}{V_1-V_2} \left(\frac{1}{4} - \varepsilon \right).
\label{dieq15}
\end{eqnarray}
The critical temperature of the selectively diluted mixed spin-1/2 and spin-1 Ising model on the honeycomb lattice can in turn be obtained from Eq. (\ref{dieq15}) after substituting there the inverse critical temperature $\beta_{\rm c} R = \ln 3$ and the critical value $\varepsilon_{\rm c} = 1/6$ of the nearest-neighbor pair correlation function of the effective spin-1/2 Ising model on a triangular lattice \cite{domb60}:
\begin{eqnarray}
\rho = \frac{1}{2} \frac{V_1^{\rm c} + V_2^{\rm c}}{V_1^{\rm c} - V_2^{\rm c}}.
\label{dicc}
\end{eqnarray}
Note that the upper index c in the critical condition (\ref{dicc}) means that the inverse critical temperature $\beta_{\rm c} = 1/(k_{\rm B} T_{\rm c})$ enters into Eq. (\ref{dieq11}) determining the functions $V_1$ and $V_2$ instead of the inverse temperature $\beta = 1/(k_{\rm B} T)$. Moreover, it can be easily verified that the critical condition (\ref{dicc}) of the selectively site diluted mixed-spin Ising model on the honeycomb lattice exactly coincides with the critical condition (\ref{mhcc}) of the undiluted mixed-spin Ising model on the honeycomb lattice on assumption that no impurities are present, i.e. the concentration of the spin-1 atoms is set to $\rho = 1$.

\begin{figure}[tb]
\begin{center}
\hspace{-1.0cm}
\includegraphics[clip,width=7.5cm]{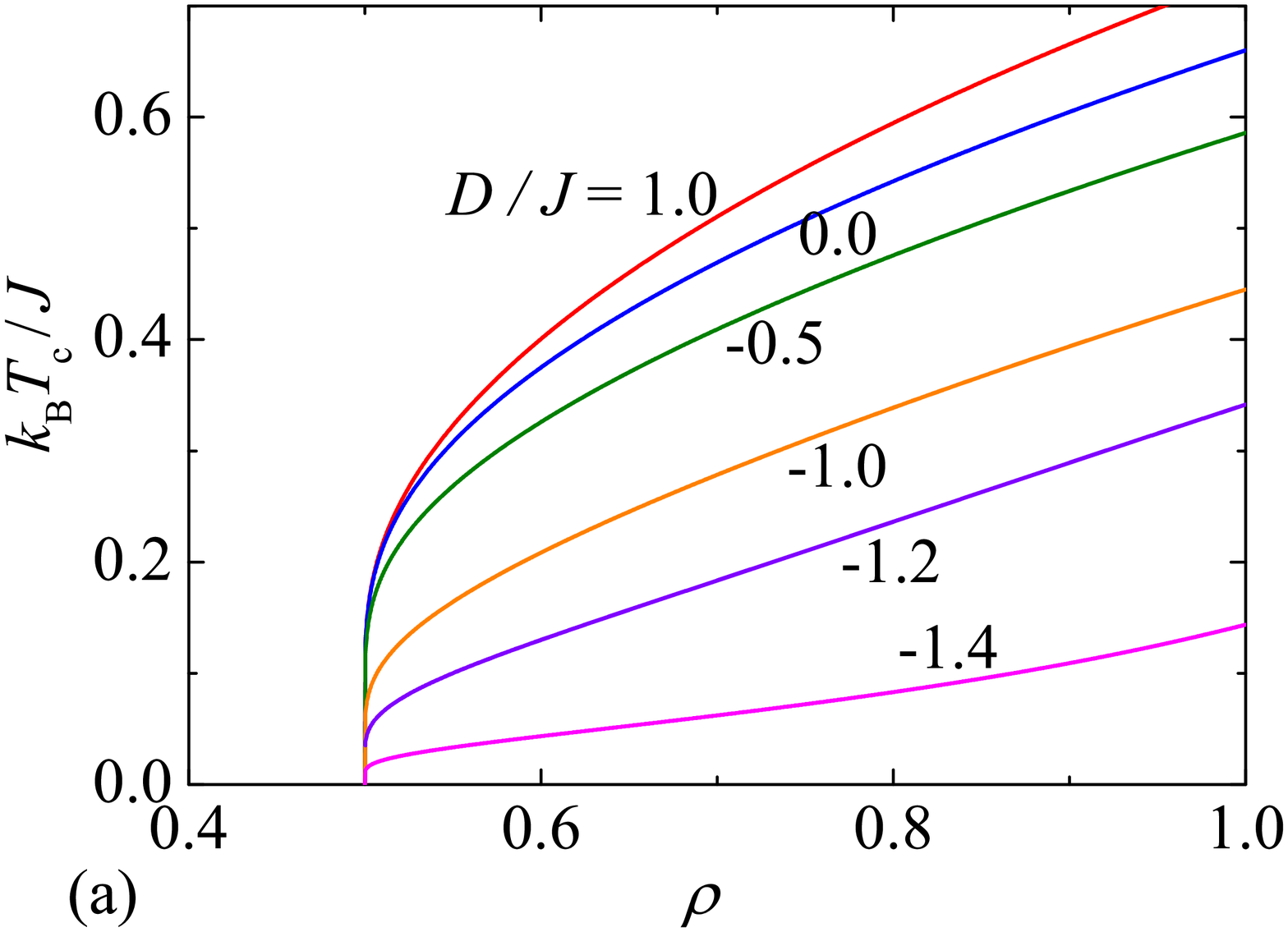}
\hspace{-1.0cm}
\includegraphics[clip,width=7.5cm]{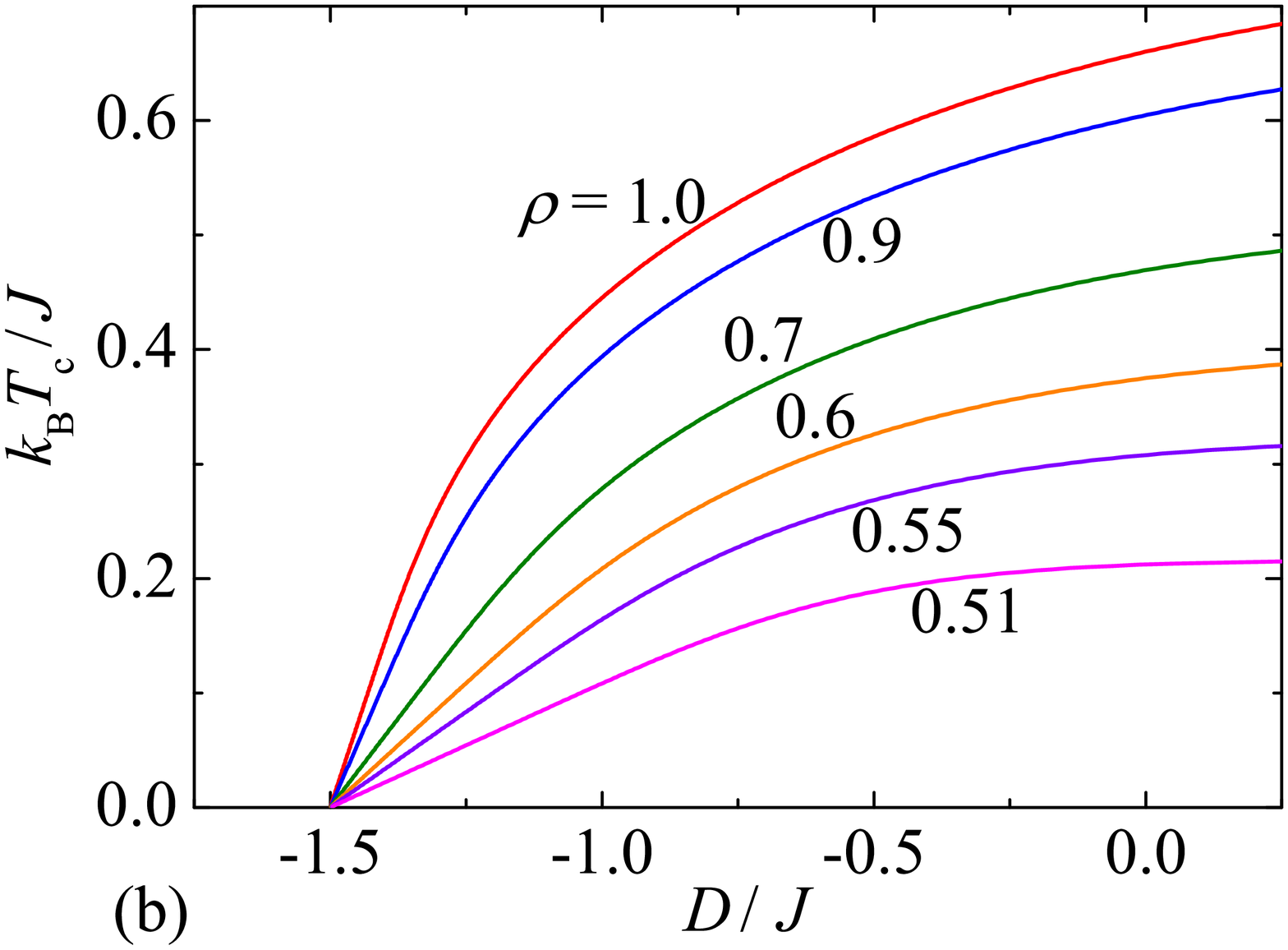}
\hspace{-1.5cm}
\vspace{-0.25cm}
\caption{(a) The critical temperature of the selectively diluted mixed spin-1/2 and spin-1 Ising model on a honeycomb lattice as a function of the concentration for several values of the uniaxial single-ion anisotropy; (b) The critical temperature of the selectively diluted mixed spin-1/2 and spin-1 Ising model on a honeycomb lattice as a function of the uniaxial single-ion anisotropy for several values of the concentration.}
\label{figmixdil}
\end{center}
\end{figure}

The numerical solution of the critical condition (\ref{dicc}) of the selectively diluted mixed spin-1/2 and spin-1 Ising model on the honeycomb lattice is presented in Fig. \ref{figmixdil}. The critical temperature is plotted in Fig. \ref{figmixdil}(a) against the concentration of the magnetic spin-1 atoms for several values of the uniaxial single-ion anisotropy. As one can see, the critical temperature monotonically decreases upon decreasing the concentration of the magnetic spin-1 atoms until it completely tends to zero at the critical concentration $\rho_{\rm c} =0.5$. This general trend remains valid independently of the uniaxial single-ion anisotropy, which has tendency to lower the critical temperature as the easy-plane (negative) single-ion anisotropy strengthens. To gain an overall insight, the critical temperature is displayed in Fig. \ref{figmixdil}(b) as a function of the uniaxial single-ion anisotropy for several values of the concentration of the magnetic spin-1 atoms. It is quite evident from this figure that the critical temperature monotonically decreases with the easy-plane single-ion anisotropy until it completely vanishes at the ground-state phase boundary $D_{\rm b}/J = -1.5$ between the ferromagnetically ordered phase and the disordered paramagnetic phase. In addition, it can be also concluded that the selective site dilution of the magnetic spin-1 atoms is responsible for a gradual reduction of the critical temperature quite similarly as the effect of the easy-plane single-ion anisotropy.

\subsection{Historical survey and future perspectives}

A great deal of research interest aimed at exactly solvable Ising models has resulted in a rather impressive list of existing literature. It is therefore beyond the scope of this work to provide a comprehensive list of all exactly solved Ising models and the selected examples are mentioned just to illustrate diversity of rigorous approaches and/or various physical motivations of for their study. It is quite clear from previous argumentations that a change of spatial dimensionality of a magnetic lattice from 1D to 2D has a dramatic effect upon phase transitions and critical phenomena. Namely, 1D Ising model cannot capture a spontaneous long-range order at any non-zero temperature in contrast to 2D Ising model, which displays an intriguing order-disorder phase transition at a finite-temperature critical point. Although the model by itself seems to be extraordinary simple, the general exact solution of the spin-1/2 Ising model on 2D lattices in a presence of the magnetic field represents unresolved problem yet. Note furthermore that this exact solution would be of fundamental importance as it would clarify dependence of the magnetization on the magnetic field and it would simultaneously enable derivation of the exact closed-form expression for the susceptibility. It has been pointed out by Istrail \cite{istr00} and Cipra \cite{cipr00} that the exact solution of any non-planar Ising model is essentially NP-complete problem, which implies that any 3D Ising model is possibly analytically intractable. In this respect, the subsequent list of fully exactly solved Ising models essentially contains only 1D and 2D lattice-statistical models.    

It is worthy to recall that the spin-1/2 Ising chain has been exactly solved in the pioneering work by Ising \cite{isin25} with the help of combinatorial approach. Afterwards, the Ising's exact results have been re-derived by a variety of other mathematical techniques as for example the transfer-matrix method due to Kramers and Wannier \cite{kram41}. To a relatively extended class of the rigorously solved 1D Ising models belong the spin-$S$ Ising chain \cite{suzu67,krin74,krin75,line79,chat84}, the spin-1/2 Ising chain accounting for further-neighbor interactions \cite{step70a,step71,dobs69,redn81}, the spin-1/2 Ising model composed of two, three or four coupled chains \cite{inou71,kalo75,fedr76,yoko89,yuri07}, the spin-1/2 Ising chains with periodically alternating interactions  \cite{ohan03} as well as the mixed-spin Ising chains \cite{urum80,geor85,cure86,sapi88,cure92,fire97,fire03,vito02,valv08,lisn13} and ladders \cite{geor92,kiss08}. It is worthy to note that Minami \cite{mina96,mina98a,mina98b} has obtained the exact solution for a rather extensive class of 1D Ising models, which involve the most of aforelisted exactly solved Ising models (including the ones with the mixed spins or alternating interactions) as a special case. 

The first complete closed-form solution for 2D Ising model has been found by Onsager when formulating the spin-1/2 Ising model on a square lattice within the transfer-matrix method and Lie algebra \cite{onsa44}. Onsager's exact solution represents one of the most significant achievements in the statistical mechanics, because it brought an important revision in the understanding of phase transitions and critical phenomena. Indeed, Onsager's exact solution has delimited a modern era of the equilibrium statistical mechanics \cite{baxt10}, since it has convincingly evidenced an outstanding phase transition accompanied with a singular behavior of thermodynamic quantities in a vicinity of the critical point. Exact results for 2D Ising model also shed light on deficiencies of approximative theories such as mean-field and effective-field methods, which mostly fail in predicting a correct behavior near a critical region. One of the most essential questions to deal with in the statistical mechanics of exactly solvable models is thus closely connected to a precise nature of singularities accompanying each phase transition. 

The main deficiency of Onsager's exact solution lies in a considerable formidability of the applied mathematical methods, which initiated intensive search for an alternative way admitting a more straightforward exact treatment of 2D Ising models. The original Onsager's solution has been slightly simplified by Kaufmann and Onsager himself by employing the theory of spinors \cite{kauf49,onsa49}. A more substantial simplification has been later achieved by taking advantage of combinatorial approaches developed by Kac, Ward and Potts \cite{kac52,pott55}, Hurst and Green \cite{hurs60}, Vdovichenko \cite{vdov64}, Jordan-Wigner fermionization by Schultz, Lieb and Mattis \cite{schu64}, the star-triangle transformation by Baxter and Enting \cite{baxt78} or more recent theories based on Grassmann variables \cite{samu80,plec85,plec88,noji98,clus09} and Clifford-Dirac algebra \cite{verg09}. Among all rigorous methods, the Pfaffian technique seems to be the most versatile method that enables an exact treatment of the spin-1/2 Ising model on any planar lattice \cite{hurs63,gree64,hurs65,mcco73}. Within this approach, the problem of solving 2D Ising model is at first reformulated as the dimer covering problem on a corresponding decorated lattice \cite{kast63} and subsequently, the relevant dimer statistics is solved by Pfaffian technique elaborated by Kasteleyn \cite{kast61}, Temperley and Fisher \cite{temp61,fish61}. 

Exact critical points of the ferromagnetic spin-1/2 Ising model on all archimedean lattices (Fig. \ref{fig:arch}) have been recently compiled by Codello \cite{code10}, whereas the relevant exact solutions were formerly derived for square \cite{onsa44}, hexagonal and triangular \cite{hout50,temp50,husi50,syoz50,newe50}, kagom\'e \cite{syoz51,liwe10}, star \cite{huck86,link87}, bathroom-tile \cite{utiy51,shin87,chen87,baxt88,hank13}, trellis \cite{thom74}, Shastry-Sutherland \cite{thom74,stre05}, bounce (ruby) \cite{link83} and square-hexagon-dodecagon \cite{link85} lattice. In addition, the ferromagnetic spin-1/2 Ising model has been rigorously examined for a variety of more complex irregular lattices such as union jack \cite{vaks66,mori86,chik87,wang87,choy87,baxt89,link89}, pentagonal \cite{urum02}, square-kagom\'e \cite{sunc06}, bow-tie \cite{stre08}, two topologically different square-hexagon \cite{link88,oitm02} and the vast number of 2D decorated lattices (see Ref.~\cite{syoz72} and references cited therein). All those exact results have furnished a rigorous proof for a strong universality in a critical behavior of the ferromagnetic spin-1/2 Ising model on 2D lattices, which is characterized through a unique set of the critical exponents.

The antiferromagnetic spin-1/2 Ising model on 2D bipartite lattices shows the essentially same critical behavior than its ferromagnetic counterpart (e.g. see foregoing discussion concerned with the critical points of the Ising square and honeycomb lattices), but the situation dramatically changes for the antiferromagnetic spin-1/2 Ising model defined on 2D non-bipartite lattices\footnote{The term bipartite lattice denotes a lattice, which can be divided into two equivalent sublattices A and B in a such way that all nearest-neighbor sites of a site from the sublattice A belong to the sublattice B and vice versa. The non-bipartite lattices are all other lattices, which cannot be divided into two equivalent sublattices under the same constraint.} when a geometric spin frustration comes into play \cite{lieb86,diep04}. Quite generally, the term \textit{geometric spin frustration} is ascribed to an incapability of spins to simultaneously satisfy all pair interactions, which is always the case when product of signs of all coupling constants along an elementary plaquette becomes negative \cite{toul77}. It is worthwhile to remark that the geometric spin frustration of 2D Ising models is at heart of many outstanding phenomena like highly degenerate ground-state manifold with a non-zero residual entropy contradicting the third law of thermodynamics  \cite{lieb86} or multiply reentrant phase transitions associated with an existence of a few successive critical points \cite{diep04}. The antiferromagnetic spin-$1/2$ Ising model on a triangular lattice is presumably the first exactly solved model with the macroscopically degenerate ground state that manifest itself through a non-zero residual (zero-point) entropy \cite{wann50}, while the spin-$1/2$ Ising model on the union jack lattice with the competing (antiferromagnetic) next-nearest-neighbor interaction is likely the first exactly solved model displaying reentrant phase transitions \cite{vaks66,mori86,chik87}. An absence of finite-temperature phase transition above the highly degenerate ground-state manifold has been also rigorously proved for the spin-$1/2$ Ising model on a kagom\'e lattice \cite{kano53,kano66}, fully frustrated square, triangular and honeycomb lattices \cite{vill77,wolf82}, brickwork lattice \cite{seze81}, chessboard lattice \cite{wozi82}, two different kinds of pentagonal lattice \cite{wald84,wald85,roja12} and triangular kagom\'e lattice \cite{zhen05,lohy08}. 

Although it could be naively expected that the macroscopic degeneracy of the ground state ultimately inhibits a spontaneous long-range order, some geometrically frustrated 2D Ising models with the highly (but not too highly) degenerate ground state may exhibit a peculiar coexistence of spontaneous long-range order with a partial disorder. In fact, the exact solutions for the spin-$1/2$ Ising model on union jack lattice \cite{vaks66,mori86,chik87}, piled-up and zig-zag domino lattices \cite{andr79}, centered honeycomb lattice \cite{diep91,diep92} and generalized kagom\'e lattice \cite{diep92,azar87,deba91} have convincingly evidenced an imperfect spontaneous order accompanied with a non-zero residual entropy. It is worth mentioning that the reentrance with a few successive critical points is also observed in most of the aforementioned exactly solved Ising models. Apart from this fact, the spin-$1/2$ Ising model on a domino lattice displays a spectacular phenomenon commonly referred to as \textit{order-by-disorder effect} \cite{vill80,bida81,zitt82}, which occurs whenever an incomplete spontaneous order is thermally induced above a disordered ground state. The fact that thermal fluctuations produce a spontaneous order above a disordered ground state by selecting the ordered state with the largest degeneracy (entropy) obviously represents a highly contra-intuitive cooperative phenomenon. 
 
A special mention also deserves the mixed spin-1/2 and spin-$S$ ($S > 1/2$) Ising model on 2D lattices, because this lattice-statistical model might be regarded as one of the most natural but still exactly soluble extensions of the simple spin-$1/2$ Ising model. The foremost reason for a theoretical investigation of the mixed-spin Ising models on 2D lattices is closely connected to an expectation of a more diverse critical behavior. Using the generalized decoration-iteration and star-triangle mapping transformations, the mixed-spin Ising model on hexagonal, diced and several decorated 2D lattices have been rigorously solved by Fisher \cite{fish59} and Yamada \cite{yama69} many years ago. In addition, the exact mapping method based on generalized algebraic transformations \cite{fish59,yama69,roja09} can be rather straightforwardly adapted in order to account for the single-ion anisotropy and/or further-neighbor interaction. The effect of uniaxial single-ion anisotropy upon a critical behavior has been thoroughly examined for the mixed-spin Ising model on honeycomb \cite{gonc85}, bathroom-tile \cite{stre06}, square-hexagon-dodecagon (the present work), two geometrically related triangles-in-triangles \cite{stre13} and several decorated 2D lattices \cite{lack98,oitm05,stre07,cano08,kiss11,jasc14}. Among other matters, the exact solution for the mixed spin-1/2 and spin-1 Ising model on the triangles-in-triangles lattices \cite{stre13} has convincingly evidenced the order-by-disorder effect when an incomplete spontaneous long-range order was invoked by thermal fluctuations just above a disordered ground state. A possible existence of reentrant phase transitions, which result from a mutual competition between the nearest-neighbor and (antiferromagnetic) further-neighbor interactions, has been examined in detail within the mixed-spin Ising model on the honeycomb \cite{gonc87,dakh00}, decorated planar \cite{mata07} and union jack (centered square) \cite{lipo95,stre06a,stre06b} lattices. It has been rigorously demonstrated that the latter mixed-spin Ising model on the union jack lattice \cite{lipo95,stre06a,stre06b} represents a valuable example of the lattice-statistical model, which exhibits a remarkable line of bicritical points with continuously varying critical exponents along with another critical line of discontinuous (first-order) phase transitions. The quantum effects introduced by the biaxial single-ion anisotropy or transverse magnetic field have also been studied within the mixed-spin Ising model on the honeycomb \cite{jasc00,stre04,jasc05a}, diced \cite{jasc05b} and some decorated 2D \cite{jasc99,stre03} lattices. 

\setcounter{equation}{0} \newpage
\section{Conclusion}
\label{conclusion}

The present article provides a brief tutorial review on applicability of the mean-field, effective-field and exact methods to the Ising and Ising-like models. To this end, the aforementioned approximate and exact theories were applied to a few selected textbook examples of lattice-statistical models such as the spin-1/2 Ising model in a longitudinal field, the spin-$S$ Blume-Capel model in a longitudinal field, the mixed-spin Ising model in a longitudinal field and the spin-$S$ Ising model in a transverse field. In spite of their conceptual simplicity the Ising and Ising-like models are  exactly tractable just on 1D or 2D lattices in a presence or absence of the external magnetic field, respectively. Besides, there exist a strong evidence that the full exact solution of the simplest spin-1/2 Ising model on non-planar lattices is presumably analytically intractable problem and hence, the approximate methods like the mean-field and effective-field theories often represent the only analytical tools to grasp qualitative features of more complex Ising-like spin systems aimed at a theoretical modeling of amorphous and disordered systems, thin films, superlattices, alloys, diluted systems and so on. However, the word of caution should be made because a danger of over-interpretation is hidden in any approximation. To overcome this possible problem, the presented exactly solved Ising models can be utilized as a useful benchmark to test accuracy and reliability of the approximate methods before applying them to more complex Ising-like spin systems for which rigorous treatment is naturally inapplicable. 
 
The mean-field theory traditionally represents the simplest but at the same time also crudest approximate method, which provides plausible results for the Ising-like models whenever spin fluctuations become negligible. This means that the mean-field approach is justified for the Ising-like models on 2D and 3D lattices with sufficiently high coordination numbers, but this theory may totally fail in predicting a correct magnetic behavior of the low-dimensional Ising models with much stronger role of spin fluctuations and short-range correlations. The non-zero critical temperature of the spin-1/2 Ising chain and the tricritical behavior of the mixed spin-1/2 and spin-1 Ising model on a honeycomb lattice thus represent obvious artifacts of the mean-field treatment (c.f. the  mean-field results with the respective exact results proving an absence of spontaneous long-range order in the spin-1/2 Ising chain and an absence of tricritical phenomenon in the mixed spin-1/2 and spin-1 Ising model on a honeycomb lattice). To compare with, the effective-field theories afford superior results with respect to the standard mean-field theory as they correctly account at least for short-range correlations. The added value of the present work lies in that it brings an insight into how to avoid the long-standing problem of this approximate method concerned with a self-consistent determination of the Helmholtz free energy, which has until now precluded a comprehensive analysis of interesting phenomena as for instance discontinuous (first-order) phase transitions accompanied with a presence of the metastable states. The foremost advantage of the effective-field method with respect to the mean-field method lies in an improved accuracy of the obtained results when keeping the computational difficulty at almost the same level (e.g. the effective-field theory correctly reproduces an absence of spontaneous long-range order in the spin-1/2 Ising chain even though it would fail to reproduce this feature for more complex 1D Ising models with greater coordination number $q>2$ like a two-leg ladder). On the other hand, the effective-field theory fails in reproducing a magnetic behavior in a vicinity of critical points quite similarly as does the mean-field theory, because it affords the same set of mean-field-like critical exponents.   

The exactly solved Ising and Ising-like models are certainly active research field in its own right, because they can be regarded as cornerstones of modern equilibrium statistical mechanics on which up-to-date theory of phase transitions and critical phenomena was built. As a matter of fact, the identical critical exponents of the ferromagnetic spin-1/2 Ising model on 2D lattices have provided a firm ground for postulating scaling and universality hypotheses. Although a substantial progress has been achieved in the area of exactly solvable Ising-like models there are still a lot of unresolved and open problems for future studies. For instance, the full exact solution of the spin-1/2 Ising model on 2D lattices taking into consideration a non-zero magnetic field would be of fundamental importance, since it would provide an exact closed-form expression for the susceptibility and it would simultaneously enable a comprehensive analysis of magnetization processes presumably accompanied with a non-trivial criticality. Similarly, there is a controversy whether or not a tricritical phenomenon may occur in the spin-$S$ Blume-Capel model on some 2D lattices, but unfortunately this model has not been exactly solved yet for any 2D lattice. While the exact solutions of the mixed-spin Ising model on three-coordinated archimedean (honeycomb, bathroom-tile, square-hexagon-dodecagon) lattices point to absence of discontinuous phase transitions, one cannot definitely rule out that the tricritical phenomenon and discontinuous phase transitions do occur in the mixed-spin Ising models on 2D lattices with a greater coordination number such as a square lattice (another yet unsolved lattice-statistical model). Bearing this in mind, one may in confidence conclude that the exactly solved Ising-like models will certainly remain topical issue in a future, because all aforementioned open problems along with many other challenging topics from this demanding but surely highly exciting research field nowadays attract considerable research interest.

\medskip
\medskip
\vspace*{1cm}
\addcontentsline{toc}{section}{Acknowledgement}
\begin{ack}
This work was financially supported by the Scientific Grant Agency of The Ministry of Education, Science, Research, and Sport of the Slovak Republic under the contract Nos. VEGA 1/0234/12 and VEGA 1/0331/15 along with the scientific grants of Slovak Research and Development Agency provided under contract Nos. APVV-0132-11 and APVV-14-0073. The authors would like to thank Dr. Taras Verkholyak for his comments and remarks on the manuscript and Dr. Viliam \v{S}tub\v{n}a for his graphical support.
\end{ack}

\newpage
\addcontentsline{toc}{section}{Appendix: Gibbs-Bogoliubov inequality}
\begin{center}
{\bf Appendix: Gibbs-Bogoliubov inequality} 
\end{center}

\begin{figure}[tb]
\begin{center}
\includegraphics[clip,width=10.0cm]{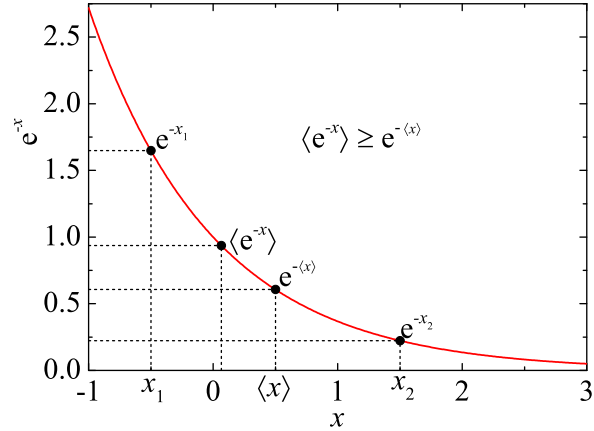}
\vspace{-0.4cm}
\caption{The convexity of the exponential function is essential for a rigorous proof of the Gibbs-Bogoliubov inequality.}
\label{figbgi}
\end{center}
\end{figure}

Let us prove a validity of the Gibbs-Bogoliubov inequality \cite{bog47,fey55,fal70}: 
\begin{eqnarray}
G \leq \Phi \equiv G_0 + \langle {\cal H} - {\cal H}_0 \rangle_0, 
\label{apagbi}
\end{eqnarray}
which can serve for an approximate calculation of the Gibbs free energy $G$ of a lattice-statistical model defined through the Hamiltonian ${\cal H}$. The upper bound for the exact Gibbs free energy $G$ is afforded by the variational estimate $\Phi$ calculated from a simpler lattice-statistical model given by the trial Hamiltonian ${\cal H}_0$ that yields the corresponding trial Gibbs free energy $G_0$. The symbol $\langle {\cal O} \rangle_{0}$ denotes canonical ensemble average of the physical quantity ${\cal O}$ calculated within the simplified lattice-statistical model:  
\begin{eqnarray}
\langle {\cal O} \rangle_{0} = \frac{1}{Z_0} \mbox{Tr} \left[ {\cal O} \exp(-\beta {\cal H}_0) \right], \qquad  Z_0 =  \mbox{Tr} \exp(-\beta {\cal H}_0). 
\label{apasym}
\end{eqnarray}
Here, the symbol Tr denotes a trace over degrees of freedom of the full lattice-statistical model, whereas the relevant partition function $Z_0$ of the simplified lattice-statistical model should be calculated exactly. The validity of the relation (\ref{apagbi}) can be proved by following an elegant proof due to Feynman \cite{fey55}. The partition function of the original lattice-statistical model is by definition given by:  
\begin{eqnarray}
Z =  \mbox{Tr} \exp(-\beta {\cal H}).
\label{apapf}
\end{eqnarray}
Now, it is advisable to decompose the total Hamiltonian of the original lattice-statistical model into a sum of two parts:
\begin{eqnarray}
{\cal H} = {\cal H}_0 + {\cal H}_1,
\label{apaham}
\end{eqnarray}
whereas the former (unperturbed) part ${\cal H}_0$ corresponds to the simplified lattice-statistical model for which the relevant calculations can be performed exactly and the latter (perturbed) part ${\cal H}_1$ involves all other (problematic) interaction terms. The ratio between the true partition function $Z$ and the trial partition function $Z_0$ can be subsequently related to a canonical ensemble average of the perturbed part of the Hamiltonian ${\cal H}_1$:
\begin{eqnarray}
\frac{Z}{Z_0} = \frac{\mbox{Tr} \exp(-\beta {\cal H})}{\mbox{Tr} \exp(-\beta {\cal H}_0)} 
= \frac{\mbox{Tr} \exp[-\beta ({\cal H}_0 + {\cal H}_1)]}{\mbox{Tr} \exp(-\beta {\cal H}_0)} = \langle \exp(-\beta {\cal H}_1) \rangle_0,
\label{apapfra}
\end{eqnarray}
which should be however calculated within the simplified lattice-statistical model. At this stage, one may take advantage of the convexity of the exponential function:
\begin{eqnarray}
\langle \exp[f(x)] \rangle \geq \exp[ \langle f(x) \rangle],
\label{apacr}
\end{eqnarray}
which holds for any real function $f(x)$ because first and second derivatives of the exponential function are always positive (see Fig. \ref{figbgi}). 
Applying the convexity relation (\ref{apacr}) to Eq. (\ref{apapfra}) one obtains the inequality:
\begin{eqnarray}
\frac{Z}{Z_0} \geq \exp(-\beta \langle {\cal H}_1  \rangle_0),
\label{apapfin}
\end{eqnarray}
which in turn gives the following relation when taking logarithm from both sides of Eq. (\ref{apapfin}):
\begin{eqnarray}
\ln Z - \ln Z_0 \geq - \beta \langle {\cal H}_1  \rangle_0.
\label{apapfina}
\end{eqnarray}
After a straightforward rearrangement of terms and substitution ${\cal H}_1 = {\cal H} - {\cal H}_0$ into Eq. (\ref{apapfina}) one directly arrives at the Gibbs-Bogoliubov inequality (\ref{apagbi}). 
The Gibbs-Bogoliubov inequality is of huge practical importance, because if one chooses the trial Hamiltonian ${\cal H}_0$ to be a simple approximation for the true Hamiltonian ${\cal H}$ of some lattice-statistical model, then it affords a rigorous upper bound for the true Gibbs free energy. Of course, it is quite natural to include into the trial Hamiltonian ${\cal H}_0$ as much interaction terms as possible from the true Hamiltonian ${\cal H}$ in order to obtain the reliable approximation of the original lattice-statistical model but still leaving the relevant calculation tractable. The simplest approximation elaborated within the Gibbs-Bogoliubov variational principle (\ref{apagbi}) lies in considering the Hamiltonian of a system of non-interacting particles as the trial Hamiltonian ${\cal H}_0$, which always provides the mean-field approximation for some interacting many-particle system. The actual interaction between the constituent particles are accordingly replaced within this approach by a fictitious external field or potential, whereas the exact Gibbs free energy $G$ represents a lower bound with respect to the right-hand-side of the Gibbs-Bogoliubov inequality (\ref{apagbi}), which may be alternatively viewed as the variational Gibbs free energy $\Phi$.

\end{document}